\documentclass[
10pt,aps,amsmath,amssymb,
reprint
,pra
,longbibliography
,notitlepage
,superscriptaddress
]{revtex4-1}
\usepackage{dcolumn}
\usepackage{bm}
\usepackage{braket}
\usepackage{tabularx}
\usepackage[table]{xcolor}
\usepackage{threeparttable}
\usepackage{graphicx}
\usepackage{mathtools}
\usepackage{multirow}
\usepackage[colorlinks=true,linkcolor=blue,urlcolor=black,citecolor=blue,bookmarksopen=true]{hyperref}
\usepackage[caption=false]{subfig}
\usepackage{physics}
\usepackage{mathrsfs}
\usepackage{xifthen}
\usepackage{array}

\usepackage{listings}
\definecolor{lightgray}{rgb}{0.95,0.95,0.95}
\definecolor{codegreen}{rgb}{0,0.6,0}
\definecolor{codegray}{rgb}{0.5,0.5,0.5}
\definecolor{codepurple}{rgb}{0.58,0,0.82}
\definecolor{backcolour}{rgb}{0.95,0.95,0.92}
\definecolor{textblue}{rgb}{.2,.2,.7}
\definecolor{textred}{rgb}{0.54,0,0}
\definecolor{textgreen}{rgb}{0,0.43,0}
\definecolor{codered}{rgb}{201,72,12}
\definecolor{calpolypomonagreen}{rgb}{0.12, 0.3, 0.17}
\definecolor{cobalt}{rgb}{0.0, 0.28, 0.67}
\lstdefinestyle{tt1}{
language={C++},
basicstyle=\linespread{0.98}\ttfamily\footnotesize,
breaklines=true,
numbers=left,
frame=single,
numberstyle=\tiny, 
stepnumber=1,
numbersep=5pt, 
tabsize=4,
commentstyle=\color{textred},   
keywordstyle=\bfseries\color{codegreen},
stringstyle=\color{textgreen},
columns=fullflexible,
keepspaces=true,
xleftmargin=\parindent,
showstringspaces=false,
otherkeywords = {True, False},
keywordstyle=[3]\color{codegreen}\bfseries,
keywords=[3]{__device__},
keywordstyle=[4]\color{cobalt}\bfseries,
keywords=[4]{ld, st, mem, qubit, gate, angle, ctrl, bit, qram, h, cx, measure, qinit, qld
warp,
block
},
escapeinside={\%*}{*)}
}

\newcommand{\pnl}{Physical and Computational Sciences, Pacific Northwest National Laboratory, Richland, WA, 99354, USA}
\newcommand{\ucsd}{Computer Science and Engineering, University of California San Diego, La Jolla, CA, 92093, USA}

\begin{document}

\title{Quantum Memory: A Missing Piece in Quantum Computing Units}
\author{Chenxu Liu}
\email{Email: chenxu.liu@pnnl.gov}
\affiliation{\pnl}

\author{Meng Wang}
\affiliation{\pnl}

\author{Samuel Stein}
\affiliation{\pnl}

\author{Yufei Ding}
\affiliation{\ucsd}

\author{Ang Li}
\email{Email: ang.li@pnnl.gov}
\affiliation{\pnl}

\begin{abstract}
Memory is an indispensable component in classical computing systems. While the development of quantum computing is still in its early stages, current quantum processing units mainly function as quantum registers. Consequently, the actual role of quantum memory in future advanced quantum computing architectures remains unclear. With the rapid scaling of qubits, it is opportune to explore the potential and feasibility of quantum memory across different substrate device technologies and application scenarios. In this paper, we provide a full design stack view of quantum memory. We start from the elementary component of a quantum memory device, quantum memory cells. We provide an abstraction to a quantum memory cell and define metrics to measure the performance of physical platforms. Combined with addressing functionality, we then review two types of quantum memory devices: random access quantum memory (RAQM) and quantum random access memory (QRAM). Building on top of these devices, quantum memory units in the computing architecture, including building a quantum memory unit, quantum cache, quantum buffer, and using QRAM for the quantum input-output module, are discussed. We further propose the programming model for the quantum memory units and discuss their possible applications. By presenting this work, we aim to attract more researchers from both the Quantum Information Science (QIS) and classical memory communities to enter this emerging and exciting area.

\end{abstract}

\maketitle

\section{Introduction}

Emergence of quantum computing and quantum networking has sparked tremendous excitement in the scientific and technological communities due to their potential to revolutionize various fields. Quantum computing harnesses the principles of quantum mechanics to perform computations exponentially faster than classical computers, offering the possibility of solving complex problems in cryptography~\cite{shor1999polynomial, gisin2002quantum}, optimization~\cite{farhi2014quantum, han2002quantum}, quantum chemistry~\cite{georgescu2014quantum, kandala2017hardware}, and finance~\cite{rebentrost2018quantum, woerner2019quantum}, etc. Furthermore, quantum networks enable the transmission of quantum information across long distances~\cite{Hensen2015, Ren2017, Chen2021_JWP, Liu2023_JWP}, facilitating secure communication~\cite{BB84, Gisin2002, Xu2020, Portmann2022} and the creation of sophisticated quantum network protocols and quantum internet~\cite{Kimble2008, Cacciapuoti2020, Cacciapuoti2020_v2, Wei2022, Illiano2022, caleffi2022distributed}. The usefulness of quantum computing systems and their connected networks lies in their ability to tackle computational challenges that are currently intractable, paving the way for significant advancements in science, industry, and society as a whole. 

One of the main goals of quantum computing research is to scale up the quantum computing systems and build a fault-tolerant large-scale quantum computer. In recent years, a lot of efforts have been demonstrated. IBM-Q demonstrates the Eagle device featuring 127 physical qubits~\cite{IBMQ2021} and the Osprey device with 433 qubits~\cite{IBMQ2022}, Google's Sycamore consists of 54 qubits~\cite{Arute2019}, Quantinuum has its `H2' device with 32 trapped ion qubits~\cite{Moses2023}, IonQ devices can hold more than 20 qubits~\cite{IonQ}, QuEra also demonstrated its 256-qubit quantum simulator~\cite{Ebadi2021, wurtz2023aquila}, etc. Although integrating thousands of qubits into a single quantum chip is possible~\cite{IBMQ2025_Roadmap} in the near future, there are still challenges in building such a large-scale fault-tolerant quantum computer along the current route. 

In the current route to reach this goal, one of the main challenges comes from the physical difficulty of integrating a huge number of physical qubits as quantum registers into a single quantum device. For example, the physical size of the quantum chip limits the number of superconducting qubits on the same chip~\cite{IBMQ2025_Roadmap, Kaushal2020}, while the electromagnetic trap size limits the number of ions living inside a single trap~\cite{Krinner2019}. Meanwhile, cross-talk also hinders fast gate operations in largely integrated quantum systems. However, using Shor's algorithm to break RSA may require thousands of logical qubits made by millions of physical qubits~\cite{gheorghiu2019benchmarking, Roetteler2017, Gidney2021}.

On the other hand, integrating a large number of quantum registers usually has a trade-off with fast and reliable gate operations on any pair of computing registers. In the ‘noisy intermediate-scale quantum’ (NISQ) era~\cite{NISQ_Preskill}, where quantum registers are made by physical qubits, the limitation is mainly reflected by slow gate operations and limited communication fidelity. For example, in superconducting devices, the coupling between qubits on different chips hinders fast and precise gate operations~\cite{IBMQ2025_Roadmap}. In trapped ion systems, increasing the number of ions in a single trap prolongs the two-qubit gate operations~\cite{Monroe2013, Kaushal2020}. In the fault-tolerant quantum computing (FTQC) era, where logical qubits are protected by quantum error correction (QEC) codes~\cite{nielsen_chuang_2010, Calderbank1996, Fowler2012, Google2023}, maintaining gate speed between remote logical qubits is even more challenging compared to the NISQ devices, due to the limited number of physical qubits allowed in a single device. Furthermore, the limited connectivity of the physical platforms requires a large number of SWAP gates to perform a gate between two remote qubits. The SWAP gate number is also proportional to the size of the quantum computing device, which further prolongs the gate operations. 

\begin{figure}
    \centering
    \includegraphics[width = 0.975 \columnwidth]{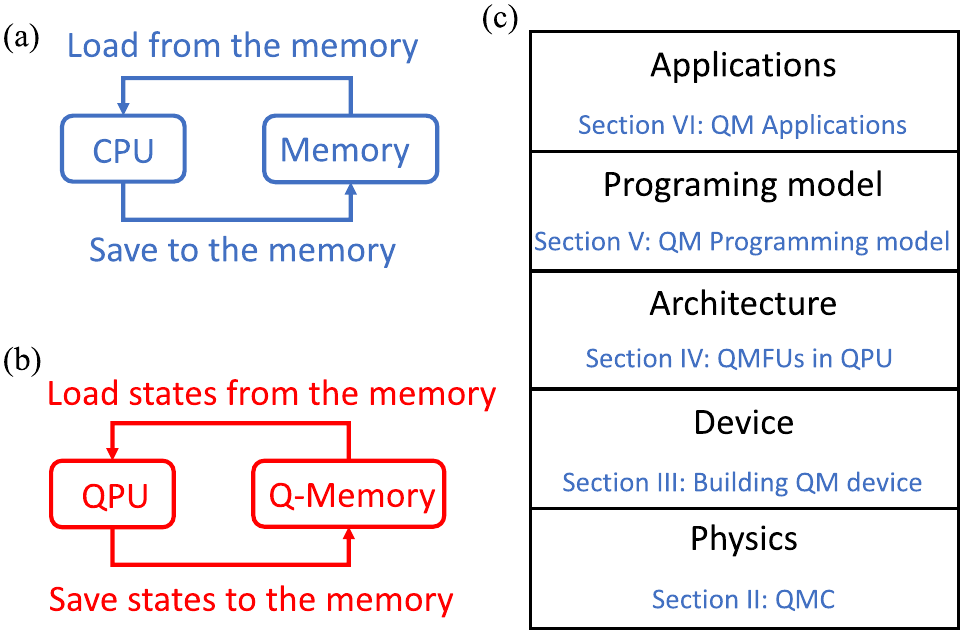}
    \caption{The usage of quantum memory and the stack of our consideration of quantum memory. In (a), we show the outline of the classical quantum computing systems, while in (b), we show our vision of quantum computing systems, where quantum memory plays the centered role in the quantum computing system. The stack of our discussion on the role of quantum memory in future quantum computing is shown in (c). This paper's structure is aligned with the design stacks.}
    \label{fig:structure_new}
\end{figure}

To resolve these issues, we take inspiration from classical computing systems, where classical memory plays the central role~\cite{hennessy2011computer,patterson2016computer}. The key insight is the CPU-memory separation in the von Neumann architecture (see Fig.~\ref{fig:structure_new}a). Similarly, in the quantum computing system design, the separation of the computing and memory devices is also desired (see Fig.~\ref{fig:structure_new}b). With this separation, future quantum computing devices can contain two main units: a quantum processing unit for computing and a quantum memory unit for information storage. The quantum processing unit contains a small number of computing registers, which can support a set of fast and reliable universal gates. The quantum memory unit, in constant, is designed for storing quantum information. It can contain a large number of quantum registers, which do not necessarily support a universal gate set. 

In addition, due to the computing-storage separation, the issue of maintaining both fast gate and large-scale integration of qubits in quantum computing devices is avoided. The quantum memory unit only needs to have reliable communication with the QPU to store and load the quantum information. By separating the requirements on computing and information storage, the QPU and the quantum memory can be realized using different physical techniques, one with fast gate operations for QPU and one with a long coherence time for quantum memory.  In addition, this design can benefit more in the FTQC systems. Because of the Eastin–Knill theorem, where a transversal universal gate set is not possible~\cite{Eastin2009}, implementing a universal QEC gate set requires techniques like magic state injection~\cite{nielsen_chuang_2010, Bravyi2005}, code deformation~\cite{Bombin2009, Horsman2012, Brown2017, Litinski2019}, code switching~\cite{Laflamme2014, Anderson2014}, etc., which further increases the complexity. However, these techniques are not necessary in quantum memory as a universal gate set is not required. It is also possible to help reduce the complexity of the QEC code itself~\cite{Terhal2015, Kovalev2013, Gidney2021_v2, bravyi2023highthreshold}.

The main focus of the current development of quantum computing systems is to build a fault-tolerant and fast quantum processing unit. The discussion and demonstration of building quantum memory devices have attracted lots of attention in the physical community~\cite{Lvovsky2009, Hedges2010, Heshami2016, Terhal2015}. However, the systematic discussion of the role of quantum memory in the quantum computing system architecture, whether a quantum memory unit should be considered, and how it should be utilized in quantum programs is still lacking. Therefore, in this paper, we attempt to follow a bottom-up manner through the design stack of quantum memory shown in Fig.~\ref{fig:structure_new}c. We not only survey the current quantum technologies and their possible usage in building quantum memory devices, but also consider how the quantum memory modules can be utilized in higher stacks, e.g., in quantum programs and software. We hope our paper can fill the gap between the physical layer and software layer of the development of quantum memory and provide useful insights for research on both frontiers. 

Specifically, at the bottom of the design stack, in Sec.~\ref{sec:QMC}, we survey the existing quantum technologies for building quantum registers and discuss their suitability of building the most elementary units of quantum memory, which is named `quantum memory cells' (QMCs). In order to unify the discussion across various physical platforms, we abstract the quantum memory cell concepts and explore the metrics that describe the essential properties of the QMCs. We then discuss quantum memory devices that are built on QMCs in Sec.~\ref{sec:QMM}. We specifically introduce two quantum memory devices, a random access quantum memory (RAQM) and a quantum random access memory (QRAM). We highlight their abstract models and their possible applications. These quantum memory devices can then be utilized to construct quantum memory modules in the future design of quantum computing architectures, which is discussed in Sec.~\ref{sec:QMFU}. We specifically give four examples, including the main quantum memory unit, quantum cache, quantum buffers, and QRAMs in quantum input-output modules. With the quantum memory modules available, we then discuss how the quantum memory can be utilized in the quantum program design. We discuss the quantum memory programming model in Sec.~\ref{sec:QMPM}. We discuss their possible application in Sec.~\ref{sec:application}. We conclude our paper in Sec.~\ref{sec:summary}.

\section{Quantum Memory Cells} \label{sec:QMC}

A quantum memory cell (QMC) is the fundamental element in quantum memory devices, analogous to classical memory cells consisting of one or a few transistors for storing a single classical bit. QMCs can be made using a quantum register or a qubit. However, they have unique requirements distinct from qubits used in quantum computing. Despite variations in physical systems for QMCs, we establish an abstract model to evaluate their performance uniformly and distinguish them from computing registers. Additionally, we introduce two metrics for quantitatively comparing different physical systems for quantum memory cells.

In Sec.~\ref{subsec:qmc_definition}, we define the QMC abstract model along with the performance metrics. Subsequently, we summarize the main results and discussions comparing various physical systems in quantum computing and quantum information processing in Sec.~\ref{subsec:qmc_compare}. A concise summary of key properties and metrics can be found in Table~\ref{tab:qmc_compare} and Fig.~\ref{fig:storage_time}. For completeness of this section, we briefly survey the physical systems one by one in the rest of the section from Sec.~\ref{subsec:sc_qubits} to Sec.~\ref{subsec:others}. These subsections are intended for readers with a particular interest in the specific quantum technologies and seeking an in-depth exploration of the references to the experiments included in our comparison. Readers who are not directly engaged with the specific physical realization or desire to focus on the broader context may choose to omit these subsections without compromising the overall coherence of the paper.

\subsection{Define a QMC} \label{subsec:qmc_definition}

A QMC can be made of a single qubit or quantum register to store one bit of quantum information. Unlike quantum computing registers, entangling gates between QMCs are not necessary. Instead, the core functionality of a QMC only includes (1) storing quantum information, (2) controlled operations to save quantum information to the QMC, and (3) load the quantum information from QMCs. 

\begin{figure}[ht]
    \centering
    \includegraphics[width = 0.55 \columnwidth]{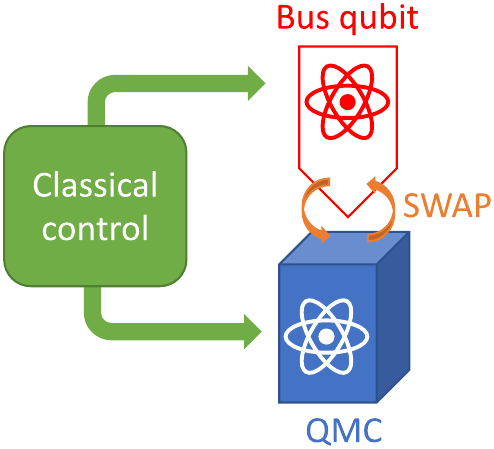}
    \caption{Demonstration of a QMC and its functionality. The QMC itself is shown as the blue cubic. It requires a bus qubit (red) which can be classically controlled to perform SWAP gates to the QMC. The read and write functionality is realized by applying a SWAP gate between a bus qubit to store and retrieve the quantum state into/out of the QMC. }
    \label{fig:qmc}
\end{figure}

The basic structure of a QMC and its related components are shown in Fig.~\ref{fig:qmc}. The QMC (blue box) has an interaction interface to transfer quantum information, which is the bus qubit (red). Due to the quantum no-cloning theorem, reading and writing (RW) processes cannot copy information and hence can only be realized using quantum operations. For example, in optical systems, RW operations can be realized by photon emission and absorption, while in gate-based systems, SWAP gates or iSWAP gates between the QMC and the bus qubit can be utilized. Using SWAP gates as an example, the RW process of a QMC can be described as 
\begin{align}
f_{\text{QMC}}(\ket{\psi}^{(\text{b})}, \ket{\phi}^{(\text{QMC})}) & = \text{SWAP}_{\text{b}, \text{QMC}} \ket{\psi}^{(\text{b})} \otimes \ket{\phi}^{(\text{QMC})} \nonumber \\
& = \ket{\phi}^{(\text{b})} \otimes \ket{\psi}^{(\text{QMC})}
\label{eq:qmc}
\end{align}
where the super-indices are for the physical qubits, `b' stands for bus qubits and SWAP is a swap gate. Although we show both qubits in pure states, if the bus (memory) qubit is already entangled with other qubits, the entanglement will be swapped to the memory (bus) qubit after the SWAP gate. 

The difference between the reading and writing processes of QMC lies in which part carries nontrivial quantum information. In the memory writing process, the bus qubit is in a useful quantum state to be stored, while in the memory reading process, the state of the QMC is important. One of the unique features of a QMC using SWAP gates as its RW operations is that reading and writing processes can be completed in a single SWAP, which is unlike the classical counterpart where an extra register and two separate operations are typically needed.

A QMC should satisfy the following requirements in terms of its functionalities.
\begin{enumerate}
    \item Quantum information storage: A QMC can store quantum information for an extended period of time. A good QMC should have a long storage time to preserve the quantum information. For NISQ devices, long storage time necessitates the physical qubit itself to have long coherence times, while in the FTQC era, it requires the logical qubit to have a small enough logical error rate. 
    \item Reading and Writing (RW) operations: A good QMC requires to have fast and accurate RW operations.
    \item Integration capability: As QMCs are employed to construct large quantum memory devices, integrating a large number of QMCs becomes essential. Therefore, a promising candidate of QMCs should have large integration capabilities. 
\end{enumerate}

There usually are tradeoffs between achieving a long storage time and fast RW operations while maintaining good integration capability. For a physical qubit, improving the coherence time involves isolating it from its environment, while fast quantum operations require strong coupling to other physical degrees of freedom to perform fast quantum gates. For a logical qubit encoded in some QEC codes, it can have smaller error rates by increasing its code distance, which may increase the number of physical qubits and complicate the coupling operations with other logical qubits. 

In order to evaluate the performance of a QMC by considering all three requirements together, we define a metric, $\alpha_{\text{in}}$, named {\it internal storage ratio}, as
\begin{equation}
    \text{Internal storage ratio: } \alpha_{\text{in}} = \frac{T_{\text{storage}}}{T_\text{RW}},
    \label{eq:internal_st}
\end{equation}
where $T_\text{storage}$ is the storage time of the QMC, while $T_{\text{RW}}$ is the time for a read or write operation. To account for the imperfection, $T_{\text{RW}}$ can be estimated by,
\begin{equation}
    T_{\text{RW}} = \tau / F_{\text{RW}} \sim \tau / p_{\text{suc}} \sim \tau / \eta,
\end{equation}
where $\tau$ is the raw gate time, $F_{\text{RW}}$ is the fidelity of the quantum gate, $p_{\text{suc}}$ is the success probability of performing the gate, while $\eta$ is the efficiency of the information storage or information retrieval. The metric $\alpha_{\text{in}}$ represents an estimation of the storage time of the QMC scaled by its RW speed. Large $\alpha_{\text{in}}$ means the QMC has faster RW operations in terms of its storage time, which is preferred. It also means that the QMC only needs to be reset after a large number of RW operations. 

Meanwhile, we consider another metric named {\it external storage ratio} as,
\begin{equation}
    \text{External storage ratio: } \alpha_{\text{ex}} = \frac{T_\text{net, storage}}{T_\text{op}} \eta,
    \label{eq:external_st}
\end{equation}
where $T_\text{op}$ is the time for a quantum operation on the possible connected computing devices, and $\eta$ is the QMC RW efficiency or fidelity. This metric measures the storage time of the QMC relative to the external essential operations, taking the imperfection of RW operations into account. Note that in Eq.~\eqref{eq:external_st}, $T_\text{net, storage}$ is the net storage time, $T_\text{net, storage} = T_\text{storage} - 2 T_\text{RW}$. Therefore, when the internal storage ratio $\alpha_\text{in} < 2$, the external storage ratio will be negative, which means that this QMC construction still needs to be further improved.

\subsection{Comparing physical platforms for QMCs} \label{subsec:qmc_compare}

\begin{table*}[tbp]
\begin{threeparttable}
\caption{\label{tab:qmc_compare} Comparison of physical techniques that can be utilized to implement quantum memory cells. We consider the physical platforms that have been surveyed in the main text. Details about our calculation and the corresponding references can be found in the main text. We consider the QMC storage time $T_\text{storage}$, possible bus qubits, the RW time $T_\text{RW}$ and its efficiency $\eta$, and the internal storage ratio $\alpha_\text{in}$ [Eq.~\eqref{eq:internal_st}], and the external storage ratio $\alpha_\text{ex}$ [Eq.~\eqref{eq:external_st}]. The transmon and fluxonium systems are used for homogenous computing systems, and hence $\alpha_\text{ex} = \alpha_\text{in}$. `MW': microwave photons, `SC': superconducting qubits. We use transmon to estimate the external storage ratio when the QMC can connect to superconducting qubits.} 
\begin{tabularx}{0.98 \textwidth}{X|X|X|X|X|X|X}
\hline
\hline
QMC & $T_\text{storage}$ & Bus & RW speed ($T_\text{RW}$) & Efficiency ($\eta$) & $\alpha_{\text{in}}$  & $\alpha_{\text{ex}}$ \\ 
\hline

Transmon  & $557.00~\mu$s  & N.A  & $40.00$~ns  & $0.998$  & $1.39 \times 10^{4}$  & \\ 
\hline
Fluxonium  & $1.48$~ms  & N.A  & $100.00$~ns  & $0.999$  & $1.48 \times 10^{4}$  & \\
\hline
MW (3D)  & $34.00$~ms  & SC  & $1000.00$~ns  & $0.994$  & $3.38 \times 10^{4}$  & $8.44 \times 10^{5}$\\ 
\hline
Trapped ions  & $300$~ms (Ca)  & Optical  & $29.94$~ns  & $0.509$  & $5.10 \times 10^{6}$  & $8.00 \times 10^{3}$\\ 
 & $5500.00$~s (Yb)  & Optical (herald)  & $16.13$~ms  & $0.901$  & $3.07 \times 10^{5}$  & $2.59 \times 10^{8}$\\ 
\hline
Neutral atoms & $800.00~\mu$s (Rb) & Optical  & $8.00~\mu$s  & $0.510$  & $51.0$  & $20.5$\\ 
& $7.90$~s (Yb) & Optical  & &  & $5.04 \times 10^{5}$  & $2.11 \times 10^{5}$\\ 
&  & Optical (expect)  & $182.48$~ns  & & $2.21 \times 10^{7}$  & $2.11 \times 10^{5}$\\ 
\hline
Atomic cloud  & $16.00$~s  & Optical  & $1.04~\mu$s  & $0.510$  & $7.84 \times 10^{6}$  & $4.27 \times 10^{5}$\\ 
 &  & Optical (expect)  &  & $0.9518$  & $1.46 \times 10^{7}$  & $7.97 \times 10^{5}$\\ 
& $800.00~\mu$s  & MW/SC  & $25.00~\mu$s  & $0.6$  & $19.2$  & $1.06 \times 10^{4}$\\ 
\hline
REIDC  & $52.9$~min  & Optical  & $400.00$~ms  & $0.0608$  & $4.83 \times 10^{2}$  & $1.01 \times 10^{7}$\\ 
&   & Optical (expect)  &  & $0.5187$  & $4.12 \times 10^{3}$  & $8.62 \times 10^{7}$\\ 
\hline
NV Nuclear spin & $12.90$~s (C)  & NV (e)  & $419.00~\mu$s  & $0.99$  & $3.05 \times 10^{4}$  &  \\ 
  & $63.00$~s (N)  &  & $389.00~\mu$s  & $0.94$  & $1.52 \times 10^{5}$  & \\ 
\hline
NV ensemble  & $200$~ns  & MW/SC  & $58.0$~ns  & $0.3742$  & $1.29$  & $<0$\\ 
& $1.80$~ms  & MW (expect)  &  &  & $1.16 \times 10^{4}$  & $1.67 \times 10^{4}$\\ 
\hline
QD  & $58.95$~ns  & SC & $23.81$~ns  & $0.80$  & $1.98$  & $<0$\\ 
 & $102.00~\mu$s  & SC (expect)  &  &  & $3.43 \times 10^{3}$  & $2.03 \times 10^{3}$\\ 
\hline
Phonons  & $130.00~\mu$s (GHz)  & MW/SC & $25.00$~ns  & $0.95$  & $4.94 \times 10^{3}$  & $3.02 \times 10^{3}$\\ 
 &  & Optical  & $714.29$~ns  & $1$ (assume)  & $1.82 \times 10^{2}$  & $6.73$\\ 
 & $100$~ms (MHz)  & Optical  & $714.29$~ns  & $1$ (assume)  & $1.40 \times 10^{5}$  & $5.24 \times 10^{3}$ \\

\hline
\hline
\end{tabularx}
\end{threeparttable}
\end{table*}

The discussion of QMCs and their abstract model applies to both NISQ- and FTQC-era quantum memory devices. However, the requirements on the physical platforms differ in these two scenarios. 

In the NISQ memory devices, both the QMCs and the computing qubits are made by single physical qubits. The storage time can be estimated by the coherence time of a single physical qubit $T_\text{coh}$. However, as the requirements of two kinds of qubits are different, in the spirit of separating the computing and memory requirements, the quantum memory devices should consist of different types of physical qubits compared to the computing devices. Reflected in our abstracted QMC model shown in Fig.~\ref{fig:qmc}, the bus qubit can be a type of qubit different from the QMC, and the SWAP gates for RW operations are implemented between two physical platforms. Therefore, the RW time $T_{\text{RW}}$ is not the two-qubit gate time typically used in characterizing computing qubits. The quality and the speed of a specific physical platform exchanging quantum information with other physical platforms are essential for building QMCs in the NISQ era. 

The rapid advancement of quantum technologies presents abundant opportunities for constructing QMCs across a diverse range of physical platforms. Nowadays, the most promising quantum substrate systems for quantum computing include superconducting qubits, microwave modes, trapped ion systems, neutral atom systems, defects and dopped ions in solid state systems, quantum dots, and mechanical and acoustical phonon systems. We examine their potential applications in building QMCs, focusing on the possibility and quality of the coupling across different physical systems.

In Table~\ref{tab:qmc_compare}, we summarize the main properties of these physical systems. We identify the bus qubits for these physical systems experimentally demonstrated by existing works, and estimate their internal and external storage ratios. Most of the physical systems can couple to photonic bus qubits. Depending on their energy scales and available quantum transitions, it is possible that QMC have optical or microwave photon interfaces.

As one of the applications of optical photonic systems is to build long-range quantum entanglements and quantum communication links, the QMCs are integrable with quantum communication devices, especially quantum repeaters~\cite{Briegel1998, Sangouard2011}. In this scenario, the main role of quantum memory is to store the quantum states to synchronize between different quantum operations. As photonic EPR pairs are commonly utilized in quantum communication protocols, and given their generation is usually slow compared to the photon measurements~\cite{Bennett1993}, photonic EPR pair generation time can be used as a time scale to measure the photonic QMC's storage time. On the other hand, photonic systems can also perform quantum computing using the measurement-based quantum computing (MBQC) scheme~\cite{MBQC_prl, MBQC_pra, MBQC_review}, where adaptive single-qubit measurements drive the computation on a pre-generated entangled resource state. In MBQC, resource state generation is one of the most significant questions, which is usually the most time-consuming operation. The resource state can also be constructed using photonic EPR pairs~\cite{FBQC}. In order to give an estimation of the external storage ratio ($\alpha_\text{ex}$) of the QMCs with optical-photon-based bus qubits, we consider using photonic EPR pair generation time as an estimate of $T_\text{op}$. Specifically, in Ref.~\cite{Zhang2021Photon}, the photon-pair generation rate can reach $52.36$~kHz ($T_\text{op} \approx 19~\mu$s).

Due to the fast gate operations between superconducting qubits, we envision that transmons and fluxonium qubits will be utilized in quantum processing units, rather than quantum memory modules (see Sec.~\ref{subsec:sc_qubits} for details). In Table~\ref{tab:qmc_compare}, we show the internal storage ratio $\alpha_\text{in}$ of superconducting qubits for reference. Superconducting qubits can strongly couple to microwave fields (see Sec.~\ref{subsec:microwave}). Therefore, the QMCs with microwave-photon-based QMCs are potentially integratable with superconducting-based QPUs. Entangling gate operations between superconducting qubits are used as the external operations when evaluating the QMC's external storage ratio. Specifically, we take the CZ gate between transmon qubits $T_\text{op} = 40$~ns with fidelity $0.998$ reported in Ref.~\cite{Marxer2023}. In Table~\ref{tab:qmc_compare}, we report possible usage of atomic clouds, spin ensembles in solid state systems, and phononic systems for QMCs interacting with microwave and superconducting qubits. However, the performance of these systems still needs to be improved to take advantage of quantum memory. 

\begin{figure}[htbp]
    \centering
    \includegraphics[width = \columnwidth]{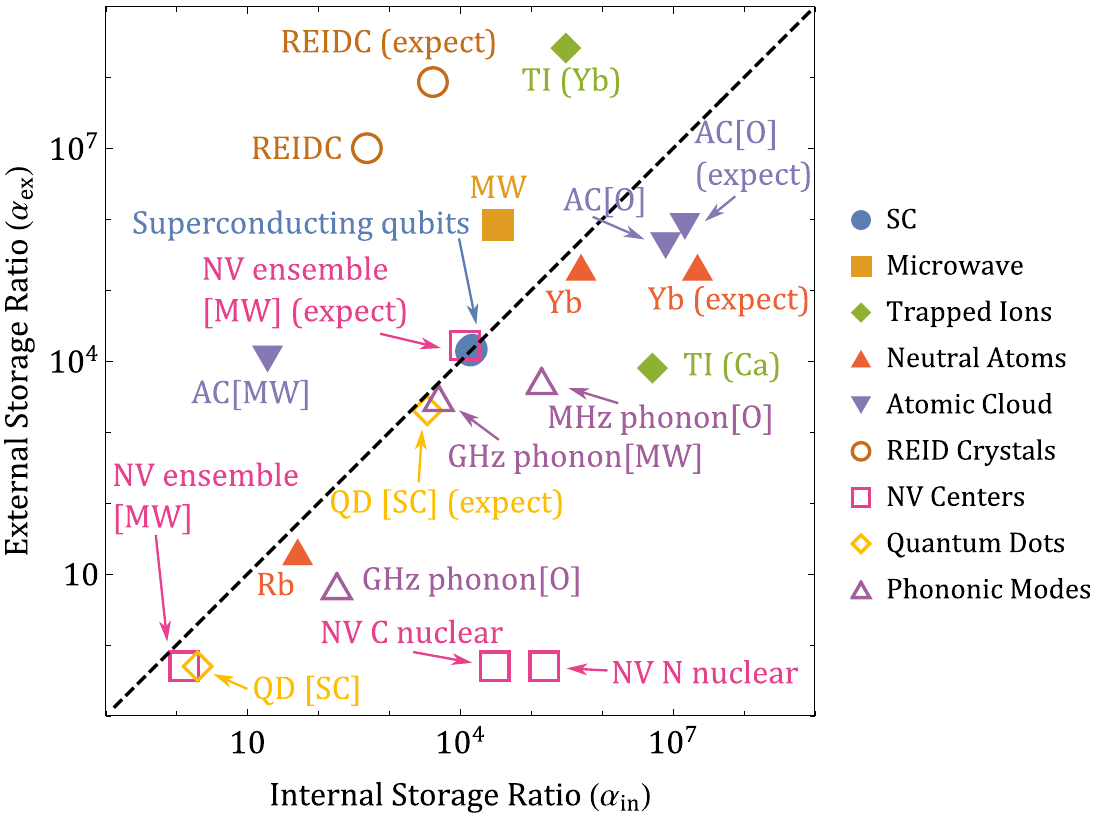}
    \caption{The internal ($\alpha_{\text{in}}$) and the external storage ratios ($\alpha_{\text{ex}}$) of QMCs built on different physical platforms. We include superconducting qubits (blue circles), microwave modes (orange squares), trapped ions (labeled as `TI', shown as green diamonds), Rubidium and Ytterbium trapped neutral atoms (reddish-orange triangles), neutral atomic clouds (labeled as `AC', purple inverted triangles), Rare-earth-ion-doped crystals (REIDC) (hollow brown circles), NV centers (hollow pink squares), quantum dots (hollow yellow diamonds), and phononic modes (hollow lavender triangles). `O' stands for connecting to optical systems, while `MW' is for connecting to microwave systems. The data points labeled as `expect' are estimated by the best performance parameters in the specific system. The detailed numbers are discussed in the corresponding subsections and summarized in Table~\ref{tab:qmc_compare}. For external storage ratios $\alpha_\text{ex} < 0$ and the external storage time of NV nuclear spin-based QMCs, we set them to be $0.5$ to be able to present them in the plot.}
    \label{fig:storage_time}
\end{figure}

In Fig.~\ref{fig:storage_time}, we plot the internal and external storage time of the QMCs built on physical systems that are discussed in the rest of the section. We plot $\alpha_{\text{in}} = \alpha_{\text{ex}}$ as the black dashed line for reference. We notice that there are two regions in Fig.~\ref{fig:storage_time}, (1) above the dashed line, and (2) below the dashed line. When a QMC point falls on the dashed line, the external operation and the RW operations take similar times, while if the point is above the line, the external operations take less time than the RW operations, and vice versa. 

Improving the QMCs can be reflected by shifting the corresponding point in the storage ratio plot (Fig.~\ref{fig:storage_time}). For example, extending the QMC coherence time can improve both internal and external storage ratios, which shifts the data point along the diagonal direction to the upper right corner. Reducing the RW time of the QMC can improve the internal storage ratio, which will push the corresponding point to the right. On the other hand, there are two methods to improve the QMCs' external storage ratio. One concentrates on the external system by accelerating external operations. The other is to increase the coupling strength and efficiency between the QMC with the external quantum system. For example, the expected trapped ion QMCs listed in Table~\ref{tab:qmc_compare} (Yb ions) are mainly from the storage time improvement, while the atomic cloud and rare-earth-ion-doped crystal QMCs are mainly from the RW operations. 

Contrary to the NISQ-era quantum memory devices, in the FTQC era, although the quantum memory and quantum computing units are made of the same species of physical qubits, the QEC codes can be different to take advantage of the QPU-memory separation. As the quantum memory does not need to support a fault-tolerant universal gate set, the QEC code on QMCs can have high thresholds and yield, e.g., using the quantum LDPC codes~\cite{bravyi2023highthreshold}. Therefore, designing FTQC-era QMCs may focus more on QEC, e.g., designing QEC codes of quantum memory and its interfaces to QPUs. We stress that the internal and external storage ratios defined in Eqs.~\eqref{eq:internal_st} and~\eqref{eq:external_st} can still quantify the performance of the FTQC QMC design. The storage time $T_\text{storage}$ can be estimated by the logical error rates, while the RW time $T_\text{RW}$ should take the gate operations between different QEC codes into consideration. 

With the focus on building QMCs, to enable fast QEC cycles on memory registers, the physical qubits should have fast and accurate measurement and gate operations between them. In addition, the physical qubits should support the topology required by the QEC codes. These requirements are exactly similar to the NISQ quantum computing qubits, and hence the discussion of physical qubits used in FTQC-era quantum memory is beyond the scope of this paper. 

\subsection{Superconducting qubits} \label{subsec:sc_qubits}

Superconducting circuit systems are one of the most promising systems for quantum computing~\cite{Blais2021, Kjaergaard2020}. Superconducting circuit systems feature strong nonlinearity provided by Josephson junctions. This nonlinearity allows for constructing quasi-atom structures capable of exhibiting quantum state manipulation, rapid initialization, quantum gate operations, and readout of the quasi-atoms' quantum states. However, the coherence time of superconducting qubits is relatively short. For example, the coherence time of transmon qubits~\cite{Koch2007, Schreier2008, Houck2008} can reach $T_2^* \approx 0.3$~ms~\cite{Place2021, Wang2022Transmon}, and be further improved to $0.557$~ms with dynamical decoupling. A fluxonium qubit~\cite{Manucharyan2009} with coherence time $T_2^*$ reaching $1.48$~ms has been reported~\cite{Somoroff2023}. On the contrary, taking the spin states of atoms and ions as an example, the coherence time can reach several seconds or even tens of minutes~\cite{Sheng2018, Bluvstein2022, Ma2022, Jenkins2022, Wang2017, Wang2021}.

On the other hand, the superconducting qubits can perform fast and high-fidelity single- and two-qubit gate operations. For example, transmon qubits can perform CZ gates in $40$~ns with a fidelity $99.8\%$~\cite{Stehlik2021, Sung2021, Marxer2023}, while a microwave-activated CZ gate between fluxonium qubits only takes $\sim 100$~ns with a fidelity $99.9\%$~\cite{Nesterov2018, Nguyen2022}. Due to these features, superconducting qubits can be leveraged as computing registers in quantum computing systems. In fact, utilizing the fast gate operations provided by superconducting qubits has been discussed in the context of hybrid quantum computing~\cite{Xiang2013, Kurizki2015, Clerk2020}. Therefore, in the NISQ-era quantum computing system design, we skip the discussion of using superconducting qubits as quantum memory cells. While in the FTQC era, where QEC is needed on quantum memory cells, the fast gate operations enable fast syndrome checking and error correction cycles, which can also be utilized as a good candidate for physical qubits to build logical quantum memory cells.

In order to compare the performance of QMCs made by other systems with a homogeneous superconducting-qubit-based quantum computing system, we based on transmon and fluxonium coherence time and gate times to estimate their storage ratio as a reference. Note that in this case, the internal and external storage ratios are essentially similar. We only focus on the internal storage ratio instead. The internal storage ratio of transmon qubits can reach $\alpha_\text{in} \approx 1.39 \times 10^4$, where we take $T_\text{RW} = 40$~ns and $F = 0.998$. The internal storage ratio of fluxonium qubits can reach $\alpha_\text{in} \approx 1.48 \times 10^4$, where we take $T_\text{RW} = 100$~ns and $F = 0.999$.

\subsection{Microwave cavities and resonators} \label{subsec:microwave}

With the development of superconducting qubits, quantum manipulation of microwave photonic states has become available, which has attracted a lot of attention recently~\cite{Gao2021}. Due to the improvement of microwave cavity fabrication techniques, microwave photon lifetimes inside a cavity keep improving, enabling the potential usage of QMCs for quantum information storage. There are two approaches to utilizing microwave resonators as QMCs: (1) using the Fock states of physical microwave photons or (2) using bosonic QEC code encoded microwave modes as qubits to store quantum information.

In the first approach, a QMC is encoded into the presence of a single photon in a cavity mode. The storage time of the QMC is largely determined by the lifetime of a photon inside the cavity~\cite{scully1997quantum}. The lifetime of 3D microwave cavities can reach $10.4$~ms~\cite{Reagor2013} to 2~s~\cite{Romanenko2020}. In this case, one choice of the bus qubit of the microwave QMCs can be a propagating microwave photon. The QMC read process is the photon emission from the microwave cavity, while the write process is the microwave photon absorption. The QMC, i.e., the microwave cavity or resonator, couples to a microwave waveguide or coaxial cable that holds the itinerant microwave photon qubit. The coupling needs to be well controlled for a high storage fidelity and RW operation efficiency~\cite{Yin2013, Wenner2014, Ibarcq2018}. The absorption efficiency can reach $99.4\%$~\cite{Wenner2014}. 

A more promising approach is to couple the microwave systems with superconducting qubits for fast computing operations. However, compared to the cavity microwave system, the superconducting qubit has a shorter coherence time, which can degrade the microwave-cavity-based QMCs. With the superconducting control qubit built in, the lifetime of the cavity photon can still reach 2~ms~\cite{Chakram2021} to $25.6$~ms~\cite{Milul2023}, while the photon coherence time can reach $34$~ms~\cite{Milul2023}, which is still significantly longer than superconducting qubits. Moreover, a superconducting bus qubit can maintain fast bus-QMC RW operations ($1~\mu$s as pointed out in Ref~\cite{Chakram2021}), which gives microwave QMCs internal storage ratio $\alpha_\text{in} \approx 3.4 \times 10^4$, where we estimate the RW fidelity as $0.994$ (estimated by the efficiency reported in Ref.~\cite{Wenner2014}) since it is not explicitly reported in Refs.~\cite{Chakram2021, Naik2018}. In addition, if the microwave-based QMC is connected with superconducting quantum computing devices, the computing operation time can be estimated by the two-qubit gate time between two transmon qubits $T_\text{op} = 40$~ns ($F = 99.8\%$), the external storage ratio can reach $\alpha_\text{ex} \approx 8.4 \times 10^5$. 

The other approach uses the quantum error correction code encoded microwave state as the QMC qubits. Recently, promising results have been shown in the microwave qubits encoded in GKP, cat, and other bosonic codes~\cite{Grimm2020, Darmawan2021, Chamberland2022, Gertler2021, Grimsmo2021, Campagne-Ibarcq2020, Hu2019}. In this case, the bus qubit needs to be a superconducting qubit, where the QMC read and write processes are equivalent to decoding and encoding the quantum information of the QMC qubit. Evaluating the performance of the logically encoded QMCs requires a detailed design of the error correction code used in the QMC and the bus qubits, which is beyond the scope of this paper.

With their multimode feature, microwave cavities are well-suited for large integration of QMCs. However, accessing different QMCs inside a single cavity is limited by the number of transmon qubits that couple to the cavity mode, typically kept low for high cavity coherence. Therefore, integrating multiple multi-mode cavities can be a promising approach~\cite{CANVAS}. However, finding the best strategy for microwave-cavity-based QMCs requires a comprehensive consideration of the memory device requirements, the cavity design, and the connectivity of the cavities, etc. 

\subsection{Trapped ions} \label{subsec:ions}

The trapped ion system is one of the popular systems not only for quantum memory but also for universal quantum computing. For a thorough review of trapped ion systems, we suggest referring to Refs.~\cite{Haffner2008, Monroe2013, Bruzewicz2019, Brown2021}. In the trapped ion systems, the quantum information is stored in the electronic spin states of the ions. The ions are trapped using radio-frequency Paul traps~\cite{Neuhauser1980} and other types of electromagnetic traps~\cite{Dehmelt1990, Chiaverini2005, Seidelin2006, Labaziewicz2008}. The quantum information can be stored in the spin states of electrons or the nuclear, which can have a long coherence time, making them suitable for quantum memory. Specifically, quantum information can either be encoded into the hyperfine levels~\cite{Harty2014, Gaebler2016, Ballance2016, Wang2017, Wang2021} or Zeeman sublevels of the same orbital~\cite{Keselman2011, Ruster2016, Hilder2022}, or other quantum states in the specific ion level structures~\cite{Bermudez2017, Toyoda2010, Pogorelov2021}. Depending on the species of the ions and the type of encoding, the coherence time of the qubits can vary significantly. For instance, in the Zeeman qubits, the coherence time can reach $300$~ms~\cite{Ruster2016}, while the hyperfine states are more isolated, and the coherence time can reach several minutes to an hour~\cite{Harty2014, Wang2017, Wang2021} ($5500$~s reported in Ref.~\cite{Wang2021}). 

Other than fast gate operations between trapped ions~\cite{Gaebler2016, Blumer2021, Moses2023, Jeon2023, Saner2023, Gaebler2016, Ballance2016, Saner2023, IonQ}, trapped ions can have strong coupling with optical cavity modes~\cite{Stute2012, Stute2013, Begley2016, Takahashi2020, Christoforou2020}. In addition, they can also be used as single-photon emitters, which can be pumped to generate entangled ion-photon pairs~\cite{Blinov2004, Bock2018, Kobel2021}. It enables trapped ions as QMCs for quantum computing systems with optical interfaces, photonic quantum computing systems, and quantum communications. In terms of using stationary photonic qubits living in the optical cavity mode as the bus qubit, the coupling strength $g/2\pi = 16.7$~MHz ($T_\text{RW} \approx 30$~ns) has been demonstrated~\cite{Christoforou2020}. With the $300$~ms long coherence time achieved in the Zeeman qubit of $\text{Ca}^+$ ions~\cite{Ruster2016}, the internal storage ratio can reach $\alpha_\text{in} \approx 5.1 \times 10^6$, where the RW efficiency is approximated by the ratio of the photon linewidth and the ion-photon coupling $\eta \approx 1 - g/\Delta \omega \approx 50.9\%$.

On the other hand, the generated entangled ion-photon pairs can be used in heralded entanglement generation schemes to create entanglement between ions with other photon emitters~\cite{Cabrillo1999, Kok2007, DLCZ, Casabone2013}. In this scenario, for a trapped-ion-based QMC, the bus qubit is another trapped ion, which is used as a single-photon emitter. Using the heralded entanglement generation scheme, the bus ion is entangled with another computing qubit and forms a Bell state~\cite{Niffenegger2020, Saha2023, Krutyanskiy2023_EG}. The bus ion can then be entangled with the QMC ion and transfer quantum information to the computing qubit. In this scheme, the RW time is determined by the bus-computing entanglement generation time, $T_\text{e-g}$, and a two-qubit gate time between ions $T_\text{Gate}$. The entanglement generation is probabilistic, whose probability is determined by the photon loss and photon detection efficiency. However, the generation process can be performed in parallel to speed up the generation time~\cite{Dhara2022}. Therefore, the internal storage ratio, in this case, depends heavily on the specific setup. 

Here we aim to furnish an approximate estimation, solely to give a qualitative understanding of the performance of ion-based QMCs in this case. In Ref.~\cite{Kobel2021}, entangled Yb ion-photon pairs can be generated with fidelity $90.1\%$ and rate $62$~Hz ($T_\text{e-g} \approx 16.1$~ms). As the two-qubit gates between ions can be implemented in $10$ to $600~\mu$s with fidelity $> 99\%$~\cite{Gaebler2016, Ballance2016, Saner2023, IonQ}, the entanglement generation time is dominating. If the Yb ions can maintain long coherence time ($5500$~s reported in Ref.~\cite{Wang2021}), the internal storage ratio can reach $\alpha_{\text{in}} \approx 3.1 \times 10^5$, where we only account the contribution of $T_\text{e-g}$ to the RW time.

Using trapped-ion-based QMCs for optical quantum computing and quantum communication, with the $300$~ms coherence time of Zeeman qubits of Ca$^{+}$~\cite{Ruster2016}, the external storage ratio can reach $\alpha_\text{ex} \approx 8.0 \times 10^3$. We estimate the RW operations based on Ref.~\cite{Christoforou2020} as above, and use photonic EPR pair generation time as the external operation time ($19~\mu$s, see Sec.~\ref{subsec:qmc_compare}). For a Yb ion-based QMCs with RW scheme as mentioned above, the external storage ratio can reach $\alpha_\text{ex} \approx 2.6 \times 10^8$ with $5500$~s storage time.

In addition, the spin degree of freedom of the ions can couple to electromagnetic fields in MHz to GHz range, which enables superconducting qubits as bus qubits to couple to superconducting quantum computing devices. Direct coupling between single microwave photons with single trapped ions is possible, but the coupling strength is estimated to be in the order of tens of Hz~\cite{Verdu2009}. The slow coupling hinges using single microwave photons as bus qubits to couple to superconducting qubits. Instead, another approach using an oscillating electric field to drive sideband transitions of ions can provide $\sim 60$~kHz coupling between an ion to superconducting qubits. Even though this coupling strength gives RW times much smaller than the coherence time of trapped ions, the RW time can be challenging as it is slower compared to superconducting qubit coherence time (see more detailed discussion in Sec.~\ref{subsec:sc_qubits}). 

Furthermore, the trapped-ion-based QMCs are suitable for large-scale integration. Since ions are naturally identical, there are no fabrication imperfections that limit the performance of individual QMCs. Specifically, commercial companies have demonstrated and made public access to 20 to 32-qubit trapped ion quantum computing units already~\cite{IonQ, Moses2023}. In addition, hundreds of ions can be trapped into a single 1D or 2D ion trap, which shows the capability to construct quantum devices with large sizes of trapped ions~\cite{Bohnet2016, Zhang2017, Kiesenhofer2023}. Unlike the trapped-ion quantum computing devices, which discourage large numbers of trapped ions in a single trap due to the hardness of driving two-qubit gates by selectively driving a single phonon mode of the trapped ion array, trapped-ion-based quantum memory devices do not require two-qubit gates between the QMCs, which releases the requirement of integrating trapped-ion QMCs. However, to reduce the RW latency, designing the structure of the quantum memories consisting of many QMCs to minimize the transporting time is an important question. 

\subsection{Neutral atoms}

Neutral atoms share several advantages with ion systems, including their intrinsic long coherence time brought by the spin degrees of freedom for information storage, the capability of precise control, and the ease of integration. However, neutral atoms have their own features, which distinguish themselves from ion-based QMCs. There are several strategies to encode quantum information into states of single atoms. Other than the hyperfine ground states of the Rydberg atoms~\cite{Sheng2018, Graham2019, Levine2019, Bluvstein2022} (named as GG qubits in Ref.~\cite{Morgado2021}), the ground state of an atom and its Rydberg excited state, i.e., a highly excited electronic state, can also be used to encode the qubit $\ket{0}$ and $\ket{1}$ states~\cite{Levine2018, Madjarov2020} (GR qubits mentioned in Ref.~\cite{Picken2019, Morgado2021}). The coherence time of the qubits varies according to the species of qubits and the trapped atoms, ranging from a few microseconds to several seconds~\cite{Levine2018, Levine2019, Norcia2019, Ebadi2021, Jenkins2022, Barnes2022, Ma2022}. 

Despite coupling to another neutral atom leveraging the ``Rydberg blockade'' effect~\cite{Lukin2001, Jaksch2000, Urban2009}, neutral atoms also have strong coupling to optical light, which makes optical interfaces (photonic bus qubits) possible. In Ref.~\cite{Langenfeld2020}, two trapped Rb atoms interacting with photonic qubits with RW efficiency $\eta^2 \approx 26\%$ has been demonstrated. With the coherence time of the Rb atoms reported in the same experiment ($800~\mu$s) and the address pulse durations ($8~\mu$s) for the RW time, the internal storage ratio is $\alpha_\text{in} \approx 51.0$. Using the QMC for optical quantum computing, where the operation time is estimated by $19~\mu$s (see Sec.~\ref{subsec:qmc_compare} for detailed discussion), the external storage ratio reaches $\alpha_\text{ex} \approx 20.5$. Note that the coherence time of other species of atoms can have a much longer coherence time. If the Yb atoms can achieve the same RW operations, the corresponding internal and external storage ratio can achieve $\alpha_\text{in} \approx 5.0 \times 10^5$ and $\alpha_\text{ex} \approx 2.1 \times 10^5$, respectively ($T_\text{coh} = 7.9$~s demonstrated in Ref.~\cite{Jenkins2022}). The strong coupling between the single atom and optical modes has also been demonstrated in experiments, where $g/2\pi = 3.2$~MHz is shown in Ref.~\cite{LiuYanxin2023}. Although in Ref.~\cite{LiuYanxin2023}, it is the Cs atoms that couple strongly to the optical modes, the RW operations of general atom-based QMCs are expected to be further improved in the future. With the coupling strength $g/2\pi = 2.74$~MHz, the RW operations take $T_{\text{RW}} = \pi/g \approx 182.5$~ns with efficiency $\eta^2 \approx 26\%$. The internal and external storage ratios can reach $\alpha_{\text{in}} \approx 2.2 \times 10^7$, and $\alpha_{\text{ex}} \approx 2.1 \times 10^5$, respectively, if the coherence time of the QMC is $7.9$~s.

In addition, a cloud of neutral atoms can also be optically controlled to serve as QMCs for optical photons. When a cloud of atoms coherently couple to the same field, the quantum interference can boost the coupling strength between the field and the collective mode of the atoms~\cite{scully1997quantum}. To control the photon absorption and emission, electromagnetically induced transparency (EIT) is one of the most commonly adopted ways to controllably absorb and emit bus optical photons, i.e., achieve RW operations. The EIT-based atomic cloud photonic quantum memories have been demonstrated in both cold and warm atomic ensembles~\cite{Hosseini2011, Dudin2013, Ding2013, Nicolas2014, Ding2015, Ding2015Nat, Parigi2015, Saunders2016, Katz2018, Hsiao2018, Jiang2019, Wang2019, Li2020, Dideriksen2021, Wang2022, Messner2023}. The EIT-based room-temperature atomic ensemble can have $\sim 0.9$~ms storage $1/e$ lifetime~\cite{Dideriksen2021, Wang2022} and in total $1$~s storage time is possible~\cite{Katz2018}, while the cold atom clouds with dynamical decoupling can extend the lifetime to $16$~s~\cite{Dudin2013}. For classical light pulse storage, the retrieval efficiency (equivalent to $\eta^2$ used in Eq.~\eqref{eq:external_st}) of the atomic quantum memory can achieve $92\%$~\cite{Hsiao2018}, while the light storage down to single-photon level has also been demonstrated~\cite{Ding2015, Ding2015Nat, Wang2019, Buser2022}. 

In the EIT-based QMCs made of atomic clouds, the RW operation is achieved by absorbing and emitting light pulses carrying the quantum information. Whenever the quantum information in the QMC is read, i.e., the light pulses are re-emitted from the optical media, the QMC gets reset. In the writing process, the writing time is determined by the speed of turning the control light pulse off. This time is compatible with the signal light pulse duration. In the cold-atom cloud system, the storage time can be estimated by the $1/e$ lifetime, where we adopt $16$~s from Ref.~\cite{Dudin2013}. We estimate the internal storage ratio using the read and write full-process efficiency $\eta^2 = 0.26$ and the control light FWHM duration $1040$~ns reported in Ref.~\cite{Dudin2013} as the RW time, which results in $\alpha_\text{in} \approx 7.8 \times 10^6$. Leveraging the QMCs for optical quantum applications ($T_\text{op} \approx 19~\mu$s), the external storage ratio is $\alpha_\text{ex} \approx 4.3 \times 10^5$. If the full-process efficiency can be improved to $0.906$ as reported in Ref.~\cite{Wang2019}, the internal and external storage ratios can be improved to $\alpha_\text{in} \approx 1.5 \times 10^7$ and $\alpha_\text{ex} \approx 8.0 \times 10^5$, respectively. 

Furthermore, similar to the ions, the spin levels of atoms can also couple to microwave photons. However, the coupling between a single atom and a single microwave photon is weak~\cite{Verdu2009, Hattermann2017}. Therefore, a cloud of atoms is leveraged to enhance the coupling strength~\cite{Rabl2006, Reversible2009, Hattermann2017, Tu2022}. Specifically, in Ref.~\cite{Hattermann2017}, an ensemble of Rb atoms coherently coupled with microwave field, enabling atomic Rabi frequency $20$~kHz. Based on the Rabi oscillation, we estimate the RW time as $25~\mu$s with efficiency $0.6$. Assuming the storage time can still reach $800~\mu$s as demonstrated in Ref.~\cite{Langenfeld2020}, the internal storage ratio is $\alpha_{\text{in}} \approx 19.2$~\footnote{We acknowledge that the hyperfine states of Rb atom used in these two works are different. Here we aim to give an estimate of the possible implementation of the QMCs.}. Suppose the microwave field can couple to transmon qubits for computing, which takes another $1~\mu$s for microwave-transmon coupling with efficiency $\eta \approx 0.994$ (see Sec.~\ref{subsec:microwave}), the external storage ratio can be $\alpha_\text{ex} \approx 1.1 \times 10^4$. Other attempts to construct coherent coupling between microwave fields in superconducting coplanar waveguides and a beam of Rydberg helium atoms have been demonstrated in experiments~\cite{Thiele2015, Garcia2019, Morgan2020}. However, limited by the coherence time and the coupling strength, high-fidelity single microwave photon level operations still need to be demonstrated, which is needed to connect the QMCs with superconducting quantum computing devices.

Similar to the trapped-ion systems, neutral-atom-based QMCs can be largely integrated, and share similar benefits with trapped-ion systems. In addition, as neutral atoms are trapped into optical lattices, which enables higher dimensional neutral atom lattices easily~\cite{YWang2015, YWang2016, Xia2015, Maller2015, Graham2019}, higher dimensional integration of neutral-atom-based QMCs into quantum memory devices is viable. Moreover, the controlled removal of the missing sites in the optical lattices~\cite{Endres2016, Bernien2017, Ebadi2021, Barnes2022} and coherent moving of the trapped atoms have been demonstrated~\cite{Bluvstein2022}, which can construct a more compact quantum memory device.

\subsection{Rare-earth-ion-doped Solid state systems}

Similar to the trapped ion systems, the quantum states of ions doped into solid-state systems can also be precisely addressed and quantum manipulated. Among different species of ions, the rare-earth ion doped (REID) solid-state system is one of the other attractive systems to build optical quantum memory~\cite{McAuslan2012, Saglamyurek2011, Clausen2011, Ledingham2012, Zhou2012, Ferguson2016, Jin2022, Ma2021_v2}, where an ensemble of doped ions are collectively manipulated. Recently, the atomic-frequency-combs-based (AFC) methods have been widely adopted in building a long-storage-time on-demand REID-based optical quantum memory~\cite{Seri2017, Kutluer2017, Laplane2017, Holzapfel2020, Businger2020, Askarani2021, Ma2021}. In the AFC-based photon absorption, a sequence of narrow control pulses is sent to a broadband optical media to carve out an equal-spacing absorption spectrum, named `frequency comb'~\cite{Afzelius2009}. The incident light can then be absorbed into the medium, exciting the medium atoms, and then be emitted after a period of time. The storage time is determined by the spectral spacing between the teeth of the frequency comb. In order to make the quantum memory on-demand, the optical excitation stored in the optical media is then converted into other excitation, e.g., a spin-wave excitation of the media~\cite{Ma2021, Askarani2021, Seri2017, Kutluer2017, Laplane2017}. This scheme also takes advantage of the long coherence time of the spin states compared to the optical excited states. Long storage time up to $52.9$~min with dynamical decoupling has been demonstrated~\cite{Ma2021}. The efficiency of $26.9\%$ has been reached for the AFC storage and retrieval. Converting the excitations to spin waves reduces the overall efficiency, causing the full-process efficiency to reduce to $\approx 7\%$ in experiments~\cite{Zhu2022}. The conditional storage fidelity can reach $99\%$~\cite{Zhou2012, Zhu2022}. Further improving the retrieval efficiency is still a challenging question of the rare-earth ion-doped solid-state optical memory.

To estimate the REID QMCs' internal and external storage ratios, we notice that the AFC technique requires a preparation step before the RW operations~\cite{Ma2021, Zhu2022}, i.e., a control light pulse is needed to prepare the medium's absorption spectrum into a frequency comb. This preparation time differs from the RW time in the definition of the internal storage ratio $\alpha_\text{in}$ (see Eq.~\eqref{eq:internal_st}). Nevertheless, we can still treat it as a time overhead to estimate the bare RW time $\tau$. Based on the experiment reported in Ref.~\cite{Ma2021}, the internal storage ratio can be estimated as $\alpha_\text{in} \approx 4.8 \times 10^2$, where we adopt $52.9$~min coherence time as the storage time, and the storage efficiency with dynamical decoupling as the RW efficiency $\eta^2 \approx 0.37\%$. As the exact time for the preparation pulses is not explicitly reported in Ref~\cite{Ma2021}, we estimate it by $400$~ms reported in Ref.~\cite{Zhu2022}. The corresponding external storage time is estimated as $\alpha_\text{ex} \approx 1.0 \times 10^7$, with photon EPR pair generation time $T_\text{op} = 19~\mu$s. This shows that the QMC has a long storage time compared to a fast EPR generation process, but a relatively short storage time compared to its own RW operations (storage preparation time overhead). One limitation is the low efficiency. Imagining the storage efficiency can be improved to $26.9\%$, which is the AFC efficiency as reported in Ref.~\cite{Zhu2022}, the internal storage time can be improved to $\alpha_{\text{in}} \approx 4.1 \times 10^3$. Accordingly, the external storage time can be $\alpha_\text{ex} \approx 8.7 \times 10^7$.

Similar to the trapped ions, the spin levels of the ions can couple to microwave fields, which makes a microwave-based bus qubit possible for quantum memory. There are several experimental attempts to use the spin states and hyperfine spin states of the ions to store microwave fields~\cite{Probst2013, Probst2015, Wolfowicz2015}. However, the coherent storing and reading out of a quantum microwave state is still lacking to fully demonstrate the feasibility of using REID crystals as a robust and reliable microwave memory. 

\subsection{Solid-state defect centers} \label{subsec:defects}

Solid-state defect centers have emerged as promising candidates for quantum memory due to their long coherence times and controllable electronic and nuclear spin states. Among all solid-state defect centers, the nitrogen-vacancy (NV) centers in diamond are one of the most promising ones due to their long spin coherence time even at room temperature~\cite{Doherty2011, Doherty2013}. The negatively charged NV centers are spin-1 systems, where the electronic spin states can be used to encode quantum information. The spin states can have $1.8$~ms lifetime in isotopically pure diamond samples without dynamical decoupling~\cite{Balasubramanian2009}. Although the nearby nucleus with nonzero spin creates a spin bath and decohere the spin states, the coherence time can be greatly extended with dynamical decoupling~\cite{Bar-Gill2013, Abobeih2018, Abobeih2022}, to $1.58$~s~\cite{Abobeih2018} for electronic spin states. 

One natural choice of bus qubits for QMCs built on NV electronic spin states can be optical photons, as NV centers can be used as single-photon emitters. The entanglement between NV center spin states and the emitted photons has been demonstrated~\cite{Togan2010}. The heralded entanglement generation between remote NV centers has been realized in experiments~\cite{Bernien2013, Pfaff2014, Hensen2015, Pompili2021, Hermans2022}. One disadvantage of this scheme is that the NV centers have a broad phonon side band~\cite{Doherty2011, Doherty2013}, which largely reduces the success probability. Another factor that decreases the success probability is the photon collection efficiency. In our analysis of using heralded entanglement generation schemes, the low success probability results in slow RW operations. To solve this problem, nano-photonic crystal structures have been utilized to Purcell enhance the photon emission to the zero-photon line and increase photon collection efficiency~\cite{Hausmann2013, Riedel2017}.

The nearby carbon and nitrogen nuclear spin levels (hyperfine levels) have extra long coherence time compared to electronic spin states~\cite{Maurer2012}. With dynamical decoupling, the coherence time can reach $63$ s~\cite{Bradley2019, Abobeih2022}. The manipulation of these nuclear levels can be achieved using NV electronic states, which enable using the nuclear spin states as QMCs~\cite{Maurer2012, Bradley2019, Pompili2021, Bartling2022, Hermans2022, Xie2023}, while using the NV electronic state as the bus qubit. The gate time between nuclear and electronic states ranges from $389~\mu$s to $1556~\mu$s~\cite{Bradley2019, Abobeih2022}. The nearby nuclear spin ensemble also has a long coherence time and is possible to store quantum information. The coherence time can reach $3.5$~ms with dynamical decoupling~\cite{WangGuoqing2023}.

With the experimental realizations of NV-center-based quantum systems shown in Ref.~\cite{Abobeih2022}, we can estimate the QMC performance. If carbon nuclear spins are used as QMCs, the storage time can be estimated by the coherence time $T = 12.9$~s with dynamical decoupling. The quantum gate between carbon nuclear spin states and the NV electronic states takes $419~\mu$s with fidelity $F_\text{gate} = 0.99$, which is used to estimate the properties of RW operations. The internal storage ratio can reach $\alpha_{\text{in}} \approx 3.05 \times 10^4$. Note that in this work, there are five carbon nuclear states used. We choose to report the largest internal storage ratio. Although the nitrogen nuclear spin states are not used as quantum memory in Ref.~\cite{Abobeih2022}, making it a QMC that can benefit from the fast gate speed ($389~\mu$s) and the long coherence time with dynamical decoupling ($63$~s). With the estimated gate fidelity $0.94$, the internal storage ratio can be $\alpha_\text{in} \approx 1.5 \times 10^6$. However, one caveat of this approach is to connect the bus qubit, i.e., the electronic state of NV centers, to other computing registers. Although universal computing can be performed by controlling the bus qubit (NV center electronic state) and a nearby nuclear spin (nitrogen in Ref.~\cite{Abobeih2022}), how to scale up the systems is still an interesting question to explore.

Except for using the quantum state of a single defect center as a QMC, an ensemble of color centers can be treated as a spin ensemble, which can coherently couple to microwave fields. It enables microwave photons as bus qubits~\cite{Diniz2011, Zhu2011, Kubo2011, Kubo2012, Heshami2014, Kalachev2019}. Similar to REID crystals and atomic clouds, optical photon storage techniques, e.g., AFC and EIT methods, can also be applied in principle. Using microwave photons as bus qubits makes the coherent coupling to microwave-connected systems possible, e.g., to superconducting qubits~\cite{Kubo2011, Kubo2012}. Direct coupling NV center ensembles with a flux qubit has also been reported in Ref.~\cite{Zhu2011}. Specifically, in the experimental demonstration in Ref.~\cite{Kubo2011}, the Ramsey measurement gives an estimate on the storage time of the NV ensembles $T_\text{storage} \approx 200$~ns. The RW of the stored state by superconducting qubits takes $58$~ns with fidelity $\eta = \sqrt{0.14}$. The internal storage ratio is $\alpha_\text{in} \approx 1.3$. Limited by the coherence of the NV ensembles, the net storage time is smaller than twice of rescaled RW time $T_\text{RW}/\eta$, so the external storage ratio $\alpha_\text{ex} < 0$. If the electronic states can be transferred to nuclear spin ensembles, where the coherence time can reach $1.8$~ms can be achieved, the internal and external storage ratio can be improved to $\alpha_\text{in} \approx 1.2 \times 10^4$ and $\alpha_\text{ex} \approx 1.7 \times 10^4$, which is compatible to homogeneous superconducting devices. To further take benefits from the NV-ensemble-based QMCs, extending its storage time and improving the RW operations is needed. 

In general, solid-state defect centers can be compactly integrated into a single crystal. The defect color centers inside the solid state systems can be nicely fabricated and implanted inside the solid crystal, which makes large-scale integration possible. In addition, other types of defect centers, e.g., SiV~\cite{Bhaskar2020, Starling2023, Stas2023}, GeV~\cite{Siyushev2017}, SnV~\cite{Debroux2021} color centers, are also under investigation to improve their coherence properties and develop new quantum manipulation techniques.

\subsection{Quantum dots} \label{subsec:qd}

Semiconductor-based quantum dot (QD) system has been attracting much attention in quantum computing and quantum information processing in recent years. Compared with other quantum systems, quantum dots can be fabricated by the well-developed deposition and lithography techniques used in the semiconductor industry. The small size of the QDs ($\sim 100$~nm) makes them easy to be integrated largely~\cite{Maurand2016, Vandersypen2017, Fedele2021, Zajac2018, Mills2022, Burkard2023}. With the available high-fidelity quantum gates, semiconductor-based quantum dots can be potentially used as one of the candidates for quantum memory cells. 

Quantum dot-based spin qubits have multiple ways to encode quantum information into the physical systems. As each quantum dot can confine an electron, which is a spin-$1/2$ particle, one natural way to encode quantum information is to use the spin state of the confined electron. The corresponding spin qubit is called `Loss-DiVincezo' (LD) qubit~\cite{Loss1998, Mills2022, Xue2022, Zajac2018, Yoneda2018, Tanttu2023, Noiri2022}. Long coherence time up to $20~\mu$s has been demonstrated in experiments~\cite{Yoneda2018}. With Hahn echo techniques, the coherence time can be extended to $100~\mu$s~\cite{Mills2022, Yoneda2018, Tanttu2023}. When there are multiple quantum dots available, especially when the tuning barriers between the quantum dots are relatively low, the electrons confined in the nearby quantum dots can couple with each other and form entangled states. The quantum information can also be encoded into the state of multiple electrons. For example, the `singlet-triplet' (ST) qubit utilizes two entangled electrons confined in two nearby quantum dots~\cite{Cerfontaine2020, Fedele2021, Takeda2020, Nichol2017}. ST qubits also show promising long spin coherence time ($\sim 2~\mu$s)~\cite{Takeda2020}. The quantum information is encoded into the singlet state and one of the three triplet states. The total spin $1/2$ states of three electrons can also be leveraged as a manifold to define a spin qubit~\cite{Weinstein2023}. As the quantum dot spin qubits in this type can realize universal quantum computing only by controlling the exchange couplings between different dots, this type of qubit is named an `exchange-only' (EO) qubit~\cite{DiVincenzo2000, Bacon2000, Burkard2023, Weinstein2023}. For a comprehensive review of the recent development of semiconductor quantum dot qubits, we suggest Ref.~\cite{Burkard2023}.

QD qubits can strongly couple to microwave fields in a superconducting resonator. The microwave field can have stronger coupling to the charge degrees of freedom~\cite{Stockklauser2017, Mi2017, Scarlino2019}, while the spin degrees of freedom of QDs can have longer coherence time~\cite{Mi2018, Landig2018, Landig2019, Samkharadze2018}. From the spectral measurements, QD charge qubits can have a strong coupling up to $\sim 119$~MHz~\cite{Stockklauser2017}, while the spin qubits can a have coupling to microwave field $52$~MHz~\cite{Landig2019}. Therefore, using microwave photons as bus qubits for a QD-based QMC is available. Moreover, as superconducting qubits can couple to resonators strongly, using the microwave field to couple superconducting qubits with QDs has also been demonstrated in experiments~\cite{Landig2019, Scarlino2019}. Therefore, a transmon qubit can also be used as a bus qubit for QD-based QMCs. Specifically, the Rabi oscillation between a transmon qubit and a QD charge qubit is experimentally demonstrated~\cite{Scarlino2019}. 

Based on Ref.~\cite{Scarlino2019}, the coherence time of QD is estimated as $59$~ns (FWHM linewidth is $2.7$~MHz). The RW operation quality can be estimated from the Rabi oscillation, where the RW time is estimated as $T_{\text{RW}} \approx 23.8$~ns, and the efficiency is estimated as $0.8$. Therefore, the internal storage ratio reaches $\alpha_\text{in} \approx 1.98$. This means further improving the coherence time of QD qubits while maintaining the strong coupling to microwave resonators and superconducting qubits is needed. Note that the rescaled single RW operation time is comparable to the storage time, the external storage ratio is negative, which means the current QD-based QMCs still need improvement to gain advantages. Suppose the coupling can be tuned such that the QMCs can still maintain good coherence time ($T_2 \approx 102~\mu$s in Ref.~\cite{Mills2022}), while the RW operations are still as good as demonstrated in Ref.~\cite{Scarlino2019}, the internal storage ratio can reach $\alpha_\text{in} \approx 3.4 \times 10^3$. Considering using this QD-based QMC for superconducting quantum computing devices, where the operation time is $T_\text{op} = 40$~ns ($F \approx 99.8\%$), the external storage ratio can be improved to $\alpha_\text{ex} \approx 2.0 \times 10^3$.

\subsection{Phononic systems} \label{subsec:phonon}

Phononic systems are also widely considered within the context of hybrid quantum systems~\cite{Xiang2013, Kurizki2015, Clerk2020} because they can interact with circuit-QED systems through the piezoelectric effect, as well as with optical systems through the optomechanical effect. This makes photonic systems an intermediary system for microwave-to-optical transduction, which has been a key focus in recent efforts to achieve long-range quantum communication between circuit-QED devices~\cite{Andrews2014, Bagci2014, Shao2019, Forsch2020, Zhong2020, Xu2021, Mirhosseini2020}. For a review of recent progress on nano-phononic systems, we refer to Refs.~\cite{Vasileiadis2021, Bachtold2022, Heinrich2021}. 

Unlike electromagnetic waves, acoustic waves require media to propagate, making phononic modes well-isolated and beneficial for maintaining a long lifetime. For example, in nano-mechanical resonators which hold phonon modes with frequency $\sim 1.4$~MHz, the single-phonon lifetime can reach $100$~s to $1000$~s~\cite{Seis2022, Beccari2022}. The coherence of the phonon modes can reach $100$~ms~\cite{Seis2022}. The lifetime of phonon modes with GHz frequencies in nano-acoustic resonators can reach $1.43$~s, while the coherence time can reach $130~\mu$s~\cite{MacCabe2020}. Therefore, the photonic modes can build QMCs where the quantum information is stored in the oscillations of these phonon modes. Fast coherent couplings between microwave modes (superconducting qubits) and the phononic modes provide necessary tools to manipulate the phononic modes in the quantum regime and show its quantum features~\cite{Teufel2011, Mirhosseini2020, Wollack2022}. An iSWAP gate between the microwave modes and phononic mode can only take $25$~ns with fidelity $0.95$~\cite{Wollack2022}. Strong dispersive couplings between the microwave and phononic modes have also been realized in experiments~\cite{Wollack2022, Lupke2022}, which can be used to couple phononic modes with microwave and optical photons as well. This makes using microwave and optical photons as the bus qubits possible. In addition, phononic QMCs can be integrated with superconducting qubits for fast quantum gate operations with the help of microwave bus qubits.

Using the phonon modes as QMCs and microwave photons as bus qubits, the internal storage time of phononic QMCs can be estimated as $\alpha_\text{in} \approx 4.9 \times 10^3$, where we consider the GHz-frequency phonon modes as the QMCs, as they naturally couple to the microwave bus modes and can be integrated with superconducting qubits. As coupling the microwave bus qubit with the superconducting qubit extends the RW time, to consider the external storage time, we take the coupling between the superconducting qubits and the microwave photon bus qubits into the consideration of phonon-based QMC RW operations ($\approx 1.04~\mu$s in total~\cite{Chakram2021}, with efficiency $\eta \approx 0.95 \times 0.994$). We then take the time for a two-qubit gate between superconducting qubits as a quantum computing operation time ($40$~ns). The external storage time can reach $\alpha_\text{ex} \approx 3.02 \times 10^{3}$.

Furthermore, the phononic modes can also couple to photonic modes via optomechanical effect, where the coherent coupling strength $g/(2\pi) \approx 700$~kHz has been demonstrated~\cite{Stockill2022}. Therefore, using optical photons as the bus qubits of a phononic-mode-based QMC is possible. In this case, the RW time of the QMC can be estimated as $0.71~\mu$s, where we consider the RW operation with unit fidelity for the estimation. The internal storage ratio can be estimated as $\alpha_\text{in} \approx 1.8 \times 10^2$ with storage time $130~\mu$s~\cite{MacCabe2020}. Using GHz phonon mode QMCs to store photonic qubits in optical quantum computing and quantum communication is also possible. With EPR pair photon generation rate at $52.36$~kHz, the external storage ratio $\alpha_\text{ex} \approx 6.7$. If the MHz phonon QMCs can have a similar RW speed with optical bus qubits, the internal and external storage ratios can reach $\alpha_\text{in} \approx 1.4 \times 10^5$ and $\alpha_\text{ex} \approx 5.2 \times 10^{7}$, which means the phonon-based QMC needs to improve its coherence time and QMC-bus coupling speed to further enhance its performance. 

Similar to microwave-cavity-based QMCs, as a mechanical membrane or acoustic resonator can support multiple phonon modes, these modes can all be used as QMCs. Therefore, the phononic QMCs can be easily integrated to form a quantum memory device. In addition, the small physical scale of the phononic systems makes them easily fit into a single dilution refrigerator, which can be integrated with superconducting quantum computing chips. 

\subsection{Others platforms} \label{subsec:others}

In addition to the physical systems we discussed above, there has been significant interest in using topologically protected states for quantum computing. One effort includes using the topological error correction codes to encode physical qubits to logical qubits, whose quantum information is topologically protected. One example is the surface code~\cite{bravyi1998quantum, Dennis2002, Fowler2012}. Using the topological error correction codes for quantum memory has been discussed in the seminal reviews Refs.~\cite{Terhal2015, Brown2012016}. This approach is not limited to any specific physical platforms. Another approach is to use the physical topological states as the basic physical qubits. In this scenario, as the physical qubit is robust to local imperfections, these physical qubits can be more robust compared to other physical qubits and reduce the QEC overhead. For example, the Majorana zero modes (MZM) localized on superconducting nanowires can be used as physical systems to encode quantum information. By braiding the MZMs, gate operations between qubits can be applied~\cite{Alicea2011}. A complimentary review of the theory and experimental realization of the MZM in the solid-state systems can be found in Ref.~\cite{Sarma2015}, while achieving couplings between Majorana qubit and superconducting qubit has also been proposed~\cite{Chirolli2022}. Reviews of topological quantum computing can be found~\cite{roy2017topological, Lahtinen2017}. Despite that the experimental realization of the MZM modes is still under some debate, it has the potential to become a promising technology to realize topologically protected quantum memory.

\section{Building Quantum Memory Devices} \label{sec:QMM}

\begin{figure}
    \centering
    \includegraphics[width = 0.6 \columnwidth]{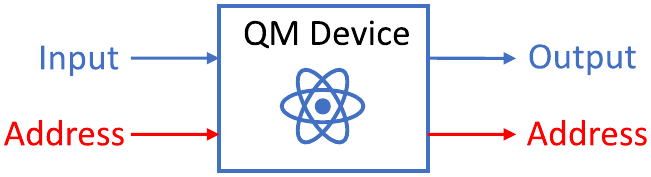}
    \caption{The abstraction of a quantum memory device. A general quantum memory device should have four ports, input, output, and two address ports. All these ports can be either classical information or quantum information.}
    \label{fig:qmd}
\end{figure}

With the development of single QMCs, how to integrate individual QMCs into a quantum memory device (QMD) is the next question to explore. Similar to the classical memory device, in order to address QMCs efficiently, it is necessary to assign addresses to QMCs such that each QMC can be accessed by its address. In Fig.~\ref{fig:qmd}, we show an abstraction of a quantum memory device. A quantum memory device should have at least three interfaces: input, output, and address ports. In the memory loading phase, the address is given, and the quantum information carried by the QMC with the given address is exported to the output port. In the writing phase, the quantum information is fetched into the QMD through the input port and saved to the QMC with the given address. In contrast to a classical memory, all the quantum memory ports can either carry classical or quantum information. The unitary nature of quantum operations requires to have an additional address port for address information. 

In order to consider different quantum memory designs and physical implementations, we define metrics to quantitatively discuss the device performance. A good quantum memory device should have fast RW speed, long storage time, and, ideally, a large scale of integration of QMCs. First of all, the internal storage time $\alpha_{\text{in}}$ discussed in Eq~\eqref{eq:internal_st} can be extended to
\begin{equation}
    \text{Storage ratio: } \alpha_{\text{QMD}} = \frac{T_\text{storage}}{T_{\text{RW}}},
\end{equation}
where $T_{\text{RW}}$ is the read and write time of the quantum memory device. Compared to the metric of QMCs, addressing the proper QMC also takes time and may dominate the RW process. We define the QMD RW time as $T_{\text{RW}} = T_{\text{addr}} + T_{\text{RW,QMC}}$, where $T_{\text{addr}}$ is the addressing time and $T_{\text{RW,QMC}}$ is the RW time for the QMCs. The numerator $T_{\text{storage}}$ is the storage time of the quantum device. For near-term implementation, where the quantum memory is not error corrected, the coherence time of the QMC qubits inside the memory device can be a good measure, and hence $T_{\text{storage}} = T_{\text{coh}} - T_{\text{RW}}$. While in the FTQC regime, the storage time can be estimated by the inverse of the logical error rate. 

On the other hand, the external storage ratio $\alpha_\text{ex}$ can also be similarly extended to describe the performance of a QMD by using QMD RW time $T_{\text{RW}}$ in Eq.~\eqref{eq:external_st}. However, to better quantify QMD's performance, especially to quantify its operation time relative to the computing operation, we consider a new metric, 
\begin{equation}
    \text{Memory latency: } \beta = \frac{T_{\text{RW}}/\eta}{T_{\text{op}}}, 
\end{equation}
where $T_{\text{op}}$ is the time for a quantum operation on the quantum computing module. This parameter $\beta$ effectively describes the latency of the QMD measured by the quantum computing speed, and hence is called the {\it memory latency}. An ideal quantum memory device should have a small latency.

In addition to these two metrics, we define another quantity named {\it addressability}, 
\begin{equation}
    \text{Addressability: } \gamma = \frac{1}{\alpha_{\text{QMD}}} \cdot \frac{N}{n} = \frac{T_{\text{RW}} N}{T_{\text{storage}} n},
\end{equation}
where $N$ is the memory capacity, i.e., the total number of QMCs integrated inside the quantum memory device, and $n$ is the number of QMCs that can be RW in parallel. The meaning of the addressability $\gamma$ is the fraction of the QMCs that can be addressed in the memory cycle allowed by the storage time of the quantum device. Ideally, we want a quantum memory device to have $\gamma < 1$. If $\gamma \gg 1$, the quantum memory device essentially integrates too many QMCs to be fully utilized, and the RW of the QMC is the bottleneck. 

\subsection{Comparison between RAQM and QRAM} \label{subsec:qram_compare}

\begin{table}[t]
\caption{\label{tab:qram_raqm} Comparison between the RAQM and QRAM. We highlight their difference and the key features of each device.} 
\begin{tabular}{l|p{1.8 cm}|p{4.2 cm}}
\hline
\hline
 & RAQM & QRAM \\
 \hline
Address Info & Classical & Quantum (encoded in the states of address qubits)\\
Addressing & Classical & Coherently routing bus qubits \\
Stored Info & Quantum & Classical or Quantum \\
\hline\hline
\end{tabular}
\end{table}

We focus on the two major types of QMDs:: (1) Random access quantum memory (RAQM), and (2) Quantum random access memory (QRAM). In this subsection, we briefly compare RAQM and QRAM in terms of their differences and applications. More detailed discussion and reviews on previous efforts of building these two devices can be found in the following subsections. 

A Random Access Quantum Memory (RAQM) is a quantum analog of classical random access memory, more specifically, a dynamical RAM in classical computer architecture. In a RAQM, many QMCs are integrated into a QMC array to store quantum information. The QMCs can be addressed individually according to their addresses. A RAQM only allows classical address information, which means the QMCs can only be addressed separately. Addressing QMCs can be realized by classical controls on the QMC array. Mapping it to the model of quantum memory shown in Fig.~\ref{fig:qmd}, the input address information is purely classical and is kept classical during the memory query. The structure of RAQM and its functionality will be discussed in detail in Sec.~\ref{subsec:raqm}.

On the other hand, a Quantum Random Access Memory (QRAM) distinguishes itself from classical RAM and RAQM by enabling coherent addressing of multiple QMCs. Coherent addressing requires a significant modification of the classical addressing techniques. Specifically, the quantum address information must be represented as quantum states of address qubits. In the abstract QMD models (Fig.~\ref{fig:qmd}), a QRAM can take both quantum input and quantum address information. The structure of a QRAM, especially its quantum addressing components, and its functionalities are discussed in Sec.~\ref{subsec:qram}. 

In Table.~\ref{tab:qram_raqm}, we highlight the differences between RAQMs and QRAMs. We notice that both RAQM and QRAM are analogous to the classical random access memory, however, in two distinct directions. As demonstrated in Sec.~\ref{subsec:raqm}, the RAQM stresses the quantum nature of the memory cells, where the quantum information can be stored and retrieved, while the random access feature is purely classical. Therefore, quantum error correction and mitigation techniques need to be implemented on the quantum memory cells to improve the information storage fidelity. On the other hand, demonstrated in Sec.~\ref{subsec:qram}, the QRAM focuses more on quantum routing to coherently address the information stored in the memory array. To improve the noise resilience, quantum error correction needs to be implemented on the quantum routing structure. 

Although a QRAM, in principle, has all the functionalities of a RAQM, we still believe RAQMs are an indispensable part of quantum computing architectures in the future. As we discussed in Sec.~\ref{subsec:qram}, if the address qubit is in a classical state, which corresponds to a single classical address, the quantum routing module in the QRAM can guide the bus qubit to the QMC deterministically. Therefore, using SWAP gates as the RW operations can read from and write to the QMC as a RAQM. However, the classical addressing in RAQMs is unnecessary to be protected by quantum error correction. Therefore, if classical addressing is sufficient for a quantum computing task, using RAQMs can greatly reduce the overhead of QEC on routing.

\newcolumntype{L}[1]{>{\raggedright\let\newline\\\arraybackslash\hspace{0pt}}m{#1}}

\begin{table}[t]
\caption{\label{tab:qmd_compare} Quantum memory devices with different requirements on the reading and writing process. In the memory reading process, the output and address information can be classical or quantum, while in the memory writing process, the input data and address data can be classical or quantum, too.  
When one bit of classical data needs to be stored in a QMC, the bus qubit needs to be prepared into $\ket{0}$ or $\ket{1}$ state and then perform the QMC writing process. When quantum data needs to be converted into a classical output, projective measurements on the state of the QMC qubit are required, which is labeled as `M' in the table. N.A. specifies the situations where the application is not clear. The blue-shaded case is purely classical, whereas the red-shaded cases attract lots of attention in the current quantum computing research (see main text for more detailed discussion).} 
\begin{ruledtabular} 
\begin{tabular}{m{1.44 cm}| m{1.44 cm} | m{1.43 cm} | m{1.44 cm}| m{2.3 cm} }
\multicolumn{2}{L{2.8 cm}|}{Reading requirement} & \multicolumn{2}{L{2.8 cm}|}{Writing requirement} & \multirow{2}{2.3 cm}{Physical Realization} \\
Output & Address & Input & Address &  \\
\hline
Classical & Classical & Classical & Classical & \cellcolor{blue!25} Classical Memory \\
Classical & Quantum & Classical & Classical & QRAM + M \\
Quantum & Classical & Classical & Classical & RAQM \\
Quantum & Quantum & Classical & Classical & \cellcolor{red!25} QRAM \\
\hline
Classical & Classical & Quantum & Classical & RAQM + M \\
Classical & Quantum & Quantum & Classical & QRAM + M \\
Quantum & Classical & Quantum & Classical & \cellcolor{red!25} RAQM \\
Quantum & Quantum & Quantum & Classical & \cellcolor{red!25} QRAM \\
\hline
\multicolumn{2}{c|}{Any} & Classical & Quantum & N.A. \\
\multicolumn{2}{c|}{Any} & Quantum & Quantum & N.A. \\
\end{tabular}
\end{ruledtabular} 
\end{table}

In Table.~\ref{tab:qmd_compare}, according to the demand of the QMD ports in the loading and writing processes, we briefly summarize the possible realizations of the QMD. Given a memory device only stores classical information in memory cells with classical address, while it is expected to read the memory cells according to a given classical address and output classical data, the memory device can be a classical memory (the first row of Table~\ref{tab:qmd_compare}). If the QMD needs to store quantum information according to classical address information in the writing process, while it loads the quantum information according to classical addresses, the QMD can be made by a RAQM (the row of `Quantum/Classical, Quantum/Classical' in Table~\ref{tab:qmd_compare}). Supposing a QMD is required to process quantum address information in the reading process, the QMD has to be a QRAM (see `Quantum output, quantum address' and `Classical output, quantum address' rows in Table~\ref{tab:qmd_compare}). If the output port needs to be connected with classical information processing modules after the memory reading process, the output information needs to be classical. As both RAQM and QRAM will output quantum states in general, the QMD is required to take measurements of the quantum states and extract classical information from the output state. Therefore, it can be constructed by a RAQM or QRAM followed by measurements. On the contrary, in the data writing process, coherent addressing of multiple QMCs inside the memory device is possible through the QRAM. However, the physical application is still unclear to our best knowledge~\cite{Liu2022quantum}.

In the rest of this section, we provide a more detailed discussion on RAQMs in Sec.~\ref{subsec:raqm} and QRAMs in Sec.~\ref{subsec:qram}.

\subsection{Random Access Quantum Memory (RAQM)} \label{subsec:raqm}

\begin{figure}[h]
    \centering
    \includegraphics[width=0.95 \columnwidth]{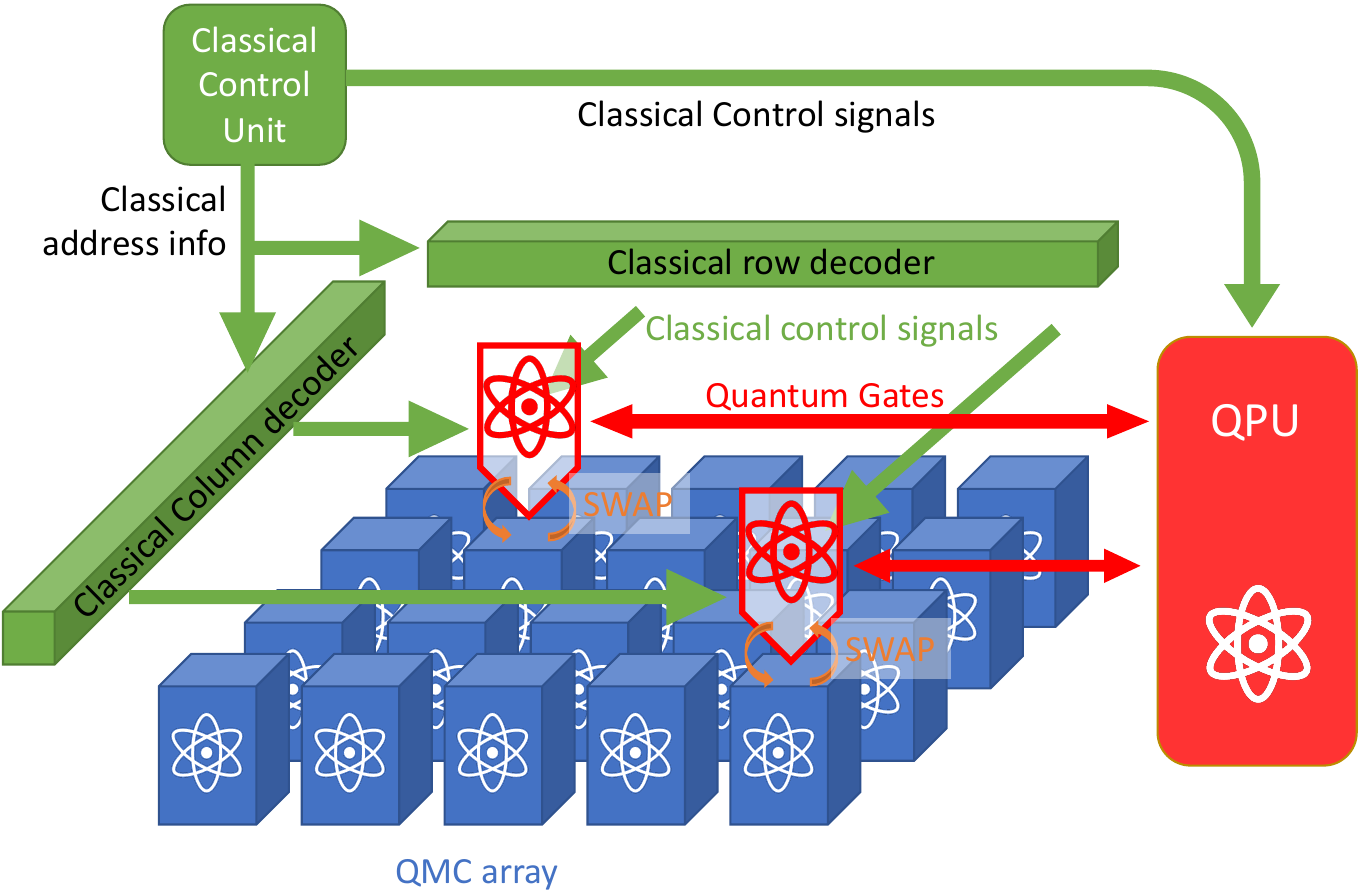}
    \caption{The structure of a RAQM. The classical addressing is achieved using the classical control and classical address decoding scheme, similar to the classical RAM construction. The memory cells are QMCs, which can store quantum information. The quantum information in the bus qubits can be addressed to the QMC with the correct address and SWAP quantum information between them. The quantum computing registers can couple with the bus qubits to perform further information processing.}
    \label{fig:RAQM}
\end{figure}

The architecture of the RAQM is illustrated in Fig.~\ref{fig:RAQM}. The classical addressing is accomplished through the classical control unit and the classical address decoding schemes. The control unit guides the bus qubit to interact with different QMCs to perform RW operations. The accessing mechanism shares similarities to a classical RAM. Other quantum registers can interact with the bus qubit to communicate with the RAQM. 

During the writing process, the bus qubit is loaded with the quantum state that needs to be stored. With the classical address given, the QMC with this address interacts with the bus qubit and performs a SWAP gate to store the quantum state into the corresponding QMC. The writing process can be formally expressed as
\begin{align}
    & W(addr, \ket{\phi})  \ket{0}_{ind=addr} \otimes \ket{\psi}_{ind \neq addr} \nonumber \\
    & \rightarrow \ket{\phi}_{ind=addr} \otimes \ket{\psi}_{ind \neq addr}
\end{align}
where $addr$ is the classical address information, $ind$ is the classical index of the QMCs in the QMC array. We assume the state of the QMC before the writing process is in state $\ket{0}$. 

During the reading process, a classical address is given. The QMC with the address qubit interacts with the bus qubit to perform a SWAP gate to swap the quantum state to the bus qubit. The reading process can be formally expressed as
\begin{align}
    & R(addr) \ket{\phi}_{ind=addr} \otimes \ket{\psi}_{ind \neq addr} \nonumber \\
    & \rightarrow \ket{\phi}_{(b)} \otimes ( \ket{0}_{ind =addr} \otimes \ket{\psi}_{ind \neq addr})
\end{align}
where we assume that the QMC with the required address $addr$ is initially in the state $\ket{\phi}$, and it is disentangled with the other qubits for simplicity. After the reading process, the bus qubit is in the state $\ket{\phi}$.

Combining both reading and writing processes, operations on RAQM can be expressed as
\begin{align}
    & F_{\text{RAQM}}(addr, \ket{\phi}^{(\text{b})}, \ket{\Psi}^{(\text{QM})}) \nonumber \\
    & = \sum_{j} \alpha_j f_\text{QMC}(\ket{\phi}^{(\text{b})}, \ket{\lambda_j}^{(\text{QM})}_{ind = addr}) \otimes \ket{\psi_j}^{\text{(QM)}}, \nonumber \\
    & = \sum_j \alpha_j \ket{\lambda_j}^{(\text{b})} \ket{\phi}^{(\text{QM})}_{ind = addr} \otimes \ket{\psi_{j}}^{(\text{QM})}, 
\end{align}
where the operation on a single QMC $f_\text{QMC}$ is shown in Eq.~\eqref{eq:qmc}.
Here, we express the quantum state of the QMC arrays in a Schmidt decomposition form
\begin{equation}
    \ket{\Psi}^{(\text{QM})} = \sum_{j} \alpha_j \ket{\lambda_j}^{(\text{QM})}_{ind = addr} \ket{\psi_j}^{(\text{QM})},
\end{equation}
where the states $\ket{\lambda_j}$ are basis states of the QMC qubit with index $ind = addr$ in the decomposition process. However, as we mentioned in Sec.~\ref{subsec:qmc_definition}, the RW process involves the exchange of entanglement between the bus qubit and the memory qubits. Generalizing to a more complicated scenario where the qubits are entangled with additional qubits is obvious.

According to the current implementation of RAQMs in photonic and microwave systems, high-quality QMCs for the memory array are critical. In addition, the following requirements also have to be fulfilled:
\begin{enumerate}
    \item Independent classical addressing of individual QMCs: The capability for independent addressing enables parallel addressing of different QMCs.
    \item Independent quantum information storage: It is necessary that the quantum state within one QMC remains decoupled from the states of other QMCs when operating separately. Ideally, the cross-talk between different QMCs should be avoided.
\end{enumerate}

The primary application of the RAQM resides in its utilization as an integrated memory device for quantum computing, a topic that will be explored further in the subsequent sections. The integration of QMCs into an array enables the storage of quantum states of a large number of qubits. Additionally, the classical random access functionality enables the storage and retrieval of quantum states in various QMCs at different times. More applications of RAQM in various quantum memory function units will be discussed in Sec.~\ref{sec:QMFU}.

\subsubsection{Experimental demonstration of RAQMs}

There have been experimental demonstrations of constructing a RAQM, ranging from storing quantum information carried by a matter qubit to a photonic pulse. A comparison between different implementations of RAQM is summarized in Table~\ref{tab:raqm_compare}. In the rest of this subsection, we briefly survey a few experimental demonstrations of RAQMs and discuss their performance.

\begin{table*}[t]
\begin{threeparttable}
\caption{\label{tab:raqm_compare} Comparison of different experimental demonstrations of RAQMs in various physical platforms. } 
\begin{tabularx}{\textwidth}{X|X|X|X|X|X|X|X|X}
    \hline \hline
   QMC array  &  Stored QI & Memory Capability ($N$) & Coherence time ($T_2$) & RW fidelity ($F$) & Efficiency ($\eta$) & $\alpha$ & $\beta$ & $\gamma$ ($n=1$)\\
   \hline
   Atomic clouds~\cite{Jiang2019, Li2020}  & photonic pulse ($\langle n \rangle \sim 0.5)$ & 210 QMCs, 105 qubits & $T_2 = 27.8~\mu$s~\cite{Jiang2019} & $F >0.9$ for $2~\mu$s storage & $\eta \sim 2\%$ to $18\%$ &  $39$ to $118$~\cite{Jiang2019} & $0.012$ to $0.037$~\cite{Jiang2019} & $1.78$ to $5.3$~\cite{Jiang2019}, $0.31$~\cite{Chang2019}\tnote{a}\\
   \hline
   Single atoms~\cite{Langenfeld2020} & photon pulse ($\langle n \rangle \sim 1)$ &  2 qubits & $T_2 = 800~\mu$s~\cite{Langenfeld2020}  & & $\eta = 26\pm3$\%\tnote{b} & $8.2$ ($2.0 \times 10^2$)\tnote{c} & $4.1$ ($0.21$)\tnote{c} & $0.20$ ($0.01$)\tnote{c}\\
   \hline
   Ensemble of Two-Level systems~\cite{OSullivan2022} & photon pulse ($\langle n \rangle \sim 100$)\tnote{d} & 4 modes demonstrated, 16 qubits in total & $T_2 = 2.0 \pm 0.2$~ms\tnote{e} & & $\eta = 0.03 \pm 0.02$ & 6.2 & 61~\tnote{f} & 2.56 \\
   \hline
   Microwave modes & Transmon states & 9 qubits & $T_2 \sim 2$ to $3$~ms\tnote{g} $T_2^* \sim 1$ to $10~\mu$s\tnote{h} & $89.0\pm2.9$\% to $96.3\pm0.7$\%$^\text{i}$\cite{Naik2018}& & $7.5$ to $415$~\tnote{h}, $2.0 \times 10^3$ to $5.9 \times 10^3$~\tnote{g} & $2.63$ to $0.508$~\tnote{h}, $12.6$ to $25.3$~\tnote{g} & $1.20$ to $2.16 \times 10^{-2}$~\tnote{h}, $0.17$ to $0.5$~\tnote{g}\\
   \hline\hline
\end{tabularx}

\begin{tablenotes}
    \item[a] This is estimated using QMC coherence and RW time parameters given in Ref.~\cite{Jiang2019} bur with only 49 QMCs integrated into the device as demonstrated in Ref.~\cite{Chang2019}, 
    \item[b] This is for the setup to reduce the cross illustration.
    \item[c] This is estimated using possible improved RW time $2~\mu$s as pointed out in Ref.~\cite{Langenfeld2020}.
    \item[d] The photon pulses are microwave pulses.
    \item[e] This is measured without dynamical decoupling sequences.
    \item[f] Connecting this RAQM with superconducting-qubit-based quantum computing device.
    \item[g] 3D microwave cavity, see Ref.~\cite{Chakram2021}.
    \item[h] 2D resonator modes, see Ref.~\cite{Naik2018}.
    \item[i] iSWAP gate fidelity from randomized benchmarking, which contains the error from the resonator modes and the transmon.
\end{tablenotes} 
\end{threeparttable}
\end{table*}

In 2017, Naik \textit{et al.} experimentally demonstrated a circuit-QED system, which can be used as a small RAQM~\cite{Naik2018}. The QMCs are made by 11 strongly coupled resonators. The modes in individual resonators are coupled to form collective modes, where 9 of them are used as the QMCs in the RAQM. The bus qubit is a transmon qubit, which can be classically controlled through microwave parametric driving to perform iSWAP gates between the transmon mode and the selective resonator mode. The iSWAP gate fidelity ranges from $95\%$ to $98.6\%$. The address is encoded in the collective mode frequencies. When a specific mode with frequency $\omega_j$ needs to be addressed, a flux modulation with frequency $\vert \omega_j - \omega_t \vert$ is applied, where $\omega_t$ is the transmon frequency. The flux modulation activates a sideband transition to implement the iSWAP gate. 

The coherent time of the cavity modes ranges from 1 to 10 $\mu$s, while the RW via the transmon-resonators mode iSWAP gate lasts $20$ to $100$~ns. Based on the device parameters reported in Ref.~\cite{Naik2018}, limited by the relatively short coherence time, the internal storage ratio $\alpha$ can vary from $7.5$ (estimated from $T_\text{RW} = 100$~ns with fidelity $F \approx 0.95$, $T_\text{storage} \approx 1~\mu$s) to $\approx 415$ ($T_\text{RW} = 20$~ns with fidelity $F \approx 0.986$, $T_\text{storage} \approx 8.46~\mu$s). The memory latency $\beta \approx 2.63$ to $0.508$, where we assume the quantum operation is a two-qubit gate time on transmon qubits, which takes $40$~ns~\cite{Marxer2023, Stehlik2021, Sung2021}. The large memory latency is due to the fast gate operations on the quantum processor relative to the RW operations. The addressability of the QM is $\gamma \approx 2.16 \times 10^{-2}$ to $1.20$, which means more QMCs can be integrated into the RAQM before the RW process becomes the bottleneck. Here, we consider that the device demonstrated cannot address different QMCs in parallel, so we take $n=1$. Note that in the worst-case scenario, the addressability is greater than unity, which means the RAQM needs to reduce the RW time further to fully appreciate the QMCs. On the other hand, introducing more RW ports while maintaining similar RW fidelity and time, which increases the parallel RW number $n$, can also reduce the addressability, e.g., to $\approx 0.6$ with $n=2$. 

Following this work, Chakram \textit{et al.} used the flute method to fabricate 3D microwave cavities, which greatly increased the coherence time of the cavity modes to $2$~ms~\cite{Chakram2021}. Similarly, the address information is also encoded into the mode frequencies. However, the transmon-cavity mode iSWAP gate is activated using an applied microwave tone with the right difference frequencies. Although the transmon-cavity mode SWAP gate time extends 
to $0.5$ - $1~\mu$s, the number of RW operations increases significantly~\cite{Chakram2021}, which can be seen from a significant increase of the internal storage time $\alpha$ to $2.0 \times 10^3$ to $5.9 \times 10^{3}$ ($T_\text{RW} = 0.5$ to $1.0~\mu$s with fidelity $F \approx 0.99$, $T_\text{storage} \approx 2$~ms and $3$~ms). However, because the RW operation becomes slower, the memory latency $\beta \approx 12.6$ to $25.3$. Due to the increase of internal storage time $\alpha$, integrating $9$ cavity modes as QMCs and addressing QMCs sequentially is still acceptable, with $\gamma \approx 1.5$ to $4.5 \times 10^{-3}$. Even with 1000 modes integrated, the addressability $\gamma \approx 0.17$ to $0.5$, which means the RAQM design is still in a good regime ($\gamma < 1$).

In addition to the circuit-QED system, RAQM is also experimentally realized in atomic cloud systems to store the quantum information carried by photonic qubits. Jiang \textit{et al.} demonstrated using Rb atom clouds as QMCs to store optical photons~\cite{Jiang2019}. They demonstrated the capability of storing $105$ dual-rail encoded photonic qubits in $210$ memory cells. The storage is achieved through electromagnetically induced transparency, while the random access feature is achieved by deflecting the control pulses to the cloud ensemble with the right address. The beam deflecting is obtained by acoustic-optical deflectors (AODs) using microwave tones.

In Ref.~\cite{Jiang2019}, the QMC coherent time is about $27.8~\mu$s. The read and write efficiency is lower than $20$\% for all memory cells (2\% to 18\% reported). Although the RW control pulse extends to $500$~ns, the memory retrieval photon pulse is emitted almost when the control pulse is applied. We take the stored photon pulse duration to estimate the RW speed of a single QMC $T_\text{RW} \approx 100$~ns. Note that the RW time of a quantum memory device should include the time cost of setting the classical address and sending the control pulse to address the QMC. Since the AOD setting time is not reported in Refs.~\cite{Jiang2019, Chang2019}, we ignore its contribution to the RW time. However, the AOD switching time is reported as $40~\mu$s in Ref.~\cite{Langenfeld2020}. If the addressing time $T_\text{addr} \approx 40~\mu$s, it will obviously dominate the RW time of QMCs, and even dominate the storage time of QMCs.  

If the addressing time is negligible, the RW time of the device is $T_{\text{RW}} \approx 100$~ns, the device internal storage time can reach $\alpha \approx 39$ (RW efficiency $\eta \approx \sqrt{0.02}$) to $118$ ($\eta \approx \sqrt{0.18}$). Using the RAQM for quantum communication, as we discussed in Sec.~\ref{subsec:qmc_compare}, the key operation is the EPR pair generation, which is estimated to $T_{\text{op}} \approx 19~\mu$s. The memory latency of the RAQM is relatively small $\beta \approx 0.012$ to $0.037$. If all QMCs inside the RAQM are addressed individually, the addressability $\gamma \approx 1.78$ to $5.3$, where we consider using 210 cells as individual QMCs. In order to efficiently use all the QMCs, addressing 2 to 6 QMCs in parallel via multiplexing is needed. This experimental setup is improved in Ref.~\cite{Chang2019}, where the control pulse is further reduced to improve RW time, and a smaller number of QMCs are integrated (49 QMCs), which makes the addressability to be $0.31$ when addressing QMCs sequentially. In addition, in Ref~\cite{Li2020}, the optical communication between two memories has been demonstrated.

On the other hand, Langenfeld \textit{et al.} demonstrates using single Rb atoms as QMCs in the RAQM~\cite{Langenfeld2020}. Although only two atoms (QMCs) are shown in the RAQM device, the combined read and write operation reaches $26$\%. Similar to the atomic cloud-based RAQMs, the random access feature is realized by guiding the control pulses to the correct atom using AODs, where the addressing time is $T_\text{addr} = 40~\mu$s. The coherence of the QMCs can reach $800~\mu$s. Although the addressing time is long, thanks to the relatively long coherence time, the internal storage time is still decent, $\alpha \approx 8.2$. However, using the RAQM to interact with a fast EPR generator makes the addressing latency not negligible, where the memory latency $\beta \approx 4.1$. The addressability of this RAQM $\gamma \approx 0.20$, which is reasonable with only two QMCs. However, if more QMCs need to be integrated into this RAQM, the slow addressing time will soon become a bottleneck. This scenario requires either suppressing the total RW time or introducing parallel addressing techniques into the RAQM RW processes. As pointed out by the authors, the addressing time is possible to be reduced to $2~\mu$s by using electro-optical deflectors. The suppression of the addressing time to $2~\mu$s can improve the RAQM performance greatly by $\alpha \approx 2.0 \times 10^2$, memory latency $\beta \approx 0.21$, and addressability $\gamma \approx 0.010$.

In addition, O'Sullivan \textit{et al.} demonstrates an echo-based scheme to store quantum information carried by photons into an ensemble of two-level atoms~\cite{OSullivan2022}. The quantum information can be encoded using chirped pulses to imprint a phase pattern to write to the RAQM. The same chirped pulse can be used to unwind the phase to read out the information. Multiple chirped pulses are used to realize the random access feature. The classical address of the memory is `labeled' by the chirped pulse. To RW of the correct QMC inside the RAQM, the corresponding chirped pulse needs to be generated and sent into the quantum memory media. The RW time of each QMC is mainly determined by the chirped pulse duration ($100~\mu$s), where the addressing time is determined by the speed of setting the chirped pulse parameters. The read or write process efficiency is $17\%$. As the addressing time is not explicitly reported in Ref.~\cite{OSullivan2022}, and it is negligible in the time sequence, we ignore the addressing time in evaluating the RAQM performance. Unlike the other RAQM experiments reported in this section, where the capacity of the RAQM is not determined by the number of physical systems that make individual QMCs, the quantum information is stored in the collective excitations of the single optical media in this experiment. Therefore, the capacity of the RAQM is determined by the number of distinct chirped pulses to access these distinct collective excitations. In the experiments, 16 modes are accessed, which give $16$ QMCs inside the RAQM. The coherent time of the QMCs is extended to $2$~ms using dynamical decoupling~\cite{OSullivan2022}. 

In terms of the performance of this RAQM, the RW time is relatively short compared to the coherence time of the quantum memory, which is shown by its internal storage time $\alpha \approx 6.2$. With the $16$ QMCs, the addressability is $\gamma \approx 2.56$, which means to address all the QMCs, multiplexing to address $n \approx 2$ QMCs in parallel is needed. However, as the bus qubit of the RAQM is microwave pulses, integrating this RAQM with either quantum communication protocols or superconducting quantum computing processors is possible. In terms of connecting this RAQM with superconducting quantum computing processors, the superconducting qubits need to interact with the microwave bus qubit, which takes another $1~\mu$s for this operation and it is part of the RW time. However, the fast gate operations between superconducting qubits make the RAQM latency a disadvantage, with $\beta \approx 61$. On the other hand, using the RAQM for quantum communication protocols requires a detailed discussion of microwave quantum communication protocols to evaluate the performance of the RAQM, which is beyond the scope of this paper. To give a rough estimate of its performance, we point out that microwave photon generation processes from superconducting qubits commonly take less than $1~\mu$s~\cite{Besse2020, Reuer2022}. Comparing it with the re-scaled RW time of the RAQM $T_\text{RW} \approx 0.24$~ms, further improving the memory latency seems necessary. On the other hand, if microwave-to-optical transduction is involved in order to convert the microwave photon to the optical domain for long-range quantum communication, due to the limited transduction efficiency, the effective microwave EPR pair generation speed is in the order of $1$~Hz to $1$~kHz levels~\cite{Delaney2022, Sahu2022, xu2021bidirectional, ang2022architectures}, which enables the latency $\beta < 1$. However, the overall performance of the RAQM and the transduction device need to be further improved to meet other requirements of efficient quantum computing and quantum information processing~\cite{ang2022architectures}.
 
\subsection{Quantum Random Access Memory (QRAM)} \label{subsec:qram}

\begin{figure*}[tb]
    \centering
    \subfloat[]{
    \includegraphics[width = 1. \columnwidth]{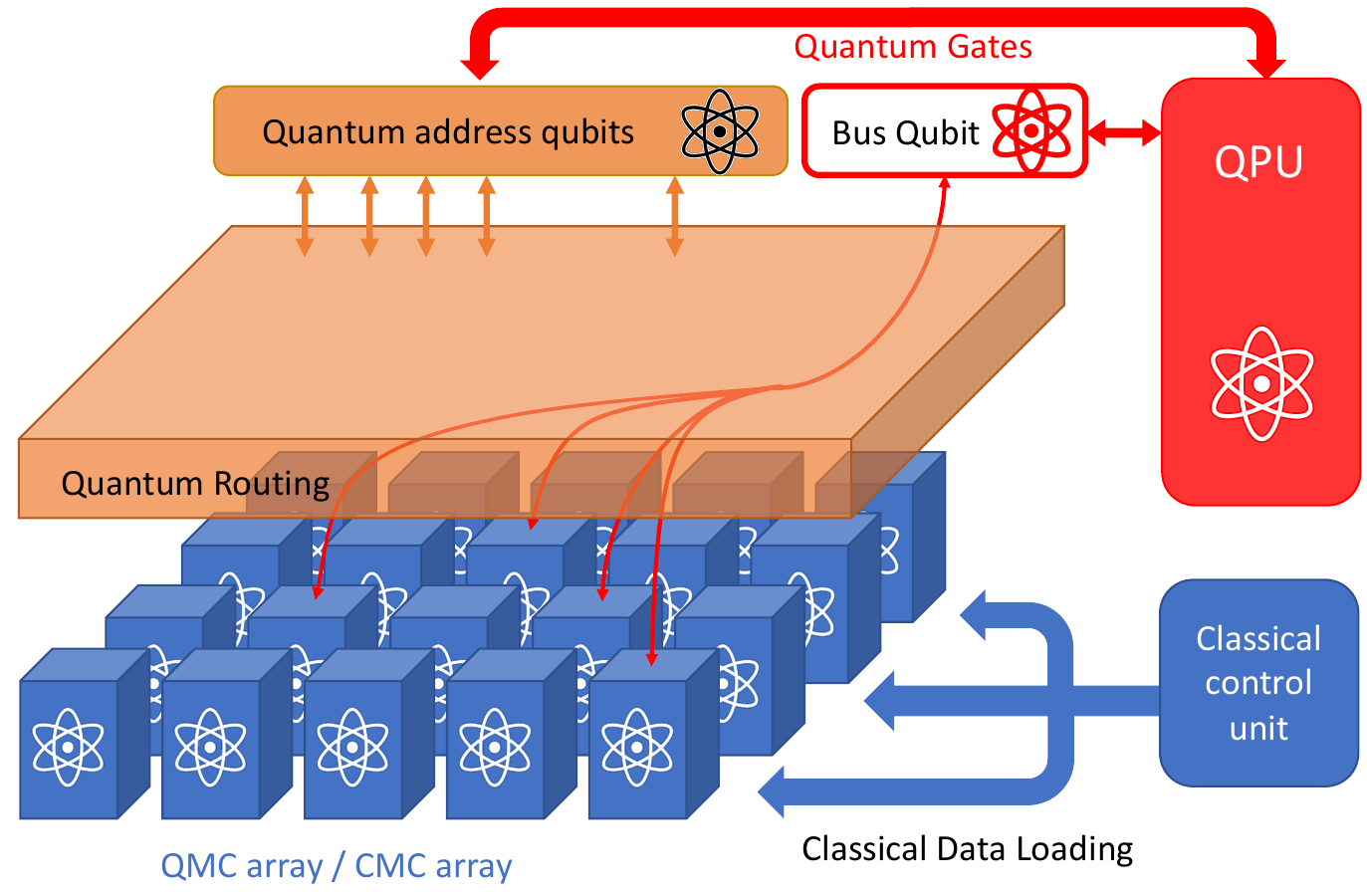}
    } \quad
    \subfloat[]{
    \includegraphics[width = 0.75 \columnwidth]{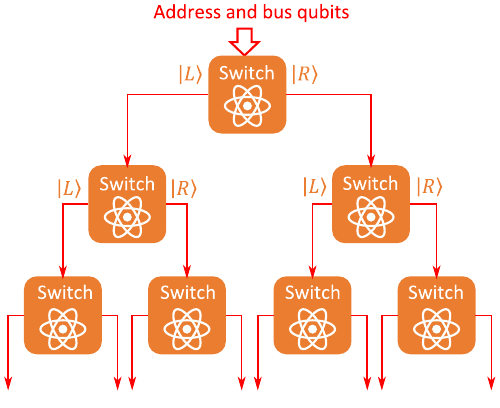}
    }
    \caption{The architecture of a QRAM is shown in (a). The key element of a QRAM is its quantum routing structure. A sketch of the quantum routing structure (up to three levels) is shown in (b). The address qubits and the bus qubits are sent through the routing systems from the top. The address qubits set the states of quantum switches. When the quantum switch is set to state $\ket{L}$ ($\ket{R}$), the next incoming qubit is routed to the left (right) child switch. }
    \label{fig:qram}
\end{figure*}

Quantum random access memory (QRAM) distinguishes itself from classical RAM and RAQM by achieving coherent addressing of the QMCs. In Fig.~\ref{fig:qram}a, we sketch the architecture of a QRAM. The key ingredient of the QRAM is the quantum routing module (see Fig.~\ref{fig:qram}b as an example). Based on the state of the address qubits, the routes of the bus qubits are coherent superposes after interacting with the quantum routing module, making the bus qubits coherently visit the corresponding QMCs and be returned as the output.

QRAMs were first proposed and designed by Giovannetti, Lloyd, and Maccone~\cite{Giovannetti2008, Giovannetti2008PRA}. Their design uses quantum routers in the quantum routing module, whose state can be set by the address qubits. After the router state is set, it coherently routes the next incoming qubit to the two different paths (see Fig.~\ref{fig:qram}b). Sending all the address qubits into the routing module carves paths that guide the bus qubits to the corresponding QMCs, which will read out the information coherently. The bus qubit is then sent out from the QRAM, and the routers are unset to return to their starting state for the next memory call. The authors named this seminal design `bucket-brigade' (BB) architecture. Unlike the direct analogy from the classical addressing structure, `fan-out' architecture, they pointed out that the BB architecture is more efficient and noise-resilient. Following this seminal work, Hong \textit{et al.} proposed an alternation of the bucket-brigade architecture by modifying the quantum switch qubits~\cite{Hong2012}.

As demonstrated in Ref.~\cite{Harrow2009}, where Harrow \textit{et al.} proposed the famous quantum linear algebra algorithms (HHL algorithm, named by the authors) to efficiently solve the inverse of a large matrix, the construction of QRAM and coherently addressing the memory content efficiently becomes indispensable for the speedup. Therefore, there is growing interest in building QRAM on different physical platforms. Hann \textit{et al.} performed a realistic analysis on how to implement QRAM based on a superconducting system and acoustic quantum memory~\cite{Hann2019}, while Chen \textit{et al.} consider using solid-state systems and photons instead~\cite{Chen2021}. Recently, Weiss {\it et al.} proposed a QRAM design using the superconducting microwave system~\cite{weiss2023qram}. In addition, based on QRAM development, the concept and the architecture of a quantum data center have been proposed in Ref.~\cite{Liu2022quantum}. Furthermore, QRAMs based on quantum random walk have also been demonstrated~\cite{Asaka2021, Asaka2023I, Asaka2023II}. In addition to the bucket-brigade architecture, there are proposals to construct QRAM with other architectures, e.g., the flip-flop QRAM~\cite{Park2019}, hybrid QRAM~\cite{Matteo2020, Alexandru2020}, etc. However, constructing a QRAM and demonstrating its performance have not been achieved in experiments yet, to the best of our knowledge.

On the other hand, the efficiency of QRAMs becomes another interesting question. In Ref.~\cite{Arunachalam2015}, Arunachalam \textit{et al.} consider the noise resilient of the coherent addressing in a bucket-brigade QRAM. They claimed that the bucket-brigade architecture is not as noise-resilient as claimed in Refs.~\cite{Giovannetti2008, Giovannetti2008PRA}, and the error scales exponentially as the number of address qubits. Contrary to this work, Hann \textit{et al.} give further analysis to the noise resilience of coherently addressing different QRAMs. They pointed out that the errors occurring on quantum switches controlled by an inactive quantum switch do not degrade the final returned states from the QRAM. Therefore, the QRAM can still be noise resilient~\cite{Hann2021}. In addition, how to improve the efficiency of the QRAM operations in various scenarios is also discussed in Refs.~\cite{chen2023efficient, Matteo2020, Alexandru2020}. The architecture for large integration of QRAM routing qubits and the memory qubits, and the possible realization using H-tree architecture are also discussed in Ref.~\cite{xu2023systems}. For recent reviews on the topic of QRAM, we refer to Refs.~\cite{HannThesis, phalak2023quantum, jaques2023qram}.

Nowadays, one of the main motivations for developing a QRAM is to realize a quantum oracle for coherently accessing the memory data,
\begin{align}
    \hat{O}_{\mathbf{x}} \sum_{j} c_j \ket{j}^{(\text{addr})} \ket{0}^{(\text{b})} = \sum_{j} c_j \ket{j}^{(\text{addr})} \ket{x_j}^{(\text{b})},
    \label{eq:oracle}
\end{align}
where $\mathbf{x}$ represents some data that needs to be accessed, $\hat{O}$ is the oracle operation. After the oracle call, the bus qubit state contains the information of $\mathbf{x}$, and it is entangled with the address qubits. Compared to a classical oracle that only allows accessing each data $\mathbf{x}$ according to the classical address, with this quantum oracle, several algorithms can be efficiently implemented with fewer oracle calls and give quantum speedup. For example, the Grover search algorithm~\cite{Grover1996}, quantum Fourier transform algorithm~\cite{nielsen_chuang_2010}, HHL algorithm for linear algebra~\cite{Haffner2008}, etc. 

This quantum oracle can be achieved by QRAM. Classical data $\mathbf{x}$ are preloaded into the QMC array according to its classical address, where the QMCs can be described by the state $\ket{\Psi^{(\text{QM})}} = \otimes_{k} \ket{x_k}$. We then prepare the address state, $\ket{\psi^{(\text{addr})}} = \sum_j c_j \ket{j}$, and initialize the bus qubit in the state $\ket{0^{(\text{b})}}$. The bus qubit and the address qubits are sent to the quantum routing module in the QRAM, which allows the bus qubit to be coherently coupled with the QMCs with address $addr = j$. To generate the output state of the quantum oracle in Eq.~\eqref{eq:oracle}, a sequence of control-NOT gates are applied to the QMC qubits and the bus qubit, which leaves the state of QMCs unchanged and not entangled with the bus qubit or address qubits. Lastly, the bus qubit and address qubits are returned by the quantum routing module.

We should stress that the above process is only one way to operate QRAM, where the information stored in each QMC is purely classical. The process disentangles the QMC qubits with the bus qubits after the CNOT gates. Furthermore, to implement the quantum oracle in Eq.~\eqref{eq:oracle}, only the reading process of the QRAM is necessary. In addition, the data reading process does not rely on SWAP gates between the bus qubit and the QMC qubit. Next, we aim to survey how QRAM can be operated and what the outcome would be. 

When operating a QRAM, a series of address qubits with the bus qubits are input into the QRAM, while after the QRAM operation, these qubits are returned. When the input address qubits are in a classical state, i.e., the state is in the computational basis, the quantum routing module will guide the bus qubit to a single QMC with the corresponding address. So, when we need to perform RW operation to the QMCs, regardless of whether the QMC state is classical or quantum, the RW operation can be achieved by a SWAP gate operation. With classical address information, RAQM can achieve the same functionality with less overhead on the quantum routing module design and operations. Therefore, if the address information is purely classical, using QRAM does not seem to be necessary.

When the address information is quantum, we have briefly discussed how to read classical data out using classical `copy' operation, which can be implemented using CNOT gates between the bus qubit and QMC qubits (or classical-controlled Pauli-X gates on the bus qubit~\cite{Hann2019, HannThesis}). However, a few questions remain unclear, e.g., what will happen when the QRAM is in the writing mode, what are the QRAM outputs when the data is quantum, etc. In Refs.~\cite{Hann2019, HannThesis} and Ref.~\cite{Liu2022quantum}, the reading and writing process of QRAMs are briefly discussed. For the completeness of our discussion and to help answer these questions, we briefly go over the reading and writing process of QRAMs, and discuss the state of the address qubits, the bus qubits, and the QMC qubits. We further assume that a single bus qubit is enough (word length is one qubit in each QMC), which can be easily generalized to cases with more bus qubits.

\begin{widetext}

\subsubsection{QRAM reading classical data} \label{subsub:qram_cr}

When the QMCs store classical information, i.e., the state of each QMC is either $\ket{0}$ or $\ket{1}$, the data stored in the QMCs can be viewed as a binary vector $\mathbf{x}$, where the $j$-th element is $x_j$, which is stored in the state of the QMC qubit with address $j$, labeled by $\ket{x_j^{(j)}}$. The address qubits are in the state $\ket{\phi^{(\text{addr}}} = \sum_j c_j \ket{j}$, where $c_j$ is the complex coefficient when the state is written in the computational basis. The bus qubit is initialized to $\ket{0^{(\text{b})}}$ state. The initial state of the bus qubit, address qubits, and the QMCs is
\begin{align}
    & \ket{\phi^{(\text{addr})}} \ket{0^{(\text{b})}} \left( \bigotimes_k \ket{x_k^{(k)}} \right)
    = \sum_j c_j \ket{j} \ket{0^{(\text{b})}} \left( \bigotimes_k \ket{x_k^{(k)}} \right)
    \label{eq:read_classical_QRAM}
\end{align}

After sending the address qubits and the bus qubit into the QRAM, according to the BB architecture, coherent paths that guide the bus qubits to the corresponding QMCs with address $j$ are activated. Mathematically, in each term of Eq.~\eqref{eq:read_classical_QRAM}, the bus qubit can interact with the QMC qubits with $j$ address. Because $\text{CNOT} \ket{x}\ket{0} = \ket{x} \ket{0 \oplus x} = \ket{x} \ket{x}$, where $x = 0, 1$, to copy the classical data out to the bus qubit, a CNOT gate with the QMC qubit as the control can be applied. After the CNOT gate, the state is transformed to
\begin{align}
    & \sum_j c_j \ket{j} \ket{0^{(\text{b})}} \left( \bigotimes_k \ket{x_k^{(k)}} \right)
    \rightarrow  \sum_j c_j \ket{j} \ket{x_j^{(\text{b})}} \left( \bigotimes_k \ket{x_k^{(k)}}  \right),
\end{align}
where QMCs are still disentangled from the bus qubit and the address qubits, while the bus and the address qubits are entangled. And the quantum oracle in Eq.~\eqref{eq:oracle} has been realized.

To be more consistent with the quantum memory we have discussed so far, we consider using a SWAP gate between the connected QMCs and the bus qubit. In this case, the QMC qubit with address $j$ is swapped with the bus qubit state, which gives
\begin{align}
    \sum_j c_j \ket{j} \ket{x_j^{(\text{b})}} \left( \ket{0^{(j)}} \bigotimes_{k \neq j} \ket{x_k^{(k)}}  \right)
    = \left[ \sum_j c_j \ket{j} \ket{x_j^{(\text{b})}} \left( \ket{0^{(j)}} \bigotimes_{\mathclap{\substack{k \in \{ addr \} \\ k \neq j}}} \ket{x_k^{(k)}}  \right) \right]
    \otimes \left( \bigotimes_{k \notin \{ addr \} } \ket{x_k^{(k)}}\right), 
    \label{eq:read_classical_qram_swap}
\end{align}
where $\ket{0^{(j)}}$ shows the QMC with the address $j$ is in the state $\ket{0}$,  $\{ addr \}$ is the set of addresses that are contained in the state of address qubits $\ket{\phi^{(\text{addr})}}$. From Eq.~\eqref{eq:read_classical_qram_swap}, the output state of the address qubits and the bus qubit is entangled with the QMC qubits that have been addressed. 

\subsubsection{QRAM reading quantum data} \label{subsub:qram_qr}

When the stored information is quantum, the state of the QMC is no longer along the computational basis. Without losing generality, we consider that the state of the QMCs can be described by $\ket{\Psi^{(\text{QM})}} = \bigotimes_{j} \ket{\psi_j^{(j)}}$, where $\ket{\psi_j^{(j)}} = m_{j,0} \ket{0^{(j)}} + m_{j,1} \ket{0^{(j)}}$, which means the QMCs do not have entanglement before the reading process. If a CNOT gate between the QMC and the bus qubit, which is in state $\ket{0^{(\text{b})}}$, is applied following the classical data readout, the resulting state is
\begin{align}
    & \text{CNOT} \ket{\psi_j^{(j)}} \ket{0^{(\text{b})}} = m_{j,0} \ket{0^{(j)} 0^{(\text{b})}} 
    +  m_{j,1} \ket{1^{(j)} 1^{(\text{b})}}, \nonumber
\end{align}
where we can no longer get the state $\ket{\psi_j^{(j)}, \psi_j^{(\text{b})}}$. Instead, the QMC qubit is entangled with the bus qubit. In fact, if the state of each individual QMCs is unknown, due to the no-cloning theorem, it is impossible to construct an operation to `copy' the QMC state to the bus qubit. 

If the QRAM is set to use the CNOT gates in its reading process, the resulting state of memory, address, and bus qubits are
\begin{align}
    & \sum_j c_j \ket{j} 
    \left(m_{j,0} \ket{0^{(j)} 0^{(\text{b})}} 
    +  m_{j,1} \ket{1^{(j)} 1^{(\text{b})}}\right)
    \left( \bigotimes_{k \neq j} \ket{\psi_k^{(k)}}  \right) \nonumber \\
    & = \left[ \sum_j c_j \ket{j} 
    \left(m_{j,0} \ket{0^{(j)} 0^{(\text{b})}} 
    +  m_{j,1} \ket{1^{(j)} 1^{(\text{b})}}\right)
    \left( \ \ \bigotimes_{\mathclap{\substack{k \in \{ addr \} \\ k \neq j}}} \ket{\psi_k^{(k)}}  \right) \right]
    \otimes \left( \bigotimes_{k \notin \{ addr \} } \ket{\psi_k^{(k)}}\right).
    \label{eq:read_quantum_qram}
\end{align}
Unlike the classical data readout, the address qubits, the bus qubit, and the addressed QMC qubits are entangled. 

If a series of SWAP gates are applied to the QMCs and the bus qubit, rather than using CNOT gates in the reading process, the state is
\begin{align}
    \sum_j c_j \ket{j} \ket{\psi_j^{(\text{b})}} \left( \ket{0^{(j)}} \bigotimes_{k \neq j} \ket{\psi_k^{(k)}}  \right)
    = \left[ \sum_j c_j \ket{j} \ket{\psi_j^{(\text{b})}} \left( \ket{0^{(j)}} \bigotimes_{\mathclap{\substack{k \in \{ addr \} \\ k \neq j}}} \ket{\psi_k^{(k)}}  \right) \right]
    \otimes \left( \bigotimes_{k \notin \{ addr \} } \ket{\psi_k^{(k)}}\right),
    \label{eq:read_quantum_qram_swap} 
\end{align}
where again, the reading process leaves the bus and the address qubits entangled with the QMC qubits. If the state stored in the QMCs is entangled, according to the entanglement pre-exists in the QMCs, more qubits in QMCs can be entangled with the bus and address qubits regardless of quantum gates used in the reading process. 

\subsubsection{Writing classical data into QRAMs}

In the memory writing process, the bus qubit is prepared in an unknown quantum state and then sent into the memory module with the address information. The memory module will save the information inside the memory media according to the address information. However, QRAMs can take coherent address information, and the bus qubit can entangle with the address qubits. Therefore, we assume the bus and the address qubits are initialized into the state $\sum_j c_j \ket{j} \ket{x_j^{(\text{b})}}$, where $x_j \in \{0, 1\}$ and then sent into the QRAM for the writing process.

Providing the QRAM memory is initialized to state $\bigotimes_k \ket{0^{(k)}}$, because the information contained in the bus qubit is classical, we can use a CNOT gate controlled by the bus qubit to copy to the QMC qubit. However, with coherent address information, the resulting state becomes,
\begin{align}
    \sum_j c_j \ket{j} \ket{x_j^{(\text{b})}} \ket{x_j^{(j)}} \left( \bigotimes_{k \neq j} \ket{0^{(k)}}  \right)
    = \left[ \sum_j c_j \ket{j} \ket{x_j^{(\text{b})}} \left( \ket{x_j^{(j)}} \bigotimes_{\mathclap{\substack{k \in \{ addr \} \\ k \neq j}}} \ket{0^{(k)}}  \right) \right]
    \otimes \left( \bigotimes_{k \notin \{ addr \} } \ket{0^{(k)}}\right). 
    \label{eq:write_classical_qram}
\end{align}
To understand this process, let us consider the bus qubit is disentangled with the address qubits at the beginning, i.e., $\ket{x_j^{(\text{b})}} = \ket{x^{(\text{b})}}$ is independent of address $j$. The state in Eq.~\eqref{eq:write_classical_qram} becomes
\begin{align}
    \ket{x^{(\text{b})}} \otimes \sum_j c_j \ket{j} \otimes \ket{x^{(j)}} \bigotimes_{k\neq j} \ket{0^{(k)}}, 
    \label{eq:write_classical_qram_simple}
\end{align}
which means the bus qubit is still disentangled from the rest of the system, and the bus qubit state is coherently saved to the QMCs with the addresses $j \in \{addr\}$. However, this is different from keeping multiple copies of $\ket{x}$ in QMCs with addresses $j \in \{ addr \}$, which results in
\begin{align}
    \ket{x^{(\text{b})}} \sum_j c_j \ket{j} \left( \bigotimes_{k \in \{addr\}} \ket{x^{(k)}}\right) \left( \bigotimes_{k \notin \{addr\}} \ket{0^{(k)}} \right), \nonumber
\end{align}
where the QMC qubits are disentangled from the rest of the system. This state is different from Eq.~\eqref{eq:write_classical_qram_simple}. Similarly, in the general case shown in Eq.~\eqref{eq:write_classical_qram}, the classical data is saved to the corresponding QMC coherently, which leaves all the qubits involved in this process entangled. This process is different from writing classical data one by one into the corresponding QMCs deterministically.

If the write operation is a SWAP gate, similar to using SWAP gates to read classical data out of a QRAM (see Sec.~\ref{subsub:qram_cr}), the writing process leaves the involved QMC qubits entangled with the address qubits. The outcome state is
\begin{align}
    \ket{0^{(\text{b})}} \sum_j c_j \ket{j}  \left( \ket{x_j^{(j)}} \bigotimes_{k \neq j} \ket{0^{(k)}} \right)
    = \ket{0^{(\text{b})}} \left[ \sum_j c_j \ket{j} \left( \ket{x_j^{(j)}} \bigotimes_{\mathclap{\substack{k \in \{ addr \} \\ k \neq j}}} \ket{0^{(k)}}  \right) \right]
    \otimes \left( \bigotimes_{k \notin \{ addr \} } \ket{0^{(k)}}\right). 
    \label{eq:write_classical_qram_swap}
\end{align}

\subsubsection{Writing quantum data into QRAMs}

When the bus qubit is prepared in a quantum state and entangled with the address qubits, i.e., the state of the bus and the address qubits are
\begin{align}
    \sum_j c_j \ket{j} \ket{x_j^{(\text{b})}} = \sum_j c_j \ket{j} \left( b_{j,0}\ket{0^{(\text{b})}} + b_{j,1} \ket{1^{(\text{b})}} \right), \nonumber
\end{align}
where the coefficients $b_{j,0}$ and $b_{j,1}$ are nonzero. In this case, the writing result is similar to the reading process, shown in Sec.~\ref{subsub:qram_qr}. If the writing operation is a CNOT gate controlled by the bus qubit, the outcome state is
\begin{align}
    & \sum_j c_j \ket{j} 
    \left(b_{j,0} \ket{0^{(\text{b})} 0^{(j)}} 
    +  b_{j,1} \ket{1^{(\text{b})} 1^{(j)}}\right)
    \left( \bigotimes_{k \neq j} \ket{0^{(k)}}  \right) \nonumber \\
    & = \left[ \sum_j c_j \ket{j} 
    \left(b_{j,0} \ket{0^{(\text{b})} 0^{(j)}} 
    +  b_{j,1} \ket{1^{(\text{b})} 1^{(j)}}\right)
    \left( \ \ \bigotimes_{\mathclap{\substack{k \in \{ addr \} \\ k \neq j}}} \ket{0^{(k)}}  \right) \right]
    \otimes \left( \bigotimes_{k \notin \{ addr \} } \ket{0^{(k)}}\right),
    \label{eq:write_quantum_qram}
\end{align}
where all the qubits involved in the process are entangled. While the writing process is performed using SWAP gates, the state is
\begin{align}
    \ket{0^{(\text{b})}} \sum_j c_j \ket{j} \left( \ket{\psi_j^{(j)}} \bigotimes_{k \neq j} \ket{0^{(k)}}  \right)
    = \ket{0^{(\text{b})}} \left[ \sum_j c_j \ket{j}  \left( \ket{0^{(j)}} \bigotimes_{\mathclap{\substack{k \in \{ addr \} \\ k \neq j}}} \ket{0^{(k)}}  \right) \right]
    \otimes \left( \bigotimes_{k \notin \{ addr \} } \ket{0^{(k)}}\right),
    \label{eq:write_quantum_qram_swap} 
\end{align}
\end{widetext}
where the bus qubit is disentangled while the other qubits involved are entangled together. 

\subsubsection{Comparison between different QRAM operation modes}

\begin{table}[t]
\caption{\label{tab:qram_mode} Comparison of different operation modes of QRAM. We consider the entanglement inside the output state. Specifically, we layout the entangled qubits, `addr' is for address qubits, `b' is for the bus qubit, `QMC' is for the QMC qubits in the memory, and `all' means all the qubits involved are entangled.} 
\begin{tabular}{p{1 cm}|p{1.8 cm}|p{1.8 cm}|p{1.8 cm}}
\hline
\hline
     & &  \multicolumn{2}{c}{RW operation} \\
\hline
     & & CNOT & SWAP \\
\hline
\multirow{2}{*}{Read} & Classical & addr, b & all \\
 & Quantum & all & all \\
\hline
\multirow{2}{*}{Write} & Classical & all & addr, QMC\\
 & Quantum & all & addr, QMC \\
\hline
\hline
\end{tabular}
\end{table}

In Table~\ref{tab:qram_mode}, we compare the four operation modes of a QRAM, where we specifically show the entanglement feature of the outcome state. We noticed that because of the coherent addressing feature, the address and the bus qubit are entangled with the quantum memory after the reading and writing queries, except in the case of reading the classical data using CNOT gates (or any other methods to copy the classical data to the bus qubit). The feature of generating entanglement between the address qubits and QMC qubits inside the memory is unique to QRAM operations, which is caused by the coherently addressing of the bus qubits to the quantum memory array. This feature can be useful to generate large-scale quantum entanglement. However, in the current stage of the quantum memory research, to our best knowledge, there is no specific usage for these operations. Instead, reading classical data coherently out of the quantum memory using CNOT-type classical copying operation can realize the quantum oracle in Eq.~\eqref{eq:oracle}, while leaving the memory disentangled with the rest of the system, which becomes the main application of QRAMs. In this sense, QRAMs can be used as a classical data quantum encoder as an I/O unit or an implementation of the quantum oracle in Eq.~\eqref{eq:oracle}, rather than traditionally believed quantum memory devices.

When a QRAM is used in this mode, the QRAM has two stages, (1) classical data loading, (2) coherent address. In stage (1), the classical data ($\mathbf{x}$) is loaded into the quantum memory (or even a classical memory module, as long as classically controlled gates on the bus qubits are available~\cite{Hann2019}), which can be represented as
\begin{align}
    W(\mathbf{x}) \ket{0}^{(\text{QM})} \rightarrow \ket{\mathbf{x}}^{(\text{QM})}.
\end{align}
In stage (2), the bus qubit is initialized at $\ket{0}$ state, and the address qubits are prepared. The bus and the address qubits are sent into the QRAM to coherently address the memory. This can be expressed by
\begin{align}
    R(\ket{addr}, \ket{0}) \ket{\mathbf{x}}^{(\text{QM})} \rightarrow \ket{addr \, \& \, \mathbf{x}} \otimes \ket{\mathbf{x}}^{(\text{QM})},
\end{align}
where the output state $\ket{addr \, \& \, \mathbf{x}}$ respects a state of both the address and the bus qubits, which respect the quantum oracle in Eq.~\eqref{eq:oracle}.

\subsubsection{Device requirements} \label{subsub:qram_require}

As there is no experimental demonstration of a working QRAM, how to build a QRAM, even how to layout different components in a QRAM, is still an open question and requires a lot of effort in material fabrication, control techniques, and device optimizations. Instead of analyzing the performance of the available quantum memory devices, we highlight a few desired properties of QRAMs to better fulfill the usage of QRAMs, i.e., coherently addressing the classical data.

When the QRAM is used to encode classical data into quantum states, which are then used in quantum algorithms to be processed, e.g., in linear algebra operations and quantum machine learning algorithms, the classical data can be enormous. In order to address the classical data efficiently, the corresponding quantum routing systems also need to be massive. This requires QRAM to have a high integration of required quantum switches and the memory qubits inside the quantum routing module. Secondly, in order to achieve the quantum advantage of the algorithms using the quantum oracle, it is necessary to have a fast oracle call. Therefore, each call of the QRAM needs to be fast such that the quantum algorithms can still outperform their classical counterparts.

In addition, depending on the specific usage of QRAM in different quantum algorithms, the latency requirement of the QRAM can be different. For example, if the QRAM is to encode classical data into a quantum state, which is then used to be processed with a deep quantum circuit, as long as the QRAM call is short enough compared to the circuit implementation time, the latency of the QRAM is acceptable. On the other side, if the quantum circuit is short, the QRAM calls need to have a small latency.

\section{Quantum Memory Functional Units in Quantum Processing Unit architecture} \label{sec:QMFU}

\begin{figure}[ht]
    \centering
    \includegraphics[width = \columnwidth]{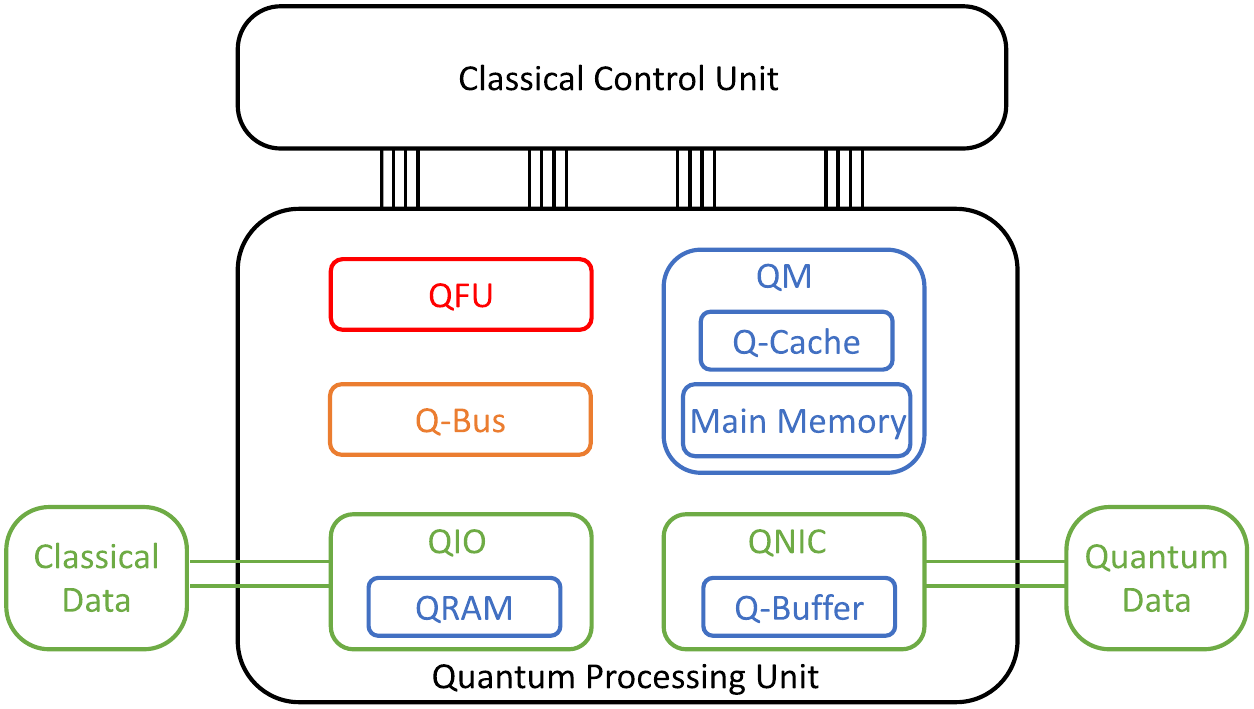}
    \caption{The proposed architecture of a quantum processing unit (QPU). We envision a QPU should include a quantum function unit (QFU), quantum memory (QM), quantum bus (Q-Bus) for communication with the QPU, quantum input-output interface (QIO) unit for an interface with classical data, and quantum network interconnect component (QNIC) for quantum data and quantum communication. A classical control unit is also necessary to control different quantum modules in the QPU. The main memory components are colored blue. }
    \label{fig:qpu_arch}
\end{figure}

Nowadays, due to the rapid development of new fabrication technology, quantum manipulation, and quantum error correction and mitigation techniques, the field of quantum computing and quantum storage has been progressing rapidly and maturing. However, current quantum computing research, especially in the realm of designing and fabricating quantum computing devices, primarily focuses on integrating more quantum registers to demonstrate their performance. Simultaneously, due to limitations imposed by physical and practical conditions, it is extremely challenging to place millions or even billions of quantum registers within the same device and maintain high connectivity. On the other hand, as we have seen from our previous sections, quantum memory techniques are increasingly mature. Therefore, it becomes possible to consider how to utilize quantum memory as an essential component within future quantum processing units (QPU).

Drawing inspiration from the architecture of classical computers, in Fig.~\ref{fig:qpu_arch}, we show the main components that can be contained in the future design of the QPU architecture. A detailed design of QPU architecture is beyond the scope of our paper, and hence we show our envision of the future QPUs without discussing the detailed designs of each functioning unit. We believe a future QPU will include the following functional units, (1) a quantum functional unit (QFU) for implementing quantum gate operations, (2) a quantum memory unit (QM) that can contain a quantum cache (Q-Cache) and a larger main quantum memory, (3) quantum bus (Q-Bus) for quantum communication within a QPU, (4) quantum input-output interface (QIO) for classical data loading, and (5) quantum network interface components (QNIC) for quantum data loading and communication. 

QFU is the central unit to implement quantum algorithms. It can contain a small number of quantum registers that support universal gate sets. When the quantum algorithm is performed, the quantum registers in QFUs implement the quantum gates required by the quantum algorithm. Limited by the size of the QFU, quantum memory is needed. The main QM and Q-Cache are two different memory modules that can store the quantum information for later usage in the QFUs, which we will discuss in Sec.~\ref{subsec:qm} and Sec.~\ref{subsec:q_cache}, respectively. As RW operations on the quantum memory devices require bus qubits, the Q-Bus is for quantum communication between different modules inside the QPU. 

The QIO and the QNIC modules are responsible for communication between the QPU and the other quantum and classical devices. QIO module provides classical interfaces with other devices. Specifically, when classical data is prepared and needs to be processed by the quantum computer, the QIO unit is responsible for encoding the classical data into quantum states. Especially, we focus on the memory device in the QIO unit, where efficiently loading classical data into the QPU can leverage QRAMs as we discussed in Sec.~\ref{subsec:qram}. The functionality of QRAMs inside the QPU architecture will be further discussed in Sec.~\ref{subsec:qram_arch}.

In the QNIC components, we consider including a quantum communication interface with other quantum devices. For example, it can enable quantum communication with other quantum sensors, which can generate quantum data for the QPU to process. It can also enable coupling with other QPUs for distributed quantum computing tasks~\cite{ang2022architectures, wu2022collcomm} 
and communication with a quantum internet for long-range quantum communication with other quantum devices. One of the key components to ensure reliable quantum communication with other devices is quantum buffers, which we will discuss further in Sec.~\ref{subsec:q_buffer}.

In the rest of the section, we direct our attention to the modules consisting of quantum memory. Particularly, we focus on the quantum memory units colored blue in Fig.~\ref{fig:qpu_arch}, which include main quantum memory, quantum cache (Q-Cache), quantum buffer (Q-Buffer), and QRAM in the QIO. We deliberate on their utilities at the architecture level and their design requirements. Specifically, we evaluate their memory qubit coherence (M.Q.C), addressing coherence (A.C), qubit integration (Q.I.), read and write parallelization (RW Para.), and operation speed (O.S.). We will discuss each of the quantum memory functional units in the following sub-sections. In Table~\ref{tab:qmfu_compare}, we give a brief comparison between different types of quantum memory units in terms of these metrics. 

\begin{table}[t]
\caption{\label{tab:qmfu_compare} Comparison between the requirements of different quantum memory modules in the architecture of QPUs. Here we focus on: memory qubit coherence (M.Q.C), addressing coherence (A.C), large qubit integration (Q.I.), read and write parallelization (RW Para.), and read and write operation speed (O.S.). We rate them in a total score 3: 1 means low requirement of the property, low capability to achieve. Seq. stands for sequential operations.} 
\begin{tabularx}{\columnwidth}{l|X|X|X|X|X}
\hline \hline
 & M.Q.C & A.C. & Q.I. & RW Para. & O.S. \\
\hline
Main QM & 3 & Classical & 3 & 3 & 1 \\
Q-Cache & 2 & Classical & 1 & 2 & 3 \\
Q Buffer& 2 to 3 & Classical & 1 & 1 & 1 to 2 \\
QRAM & Classical & 3 & 3 & Seq. & 2 to 3\\
\hline \hline
\end{tabularx}
\end{table}

\subsection{Quantum Memory Unit} \label{subsec:qm}

Analog to the role of memory in the current classical computing architecture, quantum memory is one of the central components in our architecture design. In order to expand the capability of QFU, it is necessary to use quantum memory to store quantum information carried by the qubits that are not immediately involved in the quantum operations. 

Similar to classical memory, a quantum memory unit should support the operation of reading from and writing to a QMC with a given address. Specifically, coherently addressing multiple QMCs is not necessary. As we discussed in Sec.~\ref{subsec:raqm}, RAQM can be used to realize a quantum memory unit in the QPU architecture. 

The quantum memory unit should also satisfy a few design requirements. The quantum memory unit is similar to a classical computer's main memory. As the quantum state may need to be stored in the quantum memory for an extended period of time, the quantum memory should have a low error rate to ensure the information is still authentic. Therefore, (1) the quantum memory unit should have a much longer storage time in terms of computing operation time, i.e., the device's $\beta$ metric should be large. This is the most important requirement to fulfill. (2) The quantum memory unit should have a large number of QMCs integrated into the device while maintaining low cross-talk errors, as the quantum memory unit needs to store all quantum information required in a quantum algorithm. (3) The RW operation time of the quantum memory unit should be small, ideally. Furthermore, (4) enabling addressing QMCs in parallel would be more beneficial for increasing the bandwidth of the RW operations, and simultaneously ensuring its addressability ($\gamma <1$). 

However, due to the long coherence time of the QMCs from the requirement (1), if in the NISQ era where the QMCs are not error corrected, the QMC with longer coherence time means lower coupling to the environment, which usually causes a long reading and writing time. In the FTQC era where QEC is used, suppressing the error rate to increase the storage time necessitates a large code distance and more physical qubits, which can slow down the logical SWAP gate operations. In addition, a large integration of QMCs can occupy a relatively large physical space, which makes the quantum memory unit spatially separated from the QFU. All these factors lead to extending the RW time of the quantum memory. However, to ensure the first two requirements, the requirement of latency ($\beta$) can be slightly released. 

\begin{figure}[htbp]
    \centering
    \includegraphics[width=0.6 \columnwidth]{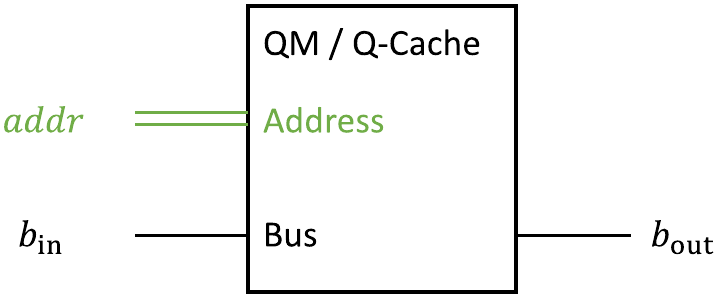}
    \caption{The interface of a quantum memory unit or a Q-Cache. The classical address information ($addr$) and quantum registers (labeled as $b_{\text{in}}$) interact with the quantum memory, while the quantum registers $b$ with the QMC with address $addr$ state is outputted, which is labeled as $b_{\text{out}}$.}
    \label{fig:qm_model}
\end{figure}

With the design requirement of the quantum memory unit, we then consider how to utilize this device in the upper stacks. The reading and writing functions are similar to the RAQM discussed in Sec.~\ref{subsec:raqm}. Specifically, the quantum memory unit should have a classical input to take in the address information. The input can take another quantum register, which carries the quantum information in the writing process while loading the quantum information from the memory in the reading process. We stress that unlike classical memory, where the reading and writing processes are unidirectional, i.e., the information is copied from and to the quantum memory, accessing the QMCs is bidirectional in the quantum case. Therefore, there is no hard distinction between the reading and writing processes, and both reading and writing processes can be represented by the model in Fig.~\ref{fig:qm_model}. However, in order to better organize the programming and highlight where the nontrivial quantum state is stored, it would be useful to have both read and write functions enabled, although the underlying physical operations are essentially the same. 

\subsection{Quantum Cache (Q-Cache)} \label{subsec:q_cache}

Quantum cache (Q-Cache) is another quantum memory functional unit inside the architecture of QPUs. According to Sec.~\ref{subsec:qm}, the quantum memory unit can have relatively long latency. In order to speed up quantum computation, analog to classical computing systems, a quantum cache (Q-Cache) can be utilized. Specifically, if the quantum information carried by a certain qubit is relatively frequently visited, instead of swapping it from and to the main quantum memory every time the operation is done, the information can be stored inside a Q-Cache, which can provide faster RW operations. 

Therefore, for the purpose of speeding up quantum computation, Q-Cache has a few design requirements. (1), the latency of the RW operations on a Q-Cache needs to be small, which is required by the operation speed of a Q-Cache. In order to achieve the low latency of the RW operations, a Q-Cache can choose bare qubits with shorter coherence time in the NISQ device, while choosing QEC codes with smaller code distance in the fault-tolerant device, compared to the ones used in the quantum memory unit. Therefore, (2) a Q-Cache may have moderately long storage times. We claim that the Q-Cache should satisfy
\begin{align}
    \alpha_{\text{ex, QC}} / \beta_{\text{QM}} = \frac{T_{\text{storage, QC}} \eta_{\text{QC}}}{T_{\text{RW, QM}}/\eta_{\text{QM}}} > r_{\text{threshold}} \sim 2,
\end{align}
where `QC' labels the properties of Q-Cache, and `QM' stands for the quantum memory unit. This means the storage time of the Q-Cache should be at least longer than storing and retrieving the quantum information from the quantum memory unit. If the quantum information carried by a qubit idles for a duration longer than twice the RW time of the main quantum memory, it would be better to be stored in the main memory, as it can experience less error. In the actual design of a QPU, the threshold value $r_\text{threshold}$ can be further optimized. In addition, to further improve the latency of the Q-Cache, the Q-Cache is expected to be located on the same chip of QFUs or nearby. Due to the spatial limitation, (3) the number of QMCs can be small. On the other hand, to increase the communication bandwidth, (4) operating RW of Q-Cache in parallel through multiple banks following the classical memory design is desired.

As the Q-Cache can also be implemented by RAQMs, the interface of a Q-Cache is similar to the quantum memory unit, which is discussed in Sec.~\ref{subsec:qm}. The interface of a Q-Cache can also be represented by Fig.~\ref{fig:qm_model}.

\subsection{Quantum Buffer} \label{subsec:q_buffer}

The quantum buffer is another quantum memory functional unit inside the QPU architecture. In the process of implementing a quantum algorithm or quantum operation, some resource states are probabilistically generated. Therefore, it is necessary to include quantum buffers to store these states and retrieve them when they are requested. The quantum buffer can be widely used in quantum communication components, especially in the QNIC shown in Fig.~\ref{fig:qpu_arch}. There are two possible applications, interfacing with quantum sensors and with quantum networks for quantum communication. Specifically, quantum sensors can prepare quantum states that encode the sensing information. The quantum states can be imported into the QPU for further processing. However, the quantum sensing process can be slow compared to the quantum computing cycles in the QFU, and various quantum sensors operate at different speeds, which necessitate quantum buffers to receive the quantum state and make them ready to be processed by the QFU. 

Another application is to use a quantum buffer to buffer information from a quantum internet. Long-range quantum communication usually relies on entanglement generation and state teleportation~\cite{Bennett1993, Briegel1998}. However, the remote entanglement generation, involving state purification and photon measurements, is probabilistic in nature. Therefore, when a qubit is successfully entangled with the remote quantum system, it can be stored in the quantum buffer for later communication use~\cite{wu2022collcomm}. Notably, the usage of quantum buffers is not limited to QNIC units. Indeed, whenever there is a need to store the probabilistically generated resource states, a quantum buffer can be utilized. One of the examples would be in the magic state distillation process, where the quantum buffer can store the generated high-fidelity magic states for later use in implementing surface code Toffoli gates. 

To fulfill these implementations, quantum buffers are required to work between two quantum systems. Without losing the generality, one of the quantum systems can be viewed as an information saver, which generates the quantum information and saves it to the quantum buffer, while the other one is the information loader, which loads the quantum information depending on its processing need. Therefore, there are two characteristic time periods, one is the time for generating the quantum state ($T_\text{g}$), which is from the information saver, while the other one is the duration of state consumption ($T_\text{c}$), which is from the information loader. Therefore, it requires (1) the external storage ratio of the quantum buffer to be long compared to the slower process, i.e.,
\begin{align}
    \alpha_{\text{ex, QB}} = \frac{T_\text{storage} \eta}{\max(T_\text{g}, T_{\text{c}})}, 
\end{align}
is a metric for a quantum buffer design and $\alpha_{\text{ex, QB}} \gg 1$ ideally. On the other hand, (2) the latency of the quantum buffer needs to be small compared to the faster process, i.e., ideally,
\begin{align}
    \beta_{\text{QB}} = \frac{T_\text{RW}/\eta}{\min(T_\text{g}, T_{\text{c}})} \ll 1.
\end{align}
The integration of the quantum buffer may not be large. It is unnecessary to buffer a huge number of quantum states, as the oldest copies can be discarded.  
Ideally, wasting quantum states is not efficient, and hence (4) we require $N \sim T_\text{g}/T_\text{c}$. The parallelization feature may not be required and depends on the requirements on the consumption side. If multiple states can be consumed simultaneously, parallelizing the reading process is needed. However, simultaneously reading from and writing to the quantum buffer should not be allowed. 

Since the metrics of quantum buffers include generation and consumption times, quantum buffers can be designed for quantum tasks differently. Notably, even for the same task, both the generation time $T_\text{g}$ and the consumption time $T_\text{c}$ can depend on the algorithms and protocols. For example, in the task of quantum communication between the current QPU and a remote quantum device, the entangled state purification protocols can have different yield and time~\cite{ang2022architectures}, which can affect $T_\text{g}$. Therefore, given quantum buffer properties, optimizing the algorithms on both generation and consumption sides to satisfy the above requirement, as well as co-design the quantum hardware and algorithms, can be interesting directions to proceed~\cite{wu2022collcomm}. 

\begin{figure}[htbp]
    \centering
    \subfloat[Q-Buffer Interface]{
    \includegraphics[width = 0.45 \columnwidth]{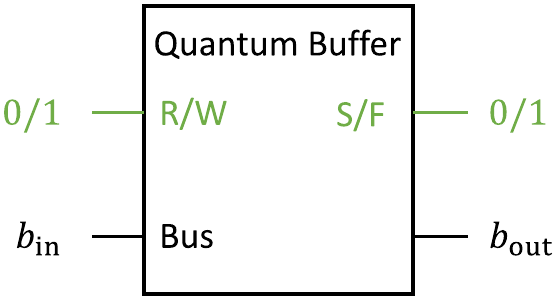}
    }\\
    \subfloat[Reading]{
    \includegraphics[width = 0.45 \columnwidth]{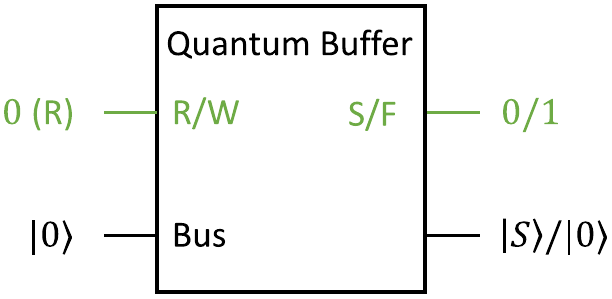}
    }
    \subfloat[Writing]{
    \includegraphics[width = 0.45 \columnwidth]{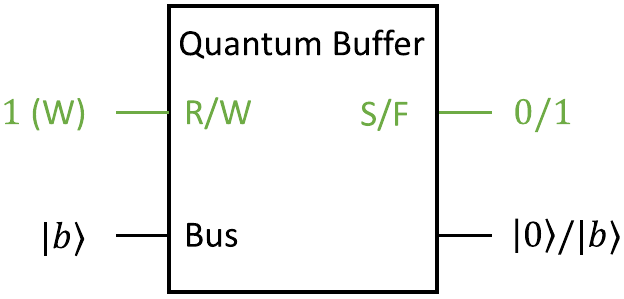}
    }
    \caption{The interfaces of the quantum buffer are shown in (a), while the reading and write processes are shown in (b) and (c), respectively.}
    \label{fig:q-buffer_model}
\end{figure}

The quantum buffer is slightly different from the memory. In most cases, a single quantum buffer stores a certain type of quantum state, while each state may have different state fidelity. The state can be constantly generated with different time intervals and can be requested from other quantum units in the other quantum function units. The random access feature is not necessary for a quantum buffer. Instead, a quantum buffer can be constructed by an array of QMCs with, for example, the first-in-first-out (FIFO) policy. In Fig.~\ref{fig:q-buffer_model}a, we show the interface of a quantum buffer. The quantum buffer can take in a single bit of classical data for the instruction of reading or writing operations. It also takes a quantum register to interact with the quantum buffer to store or retrieve the quantum information. The quantum buffer can return the bus register along with a single bit of classical data to show whether the query operation is successful. The state of the returned bus register can depend on whether the query is successful or not. 

In Fig.~\ref{fig:q-buffer_model}b, we show the reading process, which is demonstrated by the RW bit being set to be $0$. In the reading process, the state stored inside the quantum buffer is requested from other QPU modules. In the reading process, the bus register is set to be $\ket{0}$. If there are quantum states stored in the quantum buffer, which is available to be retrieved, the reading query is successful with a returning value $1$ in the output `S/F' bit. The bus register swaps the stored state out, labeled as $\ket{S}$. On the other hand, if there are no available quantum states inside the quantum buffer, the reading query fails, with S/F returning $0$. This is similar to an underflow situation in a classical buffer. If this happens, the bus register is then returned with state $\ket{0}$ without interacting with the buffer QMCs. 

The writing process of a quantum buffer is shown in Fig.~\ref{fig:q-buffer_model}c. In the writing mode, a quantum state carried by the bus register must be stored in the quantum buffer. When the quantum buffer is not full, i.e., not all quantum memory registers are used in the quantum buffer, the control unit of the quantum buffer will locate the unused memory registers, and the bus register state can be successfully stored by swapping its state to this memory register. The S/F bit will output $0$ while the bus register is set back to state $\ket{0}$, which corresponds to the original state of the memory register. On the other hand, if the quantum buffer is already full, to be consistent with the reading process, the quantum buffer will return $1$ in the S/F output bit, while holding the quantum bus register not to interact with the memory bit. This corresponds to a classical overflow situation. 

We note that although in our design shown in Fig.~\ref{fig:q-buffer_model}, we consider sequentially reading and writing quantum states of a single bus register, reading multiple registers in the same query of quantum buffer should be supported. In this case, the quantum buffer can take multiple bus registers as input, depending on its specific implementation. When the underflow or overflow situation happens, the quantum buffer should hold the reading and writing queries until all the states or storage quantum memory registers are ready.

\subsection{QRAM in QIO} \label{subsec:qram_arch}

As mentioned in Sec.~\ref{subsec:qram}, QRAMs work distinctly from the other quantum memory units. In our architecture design, QRAMs can be used in the QIO unit, where the classical data is loaded to the QRAM, while according to the algorithms, a set of address qubits are prepared to be sent into the QRAM as the address information, which coherently addressing the classical data and preparing a bus-address entangled states as the output. As we have discussed in Sec.~\ref{subsub:qram_require}, in order to efficiently obtain this goal, a few design requirements have been discussed. 

In the architecture stack, the QRAM can be built in as a general-purpose device for encoding classical data into quantum states and quantum compression of classical data. In addition, QRAMs can also be built within special function units that can perform algorithms with quantum speedups. For example, a QRAM can be built into the special function units for the Grover search algorithm~\cite{Grover1996}, which can speed up the database search. The main design requirement of QRAMs lies in having fast and reliable queries. In the framework of quantum memory devices, it is equivalent to small reading latency, where
\begin{align}
    \beta_\text{QRAM} = \frac{T_\text{R}/\eta}{T_\text{Circ}} \ll 1,
\end{align}
where $T_{\text{Circ}}$ is the time for implementing the quantum operations between two QRAM queries. On the other hand, the QRAM is designed to interact with classical data, which is usually in a large size. Therefore, to have a compact integration is greatly important. 

\begin{figure}[htbp]
    \centering
    \subfloat[QRAM Interface]{
    \includegraphics[width = 0.42 \columnwidth]{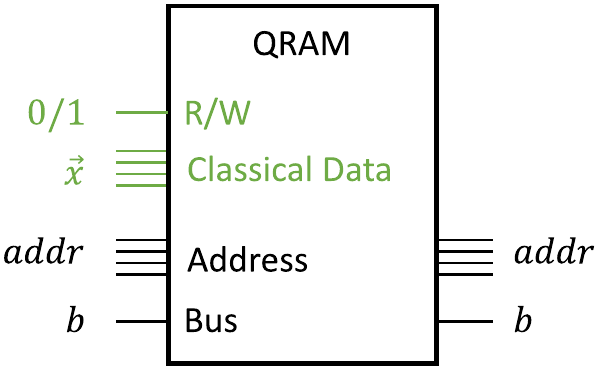}
    }\\
    \subfloat[QRAM data loading]{
    \includegraphics[width = 0.42 \columnwidth]{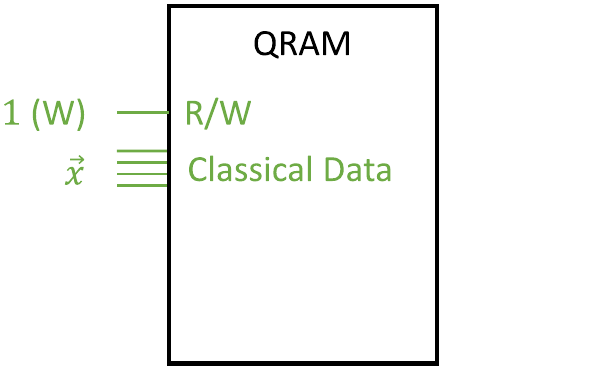}
    }
    \subfloat[QRAM query]{
    \includegraphics[width = 0.42 \columnwidth]{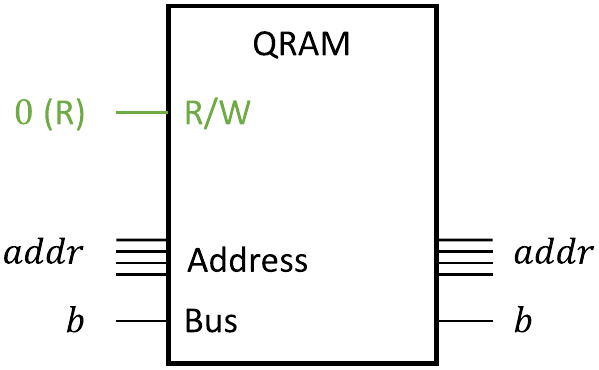}
    }
    \caption{The model of a QRAM module. The interface of a QRAM is shown in (a). The QRAM classical data loading process is shown in (b), while the QRAM query process is in (c).}
    \label{fig:qram_model}
\end{figure}

When a QRAM is used as an interface between classical data and quantum devices, in Fig.~\ref{fig:qram_model}, we show an abstracted model for a QRAM. The interface of a QRAM is shown in Fig.~\ref{fig:qram_model}a. A QRAM can have two modes: one mode is to load classical data into the memory, while the other mode is to coherently address the classical data. A QRAM can take a single bit of classical data as the mode specification (labeled as `R/W'). In addition, a QRAM can also take classical data as an input, or take a set of quantum address registers and a bus register as inputs. 

In the classical data loading mode, where the R/W classical register is set to $1$, a classical data $\vec{x}$ is sent to the QRAM. The QRAM loads the classical data $\hat{x}$ to the QRAM's memory components. In order to use the QRAM to quantum encode the classical data, or as quantum oracles shown in Eq.~\eqref{eq:oracle} in quantum algorithms, the R/W classical register is set to $0$, which shows the QRAM is in the quantum query mode. In this mode, the QRAM takes a set of quantum registers, which include address and bus registers. After a QRAM read query, the address and bus registers are returned. Their states are now changed according to the classical data $\hat{x}$ that has been stored inside the memory based on Eq.~\eqref{eq:oracle}.

\section{Quantum memory programming model} \label{sec:QMPM}

With the available quantum memory units, utilize the quantum memory modules in the future quantum programs should be available, which enables quantum-memory-aware program design. Currently, quantum programs at the assembly level are described as a quantum circuit using OpenQASM~\cite{cross2017open, qasm3}. QASM focuses on representing quantum circuits, including initialization and resetting quantum registers, applying quantum gates, performing measurements, and classically controlled gate operations. However, present quantum programs such as QASM are all centered around quantum registers, overlooking quantum memory.
Therefore, we investigate possible changes in the current gate-based quantum programming languages and APIs to incorporate quantum memory.

In order to adapt quantum assembly languages, such as QASM, for quantum memory utilities, we propose to introduce four quantum-memory-related primitives: 
\begin{itemize}
    \item[(1)] \texttt{mem} $size$. It declares the quantum memory requirements used in the corresponding program, specifying the total number of quantum memory cells ($size$). It should be used at the beginning of an assembly code. When the quantum program is executed on a quantum device, the required memory size will be passed to the quantum hardware controller to check if the hardware can support the required memory size. 
    \item[(2)] \texttt{ld} $q$ = [$addr$]. It loads the quantum state stored in the quantum memory to the corresponding quantum register or qubits. When the argument $q$ is a qubit, the state of the QMC with address $addr$ will be loaded to the qubit, while the QMC is reset. If the argument $q$ is a quantum register with $n$ qubits, the states of QMCs starting from address $addr$ are loaded to the corresponding register, while these QMCs are reset.
    \item[(3)] \texttt{st} [$addr$] = $q$. It stores the quantum state carried by the quantum register or qubit to the quantum memory cells with address \textit{addr}. Similar to the primitive \texttt{ld}, the argument $q$ can be either a qubit or a quantum register. 
    \item[(4)] \texttt{mreset} $addr$. It resets the quantum memory cell with address $addr$. If the $addr$ is missing, it will reset all the quantum memory cells declared.
\end{itemize}
With the four basic primitives to operate on quantum memory, more complex quantum memory management strategies can be implemented. The interfaces of the quantum buffer and quantum cache can be implemented as APIs, contained in quantum libraries for operating the specific quantum devices.

On the other hand, QRAM is a different type of quantum function unit distinct from quantum memory. In order to operate QRAMs in the assembly language, we proposed to include the following three primitives.
\begin{itemize}
    \item[(1)] \texttt{qram name}[$addr\_len$,$word\_len$].  It declares the QRAM needed in the program, and gives it a name \texttt{name}. The declaration requires two arguments: $addr\_len$ specifies the number of address qubits needed to address all the memory registers, and $word\_len$ specifies the word length of each memory register, i.e., the number of memory cells consisting of a single register. 
    \item[(2)] \texttt{qinit name} [$x$]. It writes classical data into the QRAM. The classical data needs to be coherently addressed via the QRAM. The argument $x$ should be a classical array of registers. The size should be compatible with the QRAM declared in the code.
    \item[(3)] \texttt{qld name}($q\_b$)[$q\_addr$]. It performs coherent addressing of the content in the QRAM with the quantum address that is represented by the state of the address qubits $q\_addr$. The quantum register $q\_b$ is also given as the bus qubit.
\end{itemize}
These three primitives provide the essential functionalities of QRAM operations, which can be utilized to construct more sophisticated quantum programs and libraries, including realizing Grover search, quantum data lookup, etc. In Table~\ref{tab:primitives}, we summarize the primitives we proposed here.

\begin{table*}[t]
\caption{\label{tab:primitives} A summary of quantum memory primitives to be included in QASM.} 
\begin{tabular}{p{2 cm}|p{4.5 cm}|l}
\hline \hline
Device & Code Format & Meaning \\
\hline
RAQM & \texttt{mem} $size$ &  declear memory usage \\
     & \texttt{ld} $q$ = [$addr$] & load quantum memory to q-register \\
     & \texttt{st} [$addr$] = $q$ & store q-register to memory \\
     & \texttt{mreset} $addr$ & reset the memory \\
\hline
QRAM & \texttt{qram name}[$addr\_len, word\_len$] & declear the usage of a QRAM \\
     & \texttt{qinit name} [$x$] & load classical data to QRAM \\
     & \texttt{qld name}($q\_b$)[$q\_addr$] & coherently address the QRAM \\
\hline \hline
\end{tabular}
\end{table*}

In addition to the primitives to initiate and operate the QRAM, as a QRAM can load classical data into its memory media, we feel it is necessary to include array operations on classical data, which is recently supported by OpenQASM 3.0~\cite{qasm3}. With the quantum memory enabled, array operations on qubits and quantum registers are also necessary. Therefore, we propose to include slice operations on quantum registers and quantum arrays. For example,
\begin{lstlisting}[style=tt1, label={code:sample}]
qubit[4] q;
mem 4;
st [0] = q[1:];
\end{lstlisting}
saves the qubit register \texttt{q[1]}, \texttt{q[2]}, and \texttt{q[3]} to memory address $0$, $1$, and $2$. 

\begin{figure*}[ht]
    \centering
    \includegraphics[width = 0.7\textwidth]{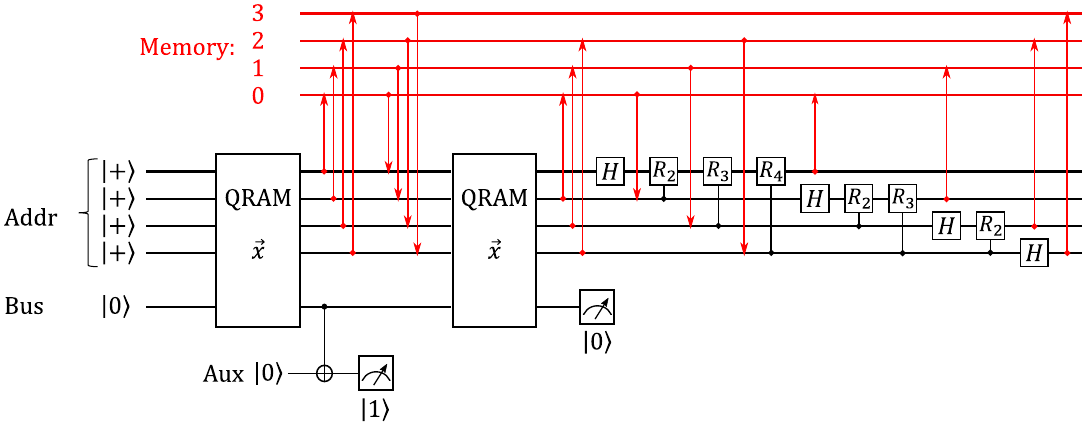}
    \caption{The amplitude encoding and QFT on the encoded state. The red arrows show the loading quantum states from and saving to the quantum memory.}
    \label{fig:example_circuit}
\end{figure*}

\begin{lstlisting}[style=tt1, caption={Code example for Fig.~\ref{fig:example_circuit}}, label={code:qft}]
OPENQASM 3;
gate cr(n) c, t {
  angle %*$\theta$*) = 2%*$\pi$*)/power(2,n)
  ctrl @ U(0, 0, %*$\theta$*)) c, t;
}

qubit[4] q;
qubit[1] b;
qubit[1] aux;
bit[2] caux;

// an example vector with binary values
bit[16] vec = [0,1,1,0,0,1,1,1,0,0,1,1,0,1,0] 

mem 4; // specify the memory requirement
qram qr[4,1]; // specify the QRAM requirement
qinit qr [vec]; // load the data into QRAM

h q; 
ldqram qr q b; // use QRAM
// save q into memory waiting for the measurement on aux qubit
st [0] = q; 

cx b aux;
measure aux -> caux[0];

if (caux[0]==1){
  ld q = [0];
  qld qr(b)[q]; // use QRAM
  
  measure b -> caux[1];
  st [0] = q[1:];
  
  int j = 0;
  for i in [0:2]{ 
    h q[i];
    j = i+1;
    while(j<4){
      if(i==0) ld q[j] = [i];
      cr(j-i+1) q[j], q[i];
      j = j+1;
    }
    st [i] = q[i];
  }
  
  h q[3]
  st [3] = q[3];
}
\end{lstlisting}

To describe the functionalities of the quantum memory and QRAM primitives, we include an example of the modified QASM code for amplitude encoding the classical data and then performing the quantum Fourier transform. The input data is a classical binary vector \texttt{vec}. The quantum circuit is shown in Fig.~\ref{fig:example_circuit}. We consider using probabilistic amplitude encoding process~\cite{HannThesis}. The encoding succeeds when the measurement on the auxiliary qubit is $\ket{1}$. The quantum Fourier transform (QFT) sequence uses control-$R_k$ gates, where 
\begin{equation}
    R_k = \left[ 
    \begin{array}{cc}
        1 & 0 \\
        0 & e^{i 2\pi/2^k}
    \end{array}
    \right],
\end{equation}
where $k$ is an integer~\cite{nielsen_chuang_2010}. The outcome state is saved in the quantum memory with addresses $0$ to $4$.

In addition to the primitives we proposed to include, we stress that including quantum memory devices can also impact the design of quantum transpilers used at higher programming levels. For example, in the current stage, a quantum algorithm is written in terms of its quantum circuit, which needs to be transpiled into a QASM code based on standardized gate operations by gate decomposition and simplification. The OpenQASM code can then be converted into instructions (like the control pulses to control the physical qubits) that can directly interact with the quantum hardware. When quantum memory is available in the architecture of quantum computing systems, a memory-aware quantum transpiler is needed to manage the quantum memory usage and optimize the performance of quantum algorithms. 

Higher-level quantum programming languages are also expected to coordinate with the presence of quantum memory devices. For example, quantum software development toolkits, such as Qiskit~\cite{Qiskit}, should include libraries that support RAQM and QRAM operations, quantum memory management, as well as using QRAMs to speed up quantum algorithms, such as the Grover search, quantum Fourier Transform, etc. With the memory-device-aware transpilers available, these libraries can be transpiled to QASM codes that can operate quantum memory units to leverage their functionalities. On top of that, a more sophisticated quantum algebra library taking advantage of quantum-memory-aware optimization and QRAMs can be fully built for general-purpose quantum computers. 

Furthermore, as addresses are introduced to operate on the quantum memory, arithmetic calculations on the quantum memory addresses should also be supported, while more sophisticated quantum data structures can be introduced into the quantum programming languages, e.g., quantum array. In addition, it is necessary to include pointers for quantum memory to enable dynamical memory allocation and efficient data access.

\section{Quantum Memory Applications} \label{sec:application}

Quantum memory is a crucial component in quantum computing systems. We explore the use of quantum memory in the architecture of a single quantum processing unit (QPU), but its potential extends far beyond that. Quantum memory can find various applications in quantum computing and quantum information processing tasks. 

In our discussion, quantum memory is usually used for saving quantum information. One example is to use the QMCs in quantum memory modules to store a multi-qubit quantum state. With a large integration of QMCs, a quantum database can be built. A huge number of QMCs can be utilized to store a large quantum state or numerous copies of intermediate-size quantum states. However, a quantum database that only consists of quantum memory for quantum information storage is not completely feasible. One limitation is from the quantum no-cloning theorem. The quantum state stored in the quantum database cannot be copied. When the quantum information is requested and measured, the information is destroyed and no longer stored in the database. Conversely, in a classical database, when the information is requested, it will not destroy the information stored in the database. Therefore, we envision that a pure quantum database is not suitable for permanent information storage. Instead, a quantum data center~\cite{Liu2022quantum} in the future can consist of both classical database and quantum memory, such that the everlasting information can be stored by a classical method in the classical memory, while the quantum memory can be used to store quantum information, either encoded from the classical information for further processing, or received from other quantum parties to be processed with the classical information. 

However, encoding classical information into quantum states can compress the data for quantum processing. For example, using amplitude encoding, a classical binary data string with length $2^N$ can be encoded into a quantum state of $N$ qubits. Fortunately, this encoding can be obtained efficiently with QRAMs~\cite{HannThesis}. After encoding the classical data into quantum states, it can then be efficiently utilized in quantum algorithms, such as quantum machine learning~\cite{Biamonte2017, Cerezo2022}. Furthermore, the quantum encoding of the classical information can also be more efficient in performing remote information processing. For example, if a client only needs to compare a large data with size $2^N$ on his/her end to a chunk of data stored in the data center, classically, the data needs to be sent for comparison. With the help of the swap test~\cite{Buhrman2001}, encoding the classical data into quantum states and sending a copy of the encoded quantum states for comparison only requires sending $N$ qubits. If the loss of the quantum network and the imperfections can be well controlled, for a given required accuracy, only polynomially many copies of the quantum state are needed. 

Furthermore, as pointed out in Ref.~\cite{Liu2022quantum}, with the help of quantum communication protocols, a quantum data center can also help improve communication privacy. In Ref.~\cite{Liu2022quantum}, Liu {\it et al.} proposed a protocol for multi-party private quantum communications, which ensures the privacy of the communication between multiple untrusted parties. 

Other than quantum data centers, quantum buffers can be used in quantum networks and quantum internet. A quantum buffer can be combined with quantum repeaters in the communication nodes, while the entangled states can be stored for quantum communication and purification. The initial application of quantum buffer in quantum networks has been discussed in Ref.~\cite{wu2022collcomm}.

Therefore, we believe that a quantum data center combined with quantum internet and quantum communication can be an excellent application example for quantum memory and QRAM devices. With these applications, we believe quantum memory research will be increasingly attractive to research in related fields.

In addition, an interesting question with the development of quantum memory devices for quantum communication is how to verify the device is authentic. In Ref.~\cite{Rosset2018}, Rosset {\it et al.} proposed a protocol to verify the quantum storage of the quantum memory devices, developed on top of the protocol in Ref.~\cite{Pusey2015}. A resource theory for quantum memory is also constructed in Ref.~\cite{Rosset2018}. This protocol has been demonstrated in experiments~\cite{Graffitti2020, Mao2020, Yu2021}. This protocol is also recently generalized to continuous-variable quantum memories~\cite{abiuso2023verification}. With the large-scale quantum memory devices available, how to verify the quantum memory and perform verification protocols more efficiently on all quantum memory devices on a quantum internet is still a valuable question for quantum memory research. 

\section{Conclusion and outlook} \label{sec:summary}

In the rapidly evolving landscape of quantum computing and quantum information science, building efficient and reliable quantum memory devices stands as a cornerstone for unlocking the full potential of quantum technologies. In this paper, we survey different aspects of quantum memory, from the physical systems for quantum memory materials to quantum memory devices, e.g., the RAQM and QRAM. We envision that the rapid development of quantum computing systems and the improvement of various physical systems enable more complicated and high-performance memory devices, which provide the opportunity to investigate the higher-stack design of quantum computing systems, especially focusing on the pivotal role of quantum memory in quantum processing units. We then discuss the quantum memory architecture design and the corresponding programming model, as well as their possible applications. Specifically, we point out the possible memory modules inside the QPU architecture, and define their software-oriented functionalities. We hope this article can present not only the status of the research in the hardware and device stacks of quantum memory to higher-stack researchers but also a view of quantum memory devices from the top stacks and show the hardware researchers the possible directions to improve their devices. We believe that it is time to start considering quantum memory and its role in quantum computing systems. We hope that our article helps to motivate people to enter this exciting field.

\vspace{2.5 mm}

\section{Acknowledgement}

This material is based upon work supported by the U.S. Department of Energy, Office of Science, National Quantum Information Science Research Centers, Co-design Center for Quantum Advantage (C2QA) under contract number DE-SC0012704, (Basic Energy Sciences, PNNL FWP 76274). The Pacific Northwest National Laboratory is operated by Battelle for the U.S. Department of Energy under Contract DE-AC05-76RL01830.

\bibliography{ref}

\begin{thebibliography}{332}%
\makeatletter
\providecommand \@ifxundefined [1]{%
 \@ifx{#1\undefined}
}%
\providecommand \@ifnum [1]{%
 \ifnum #1\expandafter \@firstoftwo
 \else \expandafter \@secondoftwo
 \fi
}%
\providecommand \@ifx [1]{%
 \ifx #1\expandafter \@firstoftwo
 \else \expandafter \@secondoftwo
 \fi
}%
\providecommand \natexlab [1]{#1}%
\providecommand \enquote  [1]{``#1''}%
\providecommand \bibnamefont  [1]{#1}%
\providecommand \bibfnamefont [1]{#1}%
\providecommand \citenamefont [1]{#1}%
\providecommand \href@noop [0]{\@secondoftwo}%
\providecommand \href [0]{\begingroup \@sanitize@url \@href}%
\providecommand \@href[1]{\@@startlink{#1}\@@href}%
\providecommand \@@href[1]{\endgroup#1\@@endlink}%
\providecommand \@sanitize@url [0]{\catcode `\\12\catcode `\$12\catcode
  `\&12\catcode `\#12\catcode `\^12\catcode `\_12\catcode `\%12\relax}%
\providecommand \@@startlink[1]{}%
\providecommand \@@endlink[0]{}%
\providecommand \url  [0]{\begingroup\@sanitize@url \@url }%
\providecommand \@url [1]{\endgroup\@href {#1}{\urlprefix }}%
\providecommand \urlprefix  [0]{URL }%
\providecommand \Eprint [0]{\href }%
\providecommand \doibase [0]{http://dx.doi.org/}%
\providecommand \selectlanguage [0]{\@gobble}%
\providecommand \bibinfo  [0]{\@secondoftwo}%
\providecommand \bibfield  [0]{\@secondoftwo}%
\providecommand \translation [1]{[#1]}%
\providecommand \BibitemOpen [0]{}%
\providecommand \bibitemStop [0]{}%
\providecommand \bibitemNoStop [0]{.\EOS\space}%
\providecommand \EOS [0]{\spacefactor3000\relax}%
\providecommand \BibitemShut  [1]{\csname bibitem#1\endcsname}%
\let\auto@bib@innerbib\@empty
\bibitem [{\citenamefont {Shor}(1999)}]{shor1999polynomial}%
  \BibitemOpen
  \bibfield  {author} {\bibinfo {author} {\bibfnamefont {Peter~W}\ \bibnamefont
  {Shor}},\ }\bibfield  {title} {\enquote {\bibinfo {title} {Polynomial-time
  algorithms for prime factorization and discrete logarithms on a quantum
  computer},}\ }\href@noop {} {\bibfield  {journal} {\bibinfo  {journal} {SIAM
  review}\ }\textbf {\bibinfo {volume} {41}},\ \bibinfo {pages} {303--332}
  (\bibinfo {year} {1999})}\BibitemShut {NoStop}%
\bibitem [{\citenamefont {Gisin}\ \emph
  {et~al.}(2002{\natexlab{a}})\citenamefont {Gisin}, \citenamefont {Ribordy},
  \citenamefont {Tittel},\ and\ \citenamefont {Zbinden}}]{gisin2002quantum}%
  \BibitemOpen
  \bibfield  {author} {\bibinfo {author} {\bibfnamefont {Nicolas}\ \bibnamefont
  {Gisin}}, \bibinfo {author} {\bibfnamefont {Gr{\'e}goire}\ \bibnamefont
  {Ribordy}}, \bibinfo {author} {\bibfnamefont {Wolfgang}\ \bibnamefont
  {Tittel}}, \ and\ \bibinfo {author} {\bibfnamefont {Hugo}\ \bibnamefont
  {Zbinden}},\ }\bibfield  {title} {\enquote {\bibinfo {title} {Quantum
  cryptography},}\ }\href@noop {} {\bibfield  {journal} {\bibinfo  {journal}
  {Reviews of modern physics}\ }\textbf {\bibinfo {volume} {74}},\ \bibinfo
  {pages} {145} (\bibinfo {year} {2002}{\natexlab{a}})}\BibitemShut {NoStop}%
\bibitem [{\citenamefont {Farhi}\ \emph {et~al.}(2014)\citenamefont {Farhi},
  \citenamefont {Goldstone},\ and\ \citenamefont {Gutmann}}]{farhi2014quantum}%
  \BibitemOpen
  \bibfield  {author} {\bibinfo {author} {\bibfnamefont {Edward}\ \bibnamefont
  {Farhi}}, \bibinfo {author} {\bibfnamefont {Jeffrey}\ \bibnamefont
  {Goldstone}}, \ and\ \bibinfo {author} {\bibfnamefont {Sam}\ \bibnamefont
  {Gutmann}},\ }\bibfield  {title} {\enquote {\bibinfo {title} {A quantum
  approximate optimization algorithm},}\ }\href@noop {} {\bibfield  {journal}
  {\bibinfo  {journal} {arXiv preprint arXiv:1411.4028}\ } (\bibinfo {year}
  {2014})}\BibitemShut {NoStop}%
\bibitem [{\citenamefont {Han}\ and\ \citenamefont
  {Kim}(2002)}]{han2002quantum}%
  \BibitemOpen
  \bibfield  {author} {\bibinfo {author} {\bibfnamefont {Kuk-Hyun}\
  \bibnamefont {Han}}\ and\ \bibinfo {author} {\bibfnamefont {Jong-Hwan}\
  \bibnamefont {Kim}},\ }\bibfield  {title} {\enquote {\bibinfo {title}
  {Quantum-inspired evolutionary algorithm for a class of combinatorial
  optimization},}\ }\href@noop {} {\bibfield  {journal} {\bibinfo  {journal}
  {IEEE transactions on evolutionary computation}\ }\textbf {\bibinfo {volume}
  {6}},\ \bibinfo {pages} {580--593} (\bibinfo {year} {2002})}\BibitemShut
  {NoStop}%
\bibitem [{\citenamefont {Georgescu}\ \emph {et~al.}(2014)\citenamefont
  {Georgescu}, \citenamefont {Ashhab},\ and\ \citenamefont
  {Nori}}]{georgescu2014quantum}%
  \BibitemOpen
  \bibfield  {author} {\bibinfo {author} {\bibfnamefont {Iulia~M}\ \bibnamefont
  {Georgescu}}, \bibinfo {author} {\bibfnamefont {Sahel}\ \bibnamefont
  {Ashhab}}, \ and\ \bibinfo {author} {\bibfnamefont {Franco}\ \bibnamefont
  {Nori}},\ }\bibfield  {title} {\enquote {\bibinfo {title} {Quantum
  simulation},}\ }\href@noop {} {\bibfield  {journal} {\bibinfo  {journal}
  {Reviews of Modern Physics}\ }\textbf {\bibinfo {volume} {86}},\ \bibinfo
  {pages} {153} (\bibinfo {year} {2014})}\BibitemShut {NoStop}%
\bibitem [{\citenamefont {Kandala}\ \emph {et~al.}(2017)\citenamefont
  {Kandala}, \citenamefont {Mezzacapo}, \citenamefont {Temme}, \citenamefont
  {Takita}, \citenamefont {Brink}, \citenamefont {Chow},\ and\ \citenamefont
  {Gambetta}}]{kandala2017hardware}%
  \BibitemOpen
  \bibfield  {author} {\bibinfo {author} {\bibfnamefont {Abhinav}\ \bibnamefont
  {Kandala}}, \bibinfo {author} {\bibfnamefont {Antonio}\ \bibnamefont
  {Mezzacapo}}, \bibinfo {author} {\bibfnamefont {Kristan}\ \bibnamefont
  {Temme}}, \bibinfo {author} {\bibfnamefont {Maika}\ \bibnamefont {Takita}},
  \bibinfo {author} {\bibfnamefont {Markus}\ \bibnamefont {Brink}}, \bibinfo
  {author} {\bibfnamefont {Jerry~M}\ \bibnamefont {Chow}}, \ and\ \bibinfo
  {author} {\bibfnamefont {Jay~M}\ \bibnamefont {Gambetta}},\ }\bibfield
  {title} {\enquote {\bibinfo {title} {Hardware-efficient variational quantum
  eigensolver for small molecules and quantum magnets},}\ }\href@noop {}
  {\bibfield  {journal} {\bibinfo  {journal} {Nature}\ }\textbf {\bibinfo
  {volume} {549}},\ \bibinfo {pages} {242--246} (\bibinfo {year}
  {2017})}\BibitemShut {NoStop}%
\bibitem [{\citenamefont {Rebentrost}\ \emph {et~al.}(2018)\citenamefont
  {Rebentrost}, \citenamefont {Gupt},\ and\ \citenamefont
  {Bromley}}]{rebentrost2018quantum}%
  \BibitemOpen
  \bibfield  {author} {\bibinfo {author} {\bibfnamefont {Patrick}\ \bibnamefont
  {Rebentrost}}, \bibinfo {author} {\bibfnamefont {Brajesh}\ \bibnamefont
  {Gupt}}, \ and\ \bibinfo {author} {\bibfnamefont {Thomas~R}\ \bibnamefont
  {Bromley}},\ }\bibfield  {title} {\enquote {\bibinfo {title} {Quantum
  computational finance: Monte carlo pricing of financial derivatives},}\
  }\href@noop {} {\bibfield  {journal} {\bibinfo  {journal} {Physical Review
  A}\ }\textbf {\bibinfo {volume} {98}},\ \bibinfo {pages} {022321} (\bibinfo
  {year} {2018})}\BibitemShut {NoStop}%
\bibitem [{\citenamefont {Woerner}\ and\ \citenamefont
  {Egger}(2019)}]{woerner2019quantum}%
  \BibitemOpen
  \bibfield  {author} {\bibinfo {author} {\bibfnamefont {Stefan}\ \bibnamefont
  {Woerner}}\ and\ \bibinfo {author} {\bibfnamefont {Daniel~J}\ \bibnamefont
  {Egger}},\ }\bibfield  {title} {\enquote {\bibinfo {title} {Quantum risk
  analysis},}\ }\href@noop {} {\bibfield  {journal} {\bibinfo  {journal} {npj
  Quantum Information}\ }\textbf {\bibinfo {volume} {5}},\ \bibinfo {pages}
  {1--8} (\bibinfo {year} {2019})}\BibitemShut {NoStop}%
\bibitem [{\citenamefont {Hensen}\ \emph {et~al.}(2015)\citenamefont {Hensen},
  \citenamefont {Bernien}, \citenamefont {Dr{\'e}au}, \citenamefont {Reiserer},
  \citenamefont {Kalb}, \citenamefont {Blok}, \citenamefont {Ruitenberg},
  \citenamefont {Vermeulen}, \citenamefont {Schouten}, \citenamefont
  {Abell{\'a}n}, \citenamefont {Amaya}, \citenamefont {Pruneri}, \citenamefont
  {Mitchell}, \citenamefont {Markham}, \citenamefont {Twitchen}, \citenamefont
  {Elkouss}, \citenamefont {Wehner}, \citenamefont {Taminiau},\ and\
  \citenamefont {Hanson}}]{Hensen2015}%
  \BibitemOpen
  \bibfield  {author} {\bibinfo {author} {\bibfnamefont {B.}~\bibnamefont
  {Hensen}}, \bibinfo {author} {\bibfnamefont {H.}~\bibnamefont {Bernien}},
  \bibinfo {author} {\bibfnamefont {A.~E.}\ \bibnamefont {Dr{\'e}au}}, \bibinfo
  {author} {\bibfnamefont {A.}~\bibnamefont {Reiserer}}, \bibinfo {author}
  {\bibfnamefont {N.}~\bibnamefont {Kalb}}, \bibinfo {author} {\bibfnamefont
  {M.~S.}\ \bibnamefont {Blok}}, \bibinfo {author} {\bibfnamefont
  {J.}~\bibnamefont {Ruitenberg}}, \bibinfo {author} {\bibfnamefont {R.~F.~L.}\
  \bibnamefont {Vermeulen}}, \bibinfo {author} {\bibfnamefont {R.~N.}\
  \bibnamefont {Schouten}}, \bibinfo {author} {\bibfnamefont {C.}~\bibnamefont
  {Abell{\'a}n}}, \bibinfo {author} {\bibfnamefont {W.}~\bibnamefont {Amaya}},
  \bibinfo {author} {\bibfnamefont {V.}~\bibnamefont {Pruneri}}, \bibinfo
  {author} {\bibfnamefont {M.~W.}\ \bibnamefont {Mitchell}}, \bibinfo {author}
  {\bibfnamefont {M.}~\bibnamefont {Markham}}, \bibinfo {author} {\bibfnamefont
  {D.~J.}\ \bibnamefont {Twitchen}}, \bibinfo {author} {\bibfnamefont
  {D.}~\bibnamefont {Elkouss}}, \bibinfo {author} {\bibfnamefont
  {S.}~\bibnamefont {Wehner}}, \bibinfo {author} {\bibfnamefont {T.~H.}\
  \bibnamefont {Taminiau}}, \ and\ \bibinfo {author} {\bibfnamefont
  {R.}~\bibnamefont {Hanson}},\ }\bibfield  {title} {\enquote {\bibinfo {title}
  {Loophole-free bell inequality violation using electron spins separated by
  1.3 kilometres},}\ }\href {\doibase 10.1038/nature15759} {\bibfield
  {journal} {\bibinfo  {journal} {Nature}\ }\textbf {\bibinfo {volume} {526}},\
  \bibinfo {pages} {682--686} (\bibinfo {year} {2015})}\BibitemShut {NoStop}%
\bibitem [{\citenamefont {Ren}\ \emph {et~al.}(2017)\citenamefont {Ren},
  \citenamefont {Xu}, \citenamefont {Yong}, \citenamefont {Zhang},
  \citenamefont {Liao}, \citenamefont {Yin}, \citenamefont {Liu}, \citenamefont
  {Cai}, \citenamefont {Yang}, \citenamefont {Li}, \citenamefont {Yang},
  \citenamefont {Han}, \citenamefont {Yao}, \citenamefont {Li}, \citenamefont
  {Wu}, \citenamefont {Wan}, \citenamefont {Liu}, \citenamefont {Liu},
  \citenamefont {Kuang}, \citenamefont {He}, \citenamefont {Shang},
  \citenamefont {Guo}, \citenamefont {Zheng}, \citenamefont {Tian},
  \citenamefont {Zhu}, \citenamefont {Liu}, \citenamefont {Lu}, \citenamefont
  {Shu}, \citenamefont {Chen}, \citenamefont {Peng}, \citenamefont {Wang},\
  and\ \citenamefont {Pan}}]{Ren2017}%
  \BibitemOpen
  \bibfield  {author} {\bibinfo {author} {\bibfnamefont {Ji-Gang}\ \bibnamefont
  {Ren}}, \bibinfo {author} {\bibfnamefont {Ping}\ \bibnamefont {Xu}}, \bibinfo
  {author} {\bibfnamefont {Hai-Lin}\ \bibnamefont {Yong}}, \bibinfo {author}
  {\bibfnamefont {Liang}\ \bibnamefont {Zhang}}, \bibinfo {author}
  {\bibfnamefont {Sheng-Kai}\ \bibnamefont {Liao}}, \bibinfo {author}
  {\bibfnamefont {Juan}\ \bibnamefont {Yin}}, \bibinfo {author} {\bibfnamefont
  {Wei-Yue}\ \bibnamefont {Liu}}, \bibinfo {author} {\bibfnamefont {Wen-Qi}\
  \bibnamefont {Cai}}, \bibinfo {author} {\bibfnamefont {Meng}\ \bibnamefont
  {Yang}}, \bibinfo {author} {\bibfnamefont {Li}~\bibnamefont {Li}}, \bibinfo
  {author} {\bibfnamefont {Kui-Xing}\ \bibnamefont {Yang}}, \bibinfo {author}
  {\bibfnamefont {Xuan}\ \bibnamefont {Han}}, \bibinfo {author} {\bibfnamefont
  {Yong-Qiang}\ \bibnamefont {Yao}}, \bibinfo {author} {\bibfnamefont
  {Ji}~\bibnamefont {Li}}, \bibinfo {author} {\bibfnamefont {Hai-Yan}\
  \bibnamefont {Wu}}, \bibinfo {author} {\bibfnamefont {Song}\ \bibnamefont
  {Wan}}, \bibinfo {author} {\bibfnamefont {Lei}\ \bibnamefont {Liu}}, \bibinfo
  {author} {\bibfnamefont {Ding-Quan}\ \bibnamefont {Liu}}, \bibinfo {author}
  {\bibfnamefont {Yao-Wu}\ \bibnamefont {Kuang}}, \bibinfo {author}
  {\bibfnamefont {Zhi-Ping}\ \bibnamefont {He}}, \bibinfo {author}
  {\bibfnamefont {Peng}\ \bibnamefont {Shang}}, \bibinfo {author}
  {\bibfnamefont {Cheng}\ \bibnamefont {Guo}}, \bibinfo {author} {\bibfnamefont
  {Ru-Hua}\ \bibnamefont {Zheng}}, \bibinfo {author} {\bibfnamefont {Kai}\
  \bibnamefont {Tian}}, \bibinfo {author} {\bibfnamefont {Zhen-Cai}\
  \bibnamefont {Zhu}}, \bibinfo {author} {\bibfnamefont {Nai-Le}\ \bibnamefont
  {Liu}}, \bibinfo {author} {\bibfnamefont {Chao-Yang}\ \bibnamefont {Lu}},
  \bibinfo {author} {\bibfnamefont {Rong}\ \bibnamefont {Shu}}, \bibinfo
  {author} {\bibfnamefont {Yu-Ao}\ \bibnamefont {Chen}}, \bibinfo {author}
  {\bibfnamefont {Cheng-Zhi}\ \bibnamefont {Peng}}, \bibinfo {author}
  {\bibfnamefont {Jian-Yu}\ \bibnamefont {Wang}}, \ and\ \bibinfo {author}
  {\bibfnamefont {Jian-Wei}\ \bibnamefont {Pan}},\ }\bibfield  {title}
  {\enquote {\bibinfo {title} {Ground-to-satellite quantum teleportation},}\
  }\href {\doibase 10.1038/nature23675} {\bibfield  {journal} {\bibinfo
  {journal} {Nature}\ }\textbf {\bibinfo {volume} {549}},\ \bibinfo {pages}
  {70--73} (\bibinfo {year} {2017})}\BibitemShut {NoStop}%
\bibitem [{\citenamefont {Chen}\ \emph
  {et~al.}(2021{\natexlab{a}})\citenamefont {Chen}, \citenamefont {Zhang},
  \citenamefont {Liu}, \citenamefont {Jiang}, \citenamefont {Zhang},
  \citenamefont {Han}, \citenamefont {Ma}, \citenamefont {Hu}, \citenamefont
  {Li}, \citenamefont {Liu}, \citenamefont {Zhou}, \citenamefont {Jiang},
  \citenamefont {Chen}, \citenamefont {Li}, \citenamefont {You}, \citenamefont
  {Wang}, \citenamefont {Wang}, \citenamefont {Zhang},\ and\ \citenamefont
  {Pan}}]{Chen2021_JWP}%
  \BibitemOpen
  \bibfield  {author} {\bibinfo {author} {\bibfnamefont {Jiu-Peng}\
  \bibnamefont {Chen}}, \bibinfo {author} {\bibfnamefont {Chi}\ \bibnamefont
  {Zhang}}, \bibinfo {author} {\bibfnamefont {Yang}\ \bibnamefont {Liu}},
  \bibinfo {author} {\bibfnamefont {Cong}\ \bibnamefont {Jiang}}, \bibinfo
  {author} {\bibfnamefont {Wei-Jun}\ \bibnamefont {Zhang}}, \bibinfo {author}
  {\bibfnamefont {Zhi-Yong}\ \bibnamefont {Han}}, \bibinfo {author}
  {\bibfnamefont {Shi-Zhao}\ \bibnamefont {Ma}}, \bibinfo {author}
  {\bibfnamefont {Xiao-Long}\ \bibnamefont {Hu}}, \bibinfo {author}
  {\bibfnamefont {Yu-Huai}\ \bibnamefont {Li}}, \bibinfo {author}
  {\bibfnamefont {Hui}\ \bibnamefont {Liu}}, \bibinfo {author} {\bibfnamefont
  {Fei}\ \bibnamefont {Zhou}}, \bibinfo {author} {\bibfnamefont {Hai-Feng}\
  \bibnamefont {Jiang}}, \bibinfo {author} {\bibfnamefont {Teng-Yun}\
  \bibnamefont {Chen}}, \bibinfo {author} {\bibfnamefont {Hao}\ \bibnamefont
  {Li}}, \bibinfo {author} {\bibfnamefont {Li-Xing}\ \bibnamefont {You}},
  \bibinfo {author} {\bibfnamefont {Zhen}\ \bibnamefont {Wang}}, \bibinfo
  {author} {\bibfnamefont {Xiang-Bin}\ \bibnamefont {Wang}}, \bibinfo {author}
  {\bibfnamefont {Qiang}\ \bibnamefont {Zhang}}, \ and\ \bibinfo {author}
  {\bibfnamefont {Jian-Wei}\ \bibnamefont {Pan}},\ }\bibfield  {title}
  {\enquote {\bibinfo {title} {Twin-field quantum key distribution over a 511
  km optical fibre linking two distant metropolitan areas},}\ }\href {\doibase
  10.1038/s41566-021-00828-5} {\bibfield  {journal} {\bibinfo  {journal}
  {Nature Photonics}\ }\textbf {\bibinfo {volume} {15}},\ \bibinfo {pages}
  {570--575} (\bibinfo {year} {2021}{\natexlab{a}})}\BibitemShut {NoStop}%
\bibitem [{\citenamefont {Liu}\ \emph {et~al.}(2023{\natexlab{a}})\citenamefont
  {Liu}, \citenamefont {Zhang}, \citenamefont {Jiang}, \citenamefont {Chen},
  \citenamefont {Zhang}, \citenamefont {Pan}, \citenamefont {Ma}, \citenamefont
  {Dong}, \citenamefont {Xiong}, \citenamefont {Zhang}, \citenamefont {Li},
  \citenamefont {Wang}, \citenamefont {Wu}, \citenamefont {Chen}, \citenamefont
  {You}, \citenamefont {Wang}, \citenamefont {Zhang},\ and\ \citenamefont
  {Pan}}]{Liu2023_JWP}%
  \BibitemOpen
  \bibfield  {author} {\bibinfo {author} {\bibfnamefont {Yang}\ \bibnamefont
  {Liu}}, \bibinfo {author} {\bibfnamefont {Wei-Jun}\ \bibnamefont {Zhang}},
  \bibinfo {author} {\bibfnamefont {Cong}\ \bibnamefont {Jiang}}, \bibinfo
  {author} {\bibfnamefont {Jiu-Peng}\ \bibnamefont {Chen}}, \bibinfo {author}
  {\bibfnamefont {Chi}\ \bibnamefont {Zhang}}, \bibinfo {author} {\bibfnamefont
  {Wen-Xin}\ \bibnamefont {Pan}}, \bibinfo {author} {\bibfnamefont
  {Di}~\bibnamefont {Ma}}, \bibinfo {author} {\bibfnamefont {Hao}\ \bibnamefont
  {Dong}}, \bibinfo {author} {\bibfnamefont {Jia-Min}\ \bibnamefont {Xiong}},
  \bibinfo {author} {\bibfnamefont {Cheng-Jun}\ \bibnamefont {Zhang}}, \bibinfo
  {author} {\bibfnamefont {Hao}\ \bibnamefont {Li}}, \bibinfo {author}
  {\bibfnamefont {Rui-Chun}\ \bibnamefont {Wang}}, \bibinfo {author}
  {\bibfnamefont {Jun}\ \bibnamefont {Wu}}, \bibinfo {author} {\bibfnamefont
  {Teng-Yun}\ \bibnamefont {Chen}}, \bibinfo {author} {\bibfnamefont {Lixing}\
  \bibnamefont {You}}, \bibinfo {author} {\bibfnamefont {Xiang-Bin}\
  \bibnamefont {Wang}}, \bibinfo {author} {\bibfnamefont {Qiang}\ \bibnamefont
  {Zhang}}, \ and\ \bibinfo {author} {\bibfnamefont {Jian-Wei}\ \bibnamefont
  {Pan}},\ }\bibfield  {title} {\enquote {\bibinfo {title} {Experimental
  twin-field quantum key distribution over 1000 km fiber distance},}\ }\href
  {\doibase 10.1103/PhysRevLett.130.210801} {\bibfield  {journal} {\bibinfo
  {journal} {Phys. Rev. Lett.}\ }\textbf {\bibinfo {volume} {130}},\ \bibinfo
  {pages} {210801} (\bibinfo {year} {2023}{\natexlab{a}})}\BibitemShut
  {NoStop}%
\bibitem [{\citenamefont {Bennett}\ and\ \citenamefont
  {Brassard}(1984)}]{BB84}%
  \BibitemOpen
  \bibfield  {author} {\bibinfo {author} {\bibfnamefont {C.~H.}\ \bibnamefont
  {Bennett}}\ and\ \bibinfo {author} {\bibfnamefont {G.}~\bibnamefont
  {Brassard}},\ }\bibfield  {title} {\enquote {\bibinfo {title} {{Quantum
  cryptography: Public key distribution and coin tossing}},}\ }in\ \href@noop
  {} {\emph {\bibinfo {booktitle} {Proceedings of IEEE International Conference
  on Computers, Systems, and Signal Processing}}}\ (\bibinfo {address}
  {India},\ \bibinfo {year} {1984})\ p.\ \bibinfo {pages} {175}\BibitemShut
  {NoStop}%
\bibitem [{\citenamefont {Gisin}\ \emph
  {et~al.}(2002{\natexlab{b}})\citenamefont {Gisin}, \citenamefont {Ribordy},
  \citenamefont {Tittel},\ and\ \citenamefont {Zbinden}}]{Gisin2002}%
  \BibitemOpen
  \bibfield  {author} {\bibinfo {author} {\bibfnamefont {Nicolas}\ \bibnamefont
  {Gisin}}, \bibinfo {author} {\bibfnamefont {Gr\'egoire}\ \bibnamefont
  {Ribordy}}, \bibinfo {author} {\bibfnamefont {Wolfgang}\ \bibnamefont
  {Tittel}}, \ and\ \bibinfo {author} {\bibfnamefont {Hugo}\ \bibnamefont
  {Zbinden}},\ }\bibfield  {title} {\enquote {\bibinfo {title} {Quantum
  cryptography},}\ }\href {\doibase 10.1103/RevModPhys.74.145} {\bibfield
  {journal} {\bibinfo  {journal} {Rev. Mod. Phys.}\ }\textbf {\bibinfo {volume}
  {74}},\ \bibinfo {pages} {145--195} (\bibinfo {year}
  {2002}{\natexlab{b}})}\BibitemShut {NoStop}%
\bibitem [{\citenamefont {Xu}\ \emph {et~al.}(2020)\citenamefont {Xu},
  \citenamefont {Ma}, \citenamefont {Zhang}, \citenamefont {Lo},\ and\
  \citenamefont {Pan}}]{Xu2020}%
  \BibitemOpen
  \bibfield  {author} {\bibinfo {author} {\bibfnamefont {Feihu}\ \bibnamefont
  {Xu}}, \bibinfo {author} {\bibfnamefont {Xiongfeng}\ \bibnamefont {Ma}},
  \bibinfo {author} {\bibfnamefont {Qiang}\ \bibnamefont {Zhang}}, \bibinfo
  {author} {\bibfnamefont {Hoi-Kwong}\ \bibnamefont {Lo}}, \ and\ \bibinfo
  {author} {\bibfnamefont {Jian-Wei}\ \bibnamefont {Pan}},\ }\bibfield  {title}
  {\enquote {\bibinfo {title} {Secure quantum key distribution with realistic
  devices},}\ }\href {\doibase 10.1103/RevModPhys.92.025002} {\bibfield
  {journal} {\bibinfo  {journal} {Rev. Mod. Phys.}\ }\textbf {\bibinfo {volume}
  {92}},\ \bibinfo {pages} {025002} (\bibinfo {year} {2020})}\BibitemShut
  {NoStop}%
\bibitem [{\citenamefont {Portmann}\ and\ \citenamefont
  {Renner}(2022)}]{Portmann2022}%
  \BibitemOpen
  \bibfield  {author} {\bibinfo {author} {\bibfnamefont {Christopher}\
  \bibnamefont {Portmann}}\ and\ \bibinfo {author} {\bibfnamefont {Renato}\
  \bibnamefont {Renner}},\ }\bibfield  {title} {\enquote {\bibinfo {title}
  {Security in quantum cryptography},}\ }\href {\doibase
  10.1103/RevModPhys.94.025008} {\bibfield  {journal} {\bibinfo  {journal}
  {Rev. Mod. Phys.}\ }\textbf {\bibinfo {volume} {94}},\ \bibinfo {pages}
  {025008} (\bibinfo {year} {2022})}\BibitemShut {NoStop}%
\bibitem [{\citenamefont {Kimble}(2008)}]{Kimble2008}%
  \BibitemOpen
  \bibfield  {author} {\bibinfo {author} {\bibfnamefont {H.~J.}\ \bibnamefont
  {Kimble}},\ }\bibfield  {title} {\enquote {\bibinfo {title} {The quantum
  internet},}\ }\href {\doibase 10.1038/nature07127} {\bibfield  {journal}
  {\bibinfo  {journal} {Nature}\ }\textbf {\bibinfo {volume} {453}},\ \bibinfo
  {pages} {1023--1030} (\bibinfo {year} {2008})}\BibitemShut {NoStop}%
\bibitem [{\citenamefont {Cacciapuoti}\ \emph
  {et~al.}(2020{\natexlab{a}})\citenamefont {Cacciapuoti}, \citenamefont
  {Caleffi}, \citenamefont {Tafuri}, \citenamefont {Cataliotti}, \citenamefont
  {Gherardini},\ and\ \citenamefont {Bianchi}}]{Cacciapuoti2020}%
  \BibitemOpen
  \bibfield  {author} {\bibinfo {author} {\bibfnamefont {Angela~Sara}\
  \bibnamefont {Cacciapuoti}}, \bibinfo {author} {\bibfnamefont {Marcello}\
  \bibnamefont {Caleffi}}, \bibinfo {author} {\bibfnamefont {Francesco}\
  \bibnamefont {Tafuri}}, \bibinfo {author} {\bibfnamefont {Francesco~Saverio}\
  \bibnamefont {Cataliotti}}, \bibinfo {author} {\bibfnamefont {Stefano}\
  \bibnamefont {Gherardini}}, \ and\ \bibinfo {author} {\bibfnamefont
  {Giuseppe}\ \bibnamefont {Bianchi}},\ }\bibfield  {title} {\enquote {\bibinfo
  {title} {Quantum internet: Networking challenges in distributed quantum
  computing},}\ }\href {\doibase 10.1109/MNET.001.1900092} {\bibfield
  {journal} {\bibinfo  {journal} {IEEE Network}\ }\textbf {\bibinfo {volume}
  {34}},\ \bibinfo {pages} {137--143} (\bibinfo {year}
  {2020}{\natexlab{a}})}\BibitemShut {NoStop}%
\bibitem [{\citenamefont {Cacciapuoti}\ \emph
  {et~al.}(2020{\natexlab{b}})\citenamefont {Cacciapuoti}, \citenamefont
  {Caleffi}, \citenamefont {Van~Meter},\ and\ \citenamefont
  {Hanzo}}]{Cacciapuoti2020_v2}%
  \BibitemOpen
  \bibfield  {author} {\bibinfo {author} {\bibfnamefont {Angela~Sara}\
  \bibnamefont {Cacciapuoti}}, \bibinfo {author} {\bibfnamefont {Marcello}\
  \bibnamefont {Caleffi}}, \bibinfo {author} {\bibfnamefont {Rodney}\
  \bibnamefont {Van~Meter}}, \ and\ \bibinfo {author} {\bibfnamefont {Lajos}\
  \bibnamefont {Hanzo}},\ }\bibfield  {title} {\enquote {\bibinfo {title} {When
  entanglement meets classical communications: Quantum teleportation for the
  quantum internet},}\ }\href {\doibase 10.1109/TCOMM.2020.2978071} {\bibfield
  {journal} {\bibinfo  {journal} {IEEE Transactions on Communications}\
  }\textbf {\bibinfo {volume} {68}},\ \bibinfo {pages} {3808--3833} (\bibinfo
  {year} {2020}{\natexlab{b}})}\BibitemShut {NoStop}%
\bibitem [{\citenamefont {Wei}\ \emph {et~al.}(2022)\citenamefont {Wei},
  \citenamefont {Jing}, \citenamefont {Zhang}, \citenamefont {Liao},
  \citenamefont {Yuan}, \citenamefont {Fan}, \citenamefont {Lyu}, \citenamefont
  {Zhou}, \citenamefont {Wang}, \citenamefont {Deng}, \citenamefont {Song},
  \citenamefont {Oblak}, \citenamefont {Guo},\ and\ \citenamefont
  {Zhou}}]{Wei2022}%
  \BibitemOpen
  \bibfield  {author} {\bibinfo {author} {\bibfnamefont {Shi-Hai}\ \bibnamefont
  {Wei}}, \bibinfo {author} {\bibfnamefont {Bo}~\bibnamefont {Jing}}, \bibinfo
  {author} {\bibfnamefont {Xue-Ying}\ \bibnamefont {Zhang}}, \bibinfo {author}
  {\bibfnamefont {Jin-Yu}\ \bibnamefont {Liao}}, \bibinfo {author}
  {\bibfnamefont {Chen-Zhi}\ \bibnamefont {Yuan}}, \bibinfo {author}
  {\bibfnamefont {Bo-Yu}\ \bibnamefont {Fan}}, \bibinfo {author} {\bibfnamefont
  {Chen}\ \bibnamefont {Lyu}}, \bibinfo {author} {\bibfnamefont {Dian-Li}\
  \bibnamefont {Zhou}}, \bibinfo {author} {\bibfnamefont {You}\ \bibnamefont
  {Wang}}, \bibinfo {author} {\bibfnamefont {Guang-Wei}\ \bibnamefont {Deng}},
  \bibinfo {author} {\bibfnamefont {Hai-Zhi}\ \bibnamefont {Song}}, \bibinfo
  {author} {\bibfnamefont {Daniel}\ \bibnamefont {Oblak}}, \bibinfo {author}
  {\bibfnamefont {Guang-Can}\ \bibnamefont {Guo}}, \ and\ \bibinfo {author}
  {\bibfnamefont {Qiang}\ \bibnamefont {Zhou}},\ }\bibfield  {title} {\enquote
  {\bibinfo {title} {Towards real-world quantum networks: A review},}\ }\href
  {\doibase https://doi.org/10.1002/lpor.202100219} {\bibfield  {journal}
  {\bibinfo  {journal} {Laser \& Photonics Reviews}\ }\textbf {\bibinfo
  {volume} {16}},\ \bibinfo {pages} {2100219} (\bibinfo {year}
  {2022})}\BibitemShut {NoStop}%
\bibitem [{\citenamefont {Illiano}\ \emph {et~al.}(2022)\citenamefont
  {Illiano}, \citenamefont {Caleffi}, \citenamefont {Manzalini},\ and\
  \citenamefont {Cacciapuoti}}]{Illiano2022}%
  \BibitemOpen
  \bibfield  {author} {\bibinfo {author} {\bibfnamefont {Jessica}\ \bibnamefont
  {Illiano}}, \bibinfo {author} {\bibfnamefont {Marcello}\ \bibnamefont
  {Caleffi}}, \bibinfo {author} {\bibfnamefont {Antonio}\ \bibnamefont
  {Manzalini}}, \ and\ \bibinfo {author} {\bibfnamefont {Angela~Sara}\
  \bibnamefont {Cacciapuoti}},\ }\bibfield  {title} {\enquote {\bibinfo {title}
  {Quantum internet protocol stack: A comprehensive survey},}\ }\href {\doibase
  https://doi.org/10.1016/j.comnet.2022.109092} {\bibfield  {journal} {\bibinfo
   {journal} {Computer Networks}\ }\textbf {\bibinfo {volume} {213}},\ \bibinfo
  {pages} {109092} (\bibinfo {year} {2022})}\BibitemShut {NoStop}%
\bibitem [{\citenamefont {Caleffi}\ \emph {et~al.}(2022)\citenamefont
  {Caleffi}, \citenamefont {Amoretti}, \citenamefont {Ferrari}, \citenamefont
  {Cuomo}, \citenamefont {Illiano}, \citenamefont {Manzalini},\ and\
  \citenamefont {Cacciapuoti}}]{caleffi2022distributed}%
  \BibitemOpen
  \bibfield  {author} {\bibinfo {author} {\bibfnamefont {Marcello}\
  \bibnamefont {Caleffi}}, \bibinfo {author} {\bibfnamefont {Michele}\
  \bibnamefont {Amoretti}}, \bibinfo {author} {\bibfnamefont {Davide}\
  \bibnamefont {Ferrari}}, \bibinfo {author} {\bibfnamefont {Daniele}\
  \bibnamefont {Cuomo}}, \bibinfo {author} {\bibfnamefont {Jessica}\
  \bibnamefont {Illiano}}, \bibinfo {author} {\bibfnamefont {Antonio}\
  \bibnamefont {Manzalini}}, \ and\ \bibinfo {author} {\bibfnamefont
  {Angela~Sara}\ \bibnamefont {Cacciapuoti}},\ }\href@noop {} {\enquote
  {\bibinfo {title} {Distributed quantum computing: a survey},}\ } (\bibinfo
  {year} {2022}),\ \Eprint {http://arxiv.org/abs/2212.10609} {arXiv:2212.10609
  [quant-ph]} \BibitemShut {NoStop}%
\bibitem [{\citenamefont {Chow}\ \emph {et~al.}(2021)\citenamefont {Chow},
  \citenamefont {Dial},\ and\ \citenamefont {Gambetta}}]{IBMQ2021}%
  \BibitemOpen
  \bibfield  {author} {\bibinfo {author} {\bibfnamefont {Jerry}\ \bibnamefont
  {Chow}}, \bibinfo {author} {\bibfnamefont {Oliver}\ \bibnamefont {Dial}}, \
  and\ \bibinfo {author} {\bibfnamefont {Jay}\ \bibnamefont {Gambetta}},\
  }\href {https://research.ibm.com/blog/127-qubit-quantum-processor-eagle}
  {\enquote {\bibinfo {title} {{IBM Quantum breaks the 100‑qubit processor
  barrier}},}\ } (\bibinfo {year} {2021})\BibitemShut {NoStop}%
\bibitem [{IBM(2022)}]{IBMQ2022}%
  \BibitemOpen
  \href
  {https://newsroom.ibm.com/2022-11-09-IBM-Unveils-400-Qubit-Plus-Quantum-Processor-and-Next-Generation-IBM-Quantum-System-Two}
  {\enquote {\bibinfo {title} {{IBM Unveils 400 Qubit-Plus Quantum Processor
  and Next-Generation IBM Quantum System Two}},}\ } (\bibinfo {year}
  {2022})\BibitemShut {NoStop}%
\bibitem [{\citenamefont {Arute}\ \emph {et~al.}(2019)\citenamefont {Arute},
  \citenamefont {Arya}, \citenamefont {Babbush}, \citenamefont {Bacon},
  \citenamefont {Bardin}, \citenamefont {Barends}, \citenamefont {Biswas},
  \citenamefont {Boixo}, \citenamefont {Brandao}, \citenamefont {Buell},
  \citenamefont {Burkett}, \citenamefont {Chen}, \citenamefont {Chen},
  \citenamefont {Chiaro}, \citenamefont {Collins}, \citenamefont {Courtney},
  \citenamefont {Dunsworth}, \citenamefont {Farhi}, \citenamefont {Foxen},
  \citenamefont {Fowler}, \citenamefont {Gidney}, \citenamefont {Giustina},
  \citenamefont {Graff}, \citenamefont {Guerin}, \citenamefont {Habegger},
  \citenamefont {Harrigan}, \citenamefont {Hartmann}, \citenamefont {Ho},
  \citenamefont {Hoffmann}, \citenamefont {Huang}, \citenamefont {Humble},
  \citenamefont {Isakov}, \citenamefont {Jeffrey}, \citenamefont {Jiang},
  \citenamefont {Kafri}, \citenamefont {Kechedzhi}, \citenamefont {Kelly},
  \citenamefont {Klimov}, \citenamefont {Knysh}, \citenamefont {Korotkov},
  \citenamefont {Kostritsa}, \citenamefont {Landhuis}, \citenamefont
  {Lindmark}, \citenamefont {Lucero}, \citenamefont {Lyakh}, \citenamefont
  {Mandr{\`a}}, \citenamefont {McClean}, \citenamefont {McEwen}, \citenamefont
  {Megrant}, \citenamefont {Mi}, \citenamefont {Michielsen}, \citenamefont
  {Mohseni}, \citenamefont {Mutus}, \citenamefont {Naaman}, \citenamefont
  {Neeley}, \citenamefont {Neill}, \citenamefont {Niu}, \citenamefont {Ostby},
  \citenamefont {Petukhov}, \citenamefont {Platt}, \citenamefont {Quintana},
  \citenamefont {Rieffel}, \citenamefont {Roushan}, \citenamefont {Rubin},
  \citenamefont {Sank}, \citenamefont {Satzinger}, \citenamefont {Smelyanskiy},
  \citenamefont {Sung}, \citenamefont {Trevithick}, \citenamefont
  {Vainsencher}, \citenamefont {Villalonga}, \citenamefont {White},
  \citenamefont {Yao}, \citenamefont {Yeh}, \citenamefont {Zalcman},
  \citenamefont {Neven},\ and\ \citenamefont {Martinis}}]{Arute2019}%
  \BibitemOpen
  \bibfield  {author} {\bibinfo {author} {\bibfnamefont {Frank}\ \bibnamefont
  {Arute}}, \bibinfo {author} {\bibfnamefont {Kunal}\ \bibnamefont {Arya}},
  \bibinfo {author} {\bibfnamefont {Ryan}\ \bibnamefont {Babbush}}, \bibinfo
  {author} {\bibfnamefont {Dave}\ \bibnamefont {Bacon}}, \bibinfo {author}
  {\bibfnamefont {Joseph~C.}\ \bibnamefont {Bardin}}, \bibinfo {author}
  {\bibfnamefont {Rami}\ \bibnamefont {Barends}}, \bibinfo {author}
  {\bibfnamefont {Rupak}\ \bibnamefont {Biswas}}, \bibinfo {author}
  {\bibfnamefont {Sergio}\ \bibnamefont {Boixo}}, \bibinfo {author}
  {\bibfnamefont {Fernando G. S.~L.}\ \bibnamefont {Brandao}}, \bibinfo
  {author} {\bibfnamefont {David~A.}\ \bibnamefont {Buell}}, \bibinfo {author}
  {\bibfnamefont {Brian}\ \bibnamefont {Burkett}}, \bibinfo {author}
  {\bibfnamefont {Yu}~\bibnamefont {Chen}}, \bibinfo {author} {\bibfnamefont
  {Zijun}\ \bibnamefont {Chen}}, \bibinfo {author} {\bibfnamefont {Ben}\
  \bibnamefont {Chiaro}}, \bibinfo {author} {\bibfnamefont {Roberto}\
  \bibnamefont {Collins}}, \bibinfo {author} {\bibfnamefont {William}\
  \bibnamefont {Courtney}}, \bibinfo {author} {\bibfnamefont {Andrew}\
  \bibnamefont {Dunsworth}}, \bibinfo {author} {\bibfnamefont {Edward}\
  \bibnamefont {Farhi}}, \bibinfo {author} {\bibfnamefont {Brooks}\
  \bibnamefont {Foxen}}, \bibinfo {author} {\bibfnamefont {Austin}\
  \bibnamefont {Fowler}}, \bibinfo {author} {\bibfnamefont {Craig}\
  \bibnamefont {Gidney}}, \bibinfo {author} {\bibfnamefont {Marissa}\
  \bibnamefont {Giustina}}, \bibinfo {author} {\bibfnamefont {Rob}\
  \bibnamefont {Graff}}, \bibinfo {author} {\bibfnamefont {Keith}\ \bibnamefont
  {Guerin}}, \bibinfo {author} {\bibfnamefont {Steve}\ \bibnamefont
  {Habegger}}, \bibinfo {author} {\bibfnamefont {Matthew~P.}\ \bibnamefont
  {Harrigan}}, \bibinfo {author} {\bibfnamefont {Michael~J.}\ \bibnamefont
  {Hartmann}}, \bibinfo {author} {\bibfnamefont {Alan}\ \bibnamefont {Ho}},
  \bibinfo {author} {\bibfnamefont {Markus}\ \bibnamefont {Hoffmann}}, \bibinfo
  {author} {\bibfnamefont {Trent}\ \bibnamefont {Huang}}, \bibinfo {author}
  {\bibfnamefont {Travis~S.}\ \bibnamefont {Humble}}, \bibinfo {author}
  {\bibfnamefont {Sergei~V.}\ \bibnamefont {Isakov}}, \bibinfo {author}
  {\bibfnamefont {Evan}\ \bibnamefont {Jeffrey}}, \bibinfo {author}
  {\bibfnamefont {Zhang}\ \bibnamefont {Jiang}}, \bibinfo {author}
  {\bibfnamefont {Dvir}\ \bibnamefont {Kafri}}, \bibinfo {author}
  {\bibfnamefont {Kostyantyn}\ \bibnamefont {Kechedzhi}}, \bibinfo {author}
  {\bibfnamefont {Julian}\ \bibnamefont {Kelly}}, \bibinfo {author}
  {\bibfnamefont {Paul~V.}\ \bibnamefont {Klimov}}, \bibinfo {author}
  {\bibfnamefont {Sergey}\ \bibnamefont {Knysh}}, \bibinfo {author}
  {\bibfnamefont {Alexander}\ \bibnamefont {Korotkov}}, \bibinfo {author}
  {\bibfnamefont {Fedor}\ \bibnamefont {Kostritsa}}, \bibinfo {author}
  {\bibfnamefont {David}\ \bibnamefont {Landhuis}}, \bibinfo {author}
  {\bibfnamefont {Mike}\ \bibnamefont {Lindmark}}, \bibinfo {author}
  {\bibfnamefont {Erik}\ \bibnamefont {Lucero}}, \bibinfo {author}
  {\bibfnamefont {Dmitry}\ \bibnamefont {Lyakh}}, \bibinfo {author}
  {\bibfnamefont {Salvatore}\ \bibnamefont {Mandr{\`a}}}, \bibinfo {author}
  {\bibfnamefont {Jarrod~R.}\ \bibnamefont {McClean}}, \bibinfo {author}
  {\bibfnamefont {Matthew}\ \bibnamefont {McEwen}}, \bibinfo {author}
  {\bibfnamefont {Anthony}\ \bibnamefont {Megrant}}, \bibinfo {author}
  {\bibfnamefont {Xiao}\ \bibnamefont {Mi}}, \bibinfo {author} {\bibfnamefont
  {Kristel}\ \bibnamefont {Michielsen}}, \bibinfo {author} {\bibfnamefont
  {Masoud}\ \bibnamefont {Mohseni}}, \bibinfo {author} {\bibfnamefont {Josh}\
  \bibnamefont {Mutus}}, \bibinfo {author} {\bibfnamefont {Ofer}\ \bibnamefont
  {Naaman}}, \bibinfo {author} {\bibfnamefont {Matthew}\ \bibnamefont
  {Neeley}}, \bibinfo {author} {\bibfnamefont {Charles}\ \bibnamefont {Neill}},
  \bibinfo {author} {\bibfnamefont {Murphy~Yuezhen}\ \bibnamefont {Niu}},
  \bibinfo {author} {\bibfnamefont {Eric}\ \bibnamefont {Ostby}}, \bibinfo
  {author} {\bibfnamefont {Andre}\ \bibnamefont {Petukhov}}, \bibinfo {author}
  {\bibfnamefont {John~C.}\ \bibnamefont {Platt}}, \bibinfo {author}
  {\bibfnamefont {Chris}\ \bibnamefont {Quintana}}, \bibinfo {author}
  {\bibfnamefont {Eleanor~G.}\ \bibnamefont {Rieffel}}, \bibinfo {author}
  {\bibfnamefont {Pedram}\ \bibnamefont {Roushan}}, \bibinfo {author}
  {\bibfnamefont {Nicholas~C.}\ \bibnamefont {Rubin}}, \bibinfo {author}
  {\bibfnamefont {Daniel}\ \bibnamefont {Sank}}, \bibinfo {author}
  {\bibfnamefont {Kevin~J.}\ \bibnamefont {Satzinger}}, \bibinfo {author}
  {\bibfnamefont {Vadim}\ \bibnamefont {Smelyanskiy}}, \bibinfo {author}
  {\bibfnamefont {Kevin~J.}\ \bibnamefont {Sung}}, \bibinfo {author}
  {\bibfnamefont {Matthew~D.}\ \bibnamefont {Trevithick}}, \bibinfo {author}
  {\bibfnamefont {Amit}\ \bibnamefont {Vainsencher}}, \bibinfo {author}
  {\bibfnamefont {Benjamin}\ \bibnamefont {Villalonga}}, \bibinfo {author}
  {\bibfnamefont {Theodore}\ \bibnamefont {White}}, \bibinfo {author}
  {\bibfnamefont {Z.~Jamie}\ \bibnamefont {Yao}}, \bibinfo {author}
  {\bibfnamefont {Ping}\ \bibnamefont {Yeh}}, \bibinfo {author} {\bibfnamefont
  {Adam}\ \bibnamefont {Zalcman}}, \bibinfo {author} {\bibfnamefont {Hartmut}\
  \bibnamefont {Neven}}, \ and\ \bibinfo {author} {\bibfnamefont {John~M.}\
  \bibnamefont {Martinis}},\ }\bibfield  {title} {\enquote {\bibinfo {title}
  {Quantum supremacy using a programmable superconducting processor},}\ }\href
  {\doibase 10.1038/s41586-019-1666-5} {\bibfield  {journal} {\bibinfo
  {journal} {Nature}\ }\textbf {\bibinfo {volume} {574}},\ \bibinfo {pages}
  {505--510} (\bibinfo {year} {2019})}\BibitemShut {NoStop}%
\bibitem [{\citenamefont {Moses}\ \emph {et~al.}(2023)\citenamefont {Moses},
  \citenamefont {Baldwin}, \citenamefont {Allman}, \citenamefont {Ancona},
  \citenamefont {Ascarrunz}, \citenamefont {Barnes}, \citenamefont
  {Bartolotta}, \citenamefont {Bjork}, \citenamefont {Blanchard}, \citenamefont
  {Bohn}, \citenamefont {Bohnet}, \citenamefont {Brown}, \citenamefont
  {Burdick}, \citenamefont {Burton}, \citenamefont {Campbell}, \citenamefont
  {au2}, \citenamefont {Carron}, \citenamefont {Chambers}, \citenamefont
  {Chan}, \citenamefont {Chen}, \citenamefont {Chernoguzov}, \citenamefont
  {Chertkov}, \citenamefont {Colina}, \citenamefont {Curtis}, \citenamefont
  {Daniel}, \citenamefont {DeCross}, \citenamefont {Deen}, \citenamefont
  {Delaney}, \citenamefont {Dreiling}, \citenamefont {Ertsgaard}, \citenamefont
  {Esposito}, \citenamefont {Estey}, \citenamefont {Fabrikant}, \citenamefont
  {Figgatt}, \citenamefont {Foltz}, \citenamefont {Foss-Feig}, \citenamefont
  {Francois}, \citenamefont {Gaebler}, \citenamefont {Gatterman}, \citenamefont
  {Gilbreth}, \citenamefont {Giles}, \citenamefont {Glynn}, \citenamefont
  {Hall}, \citenamefont {Hankin}, \citenamefont {Hansen}, \citenamefont
  {Hayes}, \citenamefont {Higashi}, \citenamefont {Hoffman}, \citenamefont
  {Horning}, \citenamefont {Hout}, \citenamefont {Jacobs}, \citenamefont
  {Johansen}, \citenamefont {Jones}, \citenamefont {Karcz}, \citenamefont
  {Klein}, \citenamefont {Lauria}, \citenamefont {Lee}, \citenamefont {Liefer},
  \citenamefont {Lytle}, \citenamefont {Lu}, \citenamefont {Lucchetti},
  \citenamefont {Malm}, \citenamefont {Matheny}, \citenamefont {Mathewson},
  \citenamefont {Mayer}, \citenamefont {Miller}, \citenamefont {Mills},
  \citenamefont {Neyenhuis}, \citenamefont {Nugent}, \citenamefont {Olson},
  \citenamefont {Parks}, \citenamefont {Price}, \citenamefont {Price},
  \citenamefont {Pugh}, \citenamefont {Ransford}, \citenamefont {Reed},
  \citenamefont {Roman}, \citenamefont {Rowe}, \citenamefont {Ryan-Anderson},
  \citenamefont {Sanders}, \citenamefont {Sedlacek}, \citenamefont {Shevchuk},
  \citenamefont {Siegfried}, \citenamefont {Skripka}, \citenamefont {Spaun},
  \citenamefont {Sprenkle}, \citenamefont {Stutz}, \citenamefont {Swallows},
  \citenamefont {Tobey}, \citenamefont {Tran}, \citenamefont {Tran},
  \citenamefont {Vogt}, \citenamefont {Volin}, \citenamefont {Walker},
  \citenamefont {Zolot},\ and\ \citenamefont {Pino}}]{Moses2023}%
  \BibitemOpen
  \bibfield  {author} {\bibinfo {author} {\bibfnamefont {S.~A.}\ \bibnamefont
  {Moses}}, \bibinfo {author} {\bibfnamefont {C.~H.}\ \bibnamefont {Baldwin}},
  \bibinfo {author} {\bibfnamefont {M.~S.}\ \bibnamefont {Allman}}, \bibinfo
  {author} {\bibfnamefont {R.}~\bibnamefont {Ancona}}, \bibinfo {author}
  {\bibfnamefont {L.}~\bibnamefont {Ascarrunz}}, \bibinfo {author}
  {\bibfnamefont {C.}~\bibnamefont {Barnes}}, \bibinfo {author} {\bibfnamefont
  {J.}~\bibnamefont {Bartolotta}}, \bibinfo {author} {\bibfnamefont
  {B.}~\bibnamefont {Bjork}}, \bibinfo {author} {\bibfnamefont
  {P.}~\bibnamefont {Blanchard}}, \bibinfo {author} {\bibfnamefont
  {M.}~\bibnamefont {Bohn}}, \bibinfo {author} {\bibfnamefont {J.~G.}\
  \bibnamefont {Bohnet}}, \bibinfo {author} {\bibfnamefont {N.~C.}\
  \bibnamefont {Brown}}, \bibinfo {author} {\bibfnamefont {N.~Q.}\ \bibnamefont
  {Burdick}}, \bibinfo {author} {\bibfnamefont {W.~C.}\ \bibnamefont {Burton}},
  \bibinfo {author} {\bibfnamefont {S.~L.}\ \bibnamefont {Campbell}}, \bibinfo
  {author} {\bibfnamefont {J.~P. Campora~III}\ \bibnamefont {au2}}, \bibinfo
  {author} {\bibfnamefont {C.}~\bibnamefont {Carron}}, \bibinfo {author}
  {\bibfnamefont {J.}~\bibnamefont {Chambers}}, \bibinfo {author}
  {\bibfnamefont {J.~W.}\ \bibnamefont {Chan}}, \bibinfo {author}
  {\bibfnamefont {Y.~H.}\ \bibnamefont {Chen}}, \bibinfo {author}
  {\bibfnamefont {A.}~\bibnamefont {Chernoguzov}}, \bibinfo {author}
  {\bibfnamefont {E.}~\bibnamefont {Chertkov}}, \bibinfo {author}
  {\bibfnamefont {J.}~\bibnamefont {Colina}}, \bibinfo {author} {\bibfnamefont
  {J.~P.}\ \bibnamefont {Curtis}}, \bibinfo {author} {\bibfnamefont
  {R.}~\bibnamefont {Daniel}}, \bibinfo {author} {\bibfnamefont
  {M.}~\bibnamefont {DeCross}}, \bibinfo {author} {\bibfnamefont
  {D.}~\bibnamefont {Deen}}, \bibinfo {author} {\bibfnamefont {C.}~\bibnamefont
  {Delaney}}, \bibinfo {author} {\bibfnamefont {J.~M.}\ \bibnamefont
  {Dreiling}}, \bibinfo {author} {\bibfnamefont {C.~T.}\ \bibnamefont
  {Ertsgaard}}, \bibinfo {author} {\bibfnamefont {J.}~\bibnamefont {Esposito}},
  \bibinfo {author} {\bibfnamefont {B.}~\bibnamefont {Estey}}, \bibinfo
  {author} {\bibfnamefont {M.}~\bibnamefont {Fabrikant}}, \bibinfo {author}
  {\bibfnamefont {C.}~\bibnamefont {Figgatt}}, \bibinfo {author} {\bibfnamefont
  {C.}~\bibnamefont {Foltz}}, \bibinfo {author} {\bibfnamefont
  {M.}~\bibnamefont {Foss-Feig}}, \bibinfo {author} {\bibfnamefont
  {D.}~\bibnamefont {Francois}}, \bibinfo {author} {\bibfnamefont {J.~P.}\
  \bibnamefont {Gaebler}}, \bibinfo {author} {\bibfnamefont {T.~M.}\
  \bibnamefont {Gatterman}}, \bibinfo {author} {\bibfnamefont {C.~N.}\
  \bibnamefont {Gilbreth}}, \bibinfo {author} {\bibfnamefont {J.}~\bibnamefont
  {Giles}}, \bibinfo {author} {\bibfnamefont {E.}~\bibnamefont {Glynn}},
  \bibinfo {author} {\bibfnamefont {A.}~\bibnamefont {Hall}}, \bibinfo {author}
  {\bibfnamefont {A.~M.}\ \bibnamefont {Hankin}}, \bibinfo {author}
  {\bibfnamefont {A.}~\bibnamefont {Hansen}}, \bibinfo {author} {\bibfnamefont
  {D.}~\bibnamefont {Hayes}}, \bibinfo {author} {\bibfnamefont
  {B.}~\bibnamefont {Higashi}}, \bibinfo {author} {\bibfnamefont {I.~M.}\
  \bibnamefont {Hoffman}}, \bibinfo {author} {\bibfnamefont {B.}~\bibnamefont
  {Horning}}, \bibinfo {author} {\bibfnamefont {J.~J.}\ \bibnamefont {Hout}},
  \bibinfo {author} {\bibfnamefont {R.}~\bibnamefont {Jacobs}}, \bibinfo
  {author} {\bibfnamefont {J.}~\bibnamefont {Johansen}}, \bibinfo {author}
  {\bibfnamefont {L.}~\bibnamefont {Jones}}, \bibinfo {author} {\bibfnamefont
  {J.}~\bibnamefont {Karcz}}, \bibinfo {author} {\bibfnamefont
  {T.}~\bibnamefont {Klein}}, \bibinfo {author} {\bibfnamefont
  {P.}~\bibnamefont {Lauria}}, \bibinfo {author} {\bibfnamefont
  {P.}~\bibnamefont {Lee}}, \bibinfo {author} {\bibfnamefont {D.}~\bibnamefont
  {Liefer}}, \bibinfo {author} {\bibfnamefont {C.}~\bibnamefont {Lytle}},
  \bibinfo {author} {\bibfnamefont {S.~T.}\ \bibnamefont {Lu}}, \bibinfo
  {author} {\bibfnamefont {D.}~\bibnamefont {Lucchetti}}, \bibinfo {author}
  {\bibfnamefont {A.}~\bibnamefont {Malm}}, \bibinfo {author} {\bibfnamefont
  {M.}~\bibnamefont {Matheny}}, \bibinfo {author} {\bibfnamefont
  {B.}~\bibnamefont {Mathewson}}, \bibinfo {author} {\bibfnamefont
  {K.}~\bibnamefont {Mayer}}, \bibinfo {author} {\bibfnamefont {D.~B.}\
  \bibnamefont {Miller}}, \bibinfo {author} {\bibfnamefont {M.}~\bibnamefont
  {Mills}}, \bibinfo {author} {\bibfnamefont {B.}~\bibnamefont {Neyenhuis}},
  \bibinfo {author} {\bibfnamefont {L.}~\bibnamefont {Nugent}}, \bibinfo
  {author} {\bibfnamefont {S.}~\bibnamefont {Olson}}, \bibinfo {author}
  {\bibfnamefont {J.}~\bibnamefont {Parks}}, \bibinfo {author} {\bibfnamefont
  {G.~N.}\ \bibnamefont {Price}}, \bibinfo {author} {\bibfnamefont
  {Z.}~\bibnamefont {Price}}, \bibinfo {author} {\bibfnamefont
  {M.}~\bibnamefont {Pugh}}, \bibinfo {author} {\bibfnamefont {A.}~\bibnamefont
  {Ransford}}, \bibinfo {author} {\bibfnamefont {A.~P.}\ \bibnamefont {Reed}},
  \bibinfo {author} {\bibfnamefont {C.}~\bibnamefont {Roman}}, \bibinfo
  {author} {\bibfnamefont {M.}~\bibnamefont {Rowe}}, \bibinfo {author}
  {\bibfnamefont {C.}~\bibnamefont {Ryan-Anderson}}, \bibinfo {author}
  {\bibfnamefont {S.}~\bibnamefont {Sanders}}, \bibinfo {author} {\bibfnamefont
  {J.}~\bibnamefont {Sedlacek}}, \bibinfo {author} {\bibfnamefont
  {P.}~\bibnamefont {Shevchuk}}, \bibinfo {author} {\bibfnamefont
  {P.}~\bibnamefont {Siegfried}}, \bibinfo {author} {\bibfnamefont
  {T.}~\bibnamefont {Skripka}}, \bibinfo {author} {\bibfnamefont
  {B.}~\bibnamefont {Spaun}}, \bibinfo {author} {\bibfnamefont {R.~T.}\
  \bibnamefont {Sprenkle}}, \bibinfo {author} {\bibfnamefont {R.~P.}\
  \bibnamefont {Stutz}}, \bibinfo {author} {\bibfnamefont {M.}~\bibnamefont
  {Swallows}}, \bibinfo {author} {\bibfnamefont {R.~I.}\ \bibnamefont {Tobey}},
  \bibinfo {author} {\bibfnamefont {A.}~\bibnamefont {Tran}}, \bibinfo {author}
  {\bibfnamefont {T.}~\bibnamefont {Tran}}, \bibinfo {author} {\bibfnamefont
  {E.}~\bibnamefont {Vogt}}, \bibinfo {author} {\bibfnamefont {C.}~\bibnamefont
  {Volin}}, \bibinfo {author} {\bibfnamefont {J.}~\bibnamefont {Walker}},
  \bibinfo {author} {\bibfnamefont {A.~M.}\ \bibnamefont {Zolot}}, \ and\
  \bibinfo {author} {\bibfnamefont {J.~M.}\ \bibnamefont {Pino}},\ }\href@noop
  {} {\enquote {\bibinfo {title} {A race track trapped-ion quantum
  processor},}\ } (\bibinfo {year} {2023}),\ \Eprint
  {http://arxiv.org/abs/2305.03828} {arXiv:2305.03828 [quant-ph]} \BibitemShut
  {NoStop}%
\bibitem [{\citenamefont {{IonQ Staff}}(2023)}]{IonQ}%
  \BibitemOpen
  \bibfield  {author} {\bibinfo {author} {\bibnamefont {{IonQ Staff}}},\
  }\href@noop {} {\enquote {\bibinfo {title} {{IonQ Aria: Practical
  Performance}},}\ }\bibinfo {howpublished}
  {\url{https://ionq.com/resources/ionq-aria-practical-performance}} (\bibinfo
  {year} {2023}),\ \bibinfo {note} {{A}ccessed: 2023-06-16}\BibitemShut
  {NoStop}%
\bibitem [{\citenamefont {Ebadi}\ \emph {et~al.}(2021)\citenamefont {Ebadi},
  \citenamefont {Wang}, \citenamefont {Levine}, \citenamefont {Keesling},
  \citenamefont {Semeghini}, \citenamefont {Omran}, \citenamefont {Bluvstein},
  \citenamefont {Samajdar}, \citenamefont {Pichler}, \citenamefont {Ho},
  \citenamefont {Choi}, \citenamefont {Sachdev}, \citenamefont {Greiner},
  \citenamefont {Vuleti{\'c}},\ and\ \citenamefont {Lukin}}]{Ebadi2021}%
  \BibitemOpen
  \bibfield  {author} {\bibinfo {author} {\bibfnamefont {Sepehr}\ \bibnamefont
  {Ebadi}}, \bibinfo {author} {\bibfnamefont {Tout~T.}\ \bibnamefont {Wang}},
  \bibinfo {author} {\bibfnamefont {Harry}\ \bibnamefont {Levine}}, \bibinfo
  {author} {\bibfnamefont {Alexander}\ \bibnamefont {Keesling}}, \bibinfo
  {author} {\bibfnamefont {Giulia}\ \bibnamefont {Semeghini}}, \bibinfo
  {author} {\bibfnamefont {Ahmed}\ \bibnamefont {Omran}}, \bibinfo {author}
  {\bibfnamefont {Dolev}\ \bibnamefont {Bluvstein}}, \bibinfo {author}
  {\bibfnamefont {Rhine}\ \bibnamefont {Samajdar}}, \bibinfo {author}
  {\bibfnamefont {Hannes}\ \bibnamefont {Pichler}}, \bibinfo {author}
  {\bibfnamefont {Wen~Wei}\ \bibnamefont {Ho}}, \bibinfo {author}
  {\bibfnamefont {Soonwon}\ \bibnamefont {Choi}}, \bibinfo {author}
  {\bibfnamefont {Subir}\ \bibnamefont {Sachdev}}, \bibinfo {author}
  {\bibfnamefont {Markus}\ \bibnamefont {Greiner}}, \bibinfo {author}
  {\bibfnamefont {Vladan}\ \bibnamefont {Vuleti{\'c}}}, \ and\ \bibinfo
  {author} {\bibfnamefont {Mikhail~D.}\ \bibnamefont {Lukin}},\ }\bibfield
  {title} {\enquote {\bibinfo {title} {Quantum phases of matter on a 256-atom
  programmable quantum simulator},}\ }\href {\doibase
  10.1038/s41586-021-03582-4} {\bibfield  {journal} {\bibinfo  {journal}
  {Nature}\ }\textbf {\bibinfo {volume} {595}},\ \bibinfo {pages} {227--232}
  (\bibinfo {year} {2021})}\BibitemShut {NoStop}%
\bibitem [{\citenamefont {Wurtz}\ \emph {et~al.}(2023)\citenamefont {Wurtz},
  \citenamefont {Bylinskii}, \citenamefont {Braverman}, \citenamefont
  {Amato-Grill}, \citenamefont {Cantu}, \citenamefont {Huber}, \citenamefont
  {Lukin}, \citenamefont {Liu}, \citenamefont {Weinberg}, \citenamefont {Long},
  \citenamefont {Wang}, \citenamefont {Gemelke},\ and\ \citenamefont
  {Keesling}}]{wurtz2023aquila}%
  \BibitemOpen
  \bibfield  {author} {\bibinfo {author} {\bibfnamefont {Jonathan}\
  \bibnamefont {Wurtz}}, \bibinfo {author} {\bibfnamefont {Alexei}\
  \bibnamefont {Bylinskii}}, \bibinfo {author} {\bibfnamefont {Boris}\
  \bibnamefont {Braverman}}, \bibinfo {author} {\bibfnamefont {Jesse}\
  \bibnamefont {Amato-Grill}}, \bibinfo {author} {\bibfnamefont {Sergio~H.}\
  \bibnamefont {Cantu}}, \bibinfo {author} {\bibfnamefont {Florian}\
  \bibnamefont {Huber}}, \bibinfo {author} {\bibfnamefont {Alexander}\
  \bibnamefont {Lukin}}, \bibinfo {author} {\bibfnamefont {Fangli}\
  \bibnamefont {Liu}}, \bibinfo {author} {\bibfnamefont {Phillip}\ \bibnamefont
  {Weinberg}}, \bibinfo {author} {\bibfnamefont {John}\ \bibnamefont {Long}},
  \bibinfo {author} {\bibfnamefont {Sheng-Tao}\ \bibnamefont {Wang}}, \bibinfo
  {author} {\bibfnamefont {Nathan}\ \bibnamefont {Gemelke}}, \ and\ \bibinfo
  {author} {\bibfnamefont {Alexander}\ \bibnamefont {Keesling}},\ }\href@noop
  {} {\enquote {\bibinfo {title} {Aquila: Quera's 256-qubit neutral-atom
  quantum computer},}\ } (\bibinfo {year} {2023}),\ \Eprint
  {http://arxiv.org/abs/2306.11727} {arXiv:2306.11727 [quant-ph]} \BibitemShut
  {NoStop}%
\bibitem [{\citenamefont {Gambetta}(2022)}]{IBMQ2025_Roadmap}%
  \BibitemOpen
  \bibfield  {author} {\bibinfo {author} {\bibfnamefont {Jay}\ \bibnamefont
  {Gambetta}},\ }\href {https://research.ibm.com/blog/ibm-quantum-roadmap-2025}
  {\enquote {\bibinfo {title} {{Expanding the IBM Quantum roadmap to anticipate
  the future of quantum-centric supercomputing}},}\ } (\bibinfo {year}
  {2022})\BibitemShut {NoStop}%
\bibitem [{\citenamefont {Kaushal}\ \emph {et~al.}(2020)\citenamefont
  {Kaushal}, \citenamefont {Lekitsch}, \citenamefont {Stahl}, \citenamefont
  {Hilder}, \citenamefont {Pijn}, \citenamefont {Schmiegelow}, \citenamefont
  {Bermudez}, \citenamefont {M{\"u}ller}, \citenamefont {Schmidt-Kaler},\ and\
  \citenamefont {Poschinger}}]{Kaushal2020}%
  \BibitemOpen
  \bibfield  {author} {\bibinfo {author} {\bibfnamefont {V.}~\bibnamefont
  {Kaushal}}, \bibinfo {author} {\bibfnamefont {B.}~\bibnamefont {Lekitsch}},
  \bibinfo {author} {\bibfnamefont {A.}~\bibnamefont {Stahl}}, \bibinfo
  {author} {\bibfnamefont {J.}~\bibnamefont {Hilder}}, \bibinfo {author}
  {\bibfnamefont {D.}~\bibnamefont {Pijn}}, \bibinfo {author} {\bibfnamefont
  {C.}~\bibnamefont {Schmiegelow}}, \bibinfo {author} {\bibfnamefont
  {A.}~\bibnamefont {Bermudez}}, \bibinfo {author} {\bibfnamefont
  {M.}~\bibnamefont {M{\"u}ller}}, \bibinfo {author} {\bibfnamefont
  {F.}~\bibnamefont {Schmidt-Kaler}}, \ and\ \bibinfo {author} {\bibfnamefont
  {U.}~\bibnamefont {Poschinger}},\ }\bibfield  {title} {\enquote {\bibinfo
  {title} {{Shuttling-based trapped-ion quantum information processing}},}\
  }\href {\doibase 10.1116/1.5126186} {\bibfield  {journal} {\bibinfo
  {journal} {AVS Quantum Science}\ }\textbf {\bibinfo {volume} {2}},\ \bibinfo
  {pages} {014101} (\bibinfo {year} {2020})}\BibitemShut {NoStop}%
\bibitem [{\citenamefont {Krinner}\ \emph {et~al.}(2019)\citenamefont
  {Krinner}, \citenamefont {Storz}, \citenamefont {Kurpiers}, \citenamefont
  {Magnard}, \citenamefont {Heinsoo}, \citenamefont {Keller}, \citenamefont
  {L{\"u}tolf}, \citenamefont {Eichler},\ and\ \citenamefont
  {Wallraff}}]{Krinner2019}%
  \BibitemOpen
  \bibfield  {author} {\bibinfo {author} {\bibfnamefont {S.}~\bibnamefont
  {Krinner}}, \bibinfo {author} {\bibfnamefont {S.}~\bibnamefont {Storz}},
  \bibinfo {author} {\bibfnamefont {P.}~\bibnamefont {Kurpiers}}, \bibinfo
  {author} {\bibfnamefont {P.}~\bibnamefont {Magnard}}, \bibinfo {author}
  {\bibfnamefont {J.}~\bibnamefont {Heinsoo}}, \bibinfo {author} {\bibfnamefont
  {R.}~\bibnamefont {Keller}}, \bibinfo {author} {\bibfnamefont
  {J.}~\bibnamefont {L{\"u}tolf}}, \bibinfo {author} {\bibfnamefont
  {C.}~\bibnamefont {Eichler}}, \ and\ \bibinfo {author} {\bibfnamefont
  {A.}~\bibnamefont {Wallraff}},\ }\bibfield  {title} {\enquote {\bibinfo
  {title} {Engineering cryogenic setups for 100-qubit scale superconducting
  circuit systems},}\ }\href {\doibase 10.1140/epjqt/s40507-019-0072-0}
  {\bibfield  {journal} {\bibinfo  {journal} {EPJ Quantum Technology}\ }\textbf
  {\bibinfo {volume} {6}},\ \bibinfo {pages} {2} (\bibinfo {year}
  {2019})}\BibitemShut {NoStop}%
\bibitem [{\citenamefont {Gheorghiu}\ and\ \citenamefont
  {Mosca}(2019)}]{gheorghiu2019benchmarking}%
  \BibitemOpen
  \bibfield  {author} {\bibinfo {author} {\bibfnamefont {Vlad}\ \bibnamefont
  {Gheorghiu}}\ and\ \bibinfo {author} {\bibfnamefont {Michele}\ \bibnamefont
  {Mosca}},\ }\href@noop {} {\enquote {\bibinfo {title} {Benchmarking the
  quantum cryptanalysis of symmetric, public-key and hash-based cryptographic
  schemes},}\ } (\bibinfo {year} {2019}),\ \Eprint
  {http://arxiv.org/abs/1902.02332} {arXiv:1902.02332 [quant-ph]} \BibitemShut
  {NoStop}%
\bibitem [{\citenamefont {Roetteler}\ \emph {et~al.}(2017)\citenamefont
  {Roetteler}, \citenamefont {Naehrig}, \citenamefont {Svore},\ and\
  \citenamefont {Lauter}}]{Roetteler2017}%
  \BibitemOpen
  \bibfield  {author} {\bibinfo {author} {\bibfnamefont {Martin}\ \bibnamefont
  {Roetteler}}, \bibinfo {author} {\bibfnamefont {Michael}\ \bibnamefont
  {Naehrig}}, \bibinfo {author} {\bibfnamefont {Krysta~M.}\ \bibnamefont
  {Svore}}, \ and\ \bibinfo {author} {\bibfnamefont {Kristin}\ \bibnamefont
  {Lauter}},\ }\bibfield  {title} {\enquote {\bibinfo {title} {Quantum resource
  estimates for computing elliptic curve discrete logarithms},}\ }in\
  \href@noop {} {\emph {\bibinfo {booktitle} {Advances in Cryptology --
  ASIACRYPT 2017}}},\ \bibinfo {editor} {edited by\ \bibinfo {editor}
  {\bibfnamefont {Tsuyoshi}\ \bibnamefont {Takagi}}\ and\ \bibinfo {editor}
  {\bibfnamefont {Thomas}\ \bibnamefont {Peyrin}}}\ (\bibinfo  {publisher}
  {Springer International Publishing},\ \bibinfo {address} {Cham},\ \bibinfo
  {year} {2017})\ pp.\ \bibinfo {pages} {241--270}\BibitemShut {NoStop}%
\bibitem [{\citenamefont {Gidney}\ and\ \citenamefont
  {Eker{\aa{}}}(2021)}]{Gidney2021}%
  \BibitemOpen
  \bibfield  {author} {\bibinfo {author} {\bibfnamefont {Craig}\ \bibnamefont
  {Gidney}}\ and\ \bibinfo {author} {\bibfnamefont {Martin}\ \bibnamefont
  {Eker{\aa{}}}},\ }\bibfield  {title} {\enquote {\bibinfo {title} {How to
  factor 2048 bit {RSA} integers in 8 hours using 20 million noisy qubits},}\
  }\href {\doibase 10.22331/q-2021-04-15-433} {\bibfield  {journal} {\bibinfo
  {journal} {{Quantum}}\ }\textbf {\bibinfo {volume} {5}},\ \bibinfo {pages}
  {433} (\bibinfo {year} {2021})}\BibitemShut {NoStop}%
\bibitem [{\citenamefont {Preskill}(2018)}]{NISQ_Preskill}%
  \BibitemOpen
  \bibfield  {author} {\bibinfo {author} {\bibfnamefont {John}\ \bibnamefont
  {Preskill}},\ }\bibfield  {title} {\enquote {\bibinfo {title} {Quantum
  {C}omputing in the {NISQ} era and beyond},}\ }\href {\doibase
  10.22331/q-2018-08-06-79} {\bibfield  {journal} {\bibinfo  {journal}
  {{Quantum}}\ }\textbf {\bibinfo {volume} {2}},\ \bibinfo {pages} {79}
  (\bibinfo {year} {2018})}\BibitemShut {NoStop}%
\bibitem [{\citenamefont {Monroe}\ and\ \citenamefont
  {Kim}(2013)}]{Monroe2013}%
  \BibitemOpen
  \bibfield  {author} {\bibinfo {author} {\bibfnamefont {C.}~\bibnamefont
  {Monroe}}\ and\ \bibinfo {author} {\bibfnamefont {J.}~\bibnamefont {Kim}},\
  }\bibfield  {title} {\enquote {\bibinfo {title} {Scaling the ion trap quantum
  processor},}\ }\href {\doibase 10.1126/science.1231298} {\bibfield  {journal}
  {\bibinfo  {journal} {Science}\ }\textbf {\bibinfo {volume} {339}},\ \bibinfo
  {pages} {1164--1169} (\bibinfo {year} {2013})}\BibitemShut {NoStop}%
\bibitem [{\citenamefont {Nielsen}\ and\ \citenamefont
  {Chuang}(2010)}]{nielsen_chuang_2010}%
  \BibitemOpen
  \bibfield  {author} {\bibinfo {author} {\bibfnamefont {Michael~A.}\
  \bibnamefont {Nielsen}}\ and\ \bibinfo {author} {\bibfnamefont {Isaac~L.}\
  \bibnamefont {Chuang}},\ }\href {\doibase 10.1017/CBO9780511976667} {\emph
  {\bibinfo {title} {Quantum Computation and Quantum Information: 10th
  Anniversary Edition}}}\ (\bibinfo  {publisher} {Cambridge University Press},\
  \bibinfo {year} {2010})\BibitemShut {NoStop}%
\bibitem [{\citenamefont {Calderbank}\ and\ \citenamefont
  {Shor}(1996)}]{Calderbank1996}%
  \BibitemOpen
  \bibfield  {author} {\bibinfo {author} {\bibfnamefont {A.~R.}\ \bibnamefont
  {Calderbank}}\ and\ \bibinfo {author} {\bibfnamefont {Peter~W.}\ \bibnamefont
  {Shor}},\ }\bibfield  {title} {\enquote {\bibinfo {title} {Good quantum
  error-correcting codes exist},}\ }\href {\doibase 10.1103/PhysRevA.54.1098}
  {\bibfield  {journal} {\bibinfo  {journal} {Phys. Rev. A}\ }\textbf {\bibinfo
  {volume} {54}},\ \bibinfo {pages} {1098--1105} (\bibinfo {year}
  {1996})}\BibitemShut {NoStop}%
\bibitem [{\citenamefont {Fowler}\ \emph {et~al.}(2012)\citenamefont {Fowler},
  \citenamefont {Mariantoni}, \citenamefont {Martinis},\ and\ \citenamefont
  {Cleland}}]{Fowler2012}%
  \BibitemOpen
  \bibfield  {author} {\bibinfo {author} {\bibfnamefont {Austin~G.}\
  \bibnamefont {Fowler}}, \bibinfo {author} {\bibfnamefont {Matteo}\
  \bibnamefont {Mariantoni}}, \bibinfo {author} {\bibfnamefont {John~M.}\
  \bibnamefont {Martinis}}, \ and\ \bibinfo {author} {\bibfnamefont
  {Andrew~N.}\ \bibnamefont {Cleland}},\ }\bibfield  {title} {\enquote
  {\bibinfo {title} {Surface codes: Towards practical large-scale quantum
  computation},}\ }\href {\doibase 10.1103/PhysRevA.86.032324} {\bibfield
  {journal} {\bibinfo  {journal} {Phys. Rev. A}\ }\textbf {\bibinfo {volume}
  {86}},\ \bibinfo {pages} {032324} (\bibinfo {year} {2012})}\BibitemShut
  {NoStop}%
\bibitem [{\citenamefont {Acharya}\ \emph {et~al.}(2023)\citenamefont
  {Acharya}, \citenamefont {Aleiner}, \citenamefont {Allen}, \citenamefont
  {Andersen}, \citenamefont {Ansmann}, \citenamefont {Arute}, \citenamefont
  {Arya}, \citenamefont {Asfaw}, \citenamefont {Atalaya}, \citenamefont
  {Babbush}, \citenamefont {Bacon}, \citenamefont {Bardin}, \citenamefont
  {Basso}, \citenamefont {Bengtsson}, \citenamefont {Boixo}, \citenamefont
  {Bortoli}, \citenamefont {Bourassa}, \citenamefont {Bovaird}, \citenamefont
  {Brill}, \citenamefont {Broughton}, \citenamefont {Buckley}, \citenamefont
  {Buell}, \citenamefont {Burger}, \citenamefont {Burkett}, \citenamefont
  {Bushnell}, \citenamefont {Chen}, \citenamefont {Chen}, \citenamefont
  {Chiaro}, \citenamefont {Cogan}, \citenamefont {Collins}, \citenamefont
  {Conner}, \citenamefont {Courtney}, \citenamefont {Crook}, \citenamefont
  {Curtin}, \citenamefont {Debroy}, \citenamefont {Del Toro~Barba},
  \citenamefont {Demura}, \citenamefont {Dunsworth}, \citenamefont {Eppens},
  \citenamefont {Erickson}, \citenamefont {Faoro}, \citenamefont {Farhi},
  \citenamefont {Fatemi}, \citenamefont {Flores~Burgos}, \citenamefont
  {Forati}, \citenamefont {Fowler}, \citenamefont {Foxen}, \citenamefont
  {Giang}, \citenamefont {Gidney}, \citenamefont {Gilboa}, \citenamefont
  {Giustina}, \citenamefont {Grajales~Dau}, \citenamefont {Gross},
  \citenamefont {Habegger}, \citenamefont {Hamilton}, \citenamefont {Harrigan},
  \citenamefont {Harrington}, \citenamefont {Higgott}, \citenamefont {Hilton},
  \citenamefont {Hoffmann}, \citenamefont {Hong}, \citenamefont {Huang},
  \citenamefont {Huff}, \citenamefont {Huggins}, \citenamefont {Ioffe},
  \citenamefont {Isakov}, \citenamefont {Iveland}, \citenamefont {Jeffrey},
  \citenamefont {Jiang}, \citenamefont {Jones}, \citenamefont {Juhas},
  \citenamefont {Kafri}, \citenamefont {Kechedzhi}, \citenamefont {Kelly},
  \citenamefont {Khattar}, \citenamefont {Khezri}, \citenamefont
  {Kieferov{\'a}}, \citenamefont {Kim}, \citenamefont {Kitaev}, \citenamefont
  {Klimov}, \citenamefont {Klots}, \citenamefont {Korotkov}, \citenamefont
  {Kostritsa}, \citenamefont {Kreikebaum}, \citenamefont {Landhuis},
  \citenamefont {Laptev}, \citenamefont {Lau}, \citenamefont {Laws},
  \citenamefont {Lee}, \citenamefont {Lee}, \citenamefont {Lester},
  \citenamefont {Lill}, \citenamefont {Liu}, \citenamefont {Locharla},
  \citenamefont {Lucero}, \citenamefont {Malone}, \citenamefont {Marshall},
  \citenamefont {Martin}, \citenamefont {McClean}, \citenamefont {McCourt},
  \citenamefont {McEwen}, \citenamefont {Megrant}, \citenamefont
  {Meurer~Costa}, \citenamefont {Mi}, \citenamefont {Miao}, \citenamefont
  {Mohseni}, \citenamefont {Montazeri}, \citenamefont {Morvan}, \citenamefont
  {Mount}, \citenamefont {Mruczkiewicz}, \citenamefont {Naaman}, \citenamefont
  {Neeley}, \citenamefont {Neill}, \citenamefont {Nersisyan}, \citenamefont
  {Neven}, \citenamefont {Newman}, \citenamefont {Ng}, \citenamefont {Nguyen},
  \citenamefont {Nguyen}, \citenamefont {Niu}, \citenamefont {O'Brien},
  \citenamefont {Opremcak}, \citenamefont {Platt}, \citenamefont {Petukhov},
  \citenamefont {Potter}, \citenamefont {Pryadko}, \citenamefont {Quintana},
  \citenamefont {Roushan}, \citenamefont {Rubin}, \citenamefont {Saei},
  \citenamefont {Sank}, \citenamefont {Sankaragomathi}, \citenamefont
  {Satzinger}, \citenamefont {Schurkus}, \citenamefont {Schuster},
  \citenamefont {Shearn}, \citenamefont {Shorter}, \citenamefont {Shvarts},
  \citenamefont {Skruzny}, \citenamefont {Smelyanskiy}, \citenamefont {Smith},
  \citenamefont {Sterling}, \citenamefont {Strain}, \citenamefont {Szalay},
  \citenamefont {Torres}, \citenamefont {Vidal}, \citenamefont {Villalonga},
  \citenamefont {Vollgraff~Heidweiller}, \citenamefont {White}, \citenamefont
  {Xing}, \citenamefont {Yao}, \citenamefont {Yeh}, \citenamefont {Yoo},
  \citenamefont {Young}, \citenamefont {Zalcman}, \citenamefont {Zhang},
  \citenamefont {Zhu},\ and\ \citenamefont {AI}}]{Google2023}%
  \BibitemOpen
  \bibfield  {author} {\bibinfo {author} {\bibfnamefont {Rajeev}\ \bibnamefont
  {Acharya}}, \bibinfo {author} {\bibfnamefont {Igor}\ \bibnamefont {Aleiner}},
  \bibinfo {author} {\bibfnamefont {Richard}\ \bibnamefont {Allen}}, \bibinfo
  {author} {\bibfnamefont {Trond~I.}\ \bibnamefont {Andersen}}, \bibinfo
  {author} {\bibfnamefont {Markus}\ \bibnamefont {Ansmann}}, \bibinfo {author}
  {\bibfnamefont {Frank}\ \bibnamefont {Arute}}, \bibinfo {author}
  {\bibfnamefont {Kunal}\ \bibnamefont {Arya}}, \bibinfo {author}
  {\bibfnamefont {Abraham}\ \bibnamefont {Asfaw}}, \bibinfo {author}
  {\bibfnamefont {Juan}\ \bibnamefont {Atalaya}}, \bibinfo {author}
  {\bibfnamefont {Ryan}\ \bibnamefont {Babbush}}, \bibinfo {author}
  {\bibfnamefont {Dave}\ \bibnamefont {Bacon}}, \bibinfo {author}
  {\bibfnamefont {Joseph~C.}\ \bibnamefont {Bardin}}, \bibinfo {author}
  {\bibfnamefont {Joao}\ \bibnamefont {Basso}}, \bibinfo {author}
  {\bibfnamefont {Andreas}\ \bibnamefont {Bengtsson}}, \bibinfo {author}
  {\bibfnamefont {Sergio}\ \bibnamefont {Boixo}}, \bibinfo {author}
  {\bibfnamefont {Gina}\ \bibnamefont {Bortoli}}, \bibinfo {author}
  {\bibfnamefont {Alexandre}\ \bibnamefont {Bourassa}}, \bibinfo {author}
  {\bibfnamefont {Jenna}\ \bibnamefont {Bovaird}}, \bibinfo {author}
  {\bibfnamefont {Leon}\ \bibnamefont {Brill}}, \bibinfo {author}
  {\bibfnamefont {Michael}\ \bibnamefont {Broughton}}, \bibinfo {author}
  {\bibfnamefont {Bob~B.}\ \bibnamefont {Buckley}}, \bibinfo {author}
  {\bibfnamefont {David~A.}\ \bibnamefont {Buell}}, \bibinfo {author}
  {\bibfnamefont {Tim}\ \bibnamefont {Burger}}, \bibinfo {author}
  {\bibfnamefont {Brian}\ \bibnamefont {Burkett}}, \bibinfo {author}
  {\bibfnamefont {Nicholas}\ \bibnamefont {Bushnell}}, \bibinfo {author}
  {\bibfnamefont {Yu}~\bibnamefont {Chen}}, \bibinfo {author} {\bibfnamefont
  {Zijun}\ \bibnamefont {Chen}}, \bibinfo {author} {\bibfnamefont {Ben}\
  \bibnamefont {Chiaro}}, \bibinfo {author} {\bibfnamefont {Josh}\ \bibnamefont
  {Cogan}}, \bibinfo {author} {\bibfnamefont {Roberto}\ \bibnamefont
  {Collins}}, \bibinfo {author} {\bibfnamefont {Paul}\ \bibnamefont {Conner}},
  \bibinfo {author} {\bibfnamefont {William}\ \bibnamefont {Courtney}},
  \bibinfo {author} {\bibfnamefont {Alexander~L.}\ \bibnamefont {Crook}},
  \bibinfo {author} {\bibfnamefont {Ben}\ \bibnamefont {Curtin}}, \bibinfo
  {author} {\bibfnamefont {Dripto~M.}\ \bibnamefont {Debroy}}, \bibinfo
  {author} {\bibfnamefont {Alexander}\ \bibnamefont {Del Toro~Barba}}, \bibinfo
  {author} {\bibfnamefont {Sean}\ \bibnamefont {Demura}}, \bibinfo {author}
  {\bibfnamefont {Andrew}\ \bibnamefont {Dunsworth}}, \bibinfo {author}
  {\bibfnamefont {Daniel}\ \bibnamefont {Eppens}}, \bibinfo {author}
  {\bibfnamefont {Catherine}\ \bibnamefont {Erickson}}, \bibinfo {author}
  {\bibfnamefont {Lara}\ \bibnamefont {Faoro}}, \bibinfo {author}
  {\bibfnamefont {Edward}\ \bibnamefont {Farhi}}, \bibinfo {author}
  {\bibfnamefont {Reza}\ \bibnamefont {Fatemi}}, \bibinfo {author}
  {\bibfnamefont {Leslie}\ \bibnamefont {Flores~Burgos}}, \bibinfo {author}
  {\bibfnamefont {Ebrahim}\ \bibnamefont {Forati}}, \bibinfo {author}
  {\bibfnamefont {Austin~G.}\ \bibnamefont {Fowler}}, \bibinfo {author}
  {\bibfnamefont {Brooks}\ \bibnamefont {Foxen}}, \bibinfo {author}
  {\bibfnamefont {William}\ \bibnamefont {Giang}}, \bibinfo {author}
  {\bibfnamefont {Craig}\ \bibnamefont {Gidney}}, \bibinfo {author}
  {\bibfnamefont {Dar}\ \bibnamefont {Gilboa}}, \bibinfo {author}
  {\bibfnamefont {Marissa}\ \bibnamefont {Giustina}}, \bibinfo {author}
  {\bibfnamefont {Alejandro}\ \bibnamefont {Grajales~Dau}}, \bibinfo {author}
  {\bibfnamefont {Jonathan~A.}\ \bibnamefont {Gross}}, \bibinfo {author}
  {\bibfnamefont {Steve}\ \bibnamefont {Habegger}}, \bibinfo {author}
  {\bibfnamefont {Michael~C.}\ \bibnamefont {Hamilton}}, \bibinfo {author}
  {\bibfnamefont {Matthew~P.}\ \bibnamefont {Harrigan}}, \bibinfo {author}
  {\bibfnamefont {Sean~D.}\ \bibnamefont {Harrington}}, \bibinfo {author}
  {\bibfnamefont {Oscar}\ \bibnamefont {Higgott}}, \bibinfo {author}
  {\bibfnamefont {Jeremy}\ \bibnamefont {Hilton}}, \bibinfo {author}
  {\bibfnamefont {Markus}\ \bibnamefont {Hoffmann}}, \bibinfo {author}
  {\bibfnamefont {Sabrina}\ \bibnamefont {Hong}}, \bibinfo {author}
  {\bibfnamefont {Trent}\ \bibnamefont {Huang}}, \bibinfo {author}
  {\bibfnamefont {Ashley}\ \bibnamefont {Huff}}, \bibinfo {author}
  {\bibfnamefont {William~J.}\ \bibnamefont {Huggins}}, \bibinfo {author}
  {\bibfnamefont {Lev~B.}\ \bibnamefont {Ioffe}}, \bibinfo {author}
  {\bibfnamefont {Sergei~V.}\ \bibnamefont {Isakov}}, \bibinfo {author}
  {\bibfnamefont {Justin}\ \bibnamefont {Iveland}}, \bibinfo {author}
  {\bibfnamefont {Evan}\ \bibnamefont {Jeffrey}}, \bibinfo {author}
  {\bibfnamefont {Zhang}\ \bibnamefont {Jiang}}, \bibinfo {author}
  {\bibfnamefont {Cody}\ \bibnamefont {Jones}}, \bibinfo {author}
  {\bibfnamefont {Pavol}\ \bibnamefont {Juhas}}, \bibinfo {author}
  {\bibfnamefont {Dvir}\ \bibnamefont {Kafri}}, \bibinfo {author}
  {\bibfnamefont {Kostyantyn}\ \bibnamefont {Kechedzhi}}, \bibinfo {author}
  {\bibfnamefont {Julian}\ \bibnamefont {Kelly}}, \bibinfo {author}
  {\bibfnamefont {Tanuj}\ \bibnamefont {Khattar}}, \bibinfo {author}
  {\bibfnamefont {Mostafa}\ \bibnamefont {Khezri}}, \bibinfo {author}
  {\bibfnamefont {M{\'a}ria}\ \bibnamefont {Kieferov{\'a}}}, \bibinfo {author}
  {\bibfnamefont {Seon}\ \bibnamefont {Kim}}, \bibinfo {author} {\bibfnamefont
  {Alexei}\ \bibnamefont {Kitaev}}, \bibinfo {author} {\bibfnamefont {Paul~V.}\
  \bibnamefont {Klimov}}, \bibinfo {author} {\bibfnamefont {Andrey~R.}\
  \bibnamefont {Klots}}, \bibinfo {author} {\bibfnamefont {Alexander~N.}\
  \bibnamefont {Korotkov}}, \bibinfo {author} {\bibfnamefont {Fedor}\
  \bibnamefont {Kostritsa}}, \bibinfo {author} {\bibfnamefont {John~Mark}\
  \bibnamefont {Kreikebaum}}, \bibinfo {author} {\bibfnamefont {David}\
  \bibnamefont {Landhuis}}, \bibinfo {author} {\bibfnamefont {Pavel}\
  \bibnamefont {Laptev}}, \bibinfo {author} {\bibfnamefont {Kim-Ming}\
  \bibnamefont {Lau}}, \bibinfo {author} {\bibfnamefont {Lily}\ \bibnamefont
  {Laws}}, \bibinfo {author} {\bibfnamefont {Joonho}\ \bibnamefont {Lee}},
  \bibinfo {author} {\bibfnamefont {Kenny}\ \bibnamefont {Lee}}, \bibinfo
  {author} {\bibfnamefont {Brian~J.}\ \bibnamefont {Lester}}, \bibinfo {author}
  {\bibfnamefont {Alexander}\ \bibnamefont {Lill}}, \bibinfo {author}
  {\bibfnamefont {Wayne}\ \bibnamefont {Liu}}, \bibinfo {author} {\bibfnamefont
  {Aditya}\ \bibnamefont {Locharla}}, \bibinfo {author} {\bibfnamefont {Erik}\
  \bibnamefont {Lucero}}, \bibinfo {author} {\bibfnamefont {Fionn~D.}\
  \bibnamefont {Malone}}, \bibinfo {author} {\bibfnamefont {Jeffrey}\
  \bibnamefont {Marshall}}, \bibinfo {author} {\bibfnamefont {Orion}\
  \bibnamefont {Martin}}, \bibinfo {author} {\bibfnamefont {Jarrod~R.}\
  \bibnamefont {McClean}}, \bibinfo {author} {\bibfnamefont {Trevor}\
  \bibnamefont {McCourt}}, \bibinfo {author} {\bibfnamefont {Matt}\
  \bibnamefont {McEwen}}, \bibinfo {author} {\bibfnamefont {Anthony}\
  \bibnamefont {Megrant}}, \bibinfo {author} {\bibfnamefont {Bernardo}\
  \bibnamefont {Meurer~Costa}}, \bibinfo {author} {\bibfnamefont {Xiao}\
  \bibnamefont {Mi}}, \bibinfo {author} {\bibfnamefont {Kevin~C.}\ \bibnamefont
  {Miao}}, \bibinfo {author} {\bibfnamefont {Masoud}\ \bibnamefont {Mohseni}},
  \bibinfo {author} {\bibfnamefont {Shirin}\ \bibnamefont {Montazeri}},
  \bibinfo {author} {\bibfnamefont {Alexis}\ \bibnamefont {Morvan}}, \bibinfo
  {author} {\bibfnamefont {Emily}\ \bibnamefont {Mount}}, \bibinfo {author}
  {\bibfnamefont {Wojciech}\ \bibnamefont {Mruczkiewicz}}, \bibinfo {author}
  {\bibfnamefont {Ofer}\ \bibnamefont {Naaman}}, \bibinfo {author}
  {\bibfnamefont {Matthew}\ \bibnamefont {Neeley}}, \bibinfo {author}
  {\bibfnamefont {Charles}\ \bibnamefont {Neill}}, \bibinfo {author}
  {\bibfnamefont {Ani}\ \bibnamefont {Nersisyan}}, \bibinfo {author}
  {\bibfnamefont {Hartmut}\ \bibnamefont {Neven}}, \bibinfo {author}
  {\bibfnamefont {Michael}\ \bibnamefont {Newman}}, \bibinfo {author}
  {\bibfnamefont {Jiun~How}\ \bibnamefont {Ng}}, \bibinfo {author}
  {\bibfnamefont {Anthony}\ \bibnamefont {Nguyen}}, \bibinfo {author}
  {\bibfnamefont {Murray}\ \bibnamefont {Nguyen}}, \bibinfo {author}
  {\bibfnamefont {Murphy~Yuezhen}\ \bibnamefont {Niu}}, \bibinfo {author}
  {\bibfnamefont {Thomas~E.}\ \bibnamefont {O'Brien}}, \bibinfo {author}
  {\bibfnamefont {Alex}\ \bibnamefont {Opremcak}}, \bibinfo {author}
  {\bibfnamefont {John}\ \bibnamefont {Platt}}, \bibinfo {author}
  {\bibfnamefont {Andre}\ \bibnamefont {Petukhov}}, \bibinfo {author}
  {\bibfnamefont {Rebecca}\ \bibnamefont {Potter}}, \bibinfo {author}
  {\bibfnamefont {Leonid~P.}\ \bibnamefont {Pryadko}}, \bibinfo {author}
  {\bibfnamefont {Chris}\ \bibnamefont {Quintana}}, \bibinfo {author}
  {\bibfnamefont {Pedram}\ \bibnamefont {Roushan}}, \bibinfo {author}
  {\bibfnamefont {Nicholas~C.}\ \bibnamefont {Rubin}}, \bibinfo {author}
  {\bibfnamefont {Negar}\ \bibnamefont {Saei}}, \bibinfo {author}
  {\bibfnamefont {Daniel}\ \bibnamefont {Sank}}, \bibinfo {author}
  {\bibfnamefont {Kannan}\ \bibnamefont {Sankaragomathi}}, \bibinfo {author}
  {\bibfnamefont {Kevin~J.}\ \bibnamefont {Satzinger}}, \bibinfo {author}
  {\bibfnamefont {Henry~F.}\ \bibnamefont {Schurkus}}, \bibinfo {author}
  {\bibfnamefont {Christopher}\ \bibnamefont {Schuster}}, \bibinfo {author}
  {\bibfnamefont {Michael~J.}\ \bibnamefont {Shearn}}, \bibinfo {author}
  {\bibfnamefont {Aaron}\ \bibnamefont {Shorter}}, \bibinfo {author}
  {\bibfnamefont {Vladimir}\ \bibnamefont {Shvarts}}, \bibinfo {author}
  {\bibfnamefont {Jindra}\ \bibnamefont {Skruzny}}, \bibinfo {author}
  {\bibfnamefont {Vadim}\ \bibnamefont {Smelyanskiy}}, \bibinfo {author}
  {\bibfnamefont {W.~Clarke}\ \bibnamefont {Smith}}, \bibinfo {author}
  {\bibfnamefont {George}\ \bibnamefont {Sterling}}, \bibinfo {author}
  {\bibfnamefont {Doug}\ \bibnamefont {Strain}}, \bibinfo {author}
  {\bibfnamefont {Marco}\ \bibnamefont {Szalay}}, \bibinfo {author}
  {\bibfnamefont {Alfredo}\ \bibnamefont {Torres}}, \bibinfo {author}
  {\bibfnamefont {Guifre}\ \bibnamefont {Vidal}}, \bibinfo {author}
  {\bibfnamefont {Benjamin}\ \bibnamefont {Villalonga}}, \bibinfo {author}
  {\bibfnamefont {Catherine}\ \bibnamefont {Vollgraff~Heidweiller}}, \bibinfo
  {author} {\bibfnamefont {Theodore}\ \bibnamefont {White}}, \bibinfo {author}
  {\bibfnamefont {Cheng}\ \bibnamefont {Xing}}, \bibinfo {author}
  {\bibfnamefont {Z.~Jamie}\ \bibnamefont {Yao}}, \bibinfo {author}
  {\bibfnamefont {Ping}\ \bibnamefont {Yeh}}, \bibinfo {author} {\bibfnamefont
  {Juhwan}\ \bibnamefont {Yoo}}, \bibinfo {author} {\bibfnamefont {Grayson}\
  \bibnamefont {Young}}, \bibinfo {author} {\bibfnamefont {Adam}\ \bibnamefont
  {Zalcman}}, \bibinfo {author} {\bibfnamefont {Yaxing}\ \bibnamefont {Zhang}},
  \bibinfo {author} {\bibfnamefont {Ningfeng}\ \bibnamefont {Zhu}}, \ and\
  \bibinfo {author} {\bibfnamefont {Google~Quantum}\ \bibnamefont {AI}},\
  }\bibfield  {title} {\enquote {\bibinfo {title} {Suppressing quantum errors
  by scaling a surface code logical qubit},}\ }\href {\doibase
  10.1038/s41586-022-05434-1} {\bibfield  {journal} {\bibinfo  {journal}
  {Nature}\ }\textbf {\bibinfo {volume} {614}},\ \bibinfo {pages} {676--681}
  (\bibinfo {year} {2023})}\BibitemShut {NoStop}%
\bibitem [{\citenamefont {Hennessy}\ and\ \citenamefont
  {Patterson}(2011)}]{hennessy2011computer}%
  \BibitemOpen
  \bibfield  {author} {\bibinfo {author} {\bibfnamefont {John~L}\ \bibnamefont
  {Hennessy}}\ and\ \bibinfo {author} {\bibfnamefont {David~A}\ \bibnamefont
  {Patterson}},\ }\href@noop {} {\emph {\bibinfo {title} {Computer
  architecture: a quantitative approach}}}\ (\bibinfo  {publisher} {Elsevier},\
  \bibinfo {year} {2011})\BibitemShut {NoStop}%
\bibitem [{\citenamefont {Patterson}\ and\ \citenamefont
  {Hennessy}(2016)}]{patterson2016computer}%
  \BibitemOpen
  \bibfield  {author} {\bibinfo {author} {\bibfnamefont {David~A}\ \bibnamefont
  {Patterson}}\ and\ \bibinfo {author} {\bibfnamefont {John~L}\ \bibnamefont
  {Hennessy}},\ }\href@noop {} {\emph {\bibinfo {title} {Computer organization
  and design ARM edition: the hardware software interface}}}\ (\bibinfo
  {publisher} {Morgan kaufmann},\ \bibinfo {year} {2016})\BibitemShut {NoStop}%
\bibitem [{\citenamefont {Eastin}\ and\ \citenamefont
  {Knill}(2009)}]{Eastin2009}%
  \BibitemOpen
  \bibfield  {author} {\bibinfo {author} {\bibfnamefont {Bryan}\ \bibnamefont
  {Eastin}}\ and\ \bibinfo {author} {\bibfnamefont {Emanuel}\ \bibnamefont
  {Knill}},\ }\bibfield  {title} {\enquote {\bibinfo {title} {Restrictions on
  transversal encoded quantum gate sets},}\ }\href {\doibase
  10.1103/PhysRevLett.102.110502} {\bibfield  {journal} {\bibinfo  {journal}
  {Phys. Rev. Lett.}\ }\textbf {\bibinfo {volume} {102}},\ \bibinfo {pages}
  {110502} (\bibinfo {year} {2009})}\BibitemShut {NoStop}%
\bibitem [{\citenamefont {Bravyi}\ and\ \citenamefont
  {Kitaev}(2005)}]{Bravyi2005}%
  \BibitemOpen
  \bibfield  {author} {\bibinfo {author} {\bibfnamefont {Sergey}\ \bibnamefont
  {Bravyi}}\ and\ \bibinfo {author} {\bibfnamefont {Alexei}\ \bibnamefont
  {Kitaev}},\ }\bibfield  {title} {\enquote {\bibinfo {title} {Universal
  quantum computation with ideal clifford gates and noisy ancillas},}\ }\href
  {\doibase 10.1103/PhysRevA.71.022316} {\bibfield  {journal} {\bibinfo
  {journal} {Phys. Rev. A}\ }\textbf {\bibinfo {volume} {71}},\ \bibinfo
  {pages} {022316} (\bibinfo {year} {2005})}\BibitemShut {NoStop}%
\bibitem [{\citenamefont {Bombin}\ and\ \citenamefont
  {Martin-Delgado}(2009)}]{Bombin2009}%
  \BibitemOpen
  \bibfield  {author} {\bibinfo {author} {\bibfnamefont {H}~\bibnamefont
  {Bombin}}\ and\ \bibinfo {author} {\bibfnamefont {M~A}\ \bibnamefont
  {Martin-Delgado}},\ }\bibfield  {title} {\enquote {\bibinfo {title} {Quantum
  measurements and gates by code deformation},}\ }\href {\doibase
  10.1088/1751-8113/42/9/095302} {\bibfield  {journal} {\bibinfo  {journal}
  {Journal of Physics A: Mathematical and Theoretical}\ }\textbf {\bibinfo
  {volume} {42}},\ \bibinfo {pages} {095302} (\bibinfo {year}
  {2009})}\BibitemShut {NoStop}%
\bibitem [{\citenamefont {Horsman}\ \emph {et~al.}(2012)\citenamefont
  {Horsman}, \citenamefont {Fowler}, \citenamefont {Devitt},\ and\
  \citenamefont {Meter}}]{Horsman2012}%
  \BibitemOpen
  \bibfield  {author} {\bibinfo {author} {\bibfnamefont {Dominic}\ \bibnamefont
  {Horsman}}, \bibinfo {author} {\bibfnamefont {Austin~G}\ \bibnamefont
  {Fowler}}, \bibinfo {author} {\bibfnamefont {Simon}\ \bibnamefont {Devitt}},
  \ and\ \bibinfo {author} {\bibfnamefont {Rodney~Van}\ \bibnamefont {Meter}},\
  }\bibfield  {title} {\enquote {\bibinfo {title} {Surface code quantum
  computing by lattice surgery},}\ }\href {\doibase
  10.1088/1367-2630/14/12/123011} {\bibfield  {journal} {\bibinfo  {journal}
  {New Journal of Physics}\ }\textbf {\bibinfo {volume} {14}},\ \bibinfo
  {pages} {123011} (\bibinfo {year} {2012})}\BibitemShut {NoStop}%
\bibitem [{\citenamefont {Brown}\ \emph {et~al.}(2017)\citenamefont {Brown},
  \citenamefont {Laubscher}, \citenamefont {Kesselring},\ and\ \citenamefont
  {Wootton}}]{Brown2017}%
  \BibitemOpen
  \bibfield  {author} {\bibinfo {author} {\bibfnamefont {Benjamin~J.}\
  \bibnamefont {Brown}}, \bibinfo {author} {\bibfnamefont {Katharina}\
  \bibnamefont {Laubscher}}, \bibinfo {author} {\bibfnamefont {Markus~S.}\
  \bibnamefont {Kesselring}}, \ and\ \bibinfo {author} {\bibfnamefont
  {James~R.}\ \bibnamefont {Wootton}},\ }\bibfield  {title} {\enquote {\bibinfo
  {title} {Poking holes and cutting corners to achieve clifford gates with the
  surface code},}\ }\href {\doibase 10.1103/PhysRevX.7.021029} {\bibfield
  {journal} {\bibinfo  {journal} {Phys. Rev. X}\ }\textbf {\bibinfo {volume}
  {7}},\ \bibinfo {pages} {021029} (\bibinfo {year} {2017})}\BibitemShut
  {NoStop}%
\bibitem [{\citenamefont {Litinski}(2019)}]{Litinski2019}%
  \BibitemOpen
  \bibfield  {author} {\bibinfo {author} {\bibfnamefont {Daniel}\ \bibnamefont
  {Litinski}},\ }\bibfield  {title} {\enquote {\bibinfo {title} {A {G}ame of
  {S}urface {C}odes: {L}arge-{S}cale {Q}uantum {C}omputing with {L}attice
  {S}urgery},}\ }\href {\doibase 10.22331/q-2019-03-05-128} {\bibfield
  {journal} {\bibinfo  {journal} {{Quantum}}\ }\textbf {\bibinfo {volume}
  {3}},\ \bibinfo {pages} {128} (\bibinfo {year} {2019})}\BibitemShut {NoStop}%
\bibitem [{\citenamefont {Jochym-O'Connor}\ and\ \citenamefont
  {Laflamme}(2014)}]{Laflamme2014}%
  \BibitemOpen
  \bibfield  {author} {\bibinfo {author} {\bibfnamefont {Tomas}\ \bibnamefont
  {Jochym-O'Connor}}\ and\ \bibinfo {author} {\bibfnamefont {Raymond}\
  \bibnamefont {Laflamme}},\ }\bibfield  {title} {\enquote {\bibinfo {title}
  {Using concatenated quantum codes for universal fault-tolerant quantum
  gates},}\ }\href {\doibase 10.1103/PhysRevLett.112.010505} {\bibfield
  {journal} {\bibinfo  {journal} {Phys. Rev. Lett.}\ }\textbf {\bibinfo
  {volume} {112}},\ \bibinfo {pages} {010505} (\bibinfo {year}
  {2014})}\BibitemShut {NoStop}%
\bibitem [{\citenamefont {Anderson}\ \emph {et~al.}(2014)\citenamefont
  {Anderson}, \citenamefont {Duclos-Cianci},\ and\ \citenamefont
  {Poulin}}]{Anderson2014}%
  \BibitemOpen
  \bibfield  {author} {\bibinfo {author} {\bibfnamefont {Jonas~T.}\
  \bibnamefont {Anderson}}, \bibinfo {author} {\bibfnamefont {Guillaume}\
  \bibnamefont {Duclos-Cianci}}, \ and\ \bibinfo {author} {\bibfnamefont
  {David}\ \bibnamefont {Poulin}},\ }\bibfield  {title} {\enquote {\bibinfo
  {title} {Fault-tolerant conversion between the steane and reed-muller quantum
  codes},}\ }\href {\doibase 10.1103/PhysRevLett.113.080501} {\bibfield
  {journal} {\bibinfo  {journal} {Phys. Rev. Lett.}\ }\textbf {\bibinfo
  {volume} {113}},\ \bibinfo {pages} {080501} (\bibinfo {year}
  {2014})}\BibitemShut {NoStop}%
\bibitem [{\citenamefont {Terhal}(2015)}]{Terhal2015}%
  \BibitemOpen
  \bibfield  {author} {\bibinfo {author} {\bibfnamefont {Barbara~M.}\
  \bibnamefont {Terhal}},\ }\bibfield  {title} {\enquote {\bibinfo {title}
  {Quantum error correction for quantum memories},}\ }\href {\doibase
  10.1103/RevModPhys.87.307} {\bibfield  {journal} {\bibinfo  {journal} {Rev.
  Mod. Phys.}\ }\textbf {\bibinfo {volume} {87}},\ \bibinfo {pages} {307--346}
  (\bibinfo {year} {2015})}\BibitemShut {NoStop}%
\bibitem [{\citenamefont {Kovalev}\ and\ \citenamefont
  {Pryadko}(2013)}]{Kovalev2013}%
  \BibitemOpen
  \bibfield  {author} {\bibinfo {author} {\bibfnamefont {Alexey~A.}\
  \bibnamefont {Kovalev}}\ and\ \bibinfo {author} {\bibfnamefont {Leonid~P.}\
  \bibnamefont {Pryadko}},\ }\bibfield  {title} {\enquote {\bibinfo {title}
  {Fault tolerance of quantum low-density parity check codes with sublinear
  distance scaling},}\ }\href {\doibase 10.1103/PhysRevA.87.020304} {\bibfield
  {journal} {\bibinfo  {journal} {Phys. Rev. A}\ }\textbf {\bibinfo {volume}
  {87}},\ \bibinfo {pages} {020304} (\bibinfo {year} {2013})}\BibitemShut
  {NoStop}%
\bibitem [{\citenamefont {Gidney}\ \emph {et~al.}(2021)\citenamefont {Gidney},
  \citenamefont {Newman}, \citenamefont {Fowler},\ and\ \citenamefont
  {Broughton}}]{Gidney2021_v2}%
  \BibitemOpen
  \bibfield  {author} {\bibinfo {author} {\bibfnamefont {Craig}\ \bibnamefont
  {Gidney}}, \bibinfo {author} {\bibfnamefont {Michael}\ \bibnamefont
  {Newman}}, \bibinfo {author} {\bibfnamefont {Austin}\ \bibnamefont {Fowler}},
  \ and\ \bibinfo {author} {\bibfnamefont {Michael}\ \bibnamefont
  {Broughton}},\ }\bibfield  {title} {\enquote {\bibinfo {title} {A
  {F}ault-{T}olerant {H}oneycomb {M}emory},}\ }\href {\doibase
  10.22331/q-2021-12-20-605} {\bibfield  {journal} {\bibinfo  {journal}
  {{Quantum}}\ }\textbf {\bibinfo {volume} {5}},\ \bibinfo {pages} {605}
  (\bibinfo {year} {2021})}\BibitemShut {NoStop}%
\bibitem [{\citenamefont {Bravyi}\ \emph {et~al.}(2023)\citenamefont {Bravyi},
  \citenamefont {Cross}, \citenamefont {Gambetta}, \citenamefont {Maslov},
  \citenamefont {Rall},\ and\ \citenamefont {Yoder}}]{bravyi2023highthreshold}%
  \BibitemOpen
  \bibfield  {author} {\bibinfo {author} {\bibfnamefont {Sergey}\ \bibnamefont
  {Bravyi}}, \bibinfo {author} {\bibfnamefont {Andrew~W.}\ \bibnamefont
  {Cross}}, \bibinfo {author} {\bibfnamefont {Jay~M.}\ \bibnamefont
  {Gambetta}}, \bibinfo {author} {\bibfnamefont {Dmitri}\ \bibnamefont
  {Maslov}}, \bibinfo {author} {\bibfnamefont {Patrick}\ \bibnamefont {Rall}},
  \ and\ \bibinfo {author} {\bibfnamefont {Theodore~J.}\ \bibnamefont
  {Yoder}},\ }\href@noop {} {\enquote {\bibinfo {title} {High-threshold and
  low-overhead fault-tolerant quantum memory},}\ } (\bibinfo {year} {2023}),\
  \Eprint {http://arxiv.org/abs/2308.07915} {arXiv:2308.07915 [quant-ph]}
  \BibitemShut {NoStop}%
\bibitem [{\citenamefont {Lvovsky}\ \emph {et~al.}(2009)\citenamefont
  {Lvovsky}, \citenamefont {Sanders},\ and\ \citenamefont
  {Tittel}}]{Lvovsky2009}%
  \BibitemOpen
  \bibfield  {author} {\bibinfo {author} {\bibfnamefont {Alexander~I.}\
  \bibnamefont {Lvovsky}}, \bibinfo {author} {\bibfnamefont {Barry~C.}\
  \bibnamefont {Sanders}}, \ and\ \bibinfo {author} {\bibfnamefont {Wolfgang}\
  \bibnamefont {Tittel}},\ }\bibfield  {title} {\enquote {\bibinfo {title}
  {Optical quantum memory},}\ }\href {\doibase 10.1038/nphoton.2009.231}
  {\bibfield  {journal} {\bibinfo  {journal} {Nature Photonics}\ }\textbf
  {\bibinfo {volume} {3}},\ \bibinfo {pages} {706--714} (\bibinfo {year}
  {2009})}\BibitemShut {NoStop}%
\bibitem [{\citenamefont {Hedges}\ \emph {et~al.}(2010)\citenamefont {Hedges},
  \citenamefont {Longdell}, \citenamefont {Li},\ and\ \citenamefont
  {Sellars}}]{Hedges2010}%
  \BibitemOpen
  \bibfield  {author} {\bibinfo {author} {\bibfnamefont {Morgan~P.}\
  \bibnamefont {Hedges}}, \bibinfo {author} {\bibfnamefont {Jevon~J.}\
  \bibnamefont {Longdell}}, \bibinfo {author} {\bibfnamefont {Yongmin}\
  \bibnamefont {Li}}, \ and\ \bibinfo {author} {\bibfnamefont {Matthew~J.}\
  \bibnamefont {Sellars}},\ }\bibfield  {title} {\enquote {\bibinfo {title}
  {Efficient quantum memory for light},}\ }\href {\doibase 10.1038/nature09081}
  {\bibfield  {journal} {\bibinfo  {journal} {Nature}\ }\textbf {\bibinfo
  {volume} {465}},\ \bibinfo {pages} {1052--1056} (\bibinfo {year}
  {2010})}\BibitemShut {NoStop}%
\bibitem [{\citenamefont {Heshami}\ \emph {et~al.}(2016)\citenamefont
  {Heshami}, \citenamefont {England}, \citenamefont {Humphreys}, \citenamefont
  {Bustard}, \citenamefont {Acosta}, \citenamefont {Nunn},\ and\ \citenamefont
  {Sussman}}]{Heshami2016}%
  \BibitemOpen
  \bibfield  {author} {\bibinfo {author} {\bibfnamefont {Khabat}\ \bibnamefont
  {Heshami}}, \bibinfo {author} {\bibfnamefont {Duncan~G.}\ \bibnamefont
  {England}}, \bibinfo {author} {\bibfnamefont {Peter~C.}\ \bibnamefont
  {Humphreys}}, \bibinfo {author} {\bibfnamefont {Philip~J.}\ \bibnamefont
  {Bustard}}, \bibinfo {author} {\bibfnamefont {Victor~M.}\ \bibnamefont
  {Acosta}}, \bibinfo {author} {\bibfnamefont {Joshua}\ \bibnamefont {Nunn}}, \
  and\ \bibinfo {author} {\bibfnamefont {Benjamin~J.}\ \bibnamefont
  {Sussman}},\ }\bibfield  {title} {\enquote {\bibinfo {title} {Quantum
  memories: emerging applications and recent advances},}\ }\href {\doibase
  10.1080/09500340.2016.1148212} {\bibfield  {journal} {\bibinfo  {journal}
  {Journal of Modern Optics}\ }\textbf {\bibinfo {volume} {63}},\ \bibinfo
  {pages} {2005--2028} (\bibinfo {year} {2016})},\ \bibinfo {note} {pMID:
  27695198}\BibitemShut {NoStop}%
\bibitem [{\citenamefont {Briegel}\ \emph {et~al.}(1998)\citenamefont
  {Briegel}, \citenamefont {D\"ur}, \citenamefont {Cirac},\ and\ \citenamefont
  {Zoller}}]{Briegel1998}%
  \BibitemOpen
  \bibfield  {author} {\bibinfo {author} {\bibfnamefont {H.-J.}\ \bibnamefont
  {Briegel}}, \bibinfo {author} {\bibfnamefont {W.}~\bibnamefont {D\"ur}},
  \bibinfo {author} {\bibfnamefont {J.~I.}\ \bibnamefont {Cirac}}, \ and\
  \bibinfo {author} {\bibfnamefont {P.}~\bibnamefont {Zoller}},\ }\bibfield
  {title} {\enquote {\bibinfo {title} {Quantum repeaters: The role of imperfect
  local operations in quantum communication},}\ }\href {\doibase
  10.1103/PhysRevLett.81.5932} {\bibfield  {journal} {\bibinfo  {journal}
  {Phys. Rev. Lett.}\ }\textbf {\bibinfo {volume} {81}},\ \bibinfo {pages}
  {5932--5935} (\bibinfo {year} {1998})}\BibitemShut {NoStop}%
\bibitem [{\citenamefont {Sangouard}\ \emph {et~al.}(2011)\citenamefont
  {Sangouard}, \citenamefont {Simon}, \citenamefont {de~Riedmatten},\ and\
  \citenamefont {Gisin}}]{Sangouard2011}%
  \BibitemOpen
  \bibfield  {author} {\bibinfo {author} {\bibfnamefont {Nicolas}\ \bibnamefont
  {Sangouard}}, \bibinfo {author} {\bibfnamefont {Christoph}\ \bibnamefont
  {Simon}}, \bibinfo {author} {\bibfnamefont {Hugues}\ \bibnamefont
  {de~Riedmatten}}, \ and\ \bibinfo {author} {\bibfnamefont {Nicolas}\
  \bibnamefont {Gisin}},\ }\bibfield  {title} {\enquote {\bibinfo {title}
  {Quantum repeaters based on atomic ensembles and linear optics},}\ }\href
  {\doibase 10.1103/RevModPhys.83.33} {\bibfield  {journal} {\bibinfo
  {journal} {Rev. Mod. Phys.}\ }\textbf {\bibinfo {volume} {83}},\ \bibinfo
  {pages} {33--80} (\bibinfo {year} {2011})}\BibitemShut {NoStop}%
\bibitem [{\citenamefont {Bennett}\ \emph {et~al.}(1993)\citenamefont
  {Bennett}, \citenamefont {Brassard}, \citenamefont {Cr\'epeau}, \citenamefont
  {Jozsa}, \citenamefont {Peres},\ and\ \citenamefont
  {Wootters}}]{Bennett1993}%
  \BibitemOpen
  \bibfield  {author} {\bibinfo {author} {\bibfnamefont {Charles~H.}\
  \bibnamefont {Bennett}}, \bibinfo {author} {\bibfnamefont {Gilles}\
  \bibnamefont {Brassard}}, \bibinfo {author} {\bibfnamefont {Claude}\
  \bibnamefont {Cr\'epeau}}, \bibinfo {author} {\bibfnamefont {Richard}\
  \bibnamefont {Jozsa}}, \bibinfo {author} {\bibfnamefont {Asher}\ \bibnamefont
  {Peres}}, \ and\ \bibinfo {author} {\bibfnamefont {William~K.}\ \bibnamefont
  {Wootters}},\ }\bibfield  {title} {\enquote {\bibinfo {title} {Teleporting an
  unknown quantum state via dual classical and einstein-podolsky-rosen
  channels},}\ }\href {\doibase 10.1103/PhysRevLett.70.1895} {\bibfield
  {journal} {\bibinfo  {journal} {Phys. Rev. Lett.}\ }\textbf {\bibinfo
  {volume} {70}},\ \bibinfo {pages} {1895--1899} (\bibinfo {year}
  {1993})}\BibitemShut {NoStop}%
\bibitem [{\citenamefont {Raussendorf}\ and\ \citenamefont
  {Briegel}(2001)}]{MBQC_prl}%
  \BibitemOpen
  \bibfield  {author} {\bibinfo {author} {\bibfnamefont {Robert}\ \bibnamefont
  {Raussendorf}}\ and\ \bibinfo {author} {\bibfnamefont {Hans~J.}\ \bibnamefont
  {Briegel}},\ }\bibfield  {title} {\enquote {\bibinfo {title} {A one-way
  quantum computer},}\ }\href {\doibase 10.1103/PhysRevLett.86.5188} {\bibfield
   {journal} {\bibinfo  {journal} {Phys. Rev. Lett.}\ }\textbf {\bibinfo
  {volume} {86}},\ \bibinfo {pages} {5188--5191} (\bibinfo {year}
  {2001})}\BibitemShut {NoStop}%
\bibitem [{\citenamefont {Raussendorf}\ \emph {et~al.}(2003)\citenamefont
  {Raussendorf}, \citenamefont {Browne},\ and\ \citenamefont
  {Briegel}}]{MBQC_pra}%
  \BibitemOpen
  \bibfield  {author} {\bibinfo {author} {\bibfnamefont {Robert}\ \bibnamefont
  {Raussendorf}}, \bibinfo {author} {\bibfnamefont {Daniel~E.}\ \bibnamefont
  {Browne}}, \ and\ \bibinfo {author} {\bibfnamefont {Hans~J.}\ \bibnamefont
  {Briegel}},\ }\bibfield  {title} {\enquote {\bibinfo {title}
  {Measurement-based quantum computation on cluster states},}\ }\href {\doibase
  10.1103/PhysRevA.68.022312} {\bibfield  {journal} {\bibinfo  {journal} {Phys.
  Rev. A}\ }\textbf {\bibinfo {volume} {68}},\ \bibinfo {pages} {022312}
  (\bibinfo {year} {2003})}\BibitemShut {NoStop}%
\bibitem [{\citenamefont {Benjamin}\ \emph {et~al.}(2009)\citenamefont
  {Benjamin}, \citenamefont {Lovett},\ and\ \citenamefont
  {Smith}}]{MBQC_review}%
  \BibitemOpen
  \bibfield  {author} {\bibinfo {author} {\bibfnamefont {S.C.}\ \bibnamefont
  {Benjamin}}, \bibinfo {author} {\bibfnamefont {B.W.}\ \bibnamefont {Lovett}},
  \ and\ \bibinfo {author} {\bibfnamefont {J.M.}\ \bibnamefont {Smith}},\
  }\bibfield  {title} {\enquote {\bibinfo {title} {Prospects for
  measurement-based quantum computing with solid state spins},}\ }\href
  {\doibase https://doi.org/10.1002/lpor.200810051} {\bibfield  {journal}
  {\bibinfo  {journal} {Laser \& Photonics Reviews}\ }\textbf {\bibinfo
  {volume} {3}},\ \bibinfo {pages} {556--574} (\bibinfo {year}
  {2009})}\BibitemShut {NoStop}%
\bibitem [{\citenamefont {Bartolucci}\ \emph {et~al.}(2021)\citenamefont
  {Bartolucci}, \citenamefont {Birchall}, \citenamefont {Bombin}, \citenamefont
  {Cable}, \citenamefont {Dawson}, \citenamefont {Gimeno-Segovia},
  \citenamefont {Johnston}, \citenamefont {Kieling}, \citenamefont {Nickerson},
  \citenamefont {Pant}, \citenamefont {Pastawski}, \citenamefont {Rudolph},\
  and\ \citenamefont {Sparrow}}]{FBQC}%
  \BibitemOpen
  \bibfield  {author} {\bibinfo {author} {\bibfnamefont {Sara}\ \bibnamefont
  {Bartolucci}}, \bibinfo {author} {\bibfnamefont {Patrick}\ \bibnamefont
  {Birchall}}, \bibinfo {author} {\bibfnamefont {Hector}\ \bibnamefont
  {Bombin}}, \bibinfo {author} {\bibfnamefont {Hugo}\ \bibnamefont {Cable}},
  \bibinfo {author} {\bibfnamefont {Chris}\ \bibnamefont {Dawson}}, \bibinfo
  {author} {\bibfnamefont {Mercedes}\ \bibnamefont {Gimeno-Segovia}}, \bibinfo
  {author} {\bibfnamefont {Eric}\ \bibnamefont {Johnston}}, \bibinfo {author}
  {\bibfnamefont {Konrad}\ \bibnamefont {Kieling}}, \bibinfo {author}
  {\bibfnamefont {Naomi}\ \bibnamefont {Nickerson}}, \bibinfo {author}
  {\bibfnamefont {Mihir}\ \bibnamefont {Pant}}, \bibinfo {author}
  {\bibfnamefont {Fernando}\ \bibnamefont {Pastawski}}, \bibinfo {author}
  {\bibfnamefont {Terry}\ \bibnamefont {Rudolph}}, \ and\ \bibinfo {author}
  {\bibfnamefont {Chris}\ \bibnamefont {Sparrow}},\ }\href@noop {} {\enquote
  {\bibinfo {title} {Fusion-based quantum computation},}\ } (\bibinfo {year}
  {2021}),\ \Eprint {http://arxiv.org/abs/2101.09310} {arXiv:2101.09310
  [quant-ph]} \BibitemShut {NoStop}%
\bibitem [{\citenamefont {Zhang}\ \emph {et~al.}(2021)\citenamefont {Zhang},
  \citenamefont {Yuan}, \citenamefont {Shen}, \citenamefont {Yu}, \citenamefont
  {Zhang}, \citenamefont {Wang}, \citenamefont {Li}, \citenamefont {Wang},
  \citenamefont {Deng}, \citenamefont {Wang}, \citenamefont {You},
  \citenamefont {Wang}, \citenamefont {Song}, \citenamefont {Guo},\ and\
  \citenamefont {Zhou}}]{Zhang2021Photon}%
  \BibitemOpen
  \bibfield  {author} {\bibinfo {author} {\bibfnamefont {Zichang}\ \bibnamefont
  {Zhang}}, \bibinfo {author} {\bibfnamefont {Chenzhi}\ \bibnamefont {Yuan}},
  \bibinfo {author} {\bibfnamefont {Si}~\bibnamefont {Shen}}, \bibinfo {author}
  {\bibfnamefont {Hao}\ \bibnamefont {Yu}}, \bibinfo {author} {\bibfnamefont
  {Ruiming}\ \bibnamefont {Zhang}}, \bibinfo {author} {\bibfnamefont {Heqing}\
  \bibnamefont {Wang}}, \bibinfo {author} {\bibfnamefont {Hao}\ \bibnamefont
  {Li}}, \bibinfo {author} {\bibfnamefont {You}\ \bibnamefont {Wang}}, \bibinfo
  {author} {\bibfnamefont {Guangwei}\ \bibnamefont {Deng}}, \bibinfo {author}
  {\bibfnamefont {Zhiming}\ \bibnamefont {Wang}}, \bibinfo {author}
  {\bibfnamefont {Lixing}\ \bibnamefont {You}}, \bibinfo {author}
  {\bibfnamefont {Zhen}\ \bibnamefont {Wang}}, \bibinfo {author} {\bibfnamefont
  {Haizhi}\ \bibnamefont {Song}}, \bibinfo {author} {\bibfnamefont {Guangcan}\
  \bibnamefont {Guo}}, \ and\ \bibinfo {author} {\bibfnamefont {Qiang}\
  \bibnamefont {Zhou}},\ }\bibfield  {title} {\enquote {\bibinfo {title}
  {High-performance quantum entanglement generation via cascaded second-order
  nonlinear processes},}\ }\href {\doibase 10.1038/s41534-021-00462-7}
  {\bibfield  {journal} {\bibinfo  {journal} {npj Quantum Information}\
  }\textbf {\bibinfo {volume} {7}},\ \bibinfo {pages} {123} (\bibinfo {year}
  {2021})}\BibitemShut {NoStop}%
\bibitem [{\citenamefont {Marxer}\ \emph {et~al.}(2023)\citenamefont {Marxer},
  \citenamefont {Veps\"al\"ainen}, \citenamefont {Jolin}, \citenamefont
  {Tuorila}, \citenamefont {Landra}, \citenamefont {Ockeloen-Korppi},
  \citenamefont {Liu}, \citenamefont {Ahonen}, \citenamefont {Auer},
  \citenamefont {Belzane}, \citenamefont {Bergholm}, \citenamefont {Chan},
  \citenamefont {Chan}, \citenamefont {Hiltunen}, \citenamefont {Hotari},
  \citenamefont {Hyypp\"a}, \citenamefont {Ikonen}, \citenamefont {Janzso},
  \citenamefont {Koistinen}, \citenamefont {Kotilahti}, \citenamefont {Li},
  \citenamefont {Luus}, \citenamefont {Papic}, \citenamefont {Partanen},
  \citenamefont {R\"abin\"a}, \citenamefont {Rosti}, \citenamefont {Savytskyi},
  \citenamefont {Sepp\"al\"a}, \citenamefont {Sevriuk}, \citenamefont {Takala},
  \citenamefont {Tarasinski}, \citenamefont {Thapa}, \citenamefont {Tosto},
  \citenamefont {Vorobeva}, \citenamefont {Yu}, \citenamefont {Tan},
  \citenamefont {Hassel}, \citenamefont {M\"ott\"onen},\ and\ \citenamefont
  {Heinsoo}}]{Marxer2023}%
  \BibitemOpen
  \bibfield  {author} {\bibinfo {author} {\bibfnamefont {Fabian}\ \bibnamefont
  {Marxer}}, \bibinfo {author} {\bibfnamefont {Antti}\ \bibnamefont
  {Veps\"al\"ainen}}, \bibinfo {author} {\bibfnamefont {Shan~W.}\ \bibnamefont
  {Jolin}}, \bibinfo {author} {\bibfnamefont {Jani}\ \bibnamefont {Tuorila}},
  \bibinfo {author} {\bibfnamefont {Alessandro}\ \bibnamefont {Landra}},
  \bibinfo {author} {\bibfnamefont {Caspar}\ \bibnamefont {Ockeloen-Korppi}},
  \bibinfo {author} {\bibfnamefont {Wei}\ \bibnamefont {Liu}}, \bibinfo
  {author} {\bibfnamefont {Olli}\ \bibnamefont {Ahonen}}, \bibinfo {author}
  {\bibfnamefont {Adrian}\ \bibnamefont {Auer}}, \bibinfo {author}
  {\bibfnamefont {Lucien}\ \bibnamefont {Belzane}}, \bibinfo {author}
  {\bibfnamefont {Ville}\ \bibnamefont {Bergholm}}, \bibinfo {author}
  {\bibfnamefont {Chun~Fai}\ \bibnamefont {Chan}}, \bibinfo {author}
  {\bibfnamefont {Kok~Wai}\ \bibnamefont {Chan}}, \bibinfo {author}
  {\bibfnamefont {Tuukka}\ \bibnamefont {Hiltunen}}, \bibinfo {author}
  {\bibfnamefont {Juho}\ \bibnamefont {Hotari}}, \bibinfo {author}
  {\bibfnamefont {Eric}\ \bibnamefont {Hyypp\"a}}, \bibinfo {author}
  {\bibfnamefont {Joni}\ \bibnamefont {Ikonen}}, \bibinfo {author}
  {\bibfnamefont {David}\ \bibnamefont {Janzso}}, \bibinfo {author}
  {\bibfnamefont {Miikka}\ \bibnamefont {Koistinen}}, \bibinfo {author}
  {\bibfnamefont {Janne}\ \bibnamefont {Kotilahti}}, \bibinfo {author}
  {\bibfnamefont {Tianyi}\ \bibnamefont {Li}}, \bibinfo {author} {\bibfnamefont
  {Jyrgen}\ \bibnamefont {Luus}}, \bibinfo {author} {\bibfnamefont {Miha}\
  \bibnamefont {Papic}}, \bibinfo {author} {\bibfnamefont {Matti}\ \bibnamefont
  {Partanen}}, \bibinfo {author} {\bibfnamefont {Jukka}\ \bibnamefont
  {R\"abin\"a}}, \bibinfo {author} {\bibfnamefont {Jari}\ \bibnamefont
  {Rosti}}, \bibinfo {author} {\bibfnamefont {Mykhailo}\ \bibnamefont
  {Savytskyi}}, \bibinfo {author} {\bibfnamefont {Marko}\ \bibnamefont
  {Sepp\"al\"a}}, \bibinfo {author} {\bibfnamefont {Vasilii}\ \bibnamefont
  {Sevriuk}}, \bibinfo {author} {\bibfnamefont {Eelis}\ \bibnamefont {Takala}},
  \bibinfo {author} {\bibfnamefont {Brian}\ \bibnamefont {Tarasinski}},
  \bibinfo {author} {\bibfnamefont {Manish~J.}\ \bibnamefont {Thapa}}, \bibinfo
  {author} {\bibfnamefont {Francesca}\ \bibnamefont {Tosto}}, \bibinfo {author}
  {\bibfnamefont {Natalia}\ \bibnamefont {Vorobeva}}, \bibinfo {author}
  {\bibfnamefont {Liuqi}\ \bibnamefont {Yu}}, \bibinfo {author} {\bibfnamefont
  {Kuan~Yen}\ \bibnamefont {Tan}}, \bibinfo {author} {\bibfnamefont {Juha}\
  \bibnamefont {Hassel}}, \bibinfo {author} {\bibfnamefont {Mikko}\
  \bibnamefont {M\"ott\"onen}}, \ and\ \bibinfo {author} {\bibfnamefont
  {Johannes}\ \bibnamefont {Heinsoo}},\ }\bibfield  {title} {\enquote {\bibinfo
  {title} {Long-distance transmon coupler with cz-gate fidelity above
  $99.8\mathrm{\%}$},}\ }\href {\doibase 10.1103/PRXQuantum.4.010314}
  {\bibfield  {journal} {\bibinfo  {journal} {PRX Quantum}\ }\textbf {\bibinfo
  {volume} {4}},\ \bibinfo {pages} {010314} (\bibinfo {year}
  {2023})}\BibitemShut {NoStop}%
\bibitem [{\citenamefont {Blais}\ \emph {et~al.}(2021)\citenamefont {Blais},
  \citenamefont {Grimsmo}, \citenamefont {Girvin},\ and\ \citenamefont
  {Wallraff}}]{Blais2021}%
  \BibitemOpen
  \bibfield  {author} {\bibinfo {author} {\bibfnamefont {Alexandre}\
  \bibnamefont {Blais}}, \bibinfo {author} {\bibfnamefont {Arne~L.}\
  \bibnamefont {Grimsmo}}, \bibinfo {author} {\bibfnamefont {S.~M.}\
  \bibnamefont {Girvin}}, \ and\ \bibinfo {author} {\bibfnamefont {Andreas}\
  \bibnamefont {Wallraff}},\ }\bibfield  {title} {\enquote {\bibinfo {title}
  {Circuit quantum electrodynamics},}\ }\href {\doibase
  10.1103/RevModPhys.93.025005} {\bibfield  {journal} {\bibinfo  {journal}
  {Rev. Mod. Phys.}\ }\textbf {\bibinfo {volume} {93}},\ \bibinfo {pages}
  {025005} (\bibinfo {year} {2021})}\BibitemShut {NoStop}%
\bibitem [{\citenamefont {Kjaergaard}\ \emph {et~al.}(2020)\citenamefont
  {Kjaergaard}, \citenamefont {Schwartz}, \citenamefont {Braum\"{u}ller},
  \citenamefont {Krantz}, \citenamefont {Wang}, \citenamefont {Gustavsson},\
  and\ \citenamefont {Oliver}}]{Kjaergaard2020}%
  \BibitemOpen
  \bibfield  {author} {\bibinfo {author} {\bibfnamefont {Morten}\ \bibnamefont
  {Kjaergaard}}, \bibinfo {author} {\bibfnamefont {Mollie~E.}\ \bibnamefont
  {Schwartz}}, \bibinfo {author} {\bibfnamefont {Jochen}\ \bibnamefont
  {Braum\"{u}ller}}, \bibinfo {author} {\bibfnamefont {Philip}\ \bibnamefont
  {Krantz}}, \bibinfo {author} {\bibfnamefont {Joel I.-J.}\ \bibnamefont
  {Wang}}, \bibinfo {author} {\bibfnamefont {Simon}\ \bibnamefont
  {Gustavsson}}, \ and\ \bibinfo {author} {\bibfnamefont {William~D.}\
  \bibnamefont {Oliver}},\ }\bibfield  {title} {\enquote {\bibinfo {title}
  {Superconducting qubits: Current state of play},}\ }\href {\doibase
  10.1146/annurev-conmatphys-031119-050605} {\bibfield  {journal} {\bibinfo
  {journal} {Annual Review of Condensed Matter Physics}\ }\textbf {\bibinfo
  {volume} {11}},\ \bibinfo {pages} {369--395} (\bibinfo {year} {2020})},\
  \Eprint
  {http://arxiv.org/abs/https://doi.org/10.1146/annurev-conmatphys-031119-050605}
  {https://doi.org/10.1146/annurev-conmatphys-031119-050605} \BibitemShut
  {NoStop}%
\bibitem [{\citenamefont {Koch}\ \emph {et~al.}(2007)\citenamefont {Koch},
  \citenamefont {Yu}, \citenamefont {Gambetta}, \citenamefont {Houck},
  \citenamefont {Schuster}, \citenamefont {Majer}, \citenamefont {Blais},
  \citenamefont {Devoret}, \citenamefont {Girvin},\ and\ \citenamefont
  {Schoelkopf}}]{Koch2007}%
  \BibitemOpen
  \bibfield  {author} {\bibinfo {author} {\bibfnamefont {Jens}\ \bibnamefont
  {Koch}}, \bibinfo {author} {\bibfnamefont {Terri~M.}\ \bibnamefont {Yu}},
  \bibinfo {author} {\bibfnamefont {Jay}\ \bibnamefont {Gambetta}}, \bibinfo
  {author} {\bibfnamefont {A.~A.}\ \bibnamefont {Houck}}, \bibinfo {author}
  {\bibfnamefont {D.~I.}\ \bibnamefont {Schuster}}, \bibinfo {author}
  {\bibfnamefont {J.}~\bibnamefont {Majer}}, \bibinfo {author} {\bibfnamefont
  {Alexandre}\ \bibnamefont {Blais}}, \bibinfo {author} {\bibfnamefont {M.~H.}\
  \bibnamefont {Devoret}}, \bibinfo {author} {\bibfnamefont {S.~M.}\
  \bibnamefont {Girvin}}, \ and\ \bibinfo {author} {\bibfnamefont {R.~J.}\
  \bibnamefont {Schoelkopf}},\ }\bibfield  {title} {\enquote {\bibinfo {title}
  {Charge-insensitive qubit design derived from the cooper pair box},}\ }\href
  {\doibase 10.1103/PhysRevA.76.042319} {\bibfield  {journal} {\bibinfo
  {journal} {Phys. Rev. A}\ }\textbf {\bibinfo {volume} {76}},\ \bibinfo
  {pages} {042319} (\bibinfo {year} {2007})}\BibitemShut {NoStop}%
\bibitem [{\citenamefont {Schreier}\ \emph {et~al.}(2008)\citenamefont
  {Schreier}, \citenamefont {Houck}, \citenamefont {Koch}, \citenamefont
  {Schuster}, \citenamefont {Johnson}, \citenamefont {Chow}, \citenamefont
  {Gambetta}, \citenamefont {Majer}, \citenamefont {Frunzio}, \citenamefont
  {Devoret}, \citenamefont {Girvin},\ and\ \citenamefont
  {Schoelkopf}}]{Schreier2008}%
  \BibitemOpen
  \bibfield  {author} {\bibinfo {author} {\bibfnamefont {J.~A.}\ \bibnamefont
  {Schreier}}, \bibinfo {author} {\bibfnamefont {A.~A.}\ \bibnamefont {Houck}},
  \bibinfo {author} {\bibfnamefont {Jens}\ \bibnamefont {Koch}}, \bibinfo
  {author} {\bibfnamefont {D.~I.}\ \bibnamefont {Schuster}}, \bibinfo {author}
  {\bibfnamefont {B.~R.}\ \bibnamefont {Johnson}}, \bibinfo {author}
  {\bibfnamefont {J.~M.}\ \bibnamefont {Chow}}, \bibinfo {author}
  {\bibfnamefont {J.~M.}\ \bibnamefont {Gambetta}}, \bibinfo {author}
  {\bibfnamefont {J.}~\bibnamefont {Majer}}, \bibinfo {author} {\bibfnamefont
  {L.}~\bibnamefont {Frunzio}}, \bibinfo {author} {\bibfnamefont {M.~H.}\
  \bibnamefont {Devoret}}, \bibinfo {author} {\bibfnamefont {S.~M.}\
  \bibnamefont {Girvin}}, \ and\ \bibinfo {author} {\bibfnamefont {R.~J.}\
  \bibnamefont {Schoelkopf}},\ }\bibfield  {title} {\enquote {\bibinfo {title}
  {Suppressing charge noise decoherence in superconducting charge qubits},}\
  }\href {\doibase 10.1103/PhysRevB.77.180502} {\bibfield  {journal} {\bibinfo
  {journal} {Phys. Rev. B}\ }\textbf {\bibinfo {volume} {77}},\ \bibinfo
  {pages} {180502} (\bibinfo {year} {2008})}\BibitemShut {NoStop}%
\bibitem [{\citenamefont {Houck}\ \emph {et~al.}(2008)\citenamefont {Houck},
  \citenamefont {Schreier}, \citenamefont {Johnson}, \citenamefont {Chow},
  \citenamefont {Koch}, \citenamefont {Gambetta}, \citenamefont {Schuster},
  \citenamefont {Frunzio}, \citenamefont {Devoret}, \citenamefont {Girvin},\
  and\ \citenamefont {Schoelkopf}}]{Houck2008}%
  \BibitemOpen
  \bibfield  {author} {\bibinfo {author} {\bibfnamefont {A.~A.}\ \bibnamefont
  {Houck}}, \bibinfo {author} {\bibfnamefont {J.~A.}\ \bibnamefont {Schreier}},
  \bibinfo {author} {\bibfnamefont {B.~R.}\ \bibnamefont {Johnson}}, \bibinfo
  {author} {\bibfnamefont {J.~M.}\ \bibnamefont {Chow}}, \bibinfo {author}
  {\bibfnamefont {Jens}\ \bibnamefont {Koch}}, \bibinfo {author} {\bibfnamefont
  {J.~M.}\ \bibnamefont {Gambetta}}, \bibinfo {author} {\bibfnamefont {D.~I.}\
  \bibnamefont {Schuster}}, \bibinfo {author} {\bibfnamefont {L.}~\bibnamefont
  {Frunzio}}, \bibinfo {author} {\bibfnamefont {M.~H.}\ \bibnamefont
  {Devoret}}, \bibinfo {author} {\bibfnamefont {S.~M.}\ \bibnamefont {Girvin}},
  \ and\ \bibinfo {author} {\bibfnamefont {R.~J.}\ \bibnamefont {Schoelkopf}},\
  }\bibfield  {title} {\enquote {\bibinfo {title} {Controlling the spontaneous
  emission of a superconducting transmon qubit},}\ }\href {\doibase
  10.1103/PhysRevLett.101.080502} {\bibfield  {journal} {\bibinfo  {journal}
  {Phys. Rev. Lett.}\ }\textbf {\bibinfo {volume} {101}},\ \bibinfo {pages}
  {080502} (\bibinfo {year} {2008})}\BibitemShut {NoStop}%
\bibitem [{\citenamefont {Place}\ \emph {et~al.}(2021)\citenamefont {Place},
  \citenamefont {Rodgers}, \citenamefont {Mundada}, \citenamefont {Smitham},
  \citenamefont {Fitzpatrick}, \citenamefont {Leng}, \citenamefont {Premkumar},
  \citenamefont {Bryon}, \citenamefont {Vrajitoarea}, \citenamefont {Sussman},
  \citenamefont {Cheng}, \citenamefont {Madhavan}, \citenamefont {Babla},
  \citenamefont {Le}, \citenamefont {Gang}, \citenamefont {J{\"a}ck},
  \citenamefont {Gyenis}, \citenamefont {Yao}, \citenamefont {Cava},
  \citenamefont {de~Leon},\ and\ \citenamefont {Houck}}]{Place2021}%
  \BibitemOpen
  \bibfield  {author} {\bibinfo {author} {\bibfnamefont {Alexander P.~M.}\
  \bibnamefont {Place}}, \bibinfo {author} {\bibfnamefont {Lila V.~H.}\
  \bibnamefont {Rodgers}}, \bibinfo {author} {\bibfnamefont {Pranav}\
  \bibnamefont {Mundada}}, \bibinfo {author} {\bibfnamefont {Basil~M.}\
  \bibnamefont {Smitham}}, \bibinfo {author} {\bibfnamefont {Mattias}\
  \bibnamefont {Fitzpatrick}}, \bibinfo {author} {\bibfnamefont {Zhaoqi}\
  \bibnamefont {Leng}}, \bibinfo {author} {\bibfnamefont {Anjali}\ \bibnamefont
  {Premkumar}}, \bibinfo {author} {\bibfnamefont {Jacob}\ \bibnamefont
  {Bryon}}, \bibinfo {author} {\bibfnamefont {Andrei}\ \bibnamefont
  {Vrajitoarea}}, \bibinfo {author} {\bibfnamefont {Sara}\ \bibnamefont
  {Sussman}}, \bibinfo {author} {\bibfnamefont {Guangming}\ \bibnamefont
  {Cheng}}, \bibinfo {author} {\bibfnamefont {Trisha}\ \bibnamefont
  {Madhavan}}, \bibinfo {author} {\bibfnamefont {Harshvardhan~K.}\ \bibnamefont
  {Babla}}, \bibinfo {author} {\bibfnamefont {Xuan~Hoang}\ \bibnamefont {Le}},
  \bibinfo {author} {\bibfnamefont {Youqi}\ \bibnamefont {Gang}}, \bibinfo
  {author} {\bibfnamefont {Berthold}\ \bibnamefont {J{\"a}ck}}, \bibinfo
  {author} {\bibfnamefont {Andr{\'a}s}\ \bibnamefont {Gyenis}}, \bibinfo
  {author} {\bibfnamefont {Nan}\ \bibnamefont {Yao}}, \bibinfo {author}
  {\bibfnamefont {Robert~J.}\ \bibnamefont {Cava}}, \bibinfo {author}
  {\bibfnamefont {Nathalie~P.}\ \bibnamefont {de~Leon}}, \ and\ \bibinfo
  {author} {\bibfnamefont {Andrew~A.}\ \bibnamefont {Houck}},\ }\bibfield
  {title} {\enquote {\bibinfo {title} {New material platform for
  superconducting transmon qubits with coherence times exceeding 0.3
  milliseconds},}\ }\href {\doibase 10.1038/s41467-021-22030-5} {\bibfield
  {journal} {\bibinfo  {journal} {Nature Communications}\ }\textbf {\bibinfo
  {volume} {12}},\ \bibinfo {pages} {1779} (\bibinfo {year}
  {2021})}\BibitemShut {NoStop}%
\bibitem [{\citenamefont {Wang}\ \emph
  {et~al.}(2022{\natexlab{a}})\citenamefont {Wang}, \citenamefont {Li},
  \citenamefont {Xu}, \citenamefont {Li}, \citenamefont {Wang}, \citenamefont
  {Yang}, \citenamefont {Mi}, \citenamefont {Liang}, \citenamefont {Su},
  \citenamefont {Yang}, \citenamefont {Wang}, \citenamefont {Wang},
  \citenamefont {Li}, \citenamefont {Chen}, \citenamefont {Li}, \citenamefont
  {Linghu}, \citenamefont {Han}, \citenamefont {Zhang}, \citenamefont {Feng},
  \citenamefont {Song}, \citenamefont {Ma}, \citenamefont {Zhang},
  \citenamefont {Wang}, \citenamefont {Zhao}, \citenamefont {Liu},
  \citenamefont {Xue}, \citenamefont {Jin},\ and\ \citenamefont
  {Yu}}]{Wang2022Transmon}%
  \BibitemOpen
  \bibfield  {author} {\bibinfo {author} {\bibfnamefont {Chenlu}\ \bibnamefont
  {Wang}}, \bibinfo {author} {\bibfnamefont {Xuegang}\ \bibnamefont {Li}},
  \bibinfo {author} {\bibfnamefont {Huikai}\ \bibnamefont {Xu}}, \bibinfo
  {author} {\bibfnamefont {Zhiyuan}\ \bibnamefont {Li}}, \bibinfo {author}
  {\bibfnamefont {Junhua}\ \bibnamefont {Wang}}, \bibinfo {author}
  {\bibfnamefont {Zhen}\ \bibnamefont {Yang}}, \bibinfo {author} {\bibfnamefont
  {Zhenyu}\ \bibnamefont {Mi}}, \bibinfo {author} {\bibfnamefont {Xuehui}\
  \bibnamefont {Liang}}, \bibinfo {author} {\bibfnamefont {Tang}\ \bibnamefont
  {Su}}, \bibinfo {author} {\bibfnamefont {Chuhong}\ \bibnamefont {Yang}},
  \bibinfo {author} {\bibfnamefont {Guangyue}\ \bibnamefont {Wang}}, \bibinfo
  {author} {\bibfnamefont {Wenyan}\ \bibnamefont {Wang}}, \bibinfo {author}
  {\bibfnamefont {Yongchao}\ \bibnamefont {Li}}, \bibinfo {author}
  {\bibfnamefont {Mo}~\bibnamefont {Chen}}, \bibinfo {author} {\bibfnamefont
  {Chengyao}\ \bibnamefont {Li}}, \bibinfo {author} {\bibfnamefont {Kehuan}\
  \bibnamefont {Linghu}}, \bibinfo {author} {\bibfnamefont {Jiaxiu}\
  \bibnamefont {Han}}, \bibinfo {author} {\bibfnamefont {Yingshan}\
  \bibnamefont {Zhang}}, \bibinfo {author} {\bibfnamefont {Yulong}\
  \bibnamefont {Feng}}, \bibinfo {author} {\bibfnamefont {Yu}~\bibnamefont
  {Song}}, \bibinfo {author} {\bibfnamefont {Teng}\ \bibnamefont {Ma}},
  \bibinfo {author} {\bibfnamefont {Jingning}\ \bibnamefont {Zhang}}, \bibinfo
  {author} {\bibfnamefont {Ruixia}\ \bibnamefont {Wang}}, \bibinfo {author}
  {\bibfnamefont {Peng}\ \bibnamefont {Zhao}}, \bibinfo {author} {\bibfnamefont
  {Weiyang}\ \bibnamefont {Liu}}, \bibinfo {author} {\bibfnamefont {Guangming}\
  \bibnamefont {Xue}}, \bibinfo {author} {\bibfnamefont {Yirong}\ \bibnamefont
  {Jin}}, \ and\ \bibinfo {author} {\bibfnamefont {Haifeng}\ \bibnamefont
  {Yu}},\ }\bibfield  {title} {\enquote {\bibinfo {title} {Towards practical
  quantum computers: transmon qubit with a lifetime approaching 0.5
  milliseconds},}\ }\href {\doibase 10.1038/s41534-021-00510-2} {\bibfield
  {journal} {\bibinfo  {journal} {npj Quantum Information}\ }\textbf {\bibinfo
  {volume} {8}},\ \bibinfo {pages} {3} (\bibinfo {year}
  {2022}{\natexlab{a}})}\BibitemShut {NoStop}%
\bibitem [{\citenamefont {Manucharyan}\ \emph {et~al.}(2009)\citenamefont
  {Manucharyan}, \citenamefont {Koch}, \citenamefont {Glazman},\ and\
  \citenamefont {Devoret}}]{Manucharyan2009}%
  \BibitemOpen
  \bibfield  {author} {\bibinfo {author} {\bibfnamefont {Vladimir~E.}\
  \bibnamefont {Manucharyan}}, \bibinfo {author} {\bibfnamefont {Jens}\
  \bibnamefont {Koch}}, \bibinfo {author} {\bibfnamefont {Leonid~I.}\
  \bibnamefont {Glazman}}, \ and\ \bibinfo {author} {\bibfnamefont {Michel~H.}\
  \bibnamefont {Devoret}},\ }\bibfield  {title} {\enquote {\bibinfo {title}
  {{Fluxonium: Single Cooper-Pair Circuit Free of Charge Offsets}},}\ }\href
  {\doibase 10.1126/science.1175552} {\bibfield  {journal} {\bibinfo  {journal}
  {Science}\ }\textbf {\bibinfo {volume} {326}},\ \bibinfo {pages} {113--116}
  (\bibinfo {year} {2009})}\BibitemShut {NoStop}%
\bibitem [{\citenamefont {Somoroff}\ \emph {et~al.}(2023)\citenamefont
  {Somoroff}, \citenamefont {Ficheux}, \citenamefont {Mencia}, \citenamefont
  {Xiong}, \citenamefont {Kuzmin},\ and\ \citenamefont
  {Manucharyan}}]{Somoroff2023}%
  \BibitemOpen
  \bibfield  {author} {\bibinfo {author} {\bibfnamefont {Aaron}\ \bibnamefont
  {Somoroff}}, \bibinfo {author} {\bibfnamefont {Quentin}\ \bibnamefont
  {Ficheux}}, \bibinfo {author} {\bibfnamefont {Raymond~A.}\ \bibnamefont
  {Mencia}}, \bibinfo {author} {\bibfnamefont {Haonan}\ \bibnamefont {Xiong}},
  \bibinfo {author} {\bibfnamefont {Roman}\ \bibnamefont {Kuzmin}}, \ and\
  \bibinfo {author} {\bibfnamefont {Vladimir~E.}\ \bibnamefont {Manucharyan}},\
  }\bibfield  {title} {\enquote {\bibinfo {title} {Millisecond coherence in a
  superconducting qubit},}\ }\href {\doibase 10.1103/PhysRevLett.130.267001}
  {\bibfield  {journal} {\bibinfo  {journal} {Phys. Rev. Lett.}\ }\textbf
  {\bibinfo {volume} {130}},\ \bibinfo {pages} {267001} (\bibinfo {year}
  {2023})}\BibitemShut {NoStop}%
\bibitem [{\citenamefont {Sheng}\ \emph {et~al.}(2018)\citenamefont {Sheng},
  \citenamefont {He}, \citenamefont {Xu}, \citenamefont {Guo}, \citenamefont
  {Wang}, \citenamefont {Xiong}, \citenamefont {Liu}, \citenamefont {Wang},\
  and\ \citenamefont {Zhan}}]{Sheng2018}%
  \BibitemOpen
  \bibfield  {author} {\bibinfo {author} {\bibfnamefont {Cheng}\ \bibnamefont
  {Sheng}}, \bibinfo {author} {\bibfnamefont {Xiaodong}\ \bibnamefont {He}},
  \bibinfo {author} {\bibfnamefont {Peng}\ \bibnamefont {Xu}}, \bibinfo
  {author} {\bibfnamefont {Ruijun}\ \bibnamefont {Guo}}, \bibinfo {author}
  {\bibfnamefont {Kunpeng}\ \bibnamefont {Wang}}, \bibinfo {author}
  {\bibfnamefont {Zongyuan}\ \bibnamefont {Xiong}}, \bibinfo {author}
  {\bibfnamefont {Min}\ \bibnamefont {Liu}}, \bibinfo {author} {\bibfnamefont
  {Jin}\ \bibnamefont {Wang}}, \ and\ \bibinfo {author} {\bibfnamefont
  {Mingsheng}\ \bibnamefont {Zhan}},\ }\bibfield  {title} {\enquote {\bibinfo
  {title} {High-fidelity single-qubit gates on neutral atoms in a
  two-dimensional magic-intensity optical dipole trap array},}\ }\href
  {\doibase 10.1103/PhysRevLett.121.240501} {\bibfield  {journal} {\bibinfo
  {journal} {Phys. Rev. Lett.}\ }\textbf {\bibinfo {volume} {121}},\ \bibinfo
  {pages} {240501} (\bibinfo {year} {2018})}\BibitemShut {NoStop}%
\bibitem [{\citenamefont {Bluvstein}\ \emph {et~al.}(2022)\citenamefont
  {Bluvstein}, \citenamefont {Levine}, \citenamefont {Semeghini}, \citenamefont
  {Wang}, \citenamefont {Ebadi}, \citenamefont {Kalinowski}, \citenamefont
  {Keesling}, \citenamefont {Maskara}, \citenamefont {Pichler}, \citenamefont
  {Greiner}, \citenamefont {Vuleti{\'c}},\ and\ \citenamefont
  {Lukin}}]{Bluvstein2022}%
  \BibitemOpen
  \bibfield  {author} {\bibinfo {author} {\bibfnamefont {Dolev}\ \bibnamefont
  {Bluvstein}}, \bibinfo {author} {\bibfnamefont {Harry}\ \bibnamefont
  {Levine}}, \bibinfo {author} {\bibfnamefont {Giulia}\ \bibnamefont
  {Semeghini}}, \bibinfo {author} {\bibfnamefont {Tout~T.}\ \bibnamefont
  {Wang}}, \bibinfo {author} {\bibfnamefont {Sepehr}\ \bibnamefont {Ebadi}},
  \bibinfo {author} {\bibfnamefont {Marcin}\ \bibnamefont {Kalinowski}},
  \bibinfo {author} {\bibfnamefont {Alexander}\ \bibnamefont {Keesling}},
  \bibinfo {author} {\bibfnamefont {Nishad}\ \bibnamefont {Maskara}}, \bibinfo
  {author} {\bibfnamefont {Hannes}\ \bibnamefont {Pichler}}, \bibinfo {author}
  {\bibfnamefont {Markus}\ \bibnamefont {Greiner}}, \bibinfo {author}
  {\bibfnamefont {Vladan}\ \bibnamefont {Vuleti{\'c}}}, \ and\ \bibinfo
  {author} {\bibfnamefont {Mikhail~D.}\ \bibnamefont {Lukin}},\ }\bibfield
  {title} {\enquote {\bibinfo {title} {A quantum processor based on coherent
  transport of entangled atom arrays},}\ }\href {\doibase
  10.1038/s41586-022-04592-6} {\bibfield  {journal} {\bibinfo  {journal}
  {Nature}\ }\textbf {\bibinfo {volume} {604}},\ \bibinfo {pages} {451--456}
  (\bibinfo {year} {2022})}\BibitemShut {NoStop}%
\bibitem [{\citenamefont {Ma}\ \emph {et~al.}(2022)\citenamefont {Ma},
  \citenamefont {Burgers}, \citenamefont {Liu}, \citenamefont {Wilson},
  \citenamefont {Zhang},\ and\ \citenamefont {Thompson}}]{Ma2022}%
  \BibitemOpen
  \bibfield  {author} {\bibinfo {author} {\bibfnamefont {Shuo}\ \bibnamefont
  {Ma}}, \bibinfo {author} {\bibfnamefont {Alex~P.}\ \bibnamefont {Burgers}},
  \bibinfo {author} {\bibfnamefont {Genyue}\ \bibnamefont {Liu}}, \bibinfo
  {author} {\bibfnamefont {Jack}\ \bibnamefont {Wilson}}, \bibinfo {author}
  {\bibfnamefont {Bichen}\ \bibnamefont {Zhang}}, \ and\ \bibinfo {author}
  {\bibfnamefont {Jeff~D.}\ \bibnamefont {Thompson}},\ }\bibfield  {title}
  {\enquote {\bibinfo {title} {Universal gate operations on nuclear spin qubits
  in an optical tweezer array of $^{171}\mathrm{Yb}$ atoms},}\ }\href {\doibase
  10.1103/PhysRevX.12.021028} {\bibfield  {journal} {\bibinfo  {journal} {Phys.
  Rev. X}\ }\textbf {\bibinfo {volume} {12}},\ \bibinfo {pages} {021028}
  (\bibinfo {year} {2022})}\BibitemShut {NoStop}%
\bibitem [{\citenamefont {Jenkins}\ \emph {et~al.}(2022)\citenamefont
  {Jenkins}, \citenamefont {Lis}, \citenamefont {Senoo}, \citenamefont
  {McGrew},\ and\ \citenamefont {Kaufman}}]{Jenkins2022}%
  \BibitemOpen
  \bibfield  {author} {\bibinfo {author} {\bibfnamefont {Alec}\ \bibnamefont
  {Jenkins}}, \bibinfo {author} {\bibfnamefont {Joanna~W.}\ \bibnamefont
  {Lis}}, \bibinfo {author} {\bibfnamefont {Aruku}\ \bibnamefont {Senoo}},
  \bibinfo {author} {\bibfnamefont {William~F.}\ \bibnamefont {McGrew}}, \ and\
  \bibinfo {author} {\bibfnamefont {Adam~M.}\ \bibnamefont {Kaufman}},\
  }\bibfield  {title} {\enquote {\bibinfo {title} {Ytterbium nuclear-spin
  qubits in an optical tweezer array},}\ }\href {\doibase
  10.1103/PhysRevX.12.021027} {\bibfield  {journal} {\bibinfo  {journal} {Phys.
  Rev. X}\ }\textbf {\bibinfo {volume} {12}},\ \bibinfo {pages} {021027}
  (\bibinfo {year} {2022})}\BibitemShut {NoStop}%
\bibitem [{\citenamefont {Wang}\ \emph {et~al.}(2017)\citenamefont {Wang},
  \citenamefont {Um}, \citenamefont {Zhang}, \citenamefont {An}, \citenamefont
  {Lyu}, \citenamefont {Zhang}, \citenamefont {Duan}, \citenamefont {Yum},\
  and\ \citenamefont {Kim}}]{Wang2017}%
  \BibitemOpen
  \bibfield  {author} {\bibinfo {author} {\bibfnamefont {Ye}~\bibnamefont
  {Wang}}, \bibinfo {author} {\bibfnamefont {Mark}\ \bibnamefont {Um}},
  \bibinfo {author} {\bibfnamefont {Junhua}\ \bibnamefont {Zhang}}, \bibinfo
  {author} {\bibfnamefont {Shuoming}\ \bibnamefont {An}}, \bibinfo {author}
  {\bibfnamefont {Ming}\ \bibnamefont {Lyu}}, \bibinfo {author} {\bibfnamefont
  {Jing-Ning}\ \bibnamefont {Zhang}}, \bibinfo {author} {\bibfnamefont {L.~M.}\
  \bibnamefont {Duan}}, \bibinfo {author} {\bibfnamefont {Dahyun}\ \bibnamefont
  {Yum}}, \ and\ \bibinfo {author} {\bibfnamefont {Kihwan}\ \bibnamefont
  {Kim}},\ }\bibfield  {title} {\enquote {\bibinfo {title} {Single-qubit
  quantum memory exceeding ten-minute coherence time},}\ }\href {\doibase
  10.1038/s41566-017-0007-1} {\bibfield  {journal} {\bibinfo  {journal} {Nature
  Photonics}\ }\textbf {\bibinfo {volume} {11}},\ \bibinfo {pages} {646--650}
  (\bibinfo {year} {2017})}\BibitemShut {NoStop}%
\bibitem [{\citenamefont {Wang}\ \emph {et~al.}(2021)\citenamefont {Wang},
  \citenamefont {Luan}, \citenamefont {Qiao}, \citenamefont {Um}, \citenamefont
  {Zhang}, \citenamefont {Wang}, \citenamefont {Yuan}, \citenamefont {Gu},
  \citenamefont {Zhang},\ and\ \citenamefont {Kim}}]{Wang2021}%
  \BibitemOpen
  \bibfield  {author} {\bibinfo {author} {\bibfnamefont {Pengfei}\ \bibnamefont
  {Wang}}, \bibinfo {author} {\bibfnamefont {Chun-Yang}\ \bibnamefont {Luan}},
  \bibinfo {author} {\bibfnamefont {Mu}~\bibnamefont {Qiao}}, \bibinfo {author}
  {\bibfnamefont {Mark}\ \bibnamefont {Um}}, \bibinfo {author} {\bibfnamefont
  {Junhua}\ \bibnamefont {Zhang}}, \bibinfo {author} {\bibfnamefont
  {Ye}~\bibnamefont {Wang}}, \bibinfo {author} {\bibfnamefont {Xiao}\
  \bibnamefont {Yuan}}, \bibinfo {author} {\bibfnamefont {Mile}\ \bibnamefont
  {Gu}}, \bibinfo {author} {\bibfnamefont {Jingning}\ \bibnamefont {Zhang}}, \
  and\ \bibinfo {author} {\bibfnamefont {Kihwan}\ \bibnamefont {Kim}},\
  }\bibfield  {title} {\enquote {\bibinfo {title} {Single ion qubit with
  estimated coherence time exceeding one hour},}\ }\href {\doibase
  10.1038/s41467-020-20330-w} {\bibfield  {journal} {\bibinfo  {journal}
  {Nature Communications}\ }\textbf {\bibinfo {volume} {12}},\ \bibinfo {pages}
  {233} (\bibinfo {year} {2021})}\BibitemShut {NoStop}%
\bibitem [{\citenamefont {Stehlik}\ \emph {et~al.}(2021)\citenamefont
  {Stehlik}, \citenamefont {Zajac}, \citenamefont {Underwood}, \citenamefont
  {Phung}, \citenamefont {Blair}, \citenamefont {Carnevale}, \citenamefont
  {Klaus}, \citenamefont {Keefe}, \citenamefont {Carniol}, \citenamefont
  {Kumph}, \citenamefont {Steffen},\ and\ \citenamefont {Dial}}]{Stehlik2021}%
  \BibitemOpen
  \bibfield  {author} {\bibinfo {author} {\bibfnamefont {J.}~\bibnamefont
  {Stehlik}}, \bibinfo {author} {\bibfnamefont {D.~M.}\ \bibnamefont {Zajac}},
  \bibinfo {author} {\bibfnamefont {D.~L.}\ \bibnamefont {Underwood}}, \bibinfo
  {author} {\bibfnamefont {T.}~\bibnamefont {Phung}}, \bibinfo {author}
  {\bibfnamefont {J.}~\bibnamefont {Blair}}, \bibinfo {author} {\bibfnamefont
  {S.}~\bibnamefont {Carnevale}}, \bibinfo {author} {\bibfnamefont
  {D.}~\bibnamefont {Klaus}}, \bibinfo {author} {\bibfnamefont {G.~A.}\
  \bibnamefont {Keefe}}, \bibinfo {author} {\bibfnamefont {A.}~\bibnamefont
  {Carniol}}, \bibinfo {author} {\bibfnamefont {M.}~\bibnamefont {Kumph}},
  \bibinfo {author} {\bibfnamefont {Matthias}\ \bibnamefont {Steffen}}, \ and\
  \bibinfo {author} {\bibfnamefont {O.~E.}\ \bibnamefont {Dial}},\ }\bibfield
  {title} {\enquote {\bibinfo {title} {Tunable coupling architecture for
  fixed-frequency transmon superconducting qubits},}\ }\href {\doibase
  10.1103/PhysRevLett.127.080505} {\bibfield  {journal} {\bibinfo  {journal}
  {Phys. Rev. Lett.}\ }\textbf {\bibinfo {volume} {127}},\ \bibinfo {pages}
  {080505} (\bibinfo {year} {2021})}\BibitemShut {NoStop}%
\bibitem [{\citenamefont {Sung}\ \emph {et~al.}(2021)\citenamefont {Sung},
  \citenamefont {Ding}, \citenamefont {Braum\"uller}, \citenamefont
  {Veps\"al\"ainen}, \citenamefont {Kannan}, \citenamefont {Kjaergaard},
  \citenamefont {Greene}, \citenamefont {Samach}, \citenamefont {McNally},
  \citenamefont {Kim}, \citenamefont {Melville}, \citenamefont {Niedzielski},
  \citenamefont {Schwartz}, \citenamefont {Yoder}, \citenamefont {Orlando},
  \citenamefont {Gustavsson},\ and\ \citenamefont {Oliver}}]{Sung2021}%
  \BibitemOpen
  \bibfield  {author} {\bibinfo {author} {\bibfnamefont {Youngkyu}\
  \bibnamefont {Sung}}, \bibinfo {author} {\bibfnamefont {Leon}\ \bibnamefont
  {Ding}}, \bibinfo {author} {\bibfnamefont {Jochen}\ \bibnamefont
  {Braum\"uller}}, \bibinfo {author} {\bibfnamefont {Antti}\ \bibnamefont
  {Veps\"al\"ainen}}, \bibinfo {author} {\bibfnamefont {Bharath}\ \bibnamefont
  {Kannan}}, \bibinfo {author} {\bibfnamefont {Morten}\ \bibnamefont
  {Kjaergaard}}, \bibinfo {author} {\bibfnamefont {Ami}\ \bibnamefont
  {Greene}}, \bibinfo {author} {\bibfnamefont {Gabriel~O.}\ \bibnamefont
  {Samach}}, \bibinfo {author} {\bibfnamefont {Chris}\ \bibnamefont {McNally}},
  \bibinfo {author} {\bibfnamefont {David}\ \bibnamefont {Kim}}, \bibinfo
  {author} {\bibfnamefont {Alexander}\ \bibnamefont {Melville}}, \bibinfo
  {author} {\bibfnamefont {Bethany~M.}\ \bibnamefont {Niedzielski}}, \bibinfo
  {author} {\bibfnamefont {Mollie~E.}\ \bibnamefont {Schwartz}}, \bibinfo
  {author} {\bibfnamefont {Jonilyn~L.}\ \bibnamefont {Yoder}}, \bibinfo
  {author} {\bibfnamefont {Terry~P.}\ \bibnamefont {Orlando}}, \bibinfo
  {author} {\bibfnamefont {Simon}\ \bibnamefont {Gustavsson}}, \ and\ \bibinfo
  {author} {\bibfnamefont {William~D.}\ \bibnamefont {Oliver}},\ }\bibfield
  {title} {\enquote {\bibinfo {title} {Realization of high-fidelity cz and
  $zz$-free iswap gates with a tunable coupler},}\ }\href {\doibase
  10.1103/PhysRevX.11.021058} {\bibfield  {journal} {\bibinfo  {journal} {Phys.
  Rev. X}\ }\textbf {\bibinfo {volume} {11}},\ \bibinfo {pages} {021058}
  (\bibinfo {year} {2021})}\BibitemShut {NoStop}%
\bibitem [{\citenamefont {Nesterov}\ \emph {et~al.}(2018)\citenamefont
  {Nesterov}, \citenamefont {Pechenezhskiy}, \citenamefont {Wang},
  \citenamefont {Manucharyan},\ and\ \citenamefont {Vavilov}}]{Nesterov2018}%
  \BibitemOpen
  \bibfield  {author} {\bibinfo {author} {\bibfnamefont {Konstantin~N.}\
  \bibnamefont {Nesterov}}, \bibinfo {author} {\bibfnamefont {Ivan~V.}\
  \bibnamefont {Pechenezhskiy}}, \bibinfo {author} {\bibfnamefont {Chen}\
  \bibnamefont {Wang}}, \bibinfo {author} {\bibfnamefont {Vladimir~E.}\
  \bibnamefont {Manucharyan}}, \ and\ \bibinfo {author} {\bibfnamefont
  {Maxim~G.}\ \bibnamefont {Vavilov}},\ }\bibfield  {title} {\enquote {\bibinfo
  {title} {{Microwave-activated controlled-$Z$ gate for fixed-frequency
  Fluxonium qubits}},}\ }\href {\doibase 10.1103/PhysRevA.98.030301} {\bibfield
   {journal} {\bibinfo  {journal} {Phys. Rev. A}\ }\textbf {\bibinfo {volume}
  {98}},\ \bibinfo {pages} {030301} (\bibinfo {year} {2018})}\BibitemShut
  {NoStop}%
\bibitem [{\citenamefont {Nguyen}\ \emph {et~al.}(2022)\citenamefont {Nguyen},
  \citenamefont {Koolstra}, \citenamefont {Kim}, \citenamefont {Morvan},
  \citenamefont {Chistolini}, \citenamefont {Singh}, \citenamefont {Nesterov},
  \citenamefont {J\"unger}, \citenamefont {Chen}, \citenamefont {Pedramrazi},
  \citenamefont {Mitchell}, \citenamefont {Kreikebaum}, \citenamefont {Puri},
  \citenamefont {Santiago},\ and\ \citenamefont {Siddiqi}}]{Nguyen2022}%
  \BibitemOpen
  \bibfield  {author} {\bibinfo {author} {\bibfnamefont {Long~B.}\ \bibnamefont
  {Nguyen}}, \bibinfo {author} {\bibfnamefont {Gerwin}\ \bibnamefont
  {Koolstra}}, \bibinfo {author} {\bibfnamefont {Yosep}\ \bibnamefont {Kim}},
  \bibinfo {author} {\bibfnamefont {Alexis}\ \bibnamefont {Morvan}}, \bibinfo
  {author} {\bibfnamefont {Trevor}\ \bibnamefont {Chistolini}}, \bibinfo
  {author} {\bibfnamefont {Shraddha}\ \bibnamefont {Singh}}, \bibinfo {author}
  {\bibfnamefont {Konstantin~N.}\ \bibnamefont {Nesterov}}, \bibinfo {author}
  {\bibfnamefont {Christian}\ \bibnamefont {J\"unger}}, \bibinfo {author}
  {\bibfnamefont {Larry}\ \bibnamefont {Chen}}, \bibinfo {author}
  {\bibfnamefont {Zahra}\ \bibnamefont {Pedramrazi}}, \bibinfo {author}
  {\bibfnamefont {Bradley~K.}\ \bibnamefont {Mitchell}}, \bibinfo {author}
  {\bibfnamefont {John~Mark}\ \bibnamefont {Kreikebaum}}, \bibinfo {author}
  {\bibfnamefont {Shruti}\ \bibnamefont {Puri}}, \bibinfo {author}
  {\bibfnamefont {David~I.}\ \bibnamefont {Santiago}}, \ and\ \bibinfo {author}
  {\bibfnamefont {Irfan}\ \bibnamefont {Siddiqi}},\ }\bibfield  {title}
  {\enquote {\bibinfo {title} {{Blueprint for a High-Performance Fluxonium
  Quantum Processor}},}\ }\href {\doibase 10.1103/PRXQuantum.3.037001}
  {\bibfield  {journal} {\bibinfo  {journal} {PRX Quantum}\ }\textbf {\bibinfo
  {volume} {3}},\ \bibinfo {pages} {037001} (\bibinfo {year}
  {2022})}\BibitemShut {NoStop}%
\bibitem [{\citenamefont {Xiang}\ \emph {et~al.}(2013)\citenamefont {Xiang},
  \citenamefont {Ashhab}, \citenamefont {You},\ and\ \citenamefont
  {Nori}}]{Xiang2013}%
  \BibitemOpen
  \bibfield  {author} {\bibinfo {author} {\bibfnamefont {Ze-Liang}\
  \bibnamefont {Xiang}}, \bibinfo {author} {\bibfnamefont {Sahel}\ \bibnamefont
  {Ashhab}}, \bibinfo {author} {\bibfnamefont {J.~Q.}\ \bibnamefont {You}}, \
  and\ \bibinfo {author} {\bibfnamefont {Franco}\ \bibnamefont {Nori}},\
  }\bibfield  {title} {\enquote {\bibinfo {title} {Hybrid quantum circuits:
  Superconducting circuits interacting with other quantum systems},}\ }\href
  {\doibase 10.1103/RevModPhys.85.623} {\bibfield  {journal} {\bibinfo
  {journal} {Rev. Mod. Phys.}\ }\textbf {\bibinfo {volume} {85}},\ \bibinfo
  {pages} {623--653} (\bibinfo {year} {2013})}\BibitemShut {NoStop}%
\bibitem [{\citenamefont {Kurizki}\ \emph {et~al.}(2015)\citenamefont
  {Kurizki}, \citenamefont {Bertet}, \citenamefont {Kubo}, \citenamefont
  {M{\o}lmer}, \citenamefont {Petrosyan}, \citenamefont {Rabl},\ and\
  \citenamefont {Schmiedmayer}}]{Kurizki2015}%
  \BibitemOpen
  \bibfield  {author} {\bibinfo {author} {\bibfnamefont {Gershon}\ \bibnamefont
  {Kurizki}}, \bibinfo {author} {\bibfnamefont {Patrice}\ \bibnamefont
  {Bertet}}, \bibinfo {author} {\bibfnamefont {Yuimaru}\ \bibnamefont {Kubo}},
  \bibinfo {author} {\bibfnamefont {Klaus}\ \bibnamefont {M{\o}lmer}}, \bibinfo
  {author} {\bibfnamefont {David}\ \bibnamefont {Petrosyan}}, \bibinfo {author}
  {\bibfnamefont {Peter}\ \bibnamefont {Rabl}}, \ and\ \bibinfo {author}
  {\bibfnamefont {J{\"o}rg}\ \bibnamefont {Schmiedmayer}},\ }\bibfield  {title}
  {\enquote {\bibinfo {title} {Quantum technologies with hybrid systems},}\
  }\href {\doibase 10.1073/pnas.1419326112} {\bibfield  {journal} {\bibinfo
  {journal} {Proceedings of the National Academy of Sciences}\ }\textbf
  {\bibinfo {volume} {112}},\ \bibinfo {pages} {3866--3873} (\bibinfo {year}
  {2015})}\BibitemShut {NoStop}%
\bibitem [{\citenamefont {Clerk}\ \emph {et~al.}(2020)\citenamefont {Clerk},
  \citenamefont {Lehnert}, \citenamefont {Bertet}, \citenamefont {Petta},\ and\
  \citenamefont {Nakamura}}]{Clerk2020}%
  \BibitemOpen
  \bibfield  {author} {\bibinfo {author} {\bibfnamefont {A.~A.}\ \bibnamefont
  {Clerk}}, \bibinfo {author} {\bibfnamefont {K.~W.}\ \bibnamefont {Lehnert}},
  \bibinfo {author} {\bibfnamefont {P.}~\bibnamefont {Bertet}}, \bibinfo
  {author} {\bibfnamefont {J.~R.}\ \bibnamefont {Petta}}, \ and\ \bibinfo
  {author} {\bibfnamefont {Y.}~\bibnamefont {Nakamura}},\ }\bibfield  {title}
  {\enquote {\bibinfo {title} {Hybrid quantum systems with circuit quantum
  electrodynamics},}\ }\href {\doibase 10.1038/s41567-020-0797-9} {\bibfield
  {journal} {\bibinfo  {journal} {Nature Physics}\ }\textbf {\bibinfo {volume}
  {16}},\ \bibinfo {pages} {257--267} (\bibinfo {year} {2020})}\BibitemShut
  {NoStop}%
\bibitem [{\citenamefont {Gao}\ \emph {et~al.}(2021)\citenamefont {Gao},
  \citenamefont {Rol}, \citenamefont {Touzard},\ and\ \citenamefont
  {Wang}}]{Gao2021}%
  \BibitemOpen
  \bibfield  {author} {\bibinfo {author} {\bibfnamefont {Yvonne~Y.}\
  \bibnamefont {Gao}}, \bibinfo {author} {\bibfnamefont {M.~Adriaan}\
  \bibnamefont {Rol}}, \bibinfo {author} {\bibfnamefont {Steven}\ \bibnamefont
  {Touzard}}, \ and\ \bibinfo {author} {\bibfnamefont {Chen}\ \bibnamefont
  {Wang}},\ }\bibfield  {title} {\enquote {\bibinfo {title} {Practical guide
  for building superconducting quantum devices},}\ }\href {\doibase
  10.1103/PRXQuantum.2.040202} {\bibfield  {journal} {\bibinfo  {journal} {PRX
  Quantum}\ }\textbf {\bibinfo {volume} {2}},\ \bibinfo {pages} {040202}
  (\bibinfo {year} {2021})}\BibitemShut {NoStop}%
\bibitem [{\citenamefont {Scully}\ and\ \citenamefont
  {Zubairy}(1997)}]{scully1997quantum}%
  \BibitemOpen
  \bibfield  {author} {\bibinfo {author} {\bibfnamefont {M.O.}\ \bibnamefont
  {Scully}}\ and\ \bibinfo {author} {\bibfnamefont {M.S.}\ \bibnamefont
  {Zubairy}},\ }\href {https://books.google.com/books?id=20ISsQCKKmQC} {\emph
  {\bibinfo {title} {Quantum Optics}}},\ Quantum Optics\ (\bibinfo  {publisher}
  {Cambridge University Press},\ \bibinfo {year} {1997})\BibitemShut {NoStop}%
\bibitem [{\citenamefont {Reagor}\ \emph {et~al.}(2013)\citenamefont {Reagor},
  \citenamefont {Paik}, \citenamefont {Catelani}, \citenamefont {Sun},
  \citenamefont {Axline}, \citenamefont {Holland}, \citenamefont {Pop},
  \citenamefont {Masluk}, \citenamefont {Brecht}, \citenamefont {Frunzio},
  \citenamefont {Devoret}, \citenamefont {Glazman},\ and\ \citenamefont
  {Schoelkopf}}]{Reagor2013}%
  \BibitemOpen
  \bibfield  {author} {\bibinfo {author} {\bibfnamefont {Matthew}\ \bibnamefont
  {Reagor}}, \bibinfo {author} {\bibfnamefont {Hanhee}\ \bibnamefont {Paik}},
  \bibinfo {author} {\bibfnamefont {Gianluigi}\ \bibnamefont {Catelani}},
  \bibinfo {author} {\bibfnamefont {Luyan}\ \bibnamefont {Sun}}, \bibinfo
  {author} {\bibfnamefont {Christopher}\ \bibnamefont {Axline}}, \bibinfo
  {author} {\bibfnamefont {Eric}\ \bibnamefont {Holland}}, \bibinfo {author}
  {\bibfnamefont {Ioan~M.}\ \bibnamefont {Pop}}, \bibinfo {author}
  {\bibfnamefont {Nicholas~A.}\ \bibnamefont {Masluk}}, \bibinfo {author}
  {\bibfnamefont {Teresa}\ \bibnamefont {Brecht}}, \bibinfo {author}
  {\bibfnamefont {Luigi}\ \bibnamefont {Frunzio}}, \bibinfo {author}
  {\bibfnamefont {Michel~H.}\ \bibnamefont {Devoret}}, \bibinfo {author}
  {\bibfnamefont {Leonid}\ \bibnamefont {Glazman}}, \ and\ \bibinfo {author}
  {\bibfnamefont {Robert~J.}\ \bibnamefont {Schoelkopf}},\ }\bibfield  {title}
  {\enquote {\bibinfo {title} {{Reaching 10 ms single photon lifetimes for
  superconducting aluminum cavities}},}\ }\href {\doibase 10.1063/1.4807015}
  {\bibfield  {journal} {\bibinfo  {journal} {Applied Physics Letters}\
  }\textbf {\bibinfo {volume} {102}} (\bibinfo {year} {2013}),\
  10.1063/1.4807015},\ \bibinfo {note} {192604}\BibitemShut {NoStop}%
\bibitem [{\citenamefont {Romanenko}\ \emph {et~al.}(2020)\citenamefont
  {Romanenko}, \citenamefont {Pilipenko}, \citenamefont {Zorzetti},
  \citenamefont {Frolov}, \citenamefont {Awida}, \citenamefont {Belomestnykh},
  \citenamefont {Posen},\ and\ \citenamefont {Grassellino}}]{Romanenko2020}%
  \BibitemOpen
  \bibfield  {author} {\bibinfo {author} {\bibfnamefont {A.}~\bibnamefont
  {Romanenko}}, \bibinfo {author} {\bibfnamefont {R.}~\bibnamefont
  {Pilipenko}}, \bibinfo {author} {\bibfnamefont {S.}~\bibnamefont {Zorzetti}},
  \bibinfo {author} {\bibfnamefont {D.}~\bibnamefont {Frolov}}, \bibinfo
  {author} {\bibfnamefont {M.}~\bibnamefont {Awida}}, \bibinfo {author}
  {\bibfnamefont {S.}~\bibnamefont {Belomestnykh}}, \bibinfo {author}
  {\bibfnamefont {S.}~\bibnamefont {Posen}}, \ and\ \bibinfo {author}
  {\bibfnamefont {A.}~\bibnamefont {Grassellino}},\ }\bibfield  {title}
  {\enquote {\bibinfo {title} {Three-dimensional superconducting resonators at
  $t < 20$ mk with photon lifetimes up to $\tau=2$ s},}\ }\href {\doibase
  10.1103/PhysRevApplied.13.034032} {\bibfield  {journal} {\bibinfo  {journal}
  {Phys. Rev. Appl.}\ }\textbf {\bibinfo {volume} {13}},\ \bibinfo {pages}
  {034032} (\bibinfo {year} {2020})}\BibitemShut {NoStop}%
\bibitem [{\citenamefont {Yin}\ \emph {et~al.}(2013)\citenamefont {Yin},
  \citenamefont {Chen}, \citenamefont {Sank}, \citenamefont {O'Malley},
  \citenamefont {White}, \citenamefont {Barends}, \citenamefont {Kelly},
  \citenamefont {Lucero}, \citenamefont {Mariantoni}, \citenamefont {Megrant},
  \citenamefont {Neill}, \citenamefont {Vainsencher}, \citenamefont {Wenner},
  \citenamefont {Korotkov}, \citenamefont {Cleland},\ and\ \citenamefont
  {Martinis}}]{Yin2013}%
  \BibitemOpen
  \bibfield  {author} {\bibinfo {author} {\bibfnamefont {Yi}~\bibnamefont
  {Yin}}, \bibinfo {author} {\bibfnamefont {Yu}~\bibnamefont {Chen}}, \bibinfo
  {author} {\bibfnamefont {Daniel}\ \bibnamefont {Sank}}, \bibinfo {author}
  {\bibfnamefont {P.~J.~J.}\ \bibnamefont {O'Malley}}, \bibinfo {author}
  {\bibfnamefont {T.~C.}\ \bibnamefont {White}}, \bibinfo {author}
  {\bibfnamefont {R.}~\bibnamefont {Barends}}, \bibinfo {author} {\bibfnamefont
  {J.}~\bibnamefont {Kelly}}, \bibinfo {author} {\bibfnamefont {Erik}\
  \bibnamefont {Lucero}}, \bibinfo {author} {\bibfnamefont {Matteo}\
  \bibnamefont {Mariantoni}}, \bibinfo {author} {\bibfnamefont
  {A.}~\bibnamefont {Megrant}}, \bibinfo {author} {\bibfnamefont
  {C.}~\bibnamefont {Neill}}, \bibinfo {author} {\bibfnamefont
  {A.}~\bibnamefont {Vainsencher}}, \bibinfo {author} {\bibfnamefont
  {J.}~\bibnamefont {Wenner}}, \bibinfo {author} {\bibfnamefont {Alexander~N.}\
  \bibnamefont {Korotkov}}, \bibinfo {author} {\bibfnamefont {A.~N.}\
  \bibnamefont {Cleland}}, \ and\ \bibinfo {author} {\bibfnamefont {John~M.}\
  \bibnamefont {Martinis}},\ }\bibfield  {title} {\enquote {\bibinfo {title}
  {Catch and release of microwave photon states},}\ }\href {\doibase
  10.1103/PhysRevLett.110.107001} {\bibfield  {journal} {\bibinfo  {journal}
  {Phys. Rev. Lett.}\ }\textbf {\bibinfo {volume} {110}},\ \bibinfo {pages}
  {107001} (\bibinfo {year} {2013})}\BibitemShut {NoStop}%
\bibitem [{\citenamefont {Wenner}\ \emph {et~al.}(2014)\citenamefont {Wenner},
  \citenamefont {Yin}, \citenamefont {Chen}, \citenamefont {Barends},
  \citenamefont {Chiaro}, \citenamefont {Jeffrey}, \citenamefont {Kelly},
  \citenamefont {Megrant}, \citenamefont {Mutus}, \citenamefont {Neill},
  \citenamefont {O'Malley}, \citenamefont {Roushan}, \citenamefont {Sank},
  \citenamefont {Vainsencher}, \citenamefont {White}, \citenamefont {Korotkov},
  \citenamefont {Cleland},\ and\ \citenamefont {Martinis}}]{Wenner2014}%
  \BibitemOpen
  \bibfield  {author} {\bibinfo {author} {\bibfnamefont {J.}~\bibnamefont
  {Wenner}}, \bibinfo {author} {\bibfnamefont {Yi}~\bibnamefont {Yin}},
  \bibinfo {author} {\bibfnamefont {Yu}~\bibnamefont {Chen}}, \bibinfo {author}
  {\bibfnamefont {R.}~\bibnamefont {Barends}}, \bibinfo {author} {\bibfnamefont
  {B.}~\bibnamefont {Chiaro}}, \bibinfo {author} {\bibfnamefont
  {E.}~\bibnamefont {Jeffrey}}, \bibinfo {author} {\bibfnamefont
  {J.}~\bibnamefont {Kelly}}, \bibinfo {author} {\bibfnamefont
  {A.}~\bibnamefont {Megrant}}, \bibinfo {author} {\bibfnamefont {J.~Y.}\
  \bibnamefont {Mutus}}, \bibinfo {author} {\bibfnamefont {C.}~\bibnamefont
  {Neill}}, \bibinfo {author} {\bibfnamefont {P.~J.~J.}\ \bibnamefont
  {O'Malley}}, \bibinfo {author} {\bibfnamefont {P.}~\bibnamefont {Roushan}},
  \bibinfo {author} {\bibfnamefont {D.}~\bibnamefont {Sank}}, \bibinfo {author}
  {\bibfnamefont {A.}~\bibnamefont {Vainsencher}}, \bibinfo {author}
  {\bibfnamefont {T.~C.}\ \bibnamefont {White}}, \bibinfo {author}
  {\bibfnamefont {Alexander~N.}\ \bibnamefont {Korotkov}}, \bibinfo {author}
  {\bibfnamefont {A.~N.}\ \bibnamefont {Cleland}}, \ and\ \bibinfo {author}
  {\bibfnamefont {John~M.}\ \bibnamefont {Martinis}},\ }\bibfield  {title}
  {\enquote {\bibinfo {title} {Catching time-reversed microwave coherent state
  photons with 99.4\% absorption efficiency},}\ }\href {\doibase
  10.1103/PhysRevLett.112.210501} {\bibfield  {journal} {\bibinfo  {journal}
  {Phys. Rev. Lett.}\ }\textbf {\bibinfo {volume} {112}},\ \bibinfo {pages}
  {210501} (\bibinfo {year} {2014})}\BibitemShut {NoStop}%
\bibitem [{\citenamefont {Campagne-Ibarcq}\ \emph {et~al.}(2018)\citenamefont
  {Campagne-Ibarcq}, \citenamefont {Zalys-Geller}, \citenamefont {Narla},
  \citenamefont {Shankar}, \citenamefont {Reinhold}, \citenamefont {Burkhart},
  \citenamefont {Axline}, \citenamefont {Pfaff}, \citenamefont {Frunzio},
  \citenamefont {Schoelkopf},\ and\ \citenamefont {Devoret}}]{Ibarcq2018}%
  \BibitemOpen
  \bibfield  {author} {\bibinfo {author} {\bibfnamefont {P.}~\bibnamefont
  {Campagne-Ibarcq}}, \bibinfo {author} {\bibfnamefont {E.}~\bibnamefont
  {Zalys-Geller}}, \bibinfo {author} {\bibfnamefont {A.}~\bibnamefont {Narla}},
  \bibinfo {author} {\bibfnamefont {S.}~\bibnamefont {Shankar}}, \bibinfo
  {author} {\bibfnamefont {P.}~\bibnamefont {Reinhold}}, \bibinfo {author}
  {\bibfnamefont {L.}~\bibnamefont {Burkhart}}, \bibinfo {author}
  {\bibfnamefont {C.}~\bibnamefont {Axline}}, \bibinfo {author} {\bibfnamefont
  {W.}~\bibnamefont {Pfaff}}, \bibinfo {author} {\bibfnamefont
  {L.}~\bibnamefont {Frunzio}}, \bibinfo {author} {\bibfnamefont {R.~J.}\
  \bibnamefont {Schoelkopf}}, \ and\ \bibinfo {author} {\bibfnamefont {M.~H.}\
  \bibnamefont {Devoret}},\ }\bibfield  {title} {\enquote {\bibinfo {title}
  {Deterministic remote entanglement of superconducting circuits through
  microwave two-photon transitions},}\ }\href {\doibase
  10.1103/PhysRevLett.120.200501} {\bibfield  {journal} {\bibinfo  {journal}
  {Phys. Rev. Lett.}\ }\textbf {\bibinfo {volume} {120}},\ \bibinfo {pages}
  {200501} (\bibinfo {year} {2018})}\BibitemShut {NoStop}%
\bibitem [{\citenamefont {Chakram}\ \emph {et~al.}(2021)\citenamefont
  {Chakram}, \citenamefont {Oriani}, \citenamefont {Naik}, \citenamefont
  {Dixit}, \citenamefont {He}, \citenamefont {Agrawal}, \citenamefont {Kwon},\
  and\ \citenamefont {Schuster}}]{Chakram2021}%
  \BibitemOpen
  \bibfield  {author} {\bibinfo {author} {\bibfnamefont {Srivatsan}\
  \bibnamefont {Chakram}}, \bibinfo {author} {\bibfnamefont {Andrew~E.}\
  \bibnamefont {Oriani}}, \bibinfo {author} {\bibfnamefont {Ravi~K.}\
  \bibnamefont {Naik}}, \bibinfo {author} {\bibfnamefont {Akash~V.}\
  \bibnamefont {Dixit}}, \bibinfo {author} {\bibfnamefont {Kevin}\ \bibnamefont
  {He}}, \bibinfo {author} {\bibfnamefont {Ankur}\ \bibnamefont {Agrawal}},
  \bibinfo {author} {\bibfnamefont {Hyeokshin}\ \bibnamefont {Kwon}}, \ and\
  \bibinfo {author} {\bibfnamefont {David~I.}\ \bibnamefont {Schuster}},\
  }\bibfield  {title} {\enquote {\bibinfo {title} {Seamless high-$q$ microwave
  cavities for multimode circuit quantum electrodynamics},}\ }\href {\doibase
  10.1103/PhysRevLett.127.107701} {\bibfield  {journal} {\bibinfo  {journal}
  {Phys. Rev. Lett.}\ }\textbf {\bibinfo {volume} {127}},\ \bibinfo {pages}
  {107701} (\bibinfo {year} {2021})}\BibitemShut {NoStop}%
\bibitem [{\citenamefont {Milul}\ \emph {et~al.}(2023)\citenamefont {Milul},
  \citenamefont {Guttel}, \citenamefont {Goldblatt}, \citenamefont {Hazanov},
  \citenamefont {Joshi}, \citenamefont {Chausovsky}, \citenamefont {Kahn},
  \citenamefont {Çiftyürek}, \citenamefont {Lafont},\ and\ \citenamefont
  {Rosenblum}}]{Milul2023}%
  \BibitemOpen
  \bibfield  {author} {\bibinfo {author} {\bibfnamefont {Ofir}\ \bibnamefont
  {Milul}}, \bibinfo {author} {\bibfnamefont {Barkay}\ \bibnamefont {Guttel}},
  \bibinfo {author} {\bibfnamefont {Uri}\ \bibnamefont {Goldblatt}}, \bibinfo
  {author} {\bibfnamefont {Sergey}\ \bibnamefont {Hazanov}}, \bibinfo {author}
  {\bibfnamefont {Lalit~M.}\ \bibnamefont {Joshi}}, \bibinfo {author}
  {\bibfnamefont {Daniel}\ \bibnamefont {Chausovsky}}, \bibinfo {author}
  {\bibfnamefont {Nitzan}\ \bibnamefont {Kahn}}, \bibinfo {author}
  {\bibfnamefont {Engin}\ \bibnamefont {Çiftyürek}}, \bibinfo {author}
  {\bibfnamefont {Fabien}\ \bibnamefont {Lafont}}, \ and\ \bibinfo {author}
  {\bibfnamefont {Serge}\ \bibnamefont {Rosenblum}},\ }\href@noop {} {\enquote
  {\bibinfo {title} {A superconducting quantum memory with tens of milliseconds
  coherence time},}\ } (\bibinfo {year} {2023}),\ \Eprint
  {http://arxiv.org/abs/2302.06442} {arXiv:2302.06442 [quant-ph]} \BibitemShut
  {NoStop}%
\bibitem [{\citenamefont {Naik}\ \emph {et~al.}(2017)\citenamefont {Naik},
  \citenamefont {Leung}, \citenamefont {Chakram}, \citenamefont {Groszkowski},
  \citenamefont {Lu}, \citenamefont {Earnest}, \citenamefont {McKay},
  \citenamefont {Koch},\ and\ \citenamefont {Schuster}}]{Naik2018}%
  \BibitemOpen
  \bibfield  {author} {\bibinfo {author} {\bibfnamefont {R.~K.}\ \bibnamefont
  {Naik}}, \bibinfo {author} {\bibfnamefont {N.}~\bibnamefont {Leung}},
  \bibinfo {author} {\bibfnamefont {S.}~\bibnamefont {Chakram}}, \bibinfo
  {author} {\bibfnamefont {Peter}\ \bibnamefont {Groszkowski}}, \bibinfo
  {author} {\bibfnamefont {Y.}~\bibnamefont {Lu}}, \bibinfo {author}
  {\bibfnamefont {N.}~\bibnamefont {Earnest}}, \bibinfo {author} {\bibfnamefont
  {D.~C.}\ \bibnamefont {McKay}}, \bibinfo {author} {\bibfnamefont {Jens}\
  \bibnamefont {Koch}}, \ and\ \bibinfo {author} {\bibfnamefont {D.~I.}\
  \bibnamefont {Schuster}},\ }\bibfield  {title} {\enquote {\bibinfo {title}
  {Random access quantum information processors using multimode circuit quantum
  electrodynamics},}\ }\href {\doibase 10.1038/s41467-017-02046-6} {\bibfield
  {journal} {\bibinfo  {journal} {Nature Communications}\ }\textbf {\bibinfo
  {volume} {8}},\ \bibinfo {pages} {1904} (\bibinfo {year} {2017})}\BibitemShut
  {NoStop}%
\bibitem [{\citenamefont {Grimm}\ \emph {et~al.}(2020)\citenamefont {Grimm},
  \citenamefont {Frattini}, \citenamefont {Puri}, \citenamefont {Mundhada},
  \citenamefont {Touzard}, \citenamefont {Mirrahimi}, \citenamefont {Girvin},
  \citenamefont {Shankar},\ and\ \citenamefont {Devoret}}]{Grimm2020}%
  \BibitemOpen
  \bibfield  {author} {\bibinfo {author} {\bibfnamefont {A.}~\bibnamefont
  {Grimm}}, \bibinfo {author} {\bibfnamefont {N.~E.}\ \bibnamefont {Frattini}},
  \bibinfo {author} {\bibfnamefont {S.}~\bibnamefont {Puri}}, \bibinfo {author}
  {\bibfnamefont {S.~O.}\ \bibnamefont {Mundhada}}, \bibinfo {author}
  {\bibfnamefont {S.}~\bibnamefont {Touzard}}, \bibinfo {author} {\bibfnamefont
  {M.}~\bibnamefont {Mirrahimi}}, \bibinfo {author} {\bibfnamefont {S.~M.}\
  \bibnamefont {Girvin}}, \bibinfo {author} {\bibfnamefont {S.}~\bibnamefont
  {Shankar}}, \ and\ \bibinfo {author} {\bibfnamefont {M.~H.}\ \bibnamefont
  {Devoret}},\ }\bibfield  {title} {\enquote {\bibinfo {title} {Stabilization
  and operation of a kerr-cat qubit},}\ }\href {\doibase
  10.1038/s41586-020-2587-z} {\bibfield  {journal} {\bibinfo  {journal}
  {Nature}\ }\textbf {\bibinfo {volume} {584}},\ \bibinfo {pages} {205--209}
  (\bibinfo {year} {2020})}\BibitemShut {NoStop}%
\bibitem [{\citenamefont {Darmawan}\ \emph {et~al.}(2021)\citenamefont
  {Darmawan}, \citenamefont {Brown}, \citenamefont {Grimsmo}, \citenamefont
  {Tuckett},\ and\ \citenamefont {Puri}}]{Darmawan2021}%
  \BibitemOpen
  \bibfield  {author} {\bibinfo {author} {\bibfnamefont {Andrew~S.}\
  \bibnamefont {Darmawan}}, \bibinfo {author} {\bibfnamefont {Benjamin~J.}\
  \bibnamefont {Brown}}, \bibinfo {author} {\bibfnamefont {Arne~L.}\
  \bibnamefont {Grimsmo}}, \bibinfo {author} {\bibfnamefont {David~K.}\
  \bibnamefont {Tuckett}}, \ and\ \bibinfo {author} {\bibfnamefont {Shruti}\
  \bibnamefont {Puri}},\ }\bibfield  {title} {\enquote {\bibinfo {title}
  {Practical quantum error correction with the xzzx code and kerr-cat
  qubits},}\ }\href {\doibase 10.1103/PRXQuantum.2.030345} {\bibfield
  {journal} {\bibinfo  {journal} {PRX Quantum}\ }\textbf {\bibinfo {volume}
  {2}},\ \bibinfo {pages} {030345} (\bibinfo {year} {2021})}\BibitemShut
  {NoStop}%
\bibitem [{\citenamefont {Chamberland}\ \emph {et~al.}(2022)\citenamefont
  {Chamberland}, \citenamefont {Noh}, \citenamefont {Arrangoiz-Arriola},
  \citenamefont {Campbell}, \citenamefont {Hann}, \citenamefont {Iverson},
  \citenamefont {Putterman}, \citenamefont {Bohdanowicz}, \citenamefont
  {Flammia}, \citenamefont {Keller}, \citenamefont {Refael}, \citenamefont
  {Preskill}, \citenamefont {Jiang}, \citenamefont {Safavi-Naeini},
  \citenamefont {Painter},\ and\ \citenamefont {Brand\~ao}}]{Chamberland2022}%
  \BibitemOpen
  \bibfield  {author} {\bibinfo {author} {\bibfnamefont {Christopher}\
  \bibnamefont {Chamberland}}, \bibinfo {author} {\bibfnamefont {Kyungjoo}\
  \bibnamefont {Noh}}, \bibinfo {author} {\bibfnamefont {Patricio}\
  \bibnamefont {Arrangoiz-Arriola}}, \bibinfo {author} {\bibfnamefont
  {Earl~T.}\ \bibnamefont {Campbell}}, \bibinfo {author} {\bibfnamefont
  {Connor~T.}\ \bibnamefont {Hann}}, \bibinfo {author} {\bibfnamefont {Joseph}\
  \bibnamefont {Iverson}}, \bibinfo {author} {\bibfnamefont {Harald}\
  \bibnamefont {Putterman}}, \bibinfo {author} {\bibfnamefont {Thomas~C.}\
  \bibnamefont {Bohdanowicz}}, \bibinfo {author} {\bibfnamefont {Steven~T.}\
  \bibnamefont {Flammia}}, \bibinfo {author} {\bibfnamefont {Andrew}\
  \bibnamefont {Keller}}, \bibinfo {author} {\bibfnamefont {Gil}\ \bibnamefont
  {Refael}}, \bibinfo {author} {\bibfnamefont {John}\ \bibnamefont {Preskill}},
  \bibinfo {author} {\bibfnamefont {Liang}\ \bibnamefont {Jiang}}, \bibinfo
  {author} {\bibfnamefont {Amir~H.}\ \bibnamefont {Safavi-Naeini}}, \bibinfo
  {author} {\bibfnamefont {Oskar}\ \bibnamefont {Painter}}, \ and\ \bibinfo
  {author} {\bibfnamefont {Fernando~G.S.L.}\ \bibnamefont {Brand\~ao}},\
  }\bibfield  {title} {\enquote {\bibinfo {title} {Building a fault-tolerant
  quantum computer using concatenated cat codes},}\ }\href {\doibase
  10.1103/PRXQuantum.3.010329} {\bibfield  {journal} {\bibinfo  {journal} {PRX
  Quantum}\ }\textbf {\bibinfo {volume} {3}},\ \bibinfo {pages} {010329}
  (\bibinfo {year} {2022})}\BibitemShut {NoStop}%
\bibitem [{\citenamefont {Gertler}\ \emph {et~al.}(2021)\citenamefont
  {Gertler}, \citenamefont {Baker}, \citenamefont {Li}, \citenamefont {Shirol},
  \citenamefont {Koch},\ and\ \citenamefont {Wang}}]{Gertler2021}%
  \BibitemOpen
  \bibfield  {author} {\bibinfo {author} {\bibfnamefont {Jeffrey~M.}\
  \bibnamefont {Gertler}}, \bibinfo {author} {\bibfnamefont {Brian}\
  \bibnamefont {Baker}}, \bibinfo {author} {\bibfnamefont {Juliang}\
  \bibnamefont {Li}}, \bibinfo {author} {\bibfnamefont {Shruti}\ \bibnamefont
  {Shirol}}, \bibinfo {author} {\bibfnamefont {Jens}\ \bibnamefont {Koch}}, \
  and\ \bibinfo {author} {\bibfnamefont {Chen}\ \bibnamefont {Wang}},\
  }\bibfield  {title} {\enquote {\bibinfo {title} {Protecting a bosonic qubit
  with autonomous quantum error correction},}\ }\href {\doibase
  10.1038/s41586-021-03257-0} {\bibfield  {journal} {\bibinfo  {journal}
  {Nature}\ }\textbf {\bibinfo {volume} {590}},\ \bibinfo {pages} {243--248}
  (\bibinfo {year} {2021})}\BibitemShut {NoStop}%
\bibitem [{\citenamefont {Grimsmo}\ and\ \citenamefont
  {Puri}(2021)}]{Grimsmo2021}%
  \BibitemOpen
  \bibfield  {author} {\bibinfo {author} {\bibfnamefont {Arne~L.}\ \bibnamefont
  {Grimsmo}}\ and\ \bibinfo {author} {\bibfnamefont {Shruti}\ \bibnamefont
  {Puri}},\ }\bibfield  {title} {\enquote {\bibinfo {title} {Quantum error
  correction with the gottesman-kitaev-preskill code},}\ }\href {\doibase
  10.1103/PRXQuantum.2.020101} {\bibfield  {journal} {\bibinfo  {journal} {PRX
  Quantum}\ }\textbf {\bibinfo {volume} {2}},\ \bibinfo {pages} {020101}
  (\bibinfo {year} {2021})}\BibitemShut {NoStop}%
\bibitem [{\citenamefont {Campagne-Ibarcq}\ \emph {et~al.}(2020)\citenamefont
  {Campagne-Ibarcq}, \citenamefont {Eickbusch}, \citenamefont {Touzard},
  \citenamefont {Zalys-Geller}, \citenamefont {Frattini}, \citenamefont
  {Sivak}, \citenamefont {Reinhold}, \citenamefont {Puri}, \citenamefont
  {Shankar}, \citenamefont {Schoelkopf}, \citenamefont {Frunzio}, \citenamefont
  {Mirrahimi},\ and\ \citenamefont {Devoret}}]{Campagne-Ibarcq2020}%
  \BibitemOpen
  \bibfield  {author} {\bibinfo {author} {\bibfnamefont {P.}~\bibnamefont
  {Campagne-Ibarcq}}, \bibinfo {author} {\bibfnamefont {A.}~\bibnamefont
  {Eickbusch}}, \bibinfo {author} {\bibfnamefont {S.}~\bibnamefont {Touzard}},
  \bibinfo {author} {\bibfnamefont {E.}~\bibnamefont {Zalys-Geller}}, \bibinfo
  {author} {\bibfnamefont {N.~E.}\ \bibnamefont {Frattini}}, \bibinfo {author}
  {\bibfnamefont {V.~V.}\ \bibnamefont {Sivak}}, \bibinfo {author}
  {\bibfnamefont {P.}~\bibnamefont {Reinhold}}, \bibinfo {author}
  {\bibfnamefont {S.}~\bibnamefont {Puri}}, \bibinfo {author} {\bibfnamefont
  {S.}~\bibnamefont {Shankar}}, \bibinfo {author} {\bibfnamefont {R.~J.}\
  \bibnamefont {Schoelkopf}}, \bibinfo {author} {\bibfnamefont
  {L.}~\bibnamefont {Frunzio}}, \bibinfo {author} {\bibfnamefont
  {M.}~\bibnamefont {Mirrahimi}}, \ and\ \bibinfo {author} {\bibfnamefont
  {M.~H.}\ \bibnamefont {Devoret}},\ }\bibfield  {title} {\enquote {\bibinfo
  {title} {Quantum error correction of a qubit encoded in grid states of an
  oscillator},}\ }\href {\doibase 10.1038/s41586-020-2603-3} {\bibfield
  {journal} {\bibinfo  {journal} {Nature}\ }\textbf {\bibinfo {volume} {584}},\
  \bibinfo {pages} {368--372} (\bibinfo {year} {2020})}\BibitemShut {NoStop}%
\bibitem [{\citenamefont {Hu}\ \emph {et~al.}(2019)\citenamefont {Hu},
  \citenamefont {Ma}, \citenamefont {Cai}, \citenamefont {Mu}, \citenamefont
  {Xu}, \citenamefont {Wang}, \citenamefont {Wu}, \citenamefont {Wang},
  \citenamefont {Song}, \citenamefont {Zou}, \citenamefont {Girvin},
  \citenamefont {Duan},\ and\ \citenamefont {Sun}}]{Hu2019}%
  \BibitemOpen
  \bibfield  {author} {\bibinfo {author} {\bibfnamefont {L.}~\bibnamefont
  {Hu}}, \bibinfo {author} {\bibfnamefont {Y.}~\bibnamefont {Ma}}, \bibinfo
  {author} {\bibfnamefont {W.}~\bibnamefont {Cai}}, \bibinfo {author}
  {\bibfnamefont {X.}~\bibnamefont {Mu}}, \bibinfo {author} {\bibfnamefont
  {Y.}~\bibnamefont {Xu}}, \bibinfo {author} {\bibfnamefont {W.}~\bibnamefont
  {Wang}}, \bibinfo {author} {\bibfnamefont {Y.}~\bibnamefont {Wu}}, \bibinfo
  {author} {\bibfnamefont {H.}~\bibnamefont {Wang}}, \bibinfo {author}
  {\bibfnamefont {Y.~P.}\ \bibnamefont {Song}}, \bibinfo {author}
  {\bibfnamefont {C.~L.}\ \bibnamefont {Zou}}, \bibinfo {author} {\bibfnamefont
  {S.~M.}\ \bibnamefont {Girvin}}, \bibinfo {author} {\bibfnamefont {L-M.}\
  \bibnamefont {Duan}}, \ and\ \bibinfo {author} {\bibfnamefont
  {L.}~\bibnamefont {Sun}},\ }\bibfield  {title} {\enquote {\bibinfo {title}
  {Quantum error correction and universal gate set operation on a binomial
  bosonic logical qubit},}\ }\href {\doibase 10.1038/s41567-018-0414-3}
  {\bibfield  {journal} {\bibinfo  {journal} {Nature Physics}\ }\textbf
  {\bibinfo {volume} {15}},\ \bibinfo {pages} {503--508} (\bibinfo {year}
  {2019})}\BibitemShut {NoStop}%
\bibitem [{\citenamefont {Stein}\ \emph {et~al.}(2023)\citenamefont {Stein},
  \citenamefont {Hua}, \citenamefont {Liu}, \citenamefont {Guinn},
  \citenamefont {Ang}, \citenamefont {Zhang}, \citenamefont {Chakram},
  \citenamefont {Ding},\ and\ \citenamefont {Li}}]{CANVAS}%
  \BibitemOpen
  \bibfield  {author} {\bibinfo {author} {\bibfnamefont {Samuel}\ \bibnamefont
  {Stein}}, \bibinfo {author} {\bibfnamefont {Fei}\ \bibnamefont {Hua}},
  \bibinfo {author} {\bibfnamefont {Chenxu}\ \bibnamefont {Liu}}, \bibinfo
  {author} {\bibfnamefont {Charles}\ \bibnamefont {Guinn}}, \bibinfo {author}
  {\bibfnamefont {James}\ \bibnamefont {Ang}}, \bibinfo {author} {\bibfnamefont
  {Eddy}\ \bibnamefont {Zhang}}, \bibinfo {author} {\bibfnamefont {Srivatsan}\
  \bibnamefont {Chakram}}, \bibinfo {author} {\bibfnamefont {Yufei}\
  \bibnamefont {Ding}}, \ and\ \bibinfo {author} {\bibfnamefont {Ang}\
  \bibnamefont {Li}},\ }\href@noop {} {\enquote {\bibinfo {title} {Canvas:
  Cavity-centric architectures for novel variational algorithm support},}\ }
  (\bibinfo {year} {2023}),\ \bibinfo {note} {in preparation}\BibitemShut
  {NoStop}%
\bibitem [{\citenamefont {H{\"a}ffner}\ \emph {et~al.}(2008)\citenamefont
  {H{\"a}ffner}, \citenamefont {Roos},\ and\ \citenamefont
  {Blatt}}]{Haffner2008}%
  \BibitemOpen
  \bibfield  {author} {\bibinfo {author} {\bibfnamefont {H.}~\bibnamefont
  {H{\"a}ffner}}, \bibinfo {author} {\bibfnamefont {C.F.}\ \bibnamefont
  {Roos}}, \ and\ \bibinfo {author} {\bibfnamefont {R.}~\bibnamefont {Blatt}},\
  }\bibfield  {title} {\enquote {\bibinfo {title} {Quantum computing with
  trapped ions},}\ }\href {\doibase
  https://doi.org/10.1016/j.physrep.2008.09.003} {\bibfield  {journal}
  {\bibinfo  {journal} {Physics Reports}\ }\textbf {\bibinfo {volume} {469}},\
  \bibinfo {pages} {155--203} (\bibinfo {year} {2008})}\BibitemShut {NoStop}%
\bibitem [{\citenamefont {Bruzewicz}\ \emph {et~al.}(2019)\citenamefont
  {Bruzewicz}, \citenamefont {Chiaverini}, \citenamefont {McConnell},\ and\
  \citenamefont {Sage}}]{Bruzewicz2019}%
  \BibitemOpen
  \bibfield  {author} {\bibinfo {author} {\bibfnamefont {Colin~D.}\
  \bibnamefont {Bruzewicz}}, \bibinfo {author} {\bibfnamefont {John}\
  \bibnamefont {Chiaverini}}, \bibinfo {author} {\bibfnamefont {Robert}\
  \bibnamefont {McConnell}}, \ and\ \bibinfo {author} {\bibfnamefont
  {Jeremy~M.}\ \bibnamefont {Sage}},\ }\bibfield  {title} {\enquote {\bibinfo
  {title} {{Trapped-ion quantum computing: Progress and challenges}},}\ }\href
  {\doibase https://doi.org/10.1063/1.5088164} {\bibfield  {journal} {\bibinfo
  {journal} {Applied Physics Reviews}\ }\textbf {\bibinfo {volume} {6}}
  (\bibinfo {year} {2019}),\ https://doi.org/10.1063/1.5088164},\ \bibinfo
  {note} {021314}\BibitemShut {NoStop}%
\bibitem [{\citenamefont {Brown}\ \emph {et~al.}(2021)\citenamefont {Brown},
  \citenamefont {Chiaverini}, \citenamefont {Sage},\ and\ \citenamefont
  {H{\"a}ffner}}]{Brown2021}%
  \BibitemOpen
  \bibfield  {author} {\bibinfo {author} {\bibfnamefont {Kenneth~R.}\
  \bibnamefont {Brown}}, \bibinfo {author} {\bibfnamefont {John}\ \bibnamefont
  {Chiaverini}}, \bibinfo {author} {\bibfnamefont {Jeremy~M.}\ \bibnamefont
  {Sage}}, \ and\ \bibinfo {author} {\bibfnamefont {Hartmut}\ \bibnamefont
  {H{\"a}ffner}},\ }\bibfield  {title} {\enquote {\bibinfo {title} {Materials
  challenges for trapped-ion quantum computers},}\ }\href {\doibase
  10.1038/s41578-021-00292-1} {\bibfield  {journal} {\bibinfo  {journal}
  {Nature Reviews Materials}\ }\textbf {\bibinfo {volume} {6}},\ \bibinfo
  {pages} {892--905} (\bibinfo {year} {2021})}\BibitemShut {NoStop}%
\bibitem [{\citenamefont {Neuhauser}\ \emph {et~al.}(1980)\citenamefont
  {Neuhauser}, \citenamefont {Hohenstatt}, \citenamefont {Toschek},\ and\
  \citenamefont {Dehmelt}}]{Neuhauser1980}%
  \BibitemOpen
  \bibfield  {author} {\bibinfo {author} {\bibfnamefont {W.}~\bibnamefont
  {Neuhauser}}, \bibinfo {author} {\bibfnamefont {M.}~\bibnamefont
  {Hohenstatt}}, \bibinfo {author} {\bibfnamefont {P.~E.}\ \bibnamefont
  {Toschek}}, \ and\ \bibinfo {author} {\bibfnamefont {H.}~\bibnamefont
  {Dehmelt}},\ }\bibfield  {title} {\enquote {\bibinfo {title} {Localized
  visible ${\mathrm{ba}}^{+}$ mono-ion oscillator},}\ }\href {\doibase
  10.1103/PhysRevA.22.1137} {\bibfield  {journal} {\bibinfo  {journal} {Phys.
  Rev. A}\ }\textbf {\bibinfo {volume} {22}},\ \bibinfo {pages} {1137--1140}
  (\bibinfo {year} {1980})}\BibitemShut {NoStop}%
\bibitem [{\citenamefont {Dehmelt}(1990)}]{Dehmelt1990}%
  \BibitemOpen
  \bibfield  {author} {\bibinfo {author} {\bibfnamefont {Hans}\ \bibnamefont
  {Dehmelt}},\ }\bibfield  {title} {\enquote {\bibinfo {title} {Experiments
  with an isolated subatomic particle at rest},}\ }\href {\doibase
  10.1103/RevModPhys.62.525} {\bibfield  {journal} {\bibinfo  {journal} {Rev.
  Mod. Phys.}\ }\textbf {\bibinfo {volume} {62}},\ \bibinfo {pages} {525--530}
  (\bibinfo {year} {1990})}\BibitemShut {NoStop}%
\bibitem [{\citenamefont {Chiaverini}\ \emph {et~al.}(2005)\citenamefont
  {Chiaverini}, \citenamefont {Blakestad}, \citenamefont {Britton},
  \citenamefont {Jost}, \citenamefont {Langer}, \citenamefont {Leibfried},
  \citenamefont {Ozeri},\ and\ \citenamefont {Wineland}}]{Chiaverini2005}%
  \BibitemOpen
  \bibfield  {author} {\bibinfo {author} {\bibfnamefont {J.}~\bibnamefont
  {Chiaverini}}, \bibinfo {author} {\bibfnamefont {R.~B.}\ \bibnamefont
  {Blakestad}}, \bibinfo {author} {\bibfnamefont {J.}~\bibnamefont {Britton}},
  \bibinfo {author} {\bibfnamefont {J.~D.}\ \bibnamefont {Jost}}, \bibinfo
  {author} {\bibfnamefont {C.}~\bibnamefont {Langer}}, \bibinfo {author}
  {\bibfnamefont {D.}~\bibnamefont {Leibfried}}, \bibinfo {author}
  {\bibfnamefont {R.}~\bibnamefont {Ozeri}}, \ and\ \bibinfo {author}
  {\bibfnamefont {D.~J.}\ \bibnamefont {Wineland}},\ }\bibfield  {title}
  {\enquote {\bibinfo {title} {Surface-electrode architecture for ion-trap
  quantum information processing},}\ }\href@noop {} {\bibfield  {journal}
  {\bibinfo  {journal} {Quantum Info. Comput.}\ }\textbf {\bibinfo {volume}
  {5}},\ \bibinfo {pages} {419–439} (\bibinfo {year} {2005})}\BibitemShut
  {NoStop}%
\bibitem [{\citenamefont {Seidelin}\ \emph {et~al.}(2006)\citenamefont
  {Seidelin}, \citenamefont {Chiaverini}, \citenamefont {Reichle},
  \citenamefont {Bollinger}, \citenamefont {Leibfried}, \citenamefont
  {Britton}, \citenamefont {Wesenberg}, \citenamefont {Blakestad},
  \citenamefont {Epstein}, \citenamefont {Hume}, \citenamefont {Itano},
  \citenamefont {Jost}, \citenamefont {Langer}, \citenamefont {Ozeri},
  \citenamefont {Shiga},\ and\ \citenamefont {Wineland}}]{Seidelin2006}%
  \BibitemOpen
  \bibfield  {author} {\bibinfo {author} {\bibfnamefont {S.}~\bibnamefont
  {Seidelin}}, \bibinfo {author} {\bibfnamefont {J.}~\bibnamefont
  {Chiaverini}}, \bibinfo {author} {\bibfnamefont {R.}~\bibnamefont {Reichle}},
  \bibinfo {author} {\bibfnamefont {J.~J.}\ \bibnamefont {Bollinger}}, \bibinfo
  {author} {\bibfnamefont {D.}~\bibnamefont {Leibfried}}, \bibinfo {author}
  {\bibfnamefont {J.}~\bibnamefont {Britton}}, \bibinfo {author} {\bibfnamefont
  {J.~H.}\ \bibnamefont {Wesenberg}}, \bibinfo {author} {\bibfnamefont {R.~B.}\
  \bibnamefont {Blakestad}}, \bibinfo {author} {\bibfnamefont {R.~J.}\
  \bibnamefont {Epstein}}, \bibinfo {author} {\bibfnamefont {D.~B.}\
  \bibnamefont {Hume}}, \bibinfo {author} {\bibfnamefont {W.~M.}\ \bibnamefont
  {Itano}}, \bibinfo {author} {\bibfnamefont {J.~D.}\ \bibnamefont {Jost}},
  \bibinfo {author} {\bibfnamefont {C.}~\bibnamefont {Langer}}, \bibinfo
  {author} {\bibfnamefont {R.}~\bibnamefont {Ozeri}}, \bibinfo {author}
  {\bibfnamefont {N.}~\bibnamefont {Shiga}}, \ and\ \bibinfo {author}
  {\bibfnamefont {D.~J.}\ \bibnamefont {Wineland}},\ }\bibfield  {title}
  {\enquote {\bibinfo {title} {{Microfabricated Surface-Electrode Ion Trap for
  Scalable Quantum Information Processing}},}\ }\href {\doibase
  10.1103/PhysRevLett.96.253003} {\bibfield  {journal} {\bibinfo  {journal}
  {Phys. Rev. Lett.}\ }\textbf {\bibinfo {volume} {96}},\ \bibinfo {pages}
  {253003} (\bibinfo {year} {2006})}\BibitemShut {NoStop}%
\bibitem [{\citenamefont {Labaziewicz}\ \emph {et~al.}(2008)\citenamefont
  {Labaziewicz}, \citenamefont {Ge}, \citenamefont {Antohi}, \citenamefont
  {Leibrandt}, \citenamefont {Brown},\ and\ \citenamefont
  {Chuang}}]{Labaziewicz2008}%
  \BibitemOpen
  \bibfield  {author} {\bibinfo {author} {\bibfnamefont {Jaroslaw}\
  \bibnamefont {Labaziewicz}}, \bibinfo {author} {\bibfnamefont {Yufei}\
  \bibnamefont {Ge}}, \bibinfo {author} {\bibfnamefont {Paul}\ \bibnamefont
  {Antohi}}, \bibinfo {author} {\bibfnamefont {David}\ \bibnamefont
  {Leibrandt}}, \bibinfo {author} {\bibfnamefont {Kenneth~R.}\ \bibnamefont
  {Brown}}, \ and\ \bibinfo {author} {\bibfnamefont {Isaac~L.}\ \bibnamefont
  {Chuang}},\ }\bibfield  {title} {\enquote {\bibinfo {title} {{Suppression of
  Heating Rates in Cryogenic Surface-Electrode Ion Traps}},}\ }\href {\doibase
  10.1103/PhysRevLett.100.013001} {\bibfield  {journal} {\bibinfo  {journal}
  {Phys. Rev. Lett.}\ }\textbf {\bibinfo {volume} {100}},\ \bibinfo {pages}
  {013001} (\bibinfo {year} {2008})}\BibitemShut {NoStop}%
\bibitem [{\citenamefont {Harty}\ \emph {et~al.}(2014)\citenamefont {Harty},
  \citenamefont {Allcock}, \citenamefont {Ballance}, \citenamefont {Guidoni},
  \citenamefont {Janacek}, \citenamefont {Linke}, \citenamefont {Stacey},\ and\
  \citenamefont {Lucas}}]{Harty2014}%
  \BibitemOpen
  \bibfield  {author} {\bibinfo {author} {\bibfnamefont {T.~P.}\ \bibnamefont
  {Harty}}, \bibinfo {author} {\bibfnamefont {D.~T.~C.}\ \bibnamefont
  {Allcock}}, \bibinfo {author} {\bibfnamefont {C.~J.}\ \bibnamefont
  {Ballance}}, \bibinfo {author} {\bibfnamefont {L.}~\bibnamefont {Guidoni}},
  \bibinfo {author} {\bibfnamefont {H.~A.}\ \bibnamefont {Janacek}}, \bibinfo
  {author} {\bibfnamefont {N.~M.}\ \bibnamefont {Linke}}, \bibinfo {author}
  {\bibfnamefont {D.~N.}\ \bibnamefont {Stacey}}, \ and\ \bibinfo {author}
  {\bibfnamefont {D.~M.}\ \bibnamefont {Lucas}},\ }\bibfield  {title} {\enquote
  {\bibinfo {title} {High-fidelity preparation, gates, memory, and readout of a
  trapped-ion quantum bit},}\ }\href {\doibase 10.1103/PhysRevLett.113.220501}
  {\bibfield  {journal} {\bibinfo  {journal} {Phys. Rev. Lett.}\ }\textbf
  {\bibinfo {volume} {113}},\ \bibinfo {pages} {220501} (\bibinfo {year}
  {2014})}\BibitemShut {NoStop}%
\bibitem [{\citenamefont {Gaebler}\ \emph {et~al.}(2016)\citenamefont
  {Gaebler}, \citenamefont {Tan}, \citenamefont {Lin}, \citenamefont {Wan},
  \citenamefont {Bowler}, \citenamefont {Keith}, \citenamefont {Glancy},
  \citenamefont {Coakley}, \citenamefont {Knill}, \citenamefont {Leibfried},\
  and\ \citenamefont {Wineland}}]{Gaebler2016}%
  \BibitemOpen
  \bibfield  {author} {\bibinfo {author} {\bibfnamefont {J.~P.}\ \bibnamefont
  {Gaebler}}, \bibinfo {author} {\bibfnamefont {T.~R.}\ \bibnamefont {Tan}},
  \bibinfo {author} {\bibfnamefont {Y.}~\bibnamefont {Lin}}, \bibinfo {author}
  {\bibfnamefont {Y.}~\bibnamefont {Wan}}, \bibinfo {author} {\bibfnamefont
  {R.}~\bibnamefont {Bowler}}, \bibinfo {author} {\bibfnamefont {A.~C.}\
  \bibnamefont {Keith}}, \bibinfo {author} {\bibfnamefont {S.}~\bibnamefont
  {Glancy}}, \bibinfo {author} {\bibfnamefont {K.}~\bibnamefont {Coakley}},
  \bibinfo {author} {\bibfnamefont {E.}~\bibnamefont {Knill}}, \bibinfo
  {author} {\bibfnamefont {D.}~\bibnamefont {Leibfried}}, \ and\ \bibinfo
  {author} {\bibfnamefont {D.~J.}\ \bibnamefont {Wineland}},\ }\bibfield
  {title} {\enquote {\bibinfo {title} {High-fidelity universal gate set for
  ${^{9}\mathrm{Be}}^{+}$ ion qubits},}\ }\href {\doibase
  10.1103/PhysRevLett.117.060505} {\bibfield  {journal} {\bibinfo  {journal}
  {Phys. Rev. Lett.}\ }\textbf {\bibinfo {volume} {117}},\ \bibinfo {pages}
  {060505} (\bibinfo {year} {2016})}\BibitemShut {NoStop}%
\bibitem [{\citenamefont {Ballance}\ \emph {et~al.}(2016)\citenamefont
  {Ballance}, \citenamefont {Harty}, \citenamefont {Linke}, \citenamefont
  {Sepiol},\ and\ \citenamefont {Lucas}}]{Ballance2016}%
  \BibitemOpen
  \bibfield  {author} {\bibinfo {author} {\bibfnamefont {C.~J.}\ \bibnamefont
  {Ballance}}, \bibinfo {author} {\bibfnamefont {T.~P.}\ \bibnamefont {Harty}},
  \bibinfo {author} {\bibfnamefont {N.~M.}\ \bibnamefont {Linke}}, \bibinfo
  {author} {\bibfnamefont {M.~A.}\ \bibnamefont {Sepiol}}, \ and\ \bibinfo
  {author} {\bibfnamefont {D.~M.}\ \bibnamefont {Lucas}},\ }\bibfield  {title}
  {\enquote {\bibinfo {title} {High-fidelity quantum logic gates using
  trapped-ion hyperfine qubits},}\ }\href {\doibase
  10.1103/PhysRevLett.117.060504} {\bibfield  {journal} {\bibinfo  {journal}
  {Phys. Rev. Lett.}\ }\textbf {\bibinfo {volume} {117}},\ \bibinfo {pages}
  {060504} (\bibinfo {year} {2016})}\BibitemShut {NoStop}%
\bibitem [{\citenamefont {Keselman}\ \emph {et~al.}(2011)\citenamefont
  {Keselman}, \citenamefont {Glickman}, \citenamefont {Akerman}, \citenamefont
  {Kotler},\ and\ \citenamefont {Ozeri}}]{Keselman2011}%
  \BibitemOpen
  \bibfield  {author} {\bibinfo {author} {\bibfnamefont {A}~\bibnamefont
  {Keselman}}, \bibinfo {author} {\bibfnamefont {Y}~\bibnamefont {Glickman}},
  \bibinfo {author} {\bibfnamefont {N}~\bibnamefont {Akerman}}, \bibinfo
  {author} {\bibfnamefont {S}~\bibnamefont {Kotler}}, \ and\ \bibinfo {author}
  {\bibfnamefont {R}~\bibnamefont {Ozeri}},\ }\bibfield  {title} {\enquote
  {\bibinfo {title} {High-fidelity state detection and tomography of a
  single-ion zeeman qubit},}\ }\href {\doibase 10.1088/1367-2630/13/7/073027}
  {\bibfield  {journal} {\bibinfo  {journal} {New Journal of Physics}\ }\textbf
  {\bibinfo {volume} {13}},\ \bibinfo {pages} {073027} (\bibinfo {year}
  {2011})}\BibitemShut {NoStop}%
\bibitem [{\citenamefont {Ruster}\ \emph {et~al.}(2016)\citenamefont {Ruster},
  \citenamefont {Schmiegelow}, \citenamefont {Kaufmann}, \citenamefont
  {Warschburger}, \citenamefont {Schmidt-Kaler},\ and\ \citenamefont
  {Poschinger}}]{Ruster2016}%
  \BibitemOpen
  \bibfield  {author} {\bibinfo {author} {\bibfnamefont {T.}~\bibnamefont
  {Ruster}}, \bibinfo {author} {\bibfnamefont {C.~T.}\ \bibnamefont
  {Schmiegelow}}, \bibinfo {author} {\bibfnamefont {H.}~\bibnamefont
  {Kaufmann}}, \bibinfo {author} {\bibfnamefont {C.}~\bibnamefont
  {Warschburger}}, \bibinfo {author} {\bibfnamefont {F.}~\bibnamefont
  {Schmidt-Kaler}}, \ and\ \bibinfo {author} {\bibfnamefont {U.~G.}\
  \bibnamefont {Poschinger}},\ }\bibfield  {title} {\enquote {\bibinfo {title}
  {A long-lived zeeman trapped-ion qubit},}\ }\href {\doibase
  10.1007/s00340-016-6527-4} {\bibfield  {journal} {\bibinfo  {journal}
  {Applied Physics B}\ }\textbf {\bibinfo {volume} {122}},\ \bibinfo {pages}
  {254} (\bibinfo {year} {2016})}\BibitemShut {NoStop}%
\bibitem [{\citenamefont {Hilder}\ \emph {et~al.}(2022)\citenamefont {Hilder},
  \citenamefont {Pijn}, \citenamefont {Onishchenko}, \citenamefont {Stahl},
  \citenamefont {Orth}, \citenamefont {Lekitsch}, \citenamefont
  {Rodriguez-Blanco}, \citenamefont {M\"uller}, \citenamefont {Schmidt-Kaler},\
  and\ \citenamefont {Poschinger}}]{Hilder2022}%
  \BibitemOpen
  \bibfield  {author} {\bibinfo {author} {\bibfnamefont {J.}~\bibnamefont
  {Hilder}}, \bibinfo {author} {\bibfnamefont {D.}~\bibnamefont {Pijn}},
  \bibinfo {author} {\bibfnamefont {O.}~\bibnamefont {Onishchenko}}, \bibinfo
  {author} {\bibfnamefont {A.}~\bibnamefont {Stahl}}, \bibinfo {author}
  {\bibfnamefont {M.}~\bibnamefont {Orth}}, \bibinfo {author} {\bibfnamefont
  {B.}~\bibnamefont {Lekitsch}}, \bibinfo {author} {\bibfnamefont
  {A.}~\bibnamefont {Rodriguez-Blanco}}, \bibinfo {author} {\bibfnamefont
  {M.}~\bibnamefont {M\"uller}}, \bibinfo {author} {\bibfnamefont
  {F.}~\bibnamefont {Schmidt-Kaler}}, \ and\ \bibinfo {author} {\bibfnamefont
  {U.~G.}\ \bibnamefont {Poschinger}},\ }\bibfield  {title} {\enquote {\bibinfo
  {title} {Fault-tolerant parity readout on a shuttling-based trapped-ion
  quantum computer},}\ }\href {\doibase 10.1103/PhysRevX.12.011032} {\bibfield
  {journal} {\bibinfo  {journal} {Phys. Rev. X}\ }\textbf {\bibinfo {volume}
  {12}},\ \bibinfo {pages} {011032} (\bibinfo {year} {2022})}\BibitemShut
  {NoStop}%
\bibitem [{\citenamefont {Bermudez}\ \emph {et~al.}(2017)\citenamefont
  {Bermudez}, \citenamefont {Xu}, \citenamefont {Nigmatullin}, \citenamefont
  {O'Gorman}, \citenamefont {Negnevitsky}, \citenamefont {Schindler},
  \citenamefont {Monz}, \citenamefont {Poschinger}, \citenamefont {Hempel},
  \citenamefont {Home}, \citenamefont {Schmidt-Kaler}, \citenamefont {Biercuk},
  \citenamefont {Blatt}, \citenamefont {Benjamin},\ and\ \citenamefont
  {M\"uller}}]{Bermudez2017}%
  \BibitemOpen
  \bibfield  {author} {\bibinfo {author} {\bibfnamefont {A.}~\bibnamefont
  {Bermudez}}, \bibinfo {author} {\bibfnamefont {X.}~\bibnamefont {Xu}},
  \bibinfo {author} {\bibfnamefont {R.}~\bibnamefont {Nigmatullin}}, \bibinfo
  {author} {\bibfnamefont {J.}~\bibnamefont {O'Gorman}}, \bibinfo {author}
  {\bibfnamefont {V.}~\bibnamefont {Negnevitsky}}, \bibinfo {author}
  {\bibfnamefont {P.}~\bibnamefont {Schindler}}, \bibinfo {author}
  {\bibfnamefont {T.}~\bibnamefont {Monz}}, \bibinfo {author} {\bibfnamefont
  {U.~G.}\ \bibnamefont {Poschinger}}, \bibinfo {author} {\bibfnamefont
  {C.}~\bibnamefont {Hempel}}, \bibinfo {author} {\bibfnamefont
  {J.}~\bibnamefont {Home}}, \bibinfo {author} {\bibfnamefont {F.}~\bibnamefont
  {Schmidt-Kaler}}, \bibinfo {author} {\bibfnamefont {M.}~\bibnamefont
  {Biercuk}}, \bibinfo {author} {\bibfnamefont {R.}~\bibnamefont {Blatt}},
  \bibinfo {author} {\bibfnamefont {S.}~\bibnamefont {Benjamin}}, \ and\
  \bibinfo {author} {\bibfnamefont {M.}~\bibnamefont {M\"uller}},\ }\bibfield
  {title} {\enquote {\bibinfo {title} {Assessing the progress of trapped-ion
  processors towards fault-tolerant quantum computation},}\ }\href {\doibase
  10.1103/PhysRevX.7.041061} {\bibfield  {journal} {\bibinfo  {journal} {Phys.
  Rev. X}\ }\textbf {\bibinfo {volume} {7}},\ \bibinfo {pages} {041061}
  (\bibinfo {year} {2017})}\BibitemShut {NoStop}%
\bibitem [{\citenamefont {Toyoda}\ \emph {et~al.}(2010)\citenamefont {Toyoda},
  \citenamefont {Haze}, \citenamefont {Yamazaki},\ and\ \citenamefont
  {Urabe}}]{Toyoda2010}%
  \BibitemOpen
  \bibfield  {author} {\bibinfo {author} {\bibfnamefont {Kenji}\ \bibnamefont
  {Toyoda}}, \bibinfo {author} {\bibfnamefont {Shinsuke}\ \bibnamefont {Haze}},
  \bibinfo {author} {\bibfnamefont {Rekishu}\ \bibnamefont {Yamazaki}}, \ and\
  \bibinfo {author} {\bibfnamefont {Shinji}\ \bibnamefont {Urabe}},\ }\bibfield
   {title} {\enquote {\bibinfo {title} {Quantum gate using qubit states
  separated by terahertz},}\ }\href {\doibase 10.1103/PhysRevA.81.032322}
  {\bibfield  {journal} {\bibinfo  {journal} {Phys. Rev. A}\ }\textbf {\bibinfo
  {volume} {81}},\ \bibinfo {pages} {032322} (\bibinfo {year}
  {2010})}\BibitemShut {NoStop}%
\bibitem [{\citenamefont {Pogorelov}\ \emph {et~al.}(2021)\citenamefont
  {Pogorelov}, \citenamefont {Feldker}, \citenamefont {Marciniak},
  \citenamefont {Postler}, \citenamefont {Jacob}, \citenamefont
  {Krieglsteiner}, \citenamefont {Podlesnic}, \citenamefont {Meth},
  \citenamefont {Negnevitsky}, \citenamefont {Stadler}, \citenamefont
  {H\"ofer}, \citenamefont {W\"achter}, \citenamefont {Lakhmanskiy},
  \citenamefont {Blatt}, \citenamefont {Schindler},\ and\ \citenamefont
  {Monz}}]{Pogorelov2021}%
  \BibitemOpen
  \bibfield  {author} {\bibinfo {author} {\bibfnamefont {I.}~\bibnamefont
  {Pogorelov}}, \bibinfo {author} {\bibfnamefont {T.}~\bibnamefont {Feldker}},
  \bibinfo {author} {\bibfnamefont {Ch.~D.}\ \bibnamefont {Marciniak}},
  \bibinfo {author} {\bibfnamefont {L.}~\bibnamefont {Postler}}, \bibinfo
  {author} {\bibfnamefont {G.}~\bibnamefont {Jacob}}, \bibinfo {author}
  {\bibfnamefont {O.}~\bibnamefont {Krieglsteiner}}, \bibinfo {author}
  {\bibfnamefont {V.}~\bibnamefont {Podlesnic}}, \bibinfo {author}
  {\bibfnamefont {M.}~\bibnamefont {Meth}}, \bibinfo {author} {\bibfnamefont
  {V.}~\bibnamefont {Negnevitsky}}, \bibinfo {author} {\bibfnamefont
  {M.}~\bibnamefont {Stadler}}, \bibinfo {author} {\bibfnamefont
  {B.}~\bibnamefont {H\"ofer}}, \bibinfo {author} {\bibfnamefont
  {C.}~\bibnamefont {W\"achter}}, \bibinfo {author} {\bibfnamefont
  {K.}~\bibnamefont {Lakhmanskiy}}, \bibinfo {author} {\bibfnamefont
  {R.}~\bibnamefont {Blatt}}, \bibinfo {author} {\bibfnamefont
  {P.}~\bibnamefont {Schindler}}, \ and\ \bibinfo {author} {\bibfnamefont
  {T.}~\bibnamefont {Monz}},\ }\bibfield  {title} {\enquote {\bibinfo {title}
  {Compact ion-trap quantum computing demonstrator},}\ }\href {\doibase
  10.1103/PRXQuantum.2.020343} {\bibfield  {journal} {\bibinfo  {journal} {PRX
  Quantum}\ }\textbf {\bibinfo {volume} {2}},\ \bibinfo {pages} {020343}
  (\bibinfo {year} {2021})}\BibitemShut {NoStop}%
\bibitem [{\citenamefont {Bl{\"u}mel}\ \emph {et~al.}(2021)\citenamefont
  {Bl{\"u}mel}, \citenamefont {Grzesiak}, \citenamefont {Pisenti},
  \citenamefont {Wright},\ and\ \citenamefont {Nam}}]{Blumer2021}%
  \BibitemOpen
  \bibfield  {author} {\bibinfo {author} {\bibfnamefont {Reinhold}\
  \bibnamefont {Bl{\"u}mel}}, \bibinfo {author} {\bibfnamefont {Nikodem}\
  \bibnamefont {Grzesiak}}, \bibinfo {author} {\bibfnamefont {Neal}\
  \bibnamefont {Pisenti}}, \bibinfo {author} {\bibfnamefont {Kenneth}\
  \bibnamefont {Wright}}, \ and\ \bibinfo {author} {\bibfnamefont {Yunseong}\
  \bibnamefont {Nam}},\ }\bibfield  {title} {\enquote {\bibinfo {title}
  {Power-optimal, stabilized entangling gate between trapped-ion qubits},}\
  }\href {\doibase 10.1038/s41534-021-00489-w} {\bibfield  {journal} {\bibinfo
  {journal} {npj Quantum Information}\ }\textbf {\bibinfo {volume} {7}},\
  \bibinfo {pages} {147} (\bibinfo {year} {2021})}\BibitemShut {NoStop}%
\bibitem [{\citenamefont {Jeon}\ \emph {et~al.}(2023)\citenamefont {Jeon},
  \citenamefont {Kang}, \citenamefont {Kim}, \citenamefont {Choi},
  \citenamefont {Kim},\ and\ \citenamefont {Kim}}]{Jeon2023}%
  \BibitemOpen
  \bibfield  {author} {\bibinfo {author} {\bibfnamefont {Honggi}\ \bibnamefont
  {Jeon}}, \bibinfo {author} {\bibfnamefont {Jiyong}\ \bibnamefont {Kang}},
  \bibinfo {author} {\bibfnamefont {Jaeun}\ \bibnamefont {Kim}}, \bibinfo
  {author} {\bibfnamefont {Wonhyeong}\ \bibnamefont {Choi}}, \bibinfo {author}
  {\bibfnamefont {Kyunghye}\ \bibnamefont {Kim}}, \ and\ \bibinfo {author}
  {\bibfnamefont {Taehyun}\ \bibnamefont {Kim}},\ }\href@noop {} {\enquote
  {\bibinfo {title} {Experimental realization of entangled coherent states in
  two-dimensional harmonic oscillators of a trapped ion},}\ } (\bibinfo {year}
  {2023}),\ \Eprint {http://arxiv.org/abs/2305.00820} {arXiv:2305.00820
  [quant-ph]} \BibitemShut {NoStop}%
\bibitem [{\citenamefont {Saner}\ \emph {et~al.}(2023)\citenamefont {Saner},
  \citenamefont {Băzăvan}, \citenamefont {Minder}, \citenamefont {Drmota},
  \citenamefont {Webb}, \citenamefont {Araneda}, \citenamefont {Srinivas},
  \citenamefont {Lucas},\ and\ \citenamefont {Ballance}}]{Saner2023}%
  \BibitemOpen
  \bibfield  {author} {\bibinfo {author} {\bibfnamefont {S.}~\bibnamefont
  {Saner}}, \bibinfo {author} {\bibfnamefont {O.}~\bibnamefont {Băzăvan}},
  \bibinfo {author} {\bibfnamefont {M.}~\bibnamefont {Minder}}, \bibinfo
  {author} {\bibfnamefont {P.}~\bibnamefont {Drmota}}, \bibinfo {author}
  {\bibfnamefont {D.~J.}\ \bibnamefont {Webb}}, \bibinfo {author}
  {\bibfnamefont {G.}~\bibnamefont {Araneda}}, \bibinfo {author} {\bibfnamefont
  {R.}~\bibnamefont {Srinivas}}, \bibinfo {author} {\bibfnamefont {D.~M.}\
  \bibnamefont {Lucas}}, \ and\ \bibinfo {author} {\bibfnamefont {C.~J.}\
  \bibnamefont {Ballance}},\ }\href@noop {} {\enquote {\bibinfo {title}
  {Breaking the entangling gate speed limit for trapped-ion qubits using a
  phase-stable standing wave},}\ } (\bibinfo {year} {2023}),\ \Eprint
  {http://arxiv.org/abs/2305.03450} {arXiv:2305.03450 [quant-ph]} \BibitemShut
  {NoStop}%
\bibitem [{\citenamefont {Stute}\ \emph {et~al.}(2012)\citenamefont {Stute},
  \citenamefont {Casabone}, \citenamefont {Schindler}, \citenamefont {Monz},
  \citenamefont {Schmidt}, \citenamefont {Brandst{\"a}tter}, \citenamefont
  {Northup},\ and\ \citenamefont {Blatt}}]{Stute2012}%
  \BibitemOpen
  \bibfield  {author} {\bibinfo {author} {\bibfnamefont {A.}~\bibnamefont
  {Stute}}, \bibinfo {author} {\bibfnamefont {B.}~\bibnamefont {Casabone}},
  \bibinfo {author} {\bibfnamefont {P.}~\bibnamefont {Schindler}}, \bibinfo
  {author} {\bibfnamefont {T.}~\bibnamefont {Monz}}, \bibinfo {author}
  {\bibfnamefont {P.~O.}\ \bibnamefont {Schmidt}}, \bibinfo {author}
  {\bibfnamefont {B.}~\bibnamefont {Brandst{\"a}tter}}, \bibinfo {author}
  {\bibfnamefont {T.~E.}\ \bibnamefont {Northup}}, \ and\ \bibinfo {author}
  {\bibfnamefont {R.}~\bibnamefont {Blatt}},\ }\bibfield  {title} {\enquote
  {\bibinfo {title} {Tunable ion--photon entanglement in an optical cavity},}\
  }\href {\doibase 10.1038/nature11120} {\bibfield  {journal} {\bibinfo
  {journal} {Nature}\ }\textbf {\bibinfo {volume} {485}},\ \bibinfo {pages}
  {482--485} (\bibinfo {year} {2012})}\BibitemShut {NoStop}%
\bibitem [{\citenamefont {Stute}\ \emph {et~al.}(2013)\citenamefont {Stute},
  \citenamefont {Casabone}, \citenamefont {Brandst{\"a}tter}, \citenamefont
  {Friebe}, \citenamefont {Northup},\ and\ \citenamefont {Blatt}}]{Stute2013}%
  \BibitemOpen
  \bibfield  {author} {\bibinfo {author} {\bibfnamefont {A.}~\bibnamefont
  {Stute}}, \bibinfo {author} {\bibfnamefont {B.}~\bibnamefont {Casabone}},
  \bibinfo {author} {\bibfnamefont {B.}~\bibnamefont {Brandst{\"a}tter}},
  \bibinfo {author} {\bibfnamefont {K.}~\bibnamefont {Friebe}}, \bibinfo
  {author} {\bibfnamefont {T.~E.}\ \bibnamefont {Northup}}, \ and\ \bibinfo
  {author} {\bibfnamefont {R.}~\bibnamefont {Blatt}},\ }\bibfield  {title}
  {\enquote {\bibinfo {title} {Quantum-state transfer from an ion to a
  photon},}\ }\href {\doibase 10.1038/nphoton.2012.358} {\bibfield  {journal}
  {\bibinfo  {journal} {Nature Photonics}\ }\textbf {\bibinfo {volume} {7}},\
  \bibinfo {pages} {219--222} (\bibinfo {year} {2013})}\BibitemShut {NoStop}%
\bibitem [{\citenamefont {Begley}\ \emph {et~al.}(2016)\citenamefont {Begley},
  \citenamefont {Vogt}, \citenamefont {Gulati}, \citenamefont {Takahashi},\
  and\ \citenamefont {Keller}}]{Begley2016}%
  \BibitemOpen
  \bibfield  {author} {\bibinfo {author} {\bibfnamefont {Stephen}\ \bibnamefont
  {Begley}}, \bibinfo {author} {\bibfnamefont {Markus}\ \bibnamefont {Vogt}},
  \bibinfo {author} {\bibfnamefont {Gurpreet~Kaur}\ \bibnamefont {Gulati}},
  \bibinfo {author} {\bibfnamefont {Hiroki}\ \bibnamefont {Takahashi}}, \ and\
  \bibinfo {author} {\bibfnamefont {Matthias}\ \bibnamefont {Keller}},\
  }\bibfield  {title} {\enquote {\bibinfo {title} {Optimized multi-ion cavity
  coupling},}\ }\href {\doibase 10.1103/PhysRevLett.116.223001} {\bibfield
  {journal} {\bibinfo  {journal} {Phys. Rev. Lett.}\ }\textbf {\bibinfo
  {volume} {116}},\ \bibinfo {pages} {223001} (\bibinfo {year}
  {2016})}\BibitemShut {NoStop}%
\bibitem [{\citenamefont {Takahashi}\ \emph {et~al.}(2020)\citenamefont
  {Takahashi}, \citenamefont {Kassa}, \citenamefont {Christoforou},\ and\
  \citenamefont {Keller}}]{Takahashi2020}%
  \BibitemOpen
  \bibfield  {author} {\bibinfo {author} {\bibfnamefont {Hiroki}\ \bibnamefont
  {Takahashi}}, \bibinfo {author} {\bibfnamefont {Ezra}\ \bibnamefont {Kassa}},
  \bibinfo {author} {\bibfnamefont {Costas}\ \bibnamefont {Christoforou}}, \
  and\ \bibinfo {author} {\bibfnamefont {Matthias}\ \bibnamefont {Keller}},\
  }\bibfield  {title} {\enquote {\bibinfo {title} {Strong coupling of a single
  ion to an optical cavity},}\ }\href {\doibase 10.1103/PhysRevLett.124.013602}
  {\bibfield  {journal} {\bibinfo  {journal} {Phys. Rev. Lett.}\ }\textbf
  {\bibinfo {volume} {124}},\ \bibinfo {pages} {013602} (\bibinfo {year}
  {2020})}\BibitemShut {NoStop}%
\bibitem [{\citenamefont {Christoforou}\ \emph {et~al.}(2020)\citenamefont
  {Christoforou}, \citenamefont {Pignot}, \citenamefont {Kassa}, \citenamefont
  {Takahashi},\ and\ \citenamefont {Keller}}]{Christoforou2020}%
  \BibitemOpen
  \bibfield  {author} {\bibinfo {author} {\bibfnamefont {Costas}\ \bibnamefont
  {Christoforou}}, \bibinfo {author} {\bibfnamefont {Corentin}\ \bibnamefont
  {Pignot}}, \bibinfo {author} {\bibfnamefont {Ezra}\ \bibnamefont {Kassa}},
  \bibinfo {author} {\bibfnamefont {Hiroki}\ \bibnamefont {Takahashi}}, \ and\
  \bibinfo {author} {\bibfnamefont {Matthias}\ \bibnamefont {Keller}},\
  }\bibfield  {title} {\enquote {\bibinfo {title} {Enhanced ion--cavity
  coupling through cavity cooling in the strong coupling regime},}\ }\href
  {\doibase 10.1038/s41598-020-72796-9} {\bibfield  {journal} {\bibinfo
  {journal} {Scientific Reports}\ }\textbf {\bibinfo {volume} {10}},\ \bibinfo
  {pages} {15693} (\bibinfo {year} {2020})}\BibitemShut {NoStop}%
\bibitem [{\citenamefont {Blinov}\ \emph {et~al.}(2004)\citenamefont {Blinov},
  \citenamefont {Moehring}, \citenamefont {Duan},\ and\ \citenamefont
  {Monroe}}]{Blinov2004}%
  \BibitemOpen
  \bibfield  {author} {\bibinfo {author} {\bibfnamefont {B.~B.}\ \bibnamefont
  {Blinov}}, \bibinfo {author} {\bibfnamefont {D.~L.}\ \bibnamefont
  {Moehring}}, \bibinfo {author} {\bibfnamefont {L.~M.}\ \bibnamefont {Duan}},
  \ and\ \bibinfo {author} {\bibfnamefont {C.}~\bibnamefont {Monroe}},\
  }\bibfield  {title} {\enquote {\bibinfo {title} {Observation of entanglement
  between a single trapped atom and a single photon},}\ }\href {\doibase
  10.1038/nature02377} {\bibfield  {journal} {\bibinfo  {journal} {Nature}\
  }\textbf {\bibinfo {volume} {428}},\ \bibinfo {pages} {153--157} (\bibinfo
  {year} {2004})}\BibitemShut {NoStop}%
\bibitem [{\citenamefont {Bock}\ \emph {et~al.}(2018)\citenamefont {Bock},
  \citenamefont {Eich}, \citenamefont {Kucera}, \citenamefont {Kreis},
  \citenamefont {Lenhard}, \citenamefont {Becher},\ and\ \citenamefont
  {Eschner}}]{Bock2018}%
  \BibitemOpen
  \bibfield  {author} {\bibinfo {author} {\bibfnamefont {Matthias}\
  \bibnamefont {Bock}}, \bibinfo {author} {\bibfnamefont {Pascal}\ \bibnamefont
  {Eich}}, \bibinfo {author} {\bibfnamefont {Stephan}\ \bibnamefont {Kucera}},
  \bibinfo {author} {\bibfnamefont {Matthias}\ \bibnamefont {Kreis}}, \bibinfo
  {author} {\bibfnamefont {Andreas}\ \bibnamefont {Lenhard}}, \bibinfo {author}
  {\bibfnamefont {Christoph}\ \bibnamefont {Becher}}, \ and\ \bibinfo {author}
  {\bibfnamefont {J{\"u}rgen}\ \bibnamefont {Eschner}},\ }\bibfield  {title}
  {\enquote {\bibinfo {title} {High-fidelity entanglement between a trapped ion
  and a telecom photon via quantum frequency conversion},}\ }\href {\doibase
  10.1038/s41467-018-04341-2} {\bibfield  {journal} {\bibinfo  {journal}
  {Nature Communications}\ }\textbf {\bibinfo {volume} {9}},\ \bibinfo {pages}
  {1998} (\bibinfo {year} {2018})}\BibitemShut {NoStop}%
\bibitem [{\citenamefont {Kobel}\ \emph {et~al.}(2021)\citenamefont {Kobel},
  \citenamefont {Breyer},\ and\ \citenamefont {K{\"o}hl}}]{Kobel2021}%
  \BibitemOpen
  \bibfield  {author} {\bibinfo {author} {\bibfnamefont {Pascal}\ \bibnamefont
  {Kobel}}, \bibinfo {author} {\bibfnamefont {Moritz}\ \bibnamefont {Breyer}},
  \ and\ \bibinfo {author} {\bibfnamefont {Michael}\ \bibnamefont {K{\"o}hl}},\
  }\bibfield  {title} {\enquote {\bibinfo {title} {Deterministic spin-photon
  entanglement from a trapped ion in a fiber fabry--perot cavity},}\ }\href
  {\doibase 10.1038/s41534-020-00338-2} {\bibfield  {journal} {\bibinfo
  {journal} {npj Quantum Information}\ }\textbf {\bibinfo {volume} {7}},\
  \bibinfo {pages} {6} (\bibinfo {year} {2021})}\BibitemShut {NoStop}%
\bibitem [{\citenamefont {Cabrillo}\ \emph {et~al.}(1999)\citenamefont
  {Cabrillo}, \citenamefont {Cirac}, \citenamefont {Garc\'{\i}a-Fern\'andez},\
  and\ \citenamefont {Zoller}}]{Cabrillo1999}%
  \BibitemOpen
  \bibfield  {author} {\bibinfo {author} {\bibfnamefont {C.}~\bibnamefont
  {Cabrillo}}, \bibinfo {author} {\bibfnamefont {J.~I.}\ \bibnamefont {Cirac}},
  \bibinfo {author} {\bibfnamefont {P.}~\bibnamefont
  {Garc\'{\i}a-Fern\'andez}}, \ and\ \bibinfo {author} {\bibfnamefont
  {P.}~\bibnamefont {Zoller}},\ }\bibfield  {title} {\enquote {\bibinfo {title}
  {Creation of entangled states of distant atoms by interference},}\ }\href
  {\doibase 10.1103/PhysRevA.59.1025} {\bibfield  {journal} {\bibinfo
  {journal} {Phys. Rev. A}\ }\textbf {\bibinfo {volume} {59}},\ \bibinfo
  {pages} {1025--1033} (\bibinfo {year} {1999})}\BibitemShut {NoStop}%
\bibitem [{\citenamefont {Kok}\ \emph {et~al.}(2007)\citenamefont {Kok},
  \citenamefont {Munro}, \citenamefont {Nemoto}, \citenamefont {Ralph},
  \citenamefont {Dowling},\ and\ \citenamefont {Milburn}}]{Kok2007}%
  \BibitemOpen
  \bibfield  {author} {\bibinfo {author} {\bibfnamefont {Pieter}\ \bibnamefont
  {Kok}}, \bibinfo {author} {\bibfnamefont {W.~J.}\ \bibnamefont {Munro}},
  \bibinfo {author} {\bibfnamefont {Kae}\ \bibnamefont {Nemoto}}, \bibinfo
  {author} {\bibfnamefont {T.~C.}\ \bibnamefont {Ralph}}, \bibinfo {author}
  {\bibfnamefont {Jonathan~P.}\ \bibnamefont {Dowling}}, \ and\ \bibinfo
  {author} {\bibfnamefont {G.~J.}\ \bibnamefont {Milburn}},\ }\bibfield
  {title} {\enquote {\bibinfo {title} {Linear optical quantum computing with
  photonic qubits},}\ }\href {\doibase 10.1103/RevModPhys.79.135} {\bibfield
  {journal} {\bibinfo  {journal} {Rev. Mod. Phys.}\ }\textbf {\bibinfo {volume}
  {79}},\ \bibinfo {pages} {135--174} (\bibinfo {year} {2007})}\BibitemShut
  {NoStop}%
\bibitem [{\citenamefont {Duan}\ \emph {et~al.}(2001)\citenamefont {Duan},
  \citenamefont {Lukin}, \citenamefont {Cirac},\ and\ \citenamefont
  {Zoller}}]{DLCZ}%
  \BibitemOpen
  \bibfield  {author} {\bibinfo {author} {\bibfnamefont {L.~M.}\ \bibnamefont
  {Duan}}, \bibinfo {author} {\bibfnamefont {M.~D.}\ \bibnamefont {Lukin}},
  \bibinfo {author} {\bibfnamefont {J.~I.}\ \bibnamefont {Cirac}}, \ and\
  \bibinfo {author} {\bibfnamefont {P.}~\bibnamefont {Zoller}},\ }\bibfield
  {title} {\enquote {\bibinfo {title} {Long-distance quantum communication with
  atomic ensembles and linear optics},}\ }\href {\doibase 10.1038/35106500}
  {\bibfield  {journal} {\bibinfo  {journal} {Nature}\ }\textbf {\bibinfo
  {volume} {414}},\ \bibinfo {pages} {413--418} (\bibinfo {year}
  {2001})}\BibitemShut {NoStop}%
\bibitem [{\citenamefont {Casabone}\ \emph {et~al.}(2013)\citenamefont
  {Casabone}, \citenamefont {Stute}, \citenamefont {Friebe}, \citenamefont
  {Brandst\"atter}, \citenamefont {Sch\"uppert}, \citenamefont {Blatt},\ and\
  \citenamefont {Northup}}]{Casabone2013}%
  \BibitemOpen
  \bibfield  {author} {\bibinfo {author} {\bibfnamefont {B.}~\bibnamefont
  {Casabone}}, \bibinfo {author} {\bibfnamefont {A.}~\bibnamefont {Stute}},
  \bibinfo {author} {\bibfnamefont {K.}~\bibnamefont {Friebe}}, \bibinfo
  {author} {\bibfnamefont {B.}~\bibnamefont {Brandst\"atter}}, \bibinfo
  {author} {\bibfnamefont {K.}~\bibnamefont {Sch\"uppert}}, \bibinfo {author}
  {\bibfnamefont {R.}~\bibnamefont {Blatt}}, \ and\ \bibinfo {author}
  {\bibfnamefont {T.~E.}\ \bibnamefont {Northup}},\ }\bibfield  {title}
  {\enquote {\bibinfo {title} {Heralded entanglement of two ions in an optical
  cavity},}\ }\href {\doibase 10.1103/PhysRevLett.111.100505} {\bibfield
  {journal} {\bibinfo  {journal} {Phys. Rev. Lett.}\ }\textbf {\bibinfo
  {volume} {111}},\ \bibinfo {pages} {100505} (\bibinfo {year}
  {2013})}\BibitemShut {NoStop}%
\bibitem [{\citenamefont {Niffenegger}\ \emph {et~al.}(2020)\citenamefont
  {Niffenegger}, \citenamefont {Stuart}, \citenamefont {Sorace-Agaskar},
  \citenamefont {Kharas}, \citenamefont {Bramhavar}, \citenamefont {Bruzewicz},
  \citenamefont {Loh}, \citenamefont {Maxson}, \citenamefont {McConnell},
  \citenamefont {Reens}, \citenamefont {West}, \citenamefont {Sage},\ and\
  \citenamefont {Chiaverini}}]{Niffenegger2020}%
  \BibitemOpen
  \bibfield  {author} {\bibinfo {author} {\bibfnamefont {R.~J.}\ \bibnamefont
  {Niffenegger}}, \bibinfo {author} {\bibfnamefont {J.}~\bibnamefont {Stuart}},
  \bibinfo {author} {\bibfnamefont {C.}~\bibnamefont {Sorace-Agaskar}},
  \bibinfo {author} {\bibfnamefont {D.}~\bibnamefont {Kharas}}, \bibinfo
  {author} {\bibfnamefont {S.}~\bibnamefont {Bramhavar}}, \bibinfo {author}
  {\bibfnamefont {C.~D.}\ \bibnamefont {Bruzewicz}}, \bibinfo {author}
  {\bibfnamefont {W.}~\bibnamefont {Loh}}, \bibinfo {author} {\bibfnamefont
  {R.~T.}\ \bibnamefont {Maxson}}, \bibinfo {author} {\bibfnamefont
  {R.}~\bibnamefont {McConnell}}, \bibinfo {author} {\bibfnamefont
  {D.}~\bibnamefont {Reens}}, \bibinfo {author} {\bibfnamefont {G.~N.}\
  \bibnamefont {West}}, \bibinfo {author} {\bibfnamefont {J.~M.}\ \bibnamefont
  {Sage}}, \ and\ \bibinfo {author} {\bibfnamefont {J.}~\bibnamefont
  {Chiaverini}},\ }\bibfield  {title} {\enquote {\bibinfo {title} {Integrated
  multi-wavelength control of an ion qubit},}\ }\href {\doibase
  10.1038/s41586-020-2811-x} {\bibfield  {journal} {\bibinfo  {journal}
  {Nature}\ }\textbf {\bibinfo {volume} {586}},\ \bibinfo {pages} {538--542}
  (\bibinfo {year} {2020})}\BibitemShut {NoStop}%
\bibitem [{\citenamefont {Saha}\ \emph {et~al.}(2023)\citenamefont {Saha},
  \citenamefont {Siverns}, \citenamefont {Hannegan}, \citenamefont {Prabhu},
  \citenamefont {Quraishi}, \citenamefont {Englund},\ and\ \citenamefont
  {Waks}}]{Saha2023}%
  \BibitemOpen
  \bibfield  {author} {\bibinfo {author} {\bibfnamefont {Uday}\ \bibnamefont
  {Saha}}, \bibinfo {author} {\bibfnamefont {James~D.}\ \bibnamefont
  {Siverns}}, \bibinfo {author} {\bibfnamefont {John}\ \bibnamefont
  {Hannegan}}, \bibinfo {author} {\bibfnamefont {Mihika}\ \bibnamefont
  {Prabhu}}, \bibinfo {author} {\bibfnamefont {Qudsia}\ \bibnamefont
  {Quraishi}}, \bibinfo {author} {\bibfnamefont {Dirk}\ \bibnamefont
  {Englund}}, \ and\ \bibinfo {author} {\bibfnamefont {Edo}\ \bibnamefont
  {Waks}},\ }\bibfield  {title} {\enquote {\bibinfo {title} {Routing single
  photons from a trapped ion using a photonic integrated circuit},}\ }\href
  {\doibase 10.1103/PhysRevApplied.19.034001} {\bibfield  {journal} {\bibinfo
  {journal} {Phys. Rev. Appl.}\ }\textbf {\bibinfo {volume} {19}},\ \bibinfo
  {pages} {034001} (\bibinfo {year} {2023})}\BibitemShut {NoStop}%
\bibitem [{\citenamefont {Krutyanskiy}\ \emph {et~al.}(2023)\citenamefont
  {Krutyanskiy}, \citenamefont {Galli}, \citenamefont {Krcmarsky},
  \citenamefont {Baier}, \citenamefont {Fioretto}, \citenamefont {Pu},
  \citenamefont {Mazloom}, \citenamefont {Sekatski}, \citenamefont {Canteri},
  \citenamefont {Teller}, \citenamefont {Schupp}, \citenamefont {Bate},
  \citenamefont {Meraner}, \citenamefont {Sangouard}, \citenamefont {Lanyon},\
  and\ \citenamefont {Northup}}]{Krutyanskiy2023_EG}%
  \BibitemOpen
  \bibfield  {author} {\bibinfo {author} {\bibfnamefont {V.}~\bibnamefont
  {Krutyanskiy}}, \bibinfo {author} {\bibfnamefont {M.}~\bibnamefont {Galli}},
  \bibinfo {author} {\bibfnamefont {V.}~\bibnamefont {Krcmarsky}}, \bibinfo
  {author} {\bibfnamefont {S.}~\bibnamefont {Baier}}, \bibinfo {author}
  {\bibfnamefont {D.~A.}\ \bibnamefont {Fioretto}}, \bibinfo {author}
  {\bibfnamefont {Y.}~\bibnamefont {Pu}}, \bibinfo {author} {\bibfnamefont
  {A.}~\bibnamefont {Mazloom}}, \bibinfo {author} {\bibfnamefont
  {P.}~\bibnamefont {Sekatski}}, \bibinfo {author} {\bibfnamefont
  {M.}~\bibnamefont {Canteri}}, \bibinfo {author} {\bibfnamefont
  {M.}~\bibnamefont {Teller}}, \bibinfo {author} {\bibfnamefont
  {J.}~\bibnamefont {Schupp}}, \bibinfo {author} {\bibfnamefont
  {J.}~\bibnamefont {Bate}}, \bibinfo {author} {\bibfnamefont {M.}~\bibnamefont
  {Meraner}}, \bibinfo {author} {\bibfnamefont {N.}~\bibnamefont {Sangouard}},
  \bibinfo {author} {\bibfnamefont {B.~P.}\ \bibnamefont {Lanyon}}, \ and\
  \bibinfo {author} {\bibfnamefont {T.~E.}\ \bibnamefont {Northup}},\
  }\bibfield  {title} {\enquote {\bibinfo {title} {Entanglement of trapped-ion
  qubits separated by 230 meters},}\ }\href {\doibase
  10.1103/PhysRevLett.130.050803} {\bibfield  {journal} {\bibinfo  {journal}
  {Phys. Rev. Lett.}\ }\textbf {\bibinfo {volume} {130}},\ \bibinfo {pages}
  {050803} (\bibinfo {year} {2023})}\BibitemShut {NoStop}%
\bibitem [{\citenamefont {Dhara}\ \emph {et~al.}(2022)\citenamefont {Dhara},
  \citenamefont {Linke}, \citenamefont {Waks}, \citenamefont {Guha},\ and\
  \citenamefont {Seshadreesan}}]{Dhara2022}%
  \BibitemOpen
  \bibfield  {author} {\bibinfo {author} {\bibfnamefont {Prajit}\ \bibnamefont
  {Dhara}}, \bibinfo {author} {\bibfnamefont {Norbert~M.}\ \bibnamefont
  {Linke}}, \bibinfo {author} {\bibfnamefont {Edo}\ \bibnamefont {Waks}},
  \bibinfo {author} {\bibfnamefont {Saikat}\ \bibnamefont {Guha}}, \ and\
  \bibinfo {author} {\bibfnamefont {Kaushik~P.}\ \bibnamefont {Seshadreesan}},\
  }\bibfield  {title} {\enquote {\bibinfo {title} {Multiplexed quantum
  repeaters based on dual-species trapped-ion systems},}\ }\href {\doibase
  10.1103/PhysRevA.105.022623} {\bibfield  {journal} {\bibinfo  {journal}
  {Phys. Rev. A}\ }\textbf {\bibinfo {volume} {105}},\ \bibinfo {pages}
  {022623} (\bibinfo {year} {2022})}\BibitemShut {NoStop}%
\bibitem [{\citenamefont {Verd\'u}\ \emph {et~al.}(2009)\citenamefont
  {Verd\'u}, \citenamefont {Zoubi}, \citenamefont {Koller}, \citenamefont
  {Majer}, \citenamefont {Ritsch},\ and\ \citenamefont
  {Schmiedmayer}}]{Verdu2009}%
  \BibitemOpen
  \bibfield  {author} {\bibinfo {author} {\bibfnamefont {J.}~\bibnamefont
  {Verd\'u}}, \bibinfo {author} {\bibfnamefont {H.}~\bibnamefont {Zoubi}},
  \bibinfo {author} {\bibfnamefont {Ch.}\ \bibnamefont {Koller}}, \bibinfo
  {author} {\bibfnamefont {J.}~\bibnamefont {Majer}}, \bibinfo {author}
  {\bibfnamefont {H.}~\bibnamefont {Ritsch}}, \ and\ \bibinfo {author}
  {\bibfnamefont {J.}~\bibnamefont {Schmiedmayer}},\ }\bibfield  {title}
  {\enquote {\bibinfo {title} {Strong magnetic coupling of an ultracold gas to
  a superconducting waveguide cavity},}\ }\href {\doibase
  10.1103/PhysRevLett.103.043603} {\bibfield  {journal} {\bibinfo  {journal}
  {Phys. Rev. Lett.}\ }\textbf {\bibinfo {volume} {103}},\ \bibinfo {pages}
  {043603} (\bibinfo {year} {2009})}\BibitemShut {NoStop}%
\bibitem [{\citenamefont {Bohnet}\ \emph {et~al.}(2016)\citenamefont {Bohnet},
  \citenamefont {Sawyer}, \citenamefont {Britton}, \citenamefont {Wall},
  \citenamefont {Rey}, \citenamefont {Foss-Feig},\ and\ \citenamefont
  {Bollinger}}]{Bohnet2016}%
  \BibitemOpen
  \bibfield  {author} {\bibinfo {author} {\bibfnamefont {Justin~G.}\
  \bibnamefont {Bohnet}}, \bibinfo {author} {\bibfnamefont {Brian~C.}\
  \bibnamefont {Sawyer}}, \bibinfo {author} {\bibfnamefont {Joseph~W.}\
  \bibnamefont {Britton}}, \bibinfo {author} {\bibfnamefont {Michael~L.}\
  \bibnamefont {Wall}}, \bibinfo {author} {\bibfnamefont {Ana~Maria}\
  \bibnamefont {Rey}}, \bibinfo {author} {\bibfnamefont {Michael}\ \bibnamefont
  {Foss-Feig}}, \ and\ \bibinfo {author} {\bibfnamefont {John~J.}\ \bibnamefont
  {Bollinger}},\ }\bibfield  {title} {\enquote {\bibinfo {title} {Quantum spin
  dynamics and entanglement generation with hundreds of trapped ions},}\ }\href
  {\doibase 10.1126/science.aad9958} {\bibfield  {journal} {\bibinfo  {journal}
  {Science}\ }\textbf {\bibinfo {volume} {352}},\ \bibinfo {pages} {1297--1301}
  (\bibinfo {year} {2016})}\BibitemShut {NoStop}%
\bibitem [{\citenamefont {Zhang}\ \emph {et~al.}(2017)\citenamefont {Zhang},
  \citenamefont {Pagano}, \citenamefont {Hess}, \citenamefont {Kyprianidis},
  \citenamefont {Becker}, \citenamefont {Kaplan}, \citenamefont {Gorshkov},
  \citenamefont {Gong},\ and\ \citenamefont {Monroe}}]{Zhang2017}%
  \BibitemOpen
  \bibfield  {author} {\bibinfo {author} {\bibfnamefont {J.}~\bibnamefont
  {Zhang}}, \bibinfo {author} {\bibfnamefont {G.}~\bibnamefont {Pagano}},
  \bibinfo {author} {\bibfnamefont {P.~W.}\ \bibnamefont {Hess}}, \bibinfo
  {author} {\bibfnamefont {A.}~\bibnamefont {Kyprianidis}}, \bibinfo {author}
  {\bibfnamefont {P.}~\bibnamefont {Becker}}, \bibinfo {author} {\bibfnamefont
  {H.}~\bibnamefont {Kaplan}}, \bibinfo {author} {\bibfnamefont {A.~V.}\
  \bibnamefont {Gorshkov}}, \bibinfo {author} {\bibfnamefont {Z.~X.}\
  \bibnamefont {Gong}}, \ and\ \bibinfo {author} {\bibfnamefont
  {C.}~\bibnamefont {Monroe}},\ }\bibfield  {title} {\enquote {\bibinfo {title}
  {Observation of a many-body dynamical phase transition with a 53-qubit
  quantum simulator},}\ }\href {\doibase 10.1038/nature24654} {\bibfield
  {journal} {\bibinfo  {journal} {Nature}\ }\textbf {\bibinfo {volume} {551}},\
  \bibinfo {pages} {601--604} (\bibinfo {year} {2017})}\BibitemShut {NoStop}%
\bibitem [{\citenamefont {Kiesenhofer}\ \emph {et~al.}(2023)\citenamefont
  {Kiesenhofer}, \citenamefont {Hainzer}, \citenamefont {Zhdanov},
  \citenamefont {Holz}, \citenamefont {Bock}, \citenamefont {Ollikainen},\ and\
  \citenamefont {Roos}}]{Kiesenhofer2023}%
  \BibitemOpen
  \bibfield  {author} {\bibinfo {author} {\bibfnamefont {Dominik}\ \bibnamefont
  {Kiesenhofer}}, \bibinfo {author} {\bibfnamefont {Helene}\ \bibnamefont
  {Hainzer}}, \bibinfo {author} {\bibfnamefont {Artem}\ \bibnamefont
  {Zhdanov}}, \bibinfo {author} {\bibfnamefont {Philip~C.}\ \bibnamefont
  {Holz}}, \bibinfo {author} {\bibfnamefont {Matthias}\ \bibnamefont {Bock}},
  \bibinfo {author} {\bibfnamefont {Tuomas}\ \bibnamefont {Ollikainen}}, \ and\
  \bibinfo {author} {\bibfnamefont {Christian~F.}\ \bibnamefont {Roos}},\
  }\bibfield  {title} {\enquote {\bibinfo {title} {Controlling two-dimensional
  coulomb crystals of more than 100 ions in a monolithic radio-frequency
  trap},}\ }\href {\doibase 10.1103/PRXQuantum.4.020317} {\bibfield  {journal}
  {\bibinfo  {journal} {PRX Quantum}\ }\textbf {\bibinfo {volume} {4}},\
  \bibinfo {pages} {020317} (\bibinfo {year} {2023})}\BibitemShut {NoStop}%
\bibitem [{\citenamefont {Graham}\ \emph {et~al.}(2019)\citenamefont {Graham},
  \citenamefont {Kwon}, \citenamefont {Grinkemeyer}, \citenamefont {Marra},
  \citenamefont {Jiang}, \citenamefont {Lichtman}, \citenamefont {Sun},
  \citenamefont {Ebert},\ and\ \citenamefont {Saffman}}]{Graham2019}%
  \BibitemOpen
  \bibfield  {author} {\bibinfo {author} {\bibfnamefont {T.~M.}\ \bibnamefont
  {Graham}}, \bibinfo {author} {\bibfnamefont {M.}~\bibnamefont {Kwon}},
  \bibinfo {author} {\bibfnamefont {B.}~\bibnamefont {Grinkemeyer}}, \bibinfo
  {author} {\bibfnamefont {Z.}~\bibnamefont {Marra}}, \bibinfo {author}
  {\bibfnamefont {X.}~\bibnamefont {Jiang}}, \bibinfo {author} {\bibfnamefont
  {M.~T.}\ \bibnamefont {Lichtman}}, \bibinfo {author} {\bibfnamefont
  {Y.}~\bibnamefont {Sun}}, \bibinfo {author} {\bibfnamefont {M.}~\bibnamefont
  {Ebert}}, \ and\ \bibinfo {author} {\bibfnamefont {M.}~\bibnamefont
  {Saffman}},\ }\bibfield  {title} {\enquote {\bibinfo {title}
  {Rydberg-mediated entanglement in a two-dimensional neutral atom qubit
  array},}\ }\href {\doibase 10.1103/PhysRevLett.123.230501} {\bibfield
  {journal} {\bibinfo  {journal} {Phys. Rev. Lett.}\ }\textbf {\bibinfo
  {volume} {123}},\ \bibinfo {pages} {230501} (\bibinfo {year}
  {2019})}\BibitemShut {NoStop}%
\bibitem [{\citenamefont {Levine}\ \emph {et~al.}(2019)\citenamefont {Levine},
  \citenamefont {Keesling}, \citenamefont {Semeghini}, \citenamefont {Omran},
  \citenamefont {Wang}, \citenamefont {Ebadi}, \citenamefont {Bernien},
  \citenamefont {Greiner}, \citenamefont {Vuleti\ifmmode~\acute{c}\else
  \'{c}\fi{}}, \citenamefont {Pichler},\ and\ \citenamefont
  {Lukin}}]{Levine2019}%
  \BibitemOpen
  \bibfield  {author} {\bibinfo {author} {\bibfnamefont {Harry}\ \bibnamefont
  {Levine}}, \bibinfo {author} {\bibfnamefont {Alexander}\ \bibnamefont
  {Keesling}}, \bibinfo {author} {\bibfnamefont {Giulia}\ \bibnamefont
  {Semeghini}}, \bibinfo {author} {\bibfnamefont {Ahmed}\ \bibnamefont
  {Omran}}, \bibinfo {author} {\bibfnamefont {Tout~T.}\ \bibnamefont {Wang}},
  \bibinfo {author} {\bibfnamefont {Sepehr}\ \bibnamefont {Ebadi}}, \bibinfo
  {author} {\bibfnamefont {Hannes}\ \bibnamefont {Bernien}}, \bibinfo {author}
  {\bibfnamefont {Markus}\ \bibnamefont {Greiner}}, \bibinfo {author}
  {\bibfnamefont {Vladan}\ \bibnamefont {Vuleti\ifmmode~\acute{c}\else
  \'{c}\fi{}}}, \bibinfo {author} {\bibfnamefont {Hannes}\ \bibnamefont
  {Pichler}}, \ and\ \bibinfo {author} {\bibfnamefont {Mikhail~D.}\
  \bibnamefont {Lukin}},\ }\bibfield  {title} {\enquote {\bibinfo {title}
  {Parallel implementation of high-fidelity multiqubit gates with neutral
  atoms},}\ }\href {\doibase 10.1103/PhysRevLett.123.170503} {\bibfield
  {journal} {\bibinfo  {journal} {Phys. Rev. Lett.}\ }\textbf {\bibinfo
  {volume} {123}},\ \bibinfo {pages} {170503} (\bibinfo {year}
  {2019})}\BibitemShut {NoStop}%
\bibitem [{\citenamefont {Morgado}\ and\ \citenamefont
  {Whitlock}(2021)}]{Morgado2021}%
  \BibitemOpen
  \bibfield  {author} {\bibinfo {author} {\bibfnamefont {M.}~\bibnamefont
  {Morgado}}\ and\ \bibinfo {author} {\bibfnamefont {S.}~\bibnamefont
  {Whitlock}},\ }\bibfield  {title} {\enquote {\bibinfo {title} {{Quantum
  simulation and computing with Rydberg-interacting qubits}},}\ }\href
  {\doibase doi.org/10.1116/5.0036562} {\bibfield  {journal} {\bibinfo
  {journal} {AVS Quantum Science}\ }\textbf {\bibinfo {volume} {3}} (\bibinfo
  {year} {2021}),\ doi.org/10.1116/5.0036562},\ \bibinfo {note}
  {023501}\BibitemShut {NoStop}%
\bibitem [{\citenamefont {Levine}\ \emph {et~al.}(2018)\citenamefont {Levine},
  \citenamefont {Keesling}, \citenamefont {Omran}, \citenamefont {Bernien},
  \citenamefont {Schwartz}, \citenamefont {Zibrov}, \citenamefont {Endres},
  \citenamefont {Greiner}, \citenamefont {Vuleti\ifmmode~\acute{c}\else
  \'{c}\fi{}},\ and\ \citenamefont {Lukin}}]{Levine2018}%
  \BibitemOpen
  \bibfield  {author} {\bibinfo {author} {\bibfnamefont {Harry}\ \bibnamefont
  {Levine}}, \bibinfo {author} {\bibfnamefont {Alexander}\ \bibnamefont
  {Keesling}}, \bibinfo {author} {\bibfnamefont {Ahmed}\ \bibnamefont {Omran}},
  \bibinfo {author} {\bibfnamefont {Hannes}\ \bibnamefont {Bernien}}, \bibinfo
  {author} {\bibfnamefont {Sylvain}\ \bibnamefont {Schwartz}}, \bibinfo
  {author} {\bibfnamefont {Alexander~S.}\ \bibnamefont {Zibrov}}, \bibinfo
  {author} {\bibfnamefont {Manuel}\ \bibnamefont {Endres}}, \bibinfo {author}
  {\bibfnamefont {Markus}\ \bibnamefont {Greiner}}, \bibinfo {author}
  {\bibfnamefont {Vladan}\ \bibnamefont {Vuleti\ifmmode~\acute{c}\else
  \'{c}\fi{}}}, \ and\ \bibinfo {author} {\bibfnamefont {Mikhail~D.}\
  \bibnamefont {Lukin}},\ }\bibfield  {title} {\enquote {\bibinfo {title}
  {High-fidelity control and entanglement of rydberg-atom qubits},}\ }\href
  {\doibase 10.1103/PhysRevLett.121.123603} {\bibfield  {journal} {\bibinfo
  {journal} {Phys. Rev. Lett.}\ }\textbf {\bibinfo {volume} {121}},\ \bibinfo
  {pages} {123603} (\bibinfo {year} {2018})}\BibitemShut {NoStop}%
\bibitem [{\citenamefont {Madjarov}\ \emph {et~al.}(2020)\citenamefont
  {Madjarov}, \citenamefont {Covey}, \citenamefont {Shaw}, \citenamefont
  {Choi}, \citenamefont {Kale}, \citenamefont {Cooper}, \citenamefont
  {Pichler}, \citenamefont {Schkolnik}, \citenamefont {Williams},\ and\
  \citenamefont {Endres}}]{Madjarov2020}%
  \BibitemOpen
  \bibfield  {author} {\bibinfo {author} {\bibfnamefont {Ivaylo~S.}\
  \bibnamefont {Madjarov}}, \bibinfo {author} {\bibfnamefont {Jacob~P.}\
  \bibnamefont {Covey}}, \bibinfo {author} {\bibfnamefont {Adam~L.}\
  \bibnamefont {Shaw}}, \bibinfo {author} {\bibfnamefont {Joonhee}\
  \bibnamefont {Choi}}, \bibinfo {author} {\bibfnamefont {Anant}\ \bibnamefont
  {Kale}}, \bibinfo {author} {\bibfnamefont {Alexandre}\ \bibnamefont
  {Cooper}}, \bibinfo {author} {\bibfnamefont {Hannes}\ \bibnamefont
  {Pichler}}, \bibinfo {author} {\bibfnamefont {Vladimir}\ \bibnamefont
  {Schkolnik}}, \bibinfo {author} {\bibfnamefont {Jason~R.}\ \bibnamefont
  {Williams}}, \ and\ \bibinfo {author} {\bibfnamefont {Manuel}\ \bibnamefont
  {Endres}},\ }\bibfield  {title} {\enquote {\bibinfo {title} {High-fidelity
  entanglement and detection of alkaline-earth rydberg atoms},}\ }\href
  {\doibase 10.1038/s41567-020-0903-z} {\bibfield  {journal} {\bibinfo
  {journal} {Nature Physics}\ }\textbf {\bibinfo {volume} {16}},\ \bibinfo
  {pages} {857--861} (\bibinfo {year} {2020})}\BibitemShut {NoStop}%
\bibitem [{\citenamefont {Picken}\ \emph {et~al.}(2018)\citenamefont {Picken},
  \citenamefont {Legaie}, \citenamefont {McDonnell},\ and\ \citenamefont
  {Pritchard}}]{Picken2019}%
  \BibitemOpen
  \bibfield  {author} {\bibinfo {author} {\bibfnamefont {C~J}\ \bibnamefont
  {Picken}}, \bibinfo {author} {\bibfnamefont {R}~\bibnamefont {Legaie}},
  \bibinfo {author} {\bibfnamefont {K}~\bibnamefont {McDonnell}}, \ and\
  \bibinfo {author} {\bibfnamefont {J~D}\ \bibnamefont {Pritchard}},\
  }\bibfield  {title} {\enquote {\bibinfo {title} {Entanglement of neutral-atom
  qubits with long ground-rydberg coherence times},}\ }\href {\doibase
  10.1088/2058-9565/aaf019} {\bibfield  {journal} {\bibinfo  {journal} {Quantum
  Science and Technology}\ }\textbf {\bibinfo {volume} {4}},\ \bibinfo {pages}
  {015011} (\bibinfo {year} {2018})}\BibitemShut {NoStop}%
\bibitem [{\citenamefont {Norcia}\ \emph {et~al.}(2019)\citenamefont {Norcia},
  \citenamefont {Young}, \citenamefont {Eckner}, \citenamefont {Oelker},
  \citenamefont {Ye},\ and\ \citenamefont {Kaufman}}]{Norcia2019}%
  \BibitemOpen
  \bibfield  {author} {\bibinfo {author} {\bibfnamefont {Matthew~A.}\
  \bibnamefont {Norcia}}, \bibinfo {author} {\bibfnamefont {Aaron~W.}\
  \bibnamefont {Young}}, \bibinfo {author} {\bibfnamefont {William~J.}\
  \bibnamefont {Eckner}}, \bibinfo {author} {\bibfnamefont {Eric}\ \bibnamefont
  {Oelker}}, \bibinfo {author} {\bibfnamefont {Jun}\ \bibnamefont {Ye}}, \ and\
  \bibinfo {author} {\bibfnamefont {Adam~M.}\ \bibnamefont {Kaufman}},\
  }\bibfield  {title} {\enquote {\bibinfo {title} {Seconds-scale coherence on
  an optical clock transition in a tweezer array},}\ }\href {\doibase
  10.1126/science.aay0644} {\bibfield  {journal} {\bibinfo  {journal}
  {Science}\ }\textbf {\bibinfo {volume} {366}},\ \bibinfo {pages} {93--97}
  (\bibinfo {year} {2019})}\BibitemShut {NoStop}%
\bibitem [{\citenamefont {Barnes}\ \emph {et~al.}(2022)\citenamefont {Barnes},
  \citenamefont {Battaglino}, \citenamefont {Bloom}, \citenamefont {Cassella},
  \citenamefont {Coxe}, \citenamefont {Crisosto}, \citenamefont {King},
  \citenamefont {Kondov}, \citenamefont {Kotru}, \citenamefont {Larsen},
  \citenamefont {Lauigan}, \citenamefont {Lester}, \citenamefont {McDonald},
  \citenamefont {Megidish}, \citenamefont {Narayanaswami}, \citenamefont
  {Nishiguchi}, \citenamefont {Notermans}, \citenamefont {Peng}, \citenamefont
  {Ryou}, \citenamefont {Wu},\ and\ \citenamefont {Yarwood}}]{Barnes2022}%
  \BibitemOpen
  \bibfield  {author} {\bibinfo {author} {\bibfnamefont {Katrina}\ \bibnamefont
  {Barnes}}, \bibinfo {author} {\bibfnamefont {Peter}\ \bibnamefont
  {Battaglino}}, \bibinfo {author} {\bibfnamefont {Benjamin~J.}\ \bibnamefont
  {Bloom}}, \bibinfo {author} {\bibfnamefont {Kayleigh}\ \bibnamefont
  {Cassella}}, \bibinfo {author} {\bibfnamefont {Robin}\ \bibnamefont {Coxe}},
  \bibinfo {author} {\bibfnamefont {Nicole}\ \bibnamefont {Crisosto}}, \bibinfo
  {author} {\bibfnamefont {Jonathan~P.}\ \bibnamefont {King}}, \bibinfo
  {author} {\bibfnamefont {Stanimir~S.}\ \bibnamefont {Kondov}}, \bibinfo
  {author} {\bibfnamefont {Krish}\ \bibnamefont {Kotru}}, \bibinfo {author}
  {\bibfnamefont {Stuart~C.}\ \bibnamefont {Larsen}}, \bibinfo {author}
  {\bibfnamefont {Joseph}\ \bibnamefont {Lauigan}}, \bibinfo {author}
  {\bibfnamefont {Brian~J.}\ \bibnamefont {Lester}}, \bibinfo {author}
  {\bibfnamefont {Mickey}\ \bibnamefont {McDonald}}, \bibinfo {author}
  {\bibfnamefont {Eli}\ \bibnamefont {Megidish}}, \bibinfo {author}
  {\bibfnamefont {Sandeep}\ \bibnamefont {Narayanaswami}}, \bibinfo {author}
  {\bibfnamefont {Ciro}\ \bibnamefont {Nishiguchi}}, \bibinfo {author}
  {\bibfnamefont {Remy}\ \bibnamefont {Notermans}}, \bibinfo {author}
  {\bibfnamefont {Lucas~S.}\ \bibnamefont {Peng}}, \bibinfo {author}
  {\bibfnamefont {Albert}\ \bibnamefont {Ryou}}, \bibinfo {author}
  {\bibfnamefont {Tsung-Yao}\ \bibnamefont {Wu}}, \ and\ \bibinfo {author}
  {\bibfnamefont {Michael}\ \bibnamefont {Yarwood}},\ }\bibfield  {title}
  {\enquote {\bibinfo {title} {Assembly and coherent control of a register of
  nuclear spin qubits},}\ }\href {\doibase 10.1038/s41467-022-29977-z}
  {\bibfield  {journal} {\bibinfo  {journal} {Nature Communications}\ }\textbf
  {\bibinfo {volume} {13}},\ \bibinfo {pages} {2779} (\bibinfo {year}
  {2022})}\BibitemShut {NoStop}%
\bibitem [{\citenamefont {Lukin}\ \emph {et~al.}(2001)\citenamefont {Lukin},
  \citenamefont {Fleischhauer}, \citenamefont {Cote}, \citenamefont {Duan},
  \citenamefont {Jaksch}, \citenamefont {Cirac},\ and\ \citenamefont
  {Zoller}}]{Lukin2001}%
  \BibitemOpen
  \bibfield  {author} {\bibinfo {author} {\bibfnamefont {M.~D.}\ \bibnamefont
  {Lukin}}, \bibinfo {author} {\bibfnamefont {M.}~\bibnamefont {Fleischhauer}},
  \bibinfo {author} {\bibfnamefont {R.}~\bibnamefont {Cote}}, \bibinfo {author}
  {\bibfnamefont {L.~M.}\ \bibnamefont {Duan}}, \bibinfo {author}
  {\bibfnamefont {D.}~\bibnamefont {Jaksch}}, \bibinfo {author} {\bibfnamefont
  {J.~I.}\ \bibnamefont {Cirac}}, \ and\ \bibinfo {author} {\bibfnamefont
  {P.}~\bibnamefont {Zoller}},\ }\bibfield  {title} {\enquote {\bibinfo {title}
  {Dipole blockade and quantum information processing in mesoscopic atomic
  ensembles},}\ }\href {\doibase 10.1103/PhysRevLett.87.037901} {\bibfield
  {journal} {\bibinfo  {journal} {Phys. Rev. Lett.}\ }\textbf {\bibinfo
  {volume} {87}},\ \bibinfo {pages} {037901} (\bibinfo {year}
  {2001})}\BibitemShut {NoStop}%
\bibitem [{\citenamefont {Jaksch}\ \emph {et~al.}(2000)\citenamefont {Jaksch},
  \citenamefont {Cirac}, \citenamefont {Zoller}, \citenamefont {Rolston},
  \citenamefont {C\^ot\'e},\ and\ \citenamefont {Lukin}}]{Jaksch2000}%
  \BibitemOpen
  \bibfield  {author} {\bibinfo {author} {\bibfnamefont {D.}~\bibnamefont
  {Jaksch}}, \bibinfo {author} {\bibfnamefont {J.~I.}\ \bibnamefont {Cirac}},
  \bibinfo {author} {\bibfnamefont {P.}~\bibnamefont {Zoller}}, \bibinfo
  {author} {\bibfnamefont {S.~L.}\ \bibnamefont {Rolston}}, \bibinfo {author}
  {\bibfnamefont {R.}~\bibnamefont {C\^ot\'e}}, \ and\ \bibinfo {author}
  {\bibfnamefont {M.~D.}\ \bibnamefont {Lukin}},\ }\bibfield  {title} {\enquote
  {\bibinfo {title} {Fast quantum gates for neutral atoms},}\ }\href {\doibase
  10.1103/PhysRevLett.85.2208} {\bibfield  {journal} {\bibinfo  {journal}
  {Phys. Rev. Lett.}\ }\textbf {\bibinfo {volume} {85}},\ \bibinfo {pages}
  {2208--2211} (\bibinfo {year} {2000})}\BibitemShut {NoStop}%
\bibitem [{\citenamefont {Urban}\ \emph {et~al.}(2009)\citenamefont {Urban},
  \citenamefont {Johnson}, \citenamefont {Henage}, \citenamefont {Isenhower},
  \citenamefont {Yavuz}, \citenamefont {Walker},\ and\ \citenamefont
  {Saffman}}]{Urban2009}%
  \BibitemOpen
  \bibfield  {author} {\bibinfo {author} {\bibfnamefont {E.}~\bibnamefont
  {Urban}}, \bibinfo {author} {\bibfnamefont {T.~A.}\ \bibnamefont {Johnson}},
  \bibinfo {author} {\bibfnamefont {T.}~\bibnamefont {Henage}}, \bibinfo
  {author} {\bibfnamefont {L.}~\bibnamefont {Isenhower}}, \bibinfo {author}
  {\bibfnamefont {D.~D.}\ \bibnamefont {Yavuz}}, \bibinfo {author}
  {\bibfnamefont {T.~G.}\ \bibnamefont {Walker}}, \ and\ \bibinfo {author}
  {\bibfnamefont {M.}~\bibnamefont {Saffman}},\ }\bibfield  {title} {\enquote
  {\bibinfo {title} {Observation of rydberg blockade between two atoms},}\
  }\href {\doibase 10.1038/nphys1178} {\bibfield  {journal} {\bibinfo
  {journal} {Nature Physics}\ }\textbf {\bibinfo {volume} {5}},\ \bibinfo
  {pages} {110--114} (\bibinfo {year} {2009})}\BibitemShut {NoStop}%
\bibitem [{\citenamefont {Langenfeld}\ \emph {et~al.}(2020)\citenamefont
  {Langenfeld}, \citenamefont {Morin}, \citenamefont {K{\"o}rber},\ and\
  \citenamefont {Rempe}}]{Langenfeld2020}%
  \BibitemOpen
  \bibfield  {author} {\bibinfo {author} {\bibfnamefont {Stefan}\ \bibnamefont
  {Langenfeld}}, \bibinfo {author} {\bibfnamefont {Olivier}\ \bibnamefont
  {Morin}}, \bibinfo {author} {\bibfnamefont {Matthias}\ \bibnamefont
  {K{\"o}rber}}, \ and\ \bibinfo {author} {\bibfnamefont {Gerhard}\
  \bibnamefont {Rempe}},\ }\bibfield  {title} {\enquote {\bibinfo {title} {A
  network-ready random-access qubits memory},}\ }\href {\doibase
  10.1038/s41534-020-00316-8} {\bibfield  {journal} {\bibinfo  {journal} {npj
  Quantum Information}\ }\textbf {\bibinfo {volume} {6}},\ \bibinfo {pages}
  {86} (\bibinfo {year} {2020})}\BibitemShut {NoStop}%
\bibitem [{\citenamefont {Liu}\ \emph {et~al.}(2023{\natexlab{b}})\citenamefont
  {Liu}, \citenamefont {Wang}, \citenamefont {Yang}, \citenamefont {Wang},
  \citenamefont {Fan}, \citenamefont {Guan}, \citenamefont {Li}, \citenamefont
  {Zhang},\ and\ \citenamefont {Zhang}}]{LiuYanxin2023}%
  \BibitemOpen
  \bibfield  {author} {\bibinfo {author} {\bibfnamefont {Yanxin}\ \bibnamefont
  {Liu}}, \bibinfo {author} {\bibfnamefont {Zhihui}\ \bibnamefont {Wang}},
  \bibinfo {author} {\bibfnamefont {Pengfei}\ \bibnamefont {Yang}}, \bibinfo
  {author} {\bibfnamefont {Qinxia}\ \bibnamefont {Wang}}, \bibinfo {author}
  {\bibfnamefont {Qing}\ \bibnamefont {Fan}}, \bibinfo {author} {\bibfnamefont
  {Shijun}\ \bibnamefont {Guan}}, \bibinfo {author} {\bibfnamefont {Gang}\
  \bibnamefont {Li}}, \bibinfo {author} {\bibfnamefont {Pengfei}\ \bibnamefont
  {Zhang}}, \ and\ \bibinfo {author} {\bibfnamefont {Tiancai}\ \bibnamefont
  {Zhang}},\ }\bibfield  {title} {\enquote {\bibinfo {title} {Realization of
  strong coupling between deterministic single-atom arrays and a high-finesse
  miniature optical cavity},}\ }\href {\doibase 10.1103/PhysRevLett.130.173601}
  {\bibfield  {journal} {\bibinfo  {journal} {Phys. Rev. Lett.}\ }\textbf
  {\bibinfo {volume} {130}},\ \bibinfo {pages} {173601} (\bibinfo {year}
  {2023}{\natexlab{b}})}\BibitemShut {NoStop}%
\bibitem [{\citenamefont {Hosseini}\ \emph {et~al.}(2011)\citenamefont
  {Hosseini}, \citenamefont {Sparkes}, \citenamefont {Campbell}, \citenamefont
  {Lam},\ and\ \citenamefont {Buchler}}]{Hosseini2011}%
  \BibitemOpen
  \bibfield  {author} {\bibinfo {author} {\bibfnamefont {M.}~\bibnamefont
  {Hosseini}}, \bibinfo {author} {\bibfnamefont {B.~M.}\ \bibnamefont
  {Sparkes}}, \bibinfo {author} {\bibfnamefont {G.}~\bibnamefont {Campbell}},
  \bibinfo {author} {\bibfnamefont {P.~K.}\ \bibnamefont {Lam}}, \ and\
  \bibinfo {author} {\bibfnamefont {B.~C.}\ \bibnamefont {Buchler}},\
  }\bibfield  {title} {\enquote {\bibinfo {title} {High efficiency coherent
  optical memory with warm rubidium vapour},}\ }\href {\doibase
  10.1038/ncomms1175} {\bibfield  {journal} {\bibinfo  {journal} {Nature
  Communications}\ }\textbf {\bibinfo {volume} {2}},\ \bibinfo {pages} {174}
  (\bibinfo {year} {2011})}\BibitemShut {NoStop}%
\bibitem [{\citenamefont {Dudin}\ \emph {et~al.}(2013)\citenamefont {Dudin},
  \citenamefont {Li},\ and\ \citenamefont {Kuzmich}}]{Dudin2013}%
  \BibitemOpen
  \bibfield  {author} {\bibinfo {author} {\bibfnamefont {Y.~O.}\ \bibnamefont
  {Dudin}}, \bibinfo {author} {\bibfnamefont {L.}~\bibnamefont {Li}}, \ and\
  \bibinfo {author} {\bibfnamefont {A.}~\bibnamefont {Kuzmich}},\ }\bibfield
  {title} {\enquote {\bibinfo {title} {Light storage on the time scale of a
  minute},}\ }\href {\doibase 10.1103/PhysRevA.87.031801} {\bibfield  {journal}
  {\bibinfo  {journal} {Phys. Rev. A}\ }\textbf {\bibinfo {volume} {87}},\
  \bibinfo {pages} {031801} (\bibinfo {year} {2013})}\BibitemShut {NoStop}%
\bibitem [{\citenamefont {Ding}\ \emph {et~al.}(2013)\citenamefont {Ding},
  \citenamefont {Zhou}, \citenamefont {Shi},\ and\ \citenamefont
  {Guo}}]{Ding2013}%
  \BibitemOpen
  \bibfield  {author} {\bibinfo {author} {\bibfnamefont {Dong-Sheng}\
  \bibnamefont {Ding}}, \bibinfo {author} {\bibfnamefont {Zhi-Yuan}\
  \bibnamefont {Zhou}}, \bibinfo {author} {\bibfnamefont {Bao-Sen}\
  \bibnamefont {Shi}}, \ and\ \bibinfo {author} {\bibfnamefont {Guang-Can}\
  \bibnamefont {Guo}},\ }\bibfield  {title} {\enquote {\bibinfo {title}
  {Single-photon-level quantum image memory based on cold atomic ensembles},}\
  }\href {https://doi.org/10.1038/ncomms3527} {\bibfield  {journal} {\bibinfo
  {journal} {Nature Communications}\ }\textbf {\bibinfo {volume} {4}},\
  \bibinfo {pages} {2527} (\bibinfo {year} {2013})}\BibitemShut {NoStop}%
\bibitem [{\citenamefont {Nicolas}\ \emph {et~al.}(2014)\citenamefont
  {Nicolas}, \citenamefont {Veissier}, \citenamefont {Giner}, \citenamefont
  {Giacobino}, \citenamefont {Maxein},\ and\ \citenamefont
  {Laurat}}]{Nicolas2014}%
  \BibitemOpen
  \bibfield  {author} {\bibinfo {author} {\bibfnamefont {A.}~\bibnamefont
  {Nicolas}}, \bibinfo {author} {\bibfnamefont {L.}~\bibnamefont {Veissier}},
  \bibinfo {author} {\bibfnamefont {L.}~\bibnamefont {Giner}}, \bibinfo
  {author} {\bibfnamefont {E.}~\bibnamefont {Giacobino}}, \bibinfo {author}
  {\bibfnamefont {D.}~\bibnamefont {Maxein}}, \ and\ \bibinfo {author}
  {\bibfnamefont {J.}~\bibnamefont {Laurat}},\ }\bibfield  {title} {\enquote
  {\bibinfo {title} {A quantum memory for orbital angular momentum photonic
  qubits},}\ }\href {\doibase 10.1038/nphoton.2013.355} {\bibfield  {journal}
  {\bibinfo  {journal} {Nature Photonics}\ }\textbf {\bibinfo {volume} {8}},\
  \bibinfo {pages} {234--238} (\bibinfo {year} {2014})}\BibitemShut {NoStop}%
\bibitem [{\citenamefont {Ding}\ \emph
  {et~al.}(2015{\natexlab{a}})\citenamefont {Ding}, \citenamefont {Zhang},
  \citenamefont {Zhou}, \citenamefont {Shi}, \citenamefont {Xiang},
  \citenamefont {Wang}, \citenamefont {Jiang}, \citenamefont {Shi},\ and\
  \citenamefont {Guo}}]{Ding2015}%
  \BibitemOpen
  \bibfield  {author} {\bibinfo {author} {\bibfnamefont {Dong-Sheng}\
  \bibnamefont {Ding}}, \bibinfo {author} {\bibfnamefont {Wei}\ \bibnamefont
  {Zhang}}, \bibinfo {author} {\bibfnamefont {Zhi-Yuan}\ \bibnamefont {Zhou}},
  \bibinfo {author} {\bibfnamefont {Shuai}\ \bibnamefont {Shi}}, \bibinfo
  {author} {\bibfnamefont {Guo-Yong}\ \bibnamefont {Xiang}}, \bibinfo {author}
  {\bibfnamefont {Xi-Shi}\ \bibnamefont {Wang}}, \bibinfo {author}
  {\bibfnamefont {Yun-Kun}\ \bibnamefont {Jiang}}, \bibinfo {author}
  {\bibfnamefont {Bao-Sen}\ \bibnamefont {Shi}}, \ and\ \bibinfo {author}
  {\bibfnamefont {Guang-Can}\ \bibnamefont {Guo}},\ }\bibfield  {title}
  {\enquote {\bibinfo {title} {Quantum storage of orbital angular momentum
  entanglement in an atomic ensemble},}\ }\href {\doibase
  10.1103/PhysRevLett.114.050502} {\bibfield  {journal} {\bibinfo  {journal}
  {Phys. Rev. Lett.}\ }\textbf {\bibinfo {volume} {114}},\ \bibinfo {pages}
  {050502} (\bibinfo {year} {2015}{\natexlab{a}})}\BibitemShut {NoStop}%
\bibitem [{\citenamefont {Ding}\ \emph
  {et~al.}(2015{\natexlab{b}})\citenamefont {Ding}, \citenamefont {Zhang},
  \citenamefont {Zhou}, \citenamefont {Shi}, \citenamefont {Shi},\ and\
  \citenamefont {Guo}}]{Ding2015Nat}%
  \BibitemOpen
  \bibfield  {author} {\bibinfo {author} {\bibfnamefont {Dong-Sheng}\
  \bibnamefont {Ding}}, \bibinfo {author} {\bibfnamefont {Wei}\ \bibnamefont
  {Zhang}}, \bibinfo {author} {\bibfnamefont {Zhi-Yuan}\ \bibnamefont {Zhou}},
  \bibinfo {author} {\bibfnamefont {Shuai}\ \bibnamefont {Shi}}, \bibinfo
  {author} {\bibfnamefont {Bao-Sen}\ \bibnamefont {Shi}}, \ and\ \bibinfo
  {author} {\bibfnamefont {Guang-Can}\ \bibnamefont {Guo}},\ }\bibfield
  {title} {\enquote {\bibinfo {title} {Raman quantum memory of photonic
  polarized entanglement},}\ }\href {\doibase 10.1038/nphoton.2015.43}
  {\bibfield  {journal} {\bibinfo  {journal} {Nature Photonics}\ }\textbf
  {\bibinfo {volume} {9}},\ \bibinfo {pages} {332--338} (\bibinfo {year}
  {2015}{\natexlab{b}})}\BibitemShut {NoStop}%
\bibitem [{\citenamefont {Parigi}\ \emph {et~al.}(2015)\citenamefont {Parigi},
  \citenamefont {D'Ambrosio}, \citenamefont {Arnold}, \citenamefont {Marrucci},
  \citenamefont {Sciarrino},\ and\ \citenamefont {Laurat}}]{Parigi2015}%
  \BibitemOpen
  \bibfield  {author} {\bibinfo {author} {\bibfnamefont {Valentina}\
  \bibnamefont {Parigi}}, \bibinfo {author} {\bibfnamefont {Vincenzo}\
  \bibnamefont {D'Ambrosio}}, \bibinfo {author} {\bibfnamefont {Christophe}\
  \bibnamefont {Arnold}}, \bibinfo {author} {\bibfnamefont {Lorenzo}\
  \bibnamefont {Marrucci}}, \bibinfo {author} {\bibfnamefont {Fabio}\
  \bibnamefont {Sciarrino}}, \ and\ \bibinfo {author} {\bibfnamefont {Julien}\
  \bibnamefont {Laurat}},\ }\bibfield  {title} {\enquote {\bibinfo {title}
  {Storage and retrieval of vector beams of light in a
  multiple-degree-of-freedom quantum memory},}\ }\href
  {https://doi.org/10.1038/ncomms8706} {\bibfield  {journal} {\bibinfo
  {journal} {Nature Communications}\ }\textbf {\bibinfo {volume} {6}},\
  \bibinfo {pages} {7706} (\bibinfo {year} {2015})}\BibitemShut {NoStop}%
\bibitem [{\citenamefont {Saunders}\ \emph {et~al.}(2016)\citenamefont
  {Saunders}, \citenamefont {Munns}, \citenamefont {Champion}, \citenamefont
  {Qiu}, \citenamefont {Kaczmarek}, \citenamefont {Poem}, \citenamefont
  {Ledingham}, \citenamefont {Walmsley},\ and\ \citenamefont
  {Nunn}}]{Saunders2016}%
  \BibitemOpen
  \bibfield  {author} {\bibinfo {author} {\bibfnamefont {D.~J.}\ \bibnamefont
  {Saunders}}, \bibinfo {author} {\bibfnamefont {J.~H.~D.}\ \bibnamefont
  {Munns}}, \bibinfo {author} {\bibfnamefont {T.~F.~M.}\ \bibnamefont
  {Champion}}, \bibinfo {author} {\bibfnamefont {C.}~\bibnamefont {Qiu}},
  \bibinfo {author} {\bibfnamefont {K.~T.}\ \bibnamefont {Kaczmarek}}, \bibinfo
  {author} {\bibfnamefont {E.}~\bibnamefont {Poem}}, \bibinfo {author}
  {\bibfnamefont {P.~M.}\ \bibnamefont {Ledingham}}, \bibinfo {author}
  {\bibfnamefont {I.~A.}\ \bibnamefont {Walmsley}}, \ and\ \bibinfo {author}
  {\bibfnamefont {J.}~\bibnamefont {Nunn}},\ }\bibfield  {title} {\enquote
  {\bibinfo {title} {Cavity-enhanced room-temperature broadband raman
  memory},}\ }\href {\doibase 10.1103/PhysRevLett.116.090501} {\bibfield
  {journal} {\bibinfo  {journal} {Phys. Rev. Lett.}\ }\textbf {\bibinfo
  {volume} {116}},\ \bibinfo {pages} {090501} (\bibinfo {year}
  {2016})}\BibitemShut {NoStop}%
\bibitem [{\citenamefont {Katz}\ and\ \citenamefont
  {Firstenberg}(2018)}]{Katz2018}%
  \BibitemOpen
  \bibfield  {author} {\bibinfo {author} {\bibfnamefont {Or}~\bibnamefont
  {Katz}}\ and\ \bibinfo {author} {\bibfnamefont {Ofer}\ \bibnamefont
  {Firstenberg}},\ }\bibfield  {title} {\enquote {\bibinfo {title} {Light
  storage for one second in room-temperature alkali vapor},}\ }\href {\doibase
  10.1038/s41467-018-04458-4} {\bibfield  {journal} {\bibinfo  {journal}
  {Nature Communications}\ }\textbf {\bibinfo {volume} {9}},\ \bibinfo {pages}
  {2074} (\bibinfo {year} {2018})}\BibitemShut {NoStop}%
\bibitem [{\citenamefont {Hsiao}\ \emph {et~al.}(2018)\citenamefont {Hsiao},
  \citenamefont {Tsai}, \citenamefont {Chen}, \citenamefont {Lin},
  \citenamefont {Hung}, \citenamefont {Lee}, \citenamefont {Chen},
  \citenamefont {Chen}, \citenamefont {Yu},\ and\ \citenamefont
  {Chen}}]{Hsiao2018}%
  \BibitemOpen
  \bibfield  {author} {\bibinfo {author} {\bibfnamefont {Ya-Fen}\ \bibnamefont
  {Hsiao}}, \bibinfo {author} {\bibfnamefont {Pin-Ju}\ \bibnamefont {Tsai}},
  \bibinfo {author} {\bibfnamefont {Hung-Shiue}\ \bibnamefont {Chen}}, \bibinfo
  {author} {\bibfnamefont {Sheng-Xiang}\ \bibnamefont {Lin}}, \bibinfo {author}
  {\bibfnamefont {Chih-Chiao}\ \bibnamefont {Hung}}, \bibinfo {author}
  {\bibfnamefont {Chih-Hsi}\ \bibnamefont {Lee}}, \bibinfo {author}
  {\bibfnamefont {Yi-Hsin}\ \bibnamefont {Chen}}, \bibinfo {author}
  {\bibfnamefont {Yong-Fan}\ \bibnamefont {Chen}}, \bibinfo {author}
  {\bibfnamefont {Ite~A.}\ \bibnamefont {Yu}}, \ and\ \bibinfo {author}
  {\bibfnamefont {Ying-Cheng}\ \bibnamefont {Chen}},\ }\bibfield  {title}
  {\enquote {\bibinfo {title} {Highly efficient coherent optical memory based
  on electromagnetically induced transparency},}\ }\href {\doibase
  10.1103/PhysRevLett.120.183602} {\bibfield  {journal} {\bibinfo  {journal}
  {Phys. Rev. Lett.}\ }\textbf {\bibinfo {volume} {120}},\ \bibinfo {pages}
  {183602} (\bibinfo {year} {2018})}\BibitemShut {NoStop}%
\bibitem [{\citenamefont {Jiang}\ \emph {et~al.}(2019)\citenamefont {Jiang},
  \citenamefont {Pu}, \citenamefont {Chang}, \citenamefont {Li}, \citenamefont
  {Zhang},\ and\ \citenamefont {Duan}}]{Jiang2019}%
  \BibitemOpen
  \bibfield  {author} {\bibinfo {author} {\bibfnamefont {N.}~\bibnamefont
  {Jiang}}, \bibinfo {author} {\bibfnamefont {Y.~F.}\ \bibnamefont {Pu}},
  \bibinfo {author} {\bibfnamefont {W.}~\bibnamefont {Chang}}, \bibinfo
  {author} {\bibfnamefont {C.}~\bibnamefont {Li}}, \bibinfo {author}
  {\bibfnamefont {S.}~\bibnamefont {Zhang}}, \ and\ \bibinfo {author}
  {\bibfnamefont {L.~M.}\ \bibnamefont {Duan}},\ }\bibfield  {title} {\enquote
  {\bibinfo {title} {Experimental realization of 105-qubit random access
  quantum memory},}\ }\href {\doibase 10.1038/s41534-019-0144-0} {\bibfield
  {journal} {\bibinfo  {journal} {npj Quantum Information}\ }\textbf {\bibinfo
  {volume} {5}},\ \bibinfo {pages} {28} (\bibinfo {year} {2019})}\BibitemShut
  {NoStop}%
\bibitem [{\citenamefont {Wang}\ \emph {et~al.}(2019)\citenamefont {Wang},
  \citenamefont {Li}, \citenamefont {Zhang}, \citenamefont {Su}, \citenamefont
  {Zhou}, \citenamefont {Liao}, \citenamefont {Du}, \citenamefont {Yan},\ and\
  \citenamefont {Zhu}}]{Wang2019}%
  \BibitemOpen
  \bibfield  {author} {\bibinfo {author} {\bibfnamefont {Yunfei}\ \bibnamefont
  {Wang}}, \bibinfo {author} {\bibfnamefont {Jianfeng}\ \bibnamefont {Li}},
  \bibinfo {author} {\bibfnamefont {Shanchao}\ \bibnamefont {Zhang}}, \bibinfo
  {author} {\bibfnamefont {Keyu}\ \bibnamefont {Su}}, \bibinfo {author}
  {\bibfnamefont {Yiru}\ \bibnamefont {Zhou}}, \bibinfo {author} {\bibfnamefont
  {Kaiyu}\ \bibnamefont {Liao}}, \bibinfo {author} {\bibfnamefont {Shengwang}\
  \bibnamefont {Du}}, \bibinfo {author} {\bibfnamefont {Hui}\ \bibnamefont
  {Yan}}, \ and\ \bibinfo {author} {\bibfnamefont {Shi-Liang}\ \bibnamefont
  {Zhu}},\ }\bibfield  {title} {\enquote {\bibinfo {title} {Efficient quantum
  memory for single-photon polarization qubits},}\ }\href {\doibase
  10.1038/s41566-019-0368-8} {\bibfield  {journal} {\bibinfo  {journal} {Nature
  Photonics}\ }\textbf {\bibinfo {volume} {13}},\ \bibinfo {pages} {346--351}
  (\bibinfo {year} {2019})}\BibitemShut {NoStop}%
\bibitem [{\citenamefont {Li}\ \emph {et~al.}(2020)\citenamefont {Li},
  \citenamefont {Jiang}, \citenamefont {Wu}, \citenamefont {Chang},
  \citenamefont {Pu}, \citenamefont {Zhang},\ and\ \citenamefont
  {Duan}}]{Li2020}%
  \BibitemOpen
  \bibfield  {author} {\bibinfo {author} {\bibfnamefont {C.}~\bibnamefont
  {Li}}, \bibinfo {author} {\bibfnamefont {N.}~\bibnamefont {Jiang}}, \bibinfo
  {author} {\bibfnamefont {Y.-K.}\ \bibnamefont {Wu}}, \bibinfo {author}
  {\bibfnamefont {W.}~\bibnamefont {Chang}}, \bibinfo {author} {\bibfnamefont
  {Y.-F.}\ \bibnamefont {Pu}}, \bibinfo {author} {\bibfnamefont
  {S.}~\bibnamefont {Zhang}}, \ and\ \bibinfo {author} {\bibfnamefont {L.-M.}\
  \bibnamefont {Duan}},\ }\bibfield  {title} {\enquote {\bibinfo {title}
  {Quantum communication between multiplexed atomic quantum memories},}\ }\href
  {\doibase 10.1103/PhysRevLett.124.240504} {\bibfield  {journal} {\bibinfo
  {journal} {Phys. Rev. Lett.}\ }\textbf {\bibinfo {volume} {124}},\ \bibinfo
  {pages} {240504} (\bibinfo {year} {2020})}\BibitemShut {NoStop}%
\bibitem [{\citenamefont {Dideriksen}\ \emph {et~al.}(2021)\citenamefont
  {Dideriksen}, \citenamefont {Schmieg}, \citenamefont {Zugenmaier},\ and\
  \citenamefont {Polzik}}]{Dideriksen2021}%
  \BibitemOpen
  \bibfield  {author} {\bibinfo {author} {\bibfnamefont {Karsten~B.}\
  \bibnamefont {Dideriksen}}, \bibinfo {author} {\bibfnamefont {Rebecca}\
  \bibnamefont {Schmieg}}, \bibinfo {author} {\bibfnamefont {Michael}\
  \bibnamefont {Zugenmaier}}, \ and\ \bibinfo {author} {\bibfnamefont
  {Eugene~S.}\ \bibnamefont {Polzik}},\ }\bibfield  {title} {\enquote {\bibinfo
  {title} {Room-temperature single-photon source with near-millisecond built-in
  memory},}\ }\href {\doibase 10.1038/s41467-021-24033-8} {\bibfield  {journal}
  {\bibinfo  {journal} {Nature Communications}\ }\textbf {\bibinfo {volume}
  {12}},\ \bibinfo {pages} {3699} (\bibinfo {year} {2021})}\BibitemShut
  {NoStop}%
\bibitem [{\citenamefont {Wang}\ \emph
  {et~al.}(2022{\natexlab{b}})\citenamefont {Wang}, \citenamefont {Craddock},
  \citenamefont {Sekelsky}, \citenamefont {Flament},\ and\ \citenamefont
  {Namazi}}]{Wang2022}%
  \BibitemOpen
  \bibfield  {author} {\bibinfo {author} {\bibfnamefont {Yang}\ \bibnamefont
  {Wang}}, \bibinfo {author} {\bibfnamefont {Alexander~N.}\ \bibnamefont
  {Craddock}}, \bibinfo {author} {\bibfnamefont {Rourke}\ \bibnamefont
  {Sekelsky}}, \bibinfo {author} {\bibfnamefont {Mael}\ \bibnamefont
  {Flament}}, \ and\ \bibinfo {author} {\bibfnamefont {Mehdi}\ \bibnamefont
  {Namazi}},\ }\bibfield  {title} {\enquote {\bibinfo {title} {Field-deployable
  quantum memory for quantum networking},}\ }\href {\doibase
  10.1103/PhysRevApplied.18.044058} {\bibfield  {journal} {\bibinfo  {journal}
  {Phys. Rev. Appl.}\ }\textbf {\bibinfo {volume} {18}},\ \bibinfo {pages}
  {044058} (\bibinfo {year} {2022}{\natexlab{b}})}\BibitemShut {NoStop}%
\bibitem [{\citenamefont {Me{\ss}ner}\ \emph {et~al.}(2023)\citenamefont
  {Me{\ss}ner}, \citenamefont {Robertson}, \citenamefont {Esguerra},
  \citenamefont {L\"{u}dge},\ and\ \citenamefont {Wolters}}]{Messner2023}%
  \BibitemOpen
  \bibfield  {author} {\bibinfo {author} {\bibfnamefont {Leon}\ \bibnamefont
  {Me{\ss}ner}}, \bibinfo {author} {\bibfnamefont {Elizabeth}\ \bibnamefont
  {Robertson}}, \bibinfo {author} {\bibfnamefont {Luisa}\ \bibnamefont
  {Esguerra}}, \bibinfo {author} {\bibfnamefont {Kathy}\ \bibnamefont
  {L\"{u}dge}}, \ and\ \bibinfo {author} {\bibfnamefont {Janik}\ \bibnamefont
  {Wolters}},\ }\bibfield  {title} {\enquote {\bibinfo {title} {Multiplexed
  random-access optical memory in warm cesium vapor},}\ }\href {\doibase
  10.1364/OE.483642} {\bibfield  {journal} {\bibinfo  {journal} {Opt. Express}\
  }\textbf {\bibinfo {volume} {31}},\ \bibinfo {pages} {10150--10158} (\bibinfo
  {year} {2023})}\BibitemShut {NoStop}%
\bibitem [{\citenamefont {Buser}\ \emph {et~al.}(2022)\citenamefont {Buser},
  \citenamefont {Mottola}, \citenamefont {Cotting}, \citenamefont {Wolters},\
  and\ \citenamefont {Treutlein}}]{Buser2022}%
  \BibitemOpen
  \bibfield  {author} {\bibinfo {author} {\bibfnamefont {Gianni}\ \bibnamefont
  {Buser}}, \bibinfo {author} {\bibfnamefont {Roberto}\ \bibnamefont
  {Mottola}}, \bibinfo {author} {\bibfnamefont {Bj\"orn}\ \bibnamefont
  {Cotting}}, \bibinfo {author} {\bibfnamefont {Janik}\ \bibnamefont
  {Wolters}}, \ and\ \bibinfo {author} {\bibfnamefont {Philipp}\ \bibnamefont
  {Treutlein}},\ }\bibfield  {title} {\enquote {\bibinfo {title} {Single-photon
  storage in a ground-state vapor cell quantum memory},}\ }\href {\doibase
  10.1103/PRXQuantum.3.020349} {\bibfield  {journal} {\bibinfo  {journal} {PRX
  Quantum}\ }\textbf {\bibinfo {volume} {3}},\ \bibinfo {pages} {020349}
  (\bibinfo {year} {2022})}\BibitemShut {NoStop}%
\bibitem [{\citenamefont {Hattermann}\ \emph {et~al.}(2017)\citenamefont
  {Hattermann}, \citenamefont {Bothner}, \citenamefont {Ley}, \citenamefont
  {Ferdinand}, \citenamefont {Wiedmaier}, \citenamefont {S{\'a}rk{\'a}ny},
  \citenamefont {Kleiner}, \citenamefont {Koelle},\ and\ \citenamefont
  {Fort{\'a}gh}}]{Hattermann2017}%
  \BibitemOpen
  \bibfield  {author} {\bibinfo {author} {\bibfnamefont {H.}~\bibnamefont
  {Hattermann}}, \bibinfo {author} {\bibfnamefont {D.}~\bibnamefont {Bothner}},
  \bibinfo {author} {\bibfnamefont {L.~Y.}\ \bibnamefont {Ley}}, \bibinfo
  {author} {\bibfnamefont {B.}~\bibnamefont {Ferdinand}}, \bibinfo {author}
  {\bibfnamefont {D.}~\bibnamefont {Wiedmaier}}, \bibinfo {author}
  {\bibfnamefont {L.}~\bibnamefont {S{\'a}rk{\'a}ny}}, \bibinfo {author}
  {\bibfnamefont {R.}~\bibnamefont {Kleiner}}, \bibinfo {author} {\bibfnamefont
  {D.}~\bibnamefont {Koelle}}, \ and\ \bibinfo {author} {\bibfnamefont
  {J.}~\bibnamefont {Fort{\'a}gh}},\ }\bibfield  {title} {\enquote {\bibinfo
  {title} {Coupling ultracold atoms to a superconducting coplanar waveguide
  resonator},}\ }\href {\doibase 10.1038/s41467-017-02439-7} {\bibfield
  {journal} {\bibinfo  {journal} {Nature Communications}\ }\textbf {\bibinfo
  {volume} {8}},\ \bibinfo {pages} {2254} (\bibinfo {year} {2017})}\BibitemShut
  {NoStop}%
\bibitem [{\citenamefont {Rabl}\ \emph {et~al.}(2006)\citenamefont {Rabl},
  \citenamefont {DeMille}, \citenamefont {Doyle}, \citenamefont {Lukin},
  \citenamefont {Schoelkopf},\ and\ \citenamefont {Zoller}}]{Rabl2006}%
  \BibitemOpen
  \bibfield  {author} {\bibinfo {author} {\bibfnamefont {P.}~\bibnamefont
  {Rabl}}, \bibinfo {author} {\bibfnamefont {D.}~\bibnamefont {DeMille}},
  \bibinfo {author} {\bibfnamefont {J.~M.}\ \bibnamefont {Doyle}}, \bibinfo
  {author} {\bibfnamefont {M.~D.}\ \bibnamefont {Lukin}}, \bibinfo {author}
  {\bibfnamefont {R.~J.}\ \bibnamefont {Schoelkopf}}, \ and\ \bibinfo {author}
  {\bibfnamefont {P.}~\bibnamefont {Zoller}},\ }\bibfield  {title} {\enquote
  {\bibinfo {title} {Hybrid quantum processors: Molecular ensembles as quantum
  memory for solid state circuits},}\ }\href {\doibase
  10.1103/PhysRevLett.97.033003} {\bibfield  {journal} {\bibinfo  {journal}
  {Phys. Rev. Lett.}\ }\textbf {\bibinfo {volume} {97}},\ \bibinfo {pages}
  {033003} (\bibinfo {year} {2006})}\BibitemShut {NoStop}%
\bibitem [{\citenamefont {Petrosyan}\ \emph {et~al.}(2009)\citenamefont
  {Petrosyan}, \citenamefont {Bensky}, \citenamefont {Kurizki}, \citenamefont
  {Mazets}, \citenamefont {Majer},\ and\ \citenamefont
  {Schmiedmayer}}]{Reversible2009}%
  \BibitemOpen
  \bibfield  {author} {\bibinfo {author} {\bibfnamefont {David}\ \bibnamefont
  {Petrosyan}}, \bibinfo {author} {\bibfnamefont {Guy}\ \bibnamefont {Bensky}},
  \bibinfo {author} {\bibfnamefont {Gershon}\ \bibnamefont {Kurizki}}, \bibinfo
  {author} {\bibfnamefont {Igor}\ \bibnamefont {Mazets}}, \bibinfo {author}
  {\bibfnamefont {Johannes}\ \bibnamefont {Majer}}, \ and\ \bibinfo {author}
  {\bibfnamefont {J\"org}\ \bibnamefont {Schmiedmayer}},\ }\bibfield  {title}
  {\enquote {\bibinfo {title} {Reversible state transfer between
  superconducting qubits and atomic ensembles},}\ }\href {\doibase
  10.1103/PhysRevA.79.040304} {\bibfield  {journal} {\bibinfo  {journal} {Phys.
  Rev. A}\ }\textbf {\bibinfo {volume} {79}},\ \bibinfo {pages} {040304}
  (\bibinfo {year} {2009})}\BibitemShut {NoStop}%
\bibitem [{\citenamefont {Tu}\ \emph {et~al.}(2022)\citenamefont {Tu},
  \citenamefont {Liao}, \citenamefont {Zhang}, \citenamefont {Liu},
  \citenamefont {Zheng}, \citenamefont {Yang}, \citenamefont {Zhang},
  \citenamefont {Yan},\ and\ \citenamefont {Zhu}}]{Tu2022}%
  \BibitemOpen
  \bibfield  {author} {\bibinfo {author} {\bibfnamefont {Hai-Tao}\ \bibnamefont
  {Tu}}, \bibinfo {author} {\bibfnamefont {Kai-Yu}\ \bibnamefont {Liao}},
  \bibinfo {author} {\bibfnamefont {Zuan-Xian}\ \bibnamefont {Zhang}}, \bibinfo
  {author} {\bibfnamefont {Xiao-Hong}\ \bibnamefont {Liu}}, \bibinfo {author}
  {\bibfnamefont {Shun-Yuan}\ \bibnamefont {Zheng}}, \bibinfo {author}
  {\bibfnamefont {Shu-Zhe}\ \bibnamefont {Yang}}, \bibinfo {author}
  {\bibfnamefont {Xin-Ding}\ \bibnamefont {Zhang}}, \bibinfo {author}
  {\bibfnamefont {Hui}\ \bibnamefont {Yan}}, \ and\ \bibinfo {author}
  {\bibfnamefont {Shi-Liang}\ \bibnamefont {Zhu}},\ }\bibfield  {title}
  {\enquote {\bibinfo {title} {High-efficiency coherent microwave-to-optics
  conversion via off-resonant scattering},}\ }\href {\doibase
  10.1038/s41566-022-00959-3} {\bibfield  {journal} {\bibinfo  {journal}
  {Nature Photonics}\ }\textbf {\bibinfo {volume} {16}},\ \bibinfo {pages}
  {291--296} (\bibinfo {year} {2022})}\BibitemShut {NoStop}%
\bibitem [{Note1()}]{Note1}%
  \BibitemOpen
  \bibinfo {note} {We acknowledge that the hyperfine states of Rb atom used in
  these two works are different. Here we aim to give an estimate of the
  possible implementation of the QMCs.}\BibitemShut {Stop}%
\bibitem [{\citenamefont {Thiele}\ \emph {et~al.}(2015)\citenamefont {Thiele},
  \citenamefont {Deiglmayr}, \citenamefont {Stammeier}, \citenamefont {Agner},
  \citenamefont {Schmutz}, \citenamefont {Merkt},\ and\ \citenamefont
  {Wallraff}}]{Thiele2015}%
  \BibitemOpen
  \bibfield  {author} {\bibinfo {author} {\bibfnamefont {T.}~\bibnamefont
  {Thiele}}, \bibinfo {author} {\bibfnamefont {J.}~\bibnamefont {Deiglmayr}},
  \bibinfo {author} {\bibfnamefont {M.}~\bibnamefont {Stammeier}}, \bibinfo
  {author} {\bibfnamefont {J.-A.}\ \bibnamefont {Agner}}, \bibinfo {author}
  {\bibfnamefont {H.}~\bibnamefont {Schmutz}}, \bibinfo {author} {\bibfnamefont
  {F.}~\bibnamefont {Merkt}}, \ and\ \bibinfo {author} {\bibfnamefont
  {A.}~\bibnamefont {Wallraff}},\ }\bibfield  {title} {\enquote {\bibinfo
  {title} {Imaging electric fields in the vicinity of cryogenic surfaces using
  rydberg atoms},}\ }\href {\doibase 10.1103/PhysRevA.92.063425} {\bibfield
  {journal} {\bibinfo  {journal} {Phys. Rev. A}\ }\textbf {\bibinfo {volume}
  {92}},\ \bibinfo {pages} {063425} (\bibinfo {year} {2015})}\BibitemShut
  {NoStop}%
\bibitem [{\citenamefont {Garcia}\ \emph {et~al.}(2019)\citenamefont {Garcia},
  \citenamefont {Stammeier}, \citenamefont {Deiglmayr}, \citenamefont {Merkt},\
  and\ \citenamefont {Wallraff}}]{Garcia2019}%
  \BibitemOpen
  \bibfield  {author} {\bibinfo {author} {\bibfnamefont {S.}~\bibnamefont
  {Garcia}}, \bibinfo {author} {\bibfnamefont {M.}~\bibnamefont {Stammeier}},
  \bibinfo {author} {\bibfnamefont {J.}~\bibnamefont {Deiglmayr}}, \bibinfo
  {author} {\bibfnamefont {F.}~\bibnamefont {Merkt}}, \ and\ \bibinfo {author}
  {\bibfnamefont {A.}~\bibnamefont {Wallraff}},\ }\bibfield  {title} {\enquote
  {\bibinfo {title} {Single-shot nondestructive detection of rydberg-atom
  ensembles by transmission measurement of a microwave cavity},}\ }\href
  {\doibase 10.1103/PhysRevLett.123.193201} {\bibfield  {journal} {\bibinfo
  {journal} {Phys. Rev. Lett.}\ }\textbf {\bibinfo {volume} {123}},\ \bibinfo
  {pages} {193201} (\bibinfo {year} {2019})}\BibitemShut {NoStop}%
\bibitem [{\citenamefont {Morgan}\ and\ \citenamefont
  {Hogan}(2020)}]{Morgan2020}%
  \BibitemOpen
  \bibfield  {author} {\bibinfo {author} {\bibfnamefont {A.~A.}\ \bibnamefont
  {Morgan}}\ and\ \bibinfo {author} {\bibfnamefont {S.~D.}\ \bibnamefont
  {Hogan}},\ }\bibfield  {title} {\enquote {\bibinfo {title} {Coupling rydberg
  atoms to microwave fields in a superconducting coplanar waveguide
  resonator},}\ }\href {\doibase 10.1103/PhysRevLett.124.193604} {\bibfield
  {journal} {\bibinfo  {journal} {Phys. Rev. Lett.}\ }\textbf {\bibinfo
  {volume} {124}},\ \bibinfo {pages} {193604} (\bibinfo {year}
  {2020})}\BibitemShut {NoStop}%
\bibitem [{\citenamefont {Wang}\ \emph {et~al.}(2015)\citenamefont {Wang},
  \citenamefont {Zhang}, \citenamefont {Corcovilos}, \citenamefont {Kumar},\
  and\ \citenamefont {Weiss}}]{YWang2015}%
  \BibitemOpen
  \bibfield  {author} {\bibinfo {author} {\bibfnamefont {Yang}\ \bibnamefont
  {Wang}}, \bibinfo {author} {\bibfnamefont {Xianli}\ \bibnamefont {Zhang}},
  \bibinfo {author} {\bibfnamefont {Theodore~A.}\ \bibnamefont {Corcovilos}},
  \bibinfo {author} {\bibfnamefont {Aishwarya}\ \bibnamefont {Kumar}}, \ and\
  \bibinfo {author} {\bibfnamefont {David~S.}\ \bibnamefont {Weiss}},\
  }\bibfield  {title} {\enquote {\bibinfo {title} {Coherent addressing of
  individual neutral atoms in a 3d optical lattice},}\ }\href {\doibase
  10.1103/PhysRevLett.115.043003} {\bibfield  {journal} {\bibinfo  {journal}
  {Phys. Rev. Lett.}\ }\textbf {\bibinfo {volume} {115}},\ \bibinfo {pages}
  {043003} (\bibinfo {year} {2015})}\BibitemShut {NoStop}%
\bibitem [{\citenamefont {Wang}\ \emph {et~al.}(2016)\citenamefont {Wang},
  \citenamefont {Kumar}, \citenamefont {Wu},\ and\ \citenamefont
  {Weiss}}]{YWang2016}%
  \BibitemOpen
  \bibfield  {author} {\bibinfo {author} {\bibfnamefont {Yang}\ \bibnamefont
  {Wang}}, \bibinfo {author} {\bibfnamefont {Aishwarya}\ \bibnamefont {Kumar}},
  \bibinfo {author} {\bibfnamefont {Tsung-Yao}\ \bibnamefont {Wu}}, \ and\
  \bibinfo {author} {\bibfnamefont {David~S.}\ \bibnamefont {Weiss}},\
  }\bibfield  {title} {\enquote {\bibinfo {title} {{Single-qubit gates based on
  targeted phase shifts in a 3D neutral atom array}},}\ }\href {\doibase
  10.1126/science.aaf2581} {\bibfield  {journal} {\bibinfo  {journal}
  {Science}\ }\textbf {\bibinfo {volume} {352}},\ \bibinfo {pages} {1562--1565}
  (\bibinfo {year} {2016})}\BibitemShut {NoStop}%
\bibitem [{\citenamefont {Xia}\ \emph {et~al.}(2015)\citenamefont {Xia},
  \citenamefont {Lichtman}, \citenamefont {Maller}, \citenamefont {Carr},
  \citenamefont {Piotrowicz}, \citenamefont {Isenhower},\ and\ \citenamefont
  {Saffman}}]{Xia2015}%
  \BibitemOpen
  \bibfield  {author} {\bibinfo {author} {\bibfnamefont {T.}~\bibnamefont
  {Xia}}, \bibinfo {author} {\bibfnamefont {M.}~\bibnamefont {Lichtman}},
  \bibinfo {author} {\bibfnamefont {K.}~\bibnamefont {Maller}}, \bibinfo
  {author} {\bibfnamefont {A.~W.}\ \bibnamefont {Carr}}, \bibinfo {author}
  {\bibfnamefont {M.~J.}\ \bibnamefont {Piotrowicz}}, \bibinfo {author}
  {\bibfnamefont {L.}~\bibnamefont {Isenhower}}, \ and\ \bibinfo {author}
  {\bibfnamefont {M.}~\bibnamefont {Saffman}},\ }\bibfield  {title} {\enquote
  {\bibinfo {title} {Randomized benchmarking of single-qubit gates in a 2d
  array of neutral-atom qubits},}\ }\href {\doibase
  10.1103/PhysRevLett.114.100503} {\bibfield  {journal} {\bibinfo  {journal}
  {Phys. Rev. Lett.}\ }\textbf {\bibinfo {volume} {114}},\ \bibinfo {pages}
  {100503} (\bibinfo {year} {2015})}\BibitemShut {NoStop}%
\bibitem [{\citenamefont {Maller}\ \emph {et~al.}(2015)\citenamefont {Maller},
  \citenamefont {Lichtman}, \citenamefont {Xia}, \citenamefont {Sun},
  \citenamefont {Piotrowicz}, \citenamefont {Carr}, \citenamefont {Isenhower},\
  and\ \citenamefont {Saffman}}]{Maller2015}%
  \BibitemOpen
  \bibfield  {author} {\bibinfo {author} {\bibfnamefont {K.~M.}\ \bibnamefont
  {Maller}}, \bibinfo {author} {\bibfnamefont {M.~T.}\ \bibnamefont
  {Lichtman}}, \bibinfo {author} {\bibfnamefont {T.}~\bibnamefont {Xia}},
  \bibinfo {author} {\bibfnamefont {Y.}~\bibnamefont {Sun}}, \bibinfo {author}
  {\bibfnamefont {M.~J.}\ \bibnamefont {Piotrowicz}}, \bibinfo {author}
  {\bibfnamefont {A.~W.}\ \bibnamefont {Carr}}, \bibinfo {author}
  {\bibfnamefont {L.}~\bibnamefont {Isenhower}}, \ and\ \bibinfo {author}
  {\bibfnamefont {M.}~\bibnamefont {Saffman}},\ }\bibfield  {title} {\enquote
  {\bibinfo {title} {Rydberg-blockade controlled-not gate and entanglement in a
  two-dimensional array of neutral-atom qubits},}\ }\href {\doibase
  10.1103/PhysRevA.92.022336} {\bibfield  {journal} {\bibinfo  {journal} {Phys.
  Rev. A}\ }\textbf {\bibinfo {volume} {92}},\ \bibinfo {pages} {022336}
  (\bibinfo {year} {2015})}\BibitemShut {NoStop}%
\bibitem [{\citenamefont {Endres}\ \emph {et~al.}(2016)\citenamefont {Endres},
  \citenamefont {Bernien}, \citenamefont {Keesling}, \citenamefont {Levine},
  \citenamefont {Anschuetz}, \citenamefont {Krajenbrink}, \citenamefont
  {Senko}, \citenamefont {Vuletic}, \citenamefont {Greiner},\ and\
  \citenamefont {Lukin}}]{Endres2016}%
  \BibitemOpen
  \bibfield  {author} {\bibinfo {author} {\bibfnamefont {Manuel}\ \bibnamefont
  {Endres}}, \bibinfo {author} {\bibfnamefont {Hannes}\ \bibnamefont
  {Bernien}}, \bibinfo {author} {\bibfnamefont {Alexander}\ \bibnamefont
  {Keesling}}, \bibinfo {author} {\bibfnamefont {Harry}\ \bibnamefont
  {Levine}}, \bibinfo {author} {\bibfnamefont {Eric~R.}\ \bibnamefont
  {Anschuetz}}, \bibinfo {author} {\bibfnamefont {Alexandre}\ \bibnamefont
  {Krajenbrink}}, \bibinfo {author} {\bibfnamefont {Crystal}\ \bibnamefont
  {Senko}}, \bibinfo {author} {\bibfnamefont {Vladan}\ \bibnamefont {Vuletic}},
  \bibinfo {author} {\bibfnamefont {Markus}\ \bibnamefont {Greiner}}, \ and\
  \bibinfo {author} {\bibfnamefont {Mikhail~D.}\ \bibnamefont {Lukin}},\
  }\bibfield  {title} {\enquote {\bibinfo {title} {Atom-by-atom assembly of
  defect-free one-dimensional cold atom arrays},}\ }\href {\doibase
  10.1126/science.aah3752} {\bibfield  {journal} {\bibinfo  {journal}
  {Science}\ }\textbf {\bibinfo {volume} {354}},\ \bibinfo {pages} {1024--1027}
  (\bibinfo {year} {2016})}\BibitemShut {NoStop}%
\bibitem [{\citenamefont {Bernien}\ \emph {et~al.}(2017)\citenamefont
  {Bernien}, \citenamefont {Schwartz}, \citenamefont {Keesling}, \citenamefont
  {Levine}, \citenamefont {Omran}, \citenamefont {Pichler}, \citenamefont
  {Choi}, \citenamefont {Zibrov}, \citenamefont {Endres}, \citenamefont
  {Greiner}, \citenamefont {Vuleti{\'c}},\ and\ \citenamefont
  {Lukin}}]{Bernien2017}%
  \BibitemOpen
  \bibfield  {author} {\bibinfo {author} {\bibfnamefont {Hannes}\ \bibnamefont
  {Bernien}}, \bibinfo {author} {\bibfnamefont {Sylvain}\ \bibnamefont
  {Schwartz}}, \bibinfo {author} {\bibfnamefont {Alexander}\ \bibnamefont
  {Keesling}}, \bibinfo {author} {\bibfnamefont {Harry}\ \bibnamefont
  {Levine}}, \bibinfo {author} {\bibfnamefont {Ahmed}\ \bibnamefont {Omran}},
  \bibinfo {author} {\bibfnamefont {Hannes}\ \bibnamefont {Pichler}}, \bibinfo
  {author} {\bibfnamefont {Soonwon}\ \bibnamefont {Choi}}, \bibinfo {author}
  {\bibfnamefont {Alexander~S.}\ \bibnamefont {Zibrov}}, \bibinfo {author}
  {\bibfnamefont {Manuel}\ \bibnamefont {Endres}}, \bibinfo {author}
  {\bibfnamefont {Markus}\ \bibnamefont {Greiner}}, \bibinfo {author}
  {\bibfnamefont {Vladan}\ \bibnamefont {Vuleti{\'c}}}, \ and\ \bibinfo
  {author} {\bibfnamefont {Mikhail~D.}\ \bibnamefont {Lukin}},\ }\bibfield
  {title} {\enquote {\bibinfo {title} {Probing many-body dynamics on a 51-atom
  quantum simulator},}\ }\href {\doibase 10.1038/nature24622} {\bibfield
  {journal} {\bibinfo  {journal} {Nature}\ }\textbf {\bibinfo {volume} {551}},\
  \bibinfo {pages} {579--584} (\bibinfo {year} {2017})}\BibitemShut {NoStop}%
\bibitem [{\citenamefont {McAuslan}\ \emph {et~al.}(2012)\citenamefont
  {McAuslan}, \citenamefont {Bartholomew}, \citenamefont {Sellars},\ and\
  \citenamefont {Longdell}}]{McAuslan2012}%
  \BibitemOpen
  \bibfield  {author} {\bibinfo {author} {\bibfnamefont {D.~L.}\ \bibnamefont
  {McAuslan}}, \bibinfo {author} {\bibfnamefont {J.~G.}\ \bibnamefont
  {Bartholomew}}, \bibinfo {author} {\bibfnamefont {M.~J.}\ \bibnamefont
  {Sellars}}, \ and\ \bibinfo {author} {\bibfnamefont {J.~J.}\ \bibnamefont
  {Longdell}},\ }\bibfield  {title} {\enquote {\bibinfo {title} {Reducing
  decoherence in optical and spin transitions in rare-earth-metal-ion--doped
  materials},}\ }\href {\doibase 10.1103/PhysRevA.85.032339} {\bibfield
  {journal} {\bibinfo  {journal} {Phys. Rev. A}\ }\textbf {\bibinfo {volume}
  {85}},\ \bibinfo {pages} {032339} (\bibinfo {year} {2012})}\BibitemShut
  {NoStop}%
\bibitem [{\citenamefont {Saglamyurek}\ \emph {et~al.}(2011)\citenamefont
  {Saglamyurek}, \citenamefont {Sinclair}, \citenamefont {Jin}, \citenamefont
  {Slater}, \citenamefont {Oblak}, \citenamefont {Bussi{\`e}res}, \citenamefont
  {George}, \citenamefont {Ricken}, \citenamefont {Sohler},\ and\ \citenamefont
  {Tittel}}]{Saglamyurek2011}%
  \BibitemOpen
  \bibfield  {author} {\bibinfo {author} {\bibfnamefont {Erhan}\ \bibnamefont
  {Saglamyurek}}, \bibinfo {author} {\bibfnamefont {Neil}\ \bibnamefont
  {Sinclair}}, \bibinfo {author} {\bibfnamefont {Jeongwan}\ \bibnamefont
  {Jin}}, \bibinfo {author} {\bibfnamefont {Joshua~A.}\ \bibnamefont {Slater}},
  \bibinfo {author} {\bibfnamefont {Daniel}\ \bibnamefont {Oblak}}, \bibinfo
  {author} {\bibfnamefont {F{\'e}lix}\ \bibnamefont {Bussi{\`e}res}}, \bibinfo
  {author} {\bibfnamefont {Mathew}\ \bibnamefont {George}}, \bibinfo {author}
  {\bibfnamefont {Raimund}\ \bibnamefont {Ricken}}, \bibinfo {author}
  {\bibfnamefont {Wolfgang}\ \bibnamefont {Sohler}}, \ and\ \bibinfo {author}
  {\bibfnamefont {Wolfgang}\ \bibnamefont {Tittel}},\ }\bibfield  {title}
  {\enquote {\bibinfo {title} {Broadband waveguide quantum memory for entangled
  photons},}\ }\href {\doibase 10.1038/nature09719} {\bibfield  {journal}
  {\bibinfo  {journal} {Nature}\ }\textbf {\bibinfo {volume} {469}},\ \bibinfo
  {pages} {512--515} (\bibinfo {year} {2011})}\BibitemShut {NoStop}%
\bibitem [{\citenamefont {Clausen}\ \emph {et~al.}(2011)\citenamefont
  {Clausen}, \citenamefont {Usmani}, \citenamefont {Bussi{\`e}res},
  \citenamefont {Sangouard}, \citenamefont {Afzelius}, \citenamefont
  {de~Riedmatten},\ and\ \citenamefont {Gisin}}]{Clausen2011}%
  \BibitemOpen
  \bibfield  {author} {\bibinfo {author} {\bibfnamefont {Christoph}\
  \bibnamefont {Clausen}}, \bibinfo {author} {\bibfnamefont {Imam}\
  \bibnamefont {Usmani}}, \bibinfo {author} {\bibfnamefont {F{\'e}lix}\
  \bibnamefont {Bussi{\`e}res}}, \bibinfo {author} {\bibfnamefont {Nicolas}\
  \bibnamefont {Sangouard}}, \bibinfo {author} {\bibfnamefont {Mikael}\
  \bibnamefont {Afzelius}}, \bibinfo {author} {\bibfnamefont {Hugues}\
  \bibnamefont {de~Riedmatten}}, \ and\ \bibinfo {author} {\bibfnamefont
  {Nicolas}\ \bibnamefont {Gisin}},\ }\bibfield  {title} {\enquote {\bibinfo
  {title} {Quantum storage of photonic entanglement in a crystal},}\ }\href
  {https://doi.org/10.1038/nature09662} {\bibfield  {journal} {\bibinfo
  {journal} {Nature}\ }\textbf {\bibinfo {volume} {469}},\ \bibinfo {pages}
  {508--511} (\bibinfo {year} {2011})}\BibitemShut {NoStop}%
\bibitem [{\citenamefont {Ledingham}\ \emph {et~al.}(2012)\citenamefont
  {Ledingham}, \citenamefont {Naylor},\ and\ \citenamefont
  {Longdell}}]{Ledingham2012}%
  \BibitemOpen
  \bibfield  {author} {\bibinfo {author} {\bibfnamefont {Patrick~M.}\
  \bibnamefont {Ledingham}}, \bibinfo {author} {\bibfnamefont {William~R.}\
  \bibnamefont {Naylor}}, \ and\ \bibinfo {author} {\bibfnamefont {Jevon~J.}\
  \bibnamefont {Longdell}},\ }\bibfield  {title} {\enquote {\bibinfo {title}
  {Experimental realization of light with time-separated correlations by
  rephasing amplified spontaneous emission},}\ }\href {\doibase
  10.1103/PhysRevLett.109.093602} {\bibfield  {journal} {\bibinfo  {journal}
  {Phys. Rev. Lett.}\ }\textbf {\bibinfo {volume} {109}},\ \bibinfo {pages}
  {093602} (\bibinfo {year} {2012})}\BibitemShut {NoStop}%
\bibitem [{\citenamefont {Zhou}\ \emph {et~al.}(2012)\citenamefont {Zhou},
  \citenamefont {Lin}, \citenamefont {Yang}, \citenamefont {Li},\ and\
  \citenamefont {Guo}}]{Zhou2012}%
  \BibitemOpen
  \bibfield  {author} {\bibinfo {author} {\bibfnamefont {Zong-Quan}\
  \bibnamefont {Zhou}}, \bibinfo {author} {\bibfnamefont {Wei-Bin}\
  \bibnamefont {Lin}}, \bibinfo {author} {\bibfnamefont {Ming}\ \bibnamefont
  {Yang}}, \bibinfo {author} {\bibfnamefont {Chuan-Feng}\ \bibnamefont {Li}}, \
  and\ \bibinfo {author} {\bibfnamefont {Guang-Can}\ \bibnamefont {Guo}},\
  }\bibfield  {title} {\enquote {\bibinfo {title} {Realization of reliable
  solid-state quantum memory for photonic polarization qubit},}\ }\href
  {\doibase 10.1103/PhysRevLett.108.190505} {\bibfield  {journal} {\bibinfo
  {journal} {Phys. Rev. Lett.}\ }\textbf {\bibinfo {volume} {108}},\ \bibinfo
  {pages} {190505} (\bibinfo {year} {2012})}\BibitemShut {NoStop}%
\bibitem [{\citenamefont {Ferguson}\ \emph {et~al.}(2016)\citenamefont
  {Ferguson}, \citenamefont {Beavan}, \citenamefont {Longdell},\ and\
  \citenamefont {Sellars}}]{Ferguson2016}%
  \BibitemOpen
  \bibfield  {author} {\bibinfo {author} {\bibfnamefont {Kate~R.}\ \bibnamefont
  {Ferguson}}, \bibinfo {author} {\bibfnamefont {Sarah~E.}\ \bibnamefont
  {Beavan}}, \bibinfo {author} {\bibfnamefont {Jevon~J.}\ \bibnamefont
  {Longdell}}, \ and\ \bibinfo {author} {\bibfnamefont {Matthew~J.}\
  \bibnamefont {Sellars}},\ }\bibfield  {title} {\enquote {\bibinfo {title}
  {Generation of light with multimode time-delayed entanglement using storage
  in a solid-state spin-wave quantum memory},}\ }\href {\doibase
  10.1103/PhysRevLett.117.020501} {\bibfield  {journal} {\bibinfo  {journal}
  {Phys. Rev. Lett.}\ }\textbf {\bibinfo {volume} {117}},\ \bibinfo {pages}
  {020501} (\bibinfo {year} {2016})}\BibitemShut {NoStop}%
\bibitem [{\citenamefont {Jin}\ \emph {et~al.}(2022)\citenamefont {Jin},
  \citenamefont {Ma}, \citenamefont {Zhou}, \citenamefont {Li},\ and\
  \citenamefont {Guo}}]{Jin2022}%
  \BibitemOpen
  \bibfield  {author} {\bibinfo {author} {\bibfnamefont {Ming}\ \bibnamefont
  {Jin}}, \bibinfo {author} {\bibfnamefont {You-Zhi}\ \bibnamefont {Ma}},
  \bibinfo {author} {\bibfnamefont {Zong-Quan}\ \bibnamefont {Zhou}}, \bibinfo
  {author} {\bibfnamefont {Chuan-Feng}\ \bibnamefont {Li}}, \ and\ \bibinfo
  {author} {\bibfnamefont {Guang-Can}\ \bibnamefont {Guo}},\ }\bibfield
  {title} {\enquote {\bibinfo {title} {A faithful solid-state spin-wave quantum
  memory for polarization qubits},}\ }\href {\doibase
  https://doi.org/10.1016/j.scib.2022.01.019} {\bibfield  {journal} {\bibinfo
  {journal} {Science Bulletin}\ }\textbf {\bibinfo {volume} {67}},\ \bibinfo
  {pages} {676--678} (\bibinfo {year} {2022})}\BibitemShut {NoStop}%
\bibitem [{\citenamefont {Ma}\ \emph {et~al.}(2021{\natexlab{a}})\citenamefont
  {Ma}, \citenamefont {Jin}, \citenamefont {Chen}, \citenamefont {Zhou},
  \citenamefont {Li},\ and\ \citenamefont {Guo}}]{Ma2021_v2}%
  \BibitemOpen
  \bibfield  {author} {\bibinfo {author} {\bibfnamefont {You-Zhi}\ \bibnamefont
  {Ma}}, \bibinfo {author} {\bibfnamefont {Ming}\ \bibnamefont {Jin}}, \bibinfo
  {author} {\bibfnamefont {Duo-Lun}\ \bibnamefont {Chen}}, \bibinfo {author}
  {\bibfnamefont {Zong-Quan}\ \bibnamefont {Zhou}}, \bibinfo {author}
  {\bibfnamefont {Chuan-Feng}\ \bibnamefont {Li}}, \ and\ \bibinfo {author}
  {\bibfnamefont {Guang-Can}\ \bibnamefont {Guo}},\ }\bibfield  {title}
  {\enquote {\bibinfo {title} {Elimination of noise in optically rephased
  photon echoes},}\ }\href {\doibase 10.1038/s41467-021-24679-4} {\bibfield
  {journal} {\bibinfo  {journal} {Nature Communications}\ }\textbf {\bibinfo
  {volume} {12}},\ \bibinfo {pages} {4378} (\bibinfo {year}
  {2021}{\natexlab{a}})}\BibitemShut {NoStop}%
\bibitem [{\citenamefont {Seri}\ \emph {et~al.}(2017)\citenamefont {Seri},
  \citenamefont {Lenhard}, \citenamefont {Riel\"ander}, \citenamefont
  {G\"undo\ifmmode~\breve{g}\else \u{g}\fi{}an}, \citenamefont {Ledingham},
  \citenamefont {Mazzera},\ and\ \citenamefont {de~Riedmatten}}]{Seri2017}%
  \BibitemOpen
  \bibfield  {author} {\bibinfo {author} {\bibfnamefont {Alessandro}\
  \bibnamefont {Seri}}, \bibinfo {author} {\bibfnamefont {Andreas}\
  \bibnamefont {Lenhard}}, \bibinfo {author} {\bibfnamefont {Daniel}\
  \bibnamefont {Riel\"ander}}, \bibinfo {author} {\bibfnamefont {Mustafa}\
  \bibnamefont {G\"undo\ifmmode~\breve{g}\else \u{g}\fi{}an}}, \bibinfo
  {author} {\bibfnamefont {Patrick~M.}\ \bibnamefont {Ledingham}}, \bibinfo
  {author} {\bibfnamefont {Margherita}\ \bibnamefont {Mazzera}}, \ and\
  \bibinfo {author} {\bibfnamefont {Hugues}\ \bibnamefont {de~Riedmatten}},\
  }\bibfield  {title} {\enquote {\bibinfo {title} {Quantum correlations between
  single telecom photons and a multimode on-demand solid-state quantum
  memory},}\ }\href {\doibase 10.1103/PhysRevX.7.021028} {\bibfield  {journal}
  {\bibinfo  {journal} {Phys. Rev. X}\ }\textbf {\bibinfo {volume} {7}},\
  \bibinfo {pages} {021028} (\bibinfo {year} {2017})}\BibitemShut {NoStop}%
\bibitem [{\citenamefont {Kutluer}\ \emph {et~al.}(2017)\citenamefont
  {Kutluer}, \citenamefont {Mazzera},\ and\ \citenamefont
  {de~Riedmatten}}]{Kutluer2017}%
  \BibitemOpen
  \bibfield  {author} {\bibinfo {author} {\bibfnamefont {Kutlu}\ \bibnamefont
  {Kutluer}}, \bibinfo {author} {\bibfnamefont {Margherita}\ \bibnamefont
  {Mazzera}}, \ and\ \bibinfo {author} {\bibfnamefont {Hugues}\ \bibnamefont
  {de~Riedmatten}},\ }\bibfield  {title} {\enquote {\bibinfo {title}
  {Solid-state source of nonclassical photon pairs with embedded multimode
  quantum memory},}\ }\href {\doibase 10.1103/PhysRevLett.118.210502}
  {\bibfield  {journal} {\bibinfo  {journal} {Phys. Rev. Lett.}\ }\textbf
  {\bibinfo {volume} {118}},\ \bibinfo {pages} {210502} (\bibinfo {year}
  {2017})}\BibitemShut {NoStop}%
\bibitem [{\citenamefont {Laplane}\ \emph {et~al.}(2017)\citenamefont
  {Laplane}, \citenamefont {Jobez}, \citenamefont {Etesse}, \citenamefont
  {Gisin},\ and\ \citenamefont {Afzelius}}]{Laplane2017}%
  \BibitemOpen
  \bibfield  {author} {\bibinfo {author} {\bibfnamefont {Cyril}\ \bibnamefont
  {Laplane}}, \bibinfo {author} {\bibfnamefont {Pierre}\ \bibnamefont {Jobez}},
  \bibinfo {author} {\bibfnamefont {Jean}\ \bibnamefont {Etesse}}, \bibinfo
  {author} {\bibfnamefont {Nicolas}\ \bibnamefont {Gisin}}, \ and\ \bibinfo
  {author} {\bibfnamefont {Mikael}\ \bibnamefont {Afzelius}},\ }\bibfield
  {title} {\enquote {\bibinfo {title} {Multimode and long-lived quantum
  correlations between photons and spins in a crystal},}\ }\href {\doibase
  10.1103/PhysRevLett.118.210501} {\bibfield  {journal} {\bibinfo  {journal}
  {Phys. Rev. Lett.}\ }\textbf {\bibinfo {volume} {118}},\ \bibinfo {pages}
  {210501} (\bibinfo {year} {2017})}\BibitemShut {NoStop}%
\bibitem [{\citenamefont {Holz\"{a}pfel}\ \emph {et~al.}(2020)\citenamefont
  {Holz\"{a}pfel}, \citenamefont {Etesse}, \citenamefont {Kaczmarek},
  \citenamefont {Tiranov}, \citenamefont {Gisin},\ and\ \citenamefont
  {Afzelius}}]{Holzapfel2020}%
  \BibitemOpen
  \bibfield  {author} {\bibinfo {author} {\bibfnamefont {Adrian}\ \bibnamefont
  {Holz\"{a}pfel}}, \bibinfo {author} {\bibfnamefont {Jean}\ \bibnamefont
  {Etesse}}, \bibinfo {author} {\bibfnamefont {Krzysztof~T}\ \bibnamefont
  {Kaczmarek}}, \bibinfo {author} {\bibfnamefont {Alexey}\ \bibnamefont
  {Tiranov}}, \bibinfo {author} {\bibfnamefont {Nicolas}\ \bibnamefont
  {Gisin}}, \ and\ \bibinfo {author} {\bibfnamefont {Mikael}\ \bibnamefont
  {Afzelius}},\ }\bibfield  {title} {\enquote {\bibinfo {title} {Optical
  storage for 0.53 s in a solid-state atomic frequency comb memory using
  dynamical decoupling},}\ }\href {\doibase 10.1088/1367-2630/ab8aac}
  {\bibfield  {journal} {\bibinfo  {journal} {New Journal of Physics}\ }\textbf
  {\bibinfo {volume} {22}},\ \bibinfo {pages} {063009} (\bibinfo {year}
  {2020})}\BibitemShut {NoStop}%
\bibitem [{\citenamefont {Businger}\ \emph {et~al.}(2020)\citenamefont
  {Businger}, \citenamefont {Tiranov}, \citenamefont {Kaczmarek}, \citenamefont
  {Welinski}, \citenamefont {Zhang}, \citenamefont {Ferrier}, \citenamefont
  {Goldner},\ and\ \citenamefont {Afzelius}}]{Businger2020}%
  \BibitemOpen
  \bibfield  {author} {\bibinfo {author} {\bibfnamefont {M.}~\bibnamefont
  {Businger}}, \bibinfo {author} {\bibfnamefont {A.}~\bibnamefont {Tiranov}},
  \bibinfo {author} {\bibfnamefont {K.~T.}\ \bibnamefont {Kaczmarek}}, \bibinfo
  {author} {\bibfnamefont {S.}~\bibnamefont {Welinski}}, \bibinfo {author}
  {\bibfnamefont {Z.}~\bibnamefont {Zhang}}, \bibinfo {author} {\bibfnamefont
  {A.}~\bibnamefont {Ferrier}}, \bibinfo {author} {\bibfnamefont
  {P.}~\bibnamefont {Goldner}}, \ and\ \bibinfo {author} {\bibfnamefont
  {M.}~\bibnamefont {Afzelius}},\ }\bibfield  {title} {\enquote {\bibinfo
  {title} {Optical spin-wave storage in a solid-state hybridized
  electron-nuclear spin ensemble},}\ }\href {\doibase
  10.1103/PhysRevLett.124.053606} {\bibfield  {journal} {\bibinfo  {journal}
  {Phys. Rev. Lett.}\ }\textbf {\bibinfo {volume} {124}},\ \bibinfo {pages}
  {053606} (\bibinfo {year} {2020})}\BibitemShut {NoStop}%
\bibitem [{\citenamefont {Askarani}\ \emph {et~al.}(2021)\citenamefont
  {Askarani}, \citenamefont {Das}, \citenamefont {Davidson}, \citenamefont
  {Amaral}, \citenamefont {Sinclair}, \citenamefont {Slater}, \citenamefont
  {Marzban}, \citenamefont {Thiel}, \citenamefont {Cone}, \citenamefont
  {Oblak},\ and\ \citenamefont {Tittel}}]{Askarani2021}%
  \BibitemOpen
  \bibfield  {author} {\bibinfo {author} {\bibfnamefont {Mohsen~Falamarzi}\
  \bibnamefont {Askarani}}, \bibinfo {author} {\bibfnamefont {Antariksha}\
  \bibnamefont {Das}}, \bibinfo {author} {\bibfnamefont {Jacob~H.}\
  \bibnamefont {Davidson}}, \bibinfo {author} {\bibfnamefont {Gustavo~C.}\
  \bibnamefont {Amaral}}, \bibinfo {author} {\bibfnamefont {Neil}\ \bibnamefont
  {Sinclair}}, \bibinfo {author} {\bibfnamefont {Joshua~A.}\ \bibnamefont
  {Slater}}, \bibinfo {author} {\bibfnamefont {Sara}\ \bibnamefont {Marzban}},
  \bibinfo {author} {\bibfnamefont {Charles~W.}\ \bibnamefont {Thiel}},
  \bibinfo {author} {\bibfnamefont {Rufus~L.}\ \bibnamefont {Cone}}, \bibinfo
  {author} {\bibfnamefont {Daniel}\ \bibnamefont {Oblak}}, \ and\ \bibinfo
  {author} {\bibfnamefont {Wolfgang}\ \bibnamefont {Tittel}},\ }\bibfield
  {title} {\enquote {\bibinfo {title} {Long-lived solid-state optical memory
  for high-rate quantum repeaters},}\ }\href {\doibase
  10.1103/PhysRevLett.127.220502} {\bibfield  {journal} {\bibinfo  {journal}
  {Phys. Rev. Lett.}\ }\textbf {\bibinfo {volume} {127}},\ \bibinfo {pages}
  {220502} (\bibinfo {year} {2021})}\BibitemShut {NoStop}%
\bibitem [{\citenamefont {Ma}\ \emph {et~al.}(2021{\natexlab{b}})\citenamefont
  {Ma}, \citenamefont {Ma}, \citenamefont {Zhou}, \citenamefont {Li},\ and\
  \citenamefont {Guo}}]{Ma2021}%
  \BibitemOpen
  \bibfield  {author} {\bibinfo {author} {\bibfnamefont {Yu}~\bibnamefont
  {Ma}}, \bibinfo {author} {\bibfnamefont {You-Zhi}\ \bibnamefont {Ma}},
  \bibinfo {author} {\bibfnamefont {Zong-Quan}\ \bibnamefont {Zhou}}, \bibinfo
  {author} {\bibfnamefont {Chuan-Feng}\ \bibnamefont {Li}}, \ and\ \bibinfo
  {author} {\bibfnamefont {Guang-Can}\ \bibnamefont {Guo}},\ }\bibfield
  {title} {\enquote {\bibinfo {title} {One-hour coherent optical storage in an
  atomic frequency comb memory},}\ }\href
  {https://doi.org/10.1038/s41467-021-22706-y} {\bibfield  {journal} {\bibinfo
  {journal} {Nature Communications}\ }\textbf {\bibinfo {volume} {12}},\
  \bibinfo {pages} {2381} (\bibinfo {year} {2021}{\natexlab{b}})}\BibitemShut
  {NoStop}%
\bibitem [{\citenamefont {Afzelius}\ \emph {et~al.}(2009)\citenamefont
  {Afzelius}, \citenamefont {Simon}, \citenamefont {de~Riedmatten},\ and\
  \citenamefont {Gisin}}]{Afzelius2009}%
  \BibitemOpen
  \bibfield  {author} {\bibinfo {author} {\bibfnamefont {Mikael}\ \bibnamefont
  {Afzelius}}, \bibinfo {author} {\bibfnamefont {Christoph}\ \bibnamefont
  {Simon}}, \bibinfo {author} {\bibfnamefont {Hugues}\ \bibnamefont
  {de~Riedmatten}}, \ and\ \bibinfo {author} {\bibfnamefont {Nicolas}\
  \bibnamefont {Gisin}},\ }\bibfield  {title} {\enquote {\bibinfo {title}
  {Multimode quantum memory based on atomic frequency combs},}\ }\href
  {\doibase 10.1103/PhysRevA.79.052329} {\bibfield  {journal} {\bibinfo
  {journal} {Phys. Rev. A}\ }\textbf {\bibinfo {volume} {79}},\ \bibinfo
  {pages} {052329} (\bibinfo {year} {2009})}\BibitemShut {NoStop}%
\bibitem [{\citenamefont {Zhu}\ \emph {et~al.}(2022)\citenamefont {Zhu},
  \citenamefont {Liu}, \citenamefont {Jin}, \citenamefont {Su}, \citenamefont
  {Liu}, \citenamefont {Li}, \citenamefont {Ye}, \citenamefont {Zhou},
  \citenamefont {Li},\ and\ \citenamefont {Guo}}]{Zhu2022}%
  \BibitemOpen
  \bibfield  {author} {\bibinfo {author} {\bibfnamefont {Tian-Xiang}\
  \bibnamefont {Zhu}}, \bibinfo {author} {\bibfnamefont {Chao}\ \bibnamefont
  {Liu}}, \bibinfo {author} {\bibfnamefont {Ming}\ \bibnamefont {Jin}},
  \bibinfo {author} {\bibfnamefont {Ming-Xu}\ \bibnamefont {Su}}, \bibinfo
  {author} {\bibfnamefont {Yu-Ping}\ \bibnamefont {Liu}}, \bibinfo {author}
  {\bibfnamefont {Wen-Juan}\ \bibnamefont {Li}}, \bibinfo {author}
  {\bibfnamefont {Yang}\ \bibnamefont {Ye}}, \bibinfo {author} {\bibfnamefont
  {Zong-Quan}\ \bibnamefont {Zhou}}, \bibinfo {author} {\bibfnamefont
  {Chuan-Feng}\ \bibnamefont {Li}}, \ and\ \bibinfo {author} {\bibfnamefont
  {Guang-Can}\ \bibnamefont {Guo}},\ }\bibfield  {title} {\enquote {\bibinfo
  {title} {On-demand integrated quantum memory for polarization qubits},}\
  }\href {\doibase 10.1103/PhysRevLett.128.180501} {\bibfield  {journal}
  {\bibinfo  {journal} {Phys. Rev. Lett.}\ }\textbf {\bibinfo {volume} {128}},\
  \bibinfo {pages} {180501} (\bibinfo {year} {2022})}\BibitemShut {NoStop}%
\bibitem [{\citenamefont {Probst}\ \emph {et~al.}(2013)\citenamefont {Probst},
  \citenamefont {Rotzinger}, \citenamefont {W\"unsch}, \citenamefont {Jung},
  \citenamefont {Jerger}, \citenamefont {Siegel}, \citenamefont {Ustinov},\
  and\ \citenamefont {Bushev}}]{Probst2013}%
  \BibitemOpen
  \bibfield  {author} {\bibinfo {author} {\bibfnamefont {S.}~\bibnamefont
  {Probst}}, \bibinfo {author} {\bibfnamefont {H.}~\bibnamefont {Rotzinger}},
  \bibinfo {author} {\bibfnamefont {S.}~\bibnamefont {W\"unsch}}, \bibinfo
  {author} {\bibfnamefont {P.}~\bibnamefont {Jung}}, \bibinfo {author}
  {\bibfnamefont {M.}~\bibnamefont {Jerger}}, \bibinfo {author} {\bibfnamefont
  {M.}~\bibnamefont {Siegel}}, \bibinfo {author} {\bibfnamefont {A.~V.}\
  \bibnamefont {Ustinov}}, \ and\ \bibinfo {author} {\bibfnamefont {P.~A.}\
  \bibnamefont {Bushev}},\ }\bibfield  {title} {\enquote {\bibinfo {title}
  {Anisotropic rare-earth spin ensemble strongly coupled to a superconducting
  resonator},}\ }\href {\doibase 10.1103/PhysRevLett.110.157001} {\bibfield
  {journal} {\bibinfo  {journal} {Phys. Rev. Lett.}\ }\textbf {\bibinfo
  {volume} {110}},\ \bibinfo {pages} {157001} (\bibinfo {year}
  {2013})}\BibitemShut {NoStop}%
\bibitem [{\citenamefont {Probst}\ \emph {et~al.}(2015)\citenamefont {Probst},
  \citenamefont {Rotzinger}, \citenamefont {Ustinov},\ and\ \citenamefont
  {Bushev}}]{Probst2015}%
  \BibitemOpen
  \bibfield  {author} {\bibinfo {author} {\bibfnamefont {S.}~\bibnamefont
  {Probst}}, \bibinfo {author} {\bibfnamefont {H.}~\bibnamefont {Rotzinger}},
  \bibinfo {author} {\bibfnamefont {A.~V.}\ \bibnamefont {Ustinov}}, \ and\
  \bibinfo {author} {\bibfnamefont {P.~A.}\ \bibnamefont {Bushev}},\ }\bibfield
   {title} {\enquote {\bibinfo {title} {Microwave multimode memory with an
  erbium spin ensemble},}\ }\href {\doibase 10.1103/PhysRevB.92.014421}
  {\bibfield  {journal} {\bibinfo  {journal} {Phys. Rev. B}\ }\textbf {\bibinfo
  {volume} {92}},\ \bibinfo {pages} {014421} (\bibinfo {year}
  {2015})}\BibitemShut {NoStop}%
\bibitem [{\citenamefont {Wolfowicz}\ \emph {et~al.}(2015)\citenamefont
  {Wolfowicz}, \citenamefont {Maier-Flaig}, \citenamefont {Marino},
  \citenamefont {Ferrier}, \citenamefont {Vezin}, \citenamefont {Morton},\ and\
  \citenamefont {Goldner}}]{Wolfowicz2015}%
  \BibitemOpen
  \bibfield  {author} {\bibinfo {author} {\bibfnamefont {Gary}\ \bibnamefont
  {Wolfowicz}}, \bibinfo {author} {\bibfnamefont {Hannes}\ \bibnamefont
  {Maier-Flaig}}, \bibinfo {author} {\bibfnamefont {Robert}\ \bibnamefont
  {Marino}}, \bibinfo {author} {\bibfnamefont {Alban}\ \bibnamefont {Ferrier}},
  \bibinfo {author} {\bibfnamefont {Herv\'e}\ \bibnamefont {Vezin}}, \bibinfo
  {author} {\bibfnamefont {John J.~L.}\ \bibnamefont {Morton}}, \ and\ \bibinfo
  {author} {\bibfnamefont {Philippe}\ \bibnamefont {Goldner}},\ }\bibfield
  {title} {\enquote {\bibinfo {title} {Coherent storage of microwave
  excitations in rare-earth nuclear spins},}\ }\href {\doibase
  10.1103/PhysRevLett.114.170503} {\bibfield  {journal} {\bibinfo  {journal}
  {Phys. Rev. Lett.}\ }\textbf {\bibinfo {volume} {114}},\ \bibinfo {pages}
  {170503} (\bibinfo {year} {2015})}\BibitemShut {NoStop}%
\bibitem [{\citenamefont {Doherty}\ \emph {et~al.}(2011)\citenamefont
  {Doherty}, \citenamefont {Manson}, \citenamefont {Delaney},\ and\
  \citenamefont {Hollenberg}}]{Doherty2011}%
  \BibitemOpen
  \bibfield  {author} {\bibinfo {author} {\bibfnamefont {M~W}\ \bibnamefont
  {Doherty}}, \bibinfo {author} {\bibfnamefont {N~B}\ \bibnamefont {Manson}},
  \bibinfo {author} {\bibfnamefont {P}~\bibnamefont {Delaney}}, \ and\ \bibinfo
  {author} {\bibfnamefont {L~C~L}\ \bibnamefont {Hollenberg}},\ }\bibfield
  {title} {\enquote {\bibinfo {title} {The negatively charged nitrogen-vacancy
  centre in diamond: the electronic solution},}\ }\href {\doibase
  10.1088/1367-2630/13/2/025019} {\bibfield  {journal} {\bibinfo  {journal}
  {New Journal of Physics}\ }\textbf {\bibinfo {volume} {13}},\ \bibinfo
  {pages} {025019} (\bibinfo {year} {2011})}\BibitemShut {NoStop}%
\bibitem [{\citenamefont {Doherty}\ \emph {et~al.}(2013)\citenamefont
  {Doherty}, \citenamefont {Manson}, \citenamefont {Delaney}, \citenamefont
  {Jelezko}, \citenamefont {Wrachtrup},\ and\ \citenamefont
  {Hollenberg}}]{Doherty2013}%
  \BibitemOpen
  \bibfield  {author} {\bibinfo {author} {\bibfnamefont {Marcus~W.}\
  \bibnamefont {Doherty}}, \bibinfo {author} {\bibfnamefont {Neil~B.}\
  \bibnamefont {Manson}}, \bibinfo {author} {\bibfnamefont {Paul}\ \bibnamefont
  {Delaney}}, \bibinfo {author} {\bibfnamefont {Fedor}\ \bibnamefont
  {Jelezko}}, \bibinfo {author} {\bibfnamefont {J{\"o}rg}\ \bibnamefont
  {Wrachtrup}}, \ and\ \bibinfo {author} {\bibfnamefont {Lloyd~C.L.}\
  \bibnamefont {Hollenberg}},\ }\bibfield  {title} {\enquote {\bibinfo {title}
  {The nitrogen-vacancy colour centre in diamond},}\ }\href {\doibase
  https://doi.org/10.1016/j.physrep.2013.02.001} {\bibfield  {journal}
  {\bibinfo  {journal} {Physics Reports}\ }\textbf {\bibinfo {volume} {528}},\
  \bibinfo {pages} {1--45} (\bibinfo {year} {2013})},\ \bibinfo {note} {the
  nitrogen-vacancy colour centre in diamond}\BibitemShut {NoStop}%
\bibitem [{\citenamefont {Balasubramanian}\ \emph {et~al.}(2009)\citenamefont
  {Balasubramanian}, \citenamefont {Neumann}, \citenamefont {Twitchen},
  \citenamefont {Markham}, \citenamefont {Kolesov}, \citenamefont {Mizuochi},
  \citenamefont {Isoya}, \citenamefont {Achard}, \citenamefont {Beck},
  \citenamefont {Tissler}, \citenamefont {Jacques}, \citenamefont {Hemmer},
  \citenamefont {Jelezko},\ and\ \citenamefont
  {Wrachtrup}}]{Balasubramanian2009}%
  \BibitemOpen
  \bibfield  {author} {\bibinfo {author} {\bibfnamefont {Gopalakrishnan}\
  \bibnamefont {Balasubramanian}}, \bibinfo {author} {\bibfnamefont {Philipp}\
  \bibnamefont {Neumann}}, \bibinfo {author} {\bibfnamefont {Daniel}\
  \bibnamefont {Twitchen}}, \bibinfo {author} {\bibfnamefont {Matthew}\
  \bibnamefont {Markham}}, \bibinfo {author} {\bibfnamefont {Roman}\
  \bibnamefont {Kolesov}}, \bibinfo {author} {\bibfnamefont {Norikazu}\
  \bibnamefont {Mizuochi}}, \bibinfo {author} {\bibfnamefont {Junichi}\
  \bibnamefont {Isoya}}, \bibinfo {author} {\bibfnamefont {Jocelyn}\
  \bibnamefont {Achard}}, \bibinfo {author} {\bibfnamefont {Johannes}\
  \bibnamefont {Beck}}, \bibinfo {author} {\bibfnamefont {Julia}\ \bibnamefont
  {Tissler}}, \bibinfo {author} {\bibfnamefont {Vincent}\ \bibnamefont
  {Jacques}}, \bibinfo {author} {\bibfnamefont {Philip~R.}\ \bibnamefont
  {Hemmer}}, \bibinfo {author} {\bibfnamefont {Fedor}\ \bibnamefont {Jelezko}},
  \ and\ \bibinfo {author} {\bibfnamefont {J{\"o}rg}\ \bibnamefont
  {Wrachtrup}},\ }\bibfield  {title} {\enquote {\bibinfo {title} {Ultralong
  spin coherence time in isotopically engineered diamond},}\ }\href {\doibase
  10.1038/nmat2420} {\bibfield  {journal} {\bibinfo  {journal} {Nature
  Materials}\ }\textbf {\bibinfo {volume} {8}},\ \bibinfo {pages} {383--387}
  (\bibinfo {year} {2009})}\BibitemShut {NoStop}%
\bibitem [{\citenamefont {Bar-Gill}\ \emph {et~al.}(2013)\citenamefont
  {Bar-Gill}, \citenamefont {Pham}, \citenamefont {Jarmola}, \citenamefont
  {Budker},\ and\ \citenamefont {Walsworth}}]{Bar-Gill2013}%
  \BibitemOpen
  \bibfield  {author} {\bibinfo {author} {\bibfnamefont {N.}~\bibnamefont
  {Bar-Gill}}, \bibinfo {author} {\bibfnamefont {L.~M.}\ \bibnamefont {Pham}},
  \bibinfo {author} {\bibfnamefont {A.}~\bibnamefont {Jarmola}}, \bibinfo
  {author} {\bibfnamefont {D.}~\bibnamefont {Budker}}, \ and\ \bibinfo {author}
  {\bibfnamefont {R.~L.}\ \bibnamefont {Walsworth}},\ }\bibfield  {title}
  {\enquote {\bibinfo {title} {Solid-state electronic spin coherence time
  approaching one second},}\ }\href {\doibase 10.1038/ncomms2771} {\bibfield
  {journal} {\bibinfo  {journal} {Nature Communications}\ }\textbf {\bibinfo
  {volume} {4}},\ \bibinfo {pages} {1743} (\bibinfo {year} {2013})}\BibitemShut
  {NoStop}%
\bibitem [{\citenamefont {Abobeih}\ \emph {et~al.}(2018)\citenamefont
  {Abobeih}, \citenamefont {Cramer}, \citenamefont {Bakker}, \citenamefont
  {Kalb}, \citenamefont {Markham}, \citenamefont {Twitchen},\ and\
  \citenamefont {Taminiau}}]{Abobeih2018}%
  \BibitemOpen
  \bibfield  {author} {\bibinfo {author} {\bibfnamefont {M.~H.}\ \bibnamefont
  {Abobeih}}, \bibinfo {author} {\bibfnamefont {J.}~\bibnamefont {Cramer}},
  \bibinfo {author} {\bibfnamefont {M.~A.}\ \bibnamefont {Bakker}}, \bibinfo
  {author} {\bibfnamefont {N.}~\bibnamefont {Kalb}}, \bibinfo {author}
  {\bibfnamefont {M.}~\bibnamefont {Markham}}, \bibinfo {author} {\bibfnamefont
  {D.~J.}\ \bibnamefont {Twitchen}}, \ and\ \bibinfo {author} {\bibfnamefont
  {T.~H.}\ \bibnamefont {Taminiau}},\ }\bibfield  {title} {\enquote {\bibinfo
  {title} {One-second coherence for a single electron spin coupled to a
  multi-qubit nuclear-spin environment},}\ }\href {\doibase
  10.1038/s41467-018-04916-z} {\bibfield  {journal} {\bibinfo  {journal}
  {Nature Communications}\ }\textbf {\bibinfo {volume} {9}},\ \bibinfo {pages}
  {2552} (\bibinfo {year} {2018})}\BibitemShut {NoStop}%
\bibitem [{\citenamefont {Abobeih}\ \emph {et~al.}(2022)\citenamefont
  {Abobeih}, \citenamefont {Wang}, \citenamefont {Randall}, \citenamefont
  {Loenen}, \citenamefont {Bradley}, \citenamefont {Markham}, \citenamefont
  {Twitchen}, \citenamefont {Terhal},\ and\ \citenamefont
  {Taminiau}}]{Abobeih2022}%
  \BibitemOpen
  \bibfield  {author} {\bibinfo {author} {\bibfnamefont {M.~H.}\ \bibnamefont
  {Abobeih}}, \bibinfo {author} {\bibfnamefont {Y.}~\bibnamefont {Wang}},
  \bibinfo {author} {\bibfnamefont {J.}~\bibnamefont {Randall}}, \bibinfo
  {author} {\bibfnamefont {S.~J.~H.}\ \bibnamefont {Loenen}}, \bibinfo {author}
  {\bibfnamefont {C.~E.}\ \bibnamefont {Bradley}}, \bibinfo {author}
  {\bibfnamefont {M.}~\bibnamefont {Markham}}, \bibinfo {author} {\bibfnamefont
  {D.~J.}\ \bibnamefont {Twitchen}}, \bibinfo {author} {\bibfnamefont {B.~M.}\
  \bibnamefont {Terhal}}, \ and\ \bibinfo {author} {\bibfnamefont {T.~H.}\
  \bibnamefont {Taminiau}},\ }\bibfield  {title} {\enquote {\bibinfo {title}
  {Fault-tolerant operation of a logical qubit in a diamond quantum
  processor},}\ }\href {\doibase 10.1038/s41586-022-04819-6} {\bibfield
  {journal} {\bibinfo  {journal} {Nature}\ }\textbf {\bibinfo {volume} {606}},\
  \bibinfo {pages} {884--889} (\bibinfo {year} {2022})}\BibitemShut {NoStop}%
\bibitem [{\citenamefont {Togan}\ \emph {et~al.}(2010)\citenamefont {Togan},
  \citenamefont {Chu}, \citenamefont {Trifonov}, \citenamefont {Jiang},
  \citenamefont {Maze}, \citenamefont {Childress}, \citenamefont {Dutt},
  \citenamefont {S{\o}rensen}, \citenamefont {Hemmer}, \citenamefont {Zibrov},\
  and\ \citenamefont {Lukin}}]{Togan2010}%
  \BibitemOpen
  \bibfield  {author} {\bibinfo {author} {\bibfnamefont {E.}~\bibnamefont
  {Togan}}, \bibinfo {author} {\bibfnamefont {Y.}~\bibnamefont {Chu}}, \bibinfo
  {author} {\bibfnamefont {A.~S.}\ \bibnamefont {Trifonov}}, \bibinfo {author}
  {\bibfnamefont {L.}~\bibnamefont {Jiang}}, \bibinfo {author} {\bibfnamefont
  {J.}~\bibnamefont {Maze}}, \bibinfo {author} {\bibfnamefont {L.}~\bibnamefont
  {Childress}}, \bibinfo {author} {\bibfnamefont {M.~V.~G.}\ \bibnamefont
  {Dutt}}, \bibinfo {author} {\bibfnamefont {A.~S.}\ \bibnamefont
  {S{\o}rensen}}, \bibinfo {author} {\bibfnamefont {P.~R.}\ \bibnamefont
  {Hemmer}}, \bibinfo {author} {\bibfnamefont {A.~S.}\ \bibnamefont {Zibrov}},
  \ and\ \bibinfo {author} {\bibfnamefont {M.~D.}\ \bibnamefont {Lukin}},\
  }\bibfield  {title} {\enquote {\bibinfo {title} {Quantum entanglement between
  an optical photon and a solid-state spin qubit},}\ }\href {\doibase
  10.1038/nature09256} {\bibfield  {journal} {\bibinfo  {journal} {Nature}\
  }\textbf {\bibinfo {volume} {466}},\ \bibinfo {pages} {730--734} (\bibinfo
  {year} {2010})}\BibitemShut {NoStop}%
\bibitem [{\citenamefont {Bernien}\ \emph {et~al.}(2013)\citenamefont
  {Bernien}, \citenamefont {Hensen}, \citenamefont {Pfaff}, \citenamefont
  {Koolstra}, \citenamefont {Blok}, \citenamefont {Robledo}, \citenamefont
  {Taminiau}, \citenamefont {Markham}, \citenamefont {Twitchen}, \citenamefont
  {Childress},\ and\ \citenamefont {Hanson}}]{Bernien2013}%
  \BibitemOpen
  \bibfield  {author} {\bibinfo {author} {\bibfnamefont {H.}~\bibnamefont
  {Bernien}}, \bibinfo {author} {\bibfnamefont {B.}~\bibnamefont {Hensen}},
  \bibinfo {author} {\bibfnamefont {W.}~\bibnamefont {Pfaff}}, \bibinfo
  {author} {\bibfnamefont {G.}~\bibnamefont {Koolstra}}, \bibinfo {author}
  {\bibfnamefont {M.~S.}\ \bibnamefont {Blok}}, \bibinfo {author}
  {\bibfnamefont {L.}~\bibnamefont {Robledo}}, \bibinfo {author} {\bibfnamefont
  {T.~H.}\ \bibnamefont {Taminiau}}, \bibinfo {author} {\bibfnamefont
  {M.}~\bibnamefont {Markham}}, \bibinfo {author} {\bibfnamefont {D.~J.}\
  \bibnamefont {Twitchen}}, \bibinfo {author} {\bibfnamefont {L.}~\bibnamefont
  {Childress}}, \ and\ \bibinfo {author} {\bibfnamefont {R.}~\bibnamefont
  {Hanson}},\ }\bibfield  {title} {\enquote {\bibinfo {title} {Heralded
  entanglement between solid-state qubits separated by three metres},}\ }\href
  {\doibase 10.1038/nature12016} {\bibfield  {journal} {\bibinfo  {journal}
  {Nature}\ }\textbf {\bibinfo {volume} {497}},\ \bibinfo {pages} {86--90}
  (\bibinfo {year} {2013})}\BibitemShut {NoStop}%
\bibitem [{\citenamefont {Pfaff}\ \emph {et~al.}(2014)\citenamefont {Pfaff},
  \citenamefont {Hensen}, \citenamefont {Bernien}, \citenamefont {van Dam},
  \citenamefont {Blok}, \citenamefont {Taminiau}, \citenamefont {Tiggelman},
  \citenamefont {Schouten}, \citenamefont {Markham}, \citenamefont {Twitchen},\
  and\ \citenamefont {Hanson}}]{Pfaff2014}%
  \BibitemOpen
  \bibfield  {author} {\bibinfo {author} {\bibfnamefont {W.}~\bibnamefont
  {Pfaff}}, \bibinfo {author} {\bibfnamefont {B.~J.}\ \bibnamefont {Hensen}},
  \bibinfo {author} {\bibfnamefont {H.}~\bibnamefont {Bernien}}, \bibinfo
  {author} {\bibfnamefont {S.~B.}\ \bibnamefont {van Dam}}, \bibinfo {author}
  {\bibfnamefont {M.~S.}\ \bibnamefont {Blok}}, \bibinfo {author}
  {\bibfnamefont {T.~H.}\ \bibnamefont {Taminiau}}, \bibinfo {author}
  {\bibfnamefont {M.~J.}\ \bibnamefont {Tiggelman}}, \bibinfo {author}
  {\bibfnamefont {R.~N.}\ \bibnamefont {Schouten}}, \bibinfo {author}
  {\bibfnamefont {M.}~\bibnamefont {Markham}}, \bibinfo {author} {\bibfnamefont
  {D.~J.}\ \bibnamefont {Twitchen}}, \ and\ \bibinfo {author} {\bibfnamefont
  {R.}~\bibnamefont {Hanson}},\ }\bibfield  {title} {\enquote {\bibinfo {title}
  {Unconditional quantum teleportation between distant solid-state quantum
  bits},}\ }\href {\doibase 10.1126/science.1253512} {\bibfield  {journal}
  {\bibinfo  {journal} {Science}\ }\textbf {\bibinfo {volume} {345}},\ \bibinfo
  {pages} {532--535} (\bibinfo {year} {2014})}\BibitemShut {NoStop}%
\bibitem [{\citenamefont {Pompili}\ \emph {et~al.}(2021)\citenamefont
  {Pompili}, \citenamefont {Hermans}, \citenamefont {Baier}, \citenamefont
  {Beukers}, \citenamefont {Humphreys}, \citenamefont {Schouten}, \citenamefont
  {Vermeulen}, \citenamefont {Tiggelman}, \citenamefont {dos Santos~Martins},
  \citenamefont {Dirkse}, \citenamefont {Wehner},\ and\ \citenamefont
  {Hanson}}]{Pompili2021}%
  \BibitemOpen
  \bibfield  {author} {\bibinfo {author} {\bibfnamefont {M.}~\bibnamefont
  {Pompili}}, \bibinfo {author} {\bibfnamefont {S.~L.~N.}\ \bibnamefont
  {Hermans}}, \bibinfo {author} {\bibfnamefont {S.}~\bibnamefont {Baier}},
  \bibinfo {author} {\bibfnamefont {H.~K.~C.}\ \bibnamefont {Beukers}},
  \bibinfo {author} {\bibfnamefont {P.~C.}\ \bibnamefont {Humphreys}}, \bibinfo
  {author} {\bibfnamefont {R.~N.}\ \bibnamefont {Schouten}}, \bibinfo {author}
  {\bibfnamefont {R.~F.~L.}\ \bibnamefont {Vermeulen}}, \bibinfo {author}
  {\bibfnamefont {M.~J.}\ \bibnamefont {Tiggelman}}, \bibinfo {author}
  {\bibfnamefont {L.}~\bibnamefont {dos Santos~Martins}}, \bibinfo {author}
  {\bibfnamefont {B.}~\bibnamefont {Dirkse}}, \bibinfo {author} {\bibfnamefont
  {S.}~\bibnamefont {Wehner}}, \ and\ \bibinfo {author} {\bibfnamefont
  {R.}~\bibnamefont {Hanson}},\ }\bibfield  {title} {\enquote {\bibinfo {title}
  {{Realization of a multinode quantum network of remote solid-state
  qubits}},}\ }\href {\doibase 10.1126/science.abg1919} {\bibfield  {journal}
  {\bibinfo  {journal} {Science}\ }\textbf {\bibinfo {volume} {372}},\ \bibinfo
  {pages} {259--264} (\bibinfo {year} {2021})}\BibitemShut {NoStop}%
\bibitem [{\citenamefont {Hermans}\ \emph {et~al.}(2022)\citenamefont
  {Hermans}, \citenamefont {Pompili}, \citenamefont {Beukers}, \citenamefont
  {Baier}, \citenamefont {Borregaard},\ and\ \citenamefont
  {Hanson}}]{Hermans2022}%
  \BibitemOpen
  \bibfield  {author} {\bibinfo {author} {\bibfnamefont {S.~L.~N.}\
  \bibnamefont {Hermans}}, \bibinfo {author} {\bibfnamefont {M.}~\bibnamefont
  {Pompili}}, \bibinfo {author} {\bibfnamefont {H.~K.~C.}\ \bibnamefont
  {Beukers}}, \bibinfo {author} {\bibfnamefont {S.}~\bibnamefont {Baier}},
  \bibinfo {author} {\bibfnamefont {J.}~\bibnamefont {Borregaard}}, \ and\
  \bibinfo {author} {\bibfnamefont {R.}~\bibnamefont {Hanson}},\ }\bibfield
  {title} {\enquote {\bibinfo {title} {Qubit teleportation between
  non-neighbouring nodes in a quantum network},}\ }\href {\doibase
  10.1038/s41586-022-04697-y} {\bibfield  {journal} {\bibinfo  {journal}
  {Nature}\ }\textbf {\bibinfo {volume} {605}},\ \bibinfo {pages} {663--668}
  (\bibinfo {year} {2022})}\BibitemShut {NoStop}%
\bibitem [{\citenamefont {Hausmann}\ \emph {et~al.}(2013)\citenamefont
  {Hausmann}, \citenamefont {Shields}, \citenamefont {Quan}, \citenamefont
  {Chu}, \citenamefont {de~Leon}, \citenamefont {Evans}, \citenamefont {Burek},
  \citenamefont {Zibrov}, \citenamefont {Markham}, \citenamefont {Twitchen},
  \citenamefont {Park}, \citenamefont {Lukin},\ and\ \citenamefont {Lonc{\v
  a}r}}]{Hausmann2013}%
  \BibitemOpen
  \bibfield  {author} {\bibinfo {author} {\bibfnamefont {B.~J.~M.}\
  \bibnamefont {Hausmann}}, \bibinfo {author} {\bibfnamefont {B.~J.}\
  \bibnamefont {Shields}}, \bibinfo {author} {\bibfnamefont {Q.}~\bibnamefont
  {Quan}}, \bibinfo {author} {\bibfnamefont {Y.}~\bibnamefont {Chu}}, \bibinfo
  {author} {\bibfnamefont {N.~P.}\ \bibnamefont {de~Leon}}, \bibinfo {author}
  {\bibfnamefont {R.}~\bibnamefont {Evans}}, \bibinfo {author} {\bibfnamefont
  {M.~J.}\ \bibnamefont {Burek}}, \bibinfo {author} {\bibfnamefont {A.~S.}\
  \bibnamefont {Zibrov}}, \bibinfo {author} {\bibfnamefont {M.}~\bibnamefont
  {Markham}}, \bibinfo {author} {\bibfnamefont {D.~J.}\ \bibnamefont
  {Twitchen}}, \bibinfo {author} {\bibfnamefont {H.}~\bibnamefont {Park}},
  \bibinfo {author} {\bibfnamefont {M.~D.}\ \bibnamefont {Lukin}}, \ and\
  \bibinfo {author} {\bibfnamefont {M.}~\bibnamefont {Lonc{\v a}r}},\
  }\bibfield  {title} {\enquote {\bibinfo {title} {Coupling of nv centers to
  photonic crystal nanobeams in diamond},}\ }\href {\doibase 10.1021/nl402174g}
  {\bibfield  {journal} {\bibinfo  {journal} {Nano Letters}\ }\textbf {\bibinfo
  {volume} {13}},\ \bibinfo {pages} {5791--5796} (\bibinfo {year}
  {2013})}\BibitemShut {NoStop}%
\bibitem [{\citenamefont {Riedel}\ \emph {et~al.}(2017)\citenamefont {Riedel},
  \citenamefont {S\"ollner}, \citenamefont {Shields}, \citenamefont
  {Starosielec}, \citenamefont {Appel}, \citenamefont {Neu}, \citenamefont
  {Maletinsky},\ and\ \citenamefont {Warburton}}]{Riedel2017}%
  \BibitemOpen
  \bibfield  {author} {\bibinfo {author} {\bibfnamefont {Daniel}\ \bibnamefont
  {Riedel}}, \bibinfo {author} {\bibfnamefont {Immo}\ \bibnamefont
  {S\"ollner}}, \bibinfo {author} {\bibfnamefont {Brendan~J.}\ \bibnamefont
  {Shields}}, \bibinfo {author} {\bibfnamefont {Sebastian}\ \bibnamefont
  {Starosielec}}, \bibinfo {author} {\bibfnamefont {Patrick}\ \bibnamefont
  {Appel}}, \bibinfo {author} {\bibfnamefont {Elke}\ \bibnamefont {Neu}},
  \bibinfo {author} {\bibfnamefont {Patrick}\ \bibnamefont {Maletinsky}}, \
  and\ \bibinfo {author} {\bibfnamefont {Richard~J.}\ \bibnamefont
  {Warburton}},\ }\bibfield  {title} {\enquote {\bibinfo {title} {Deterministic
  enhancement of coherent photon generation from a nitrogen-vacancy center in
  ultrapure diamond},}\ }\href {\doibase 10.1103/PhysRevX.7.031040} {\bibfield
  {journal} {\bibinfo  {journal} {Phys. Rev. X}\ }\textbf {\bibinfo {volume}
  {7}},\ \bibinfo {pages} {031040} (\bibinfo {year} {2017})}\BibitemShut
  {NoStop}%
\bibitem [{\citenamefont {Maurer}\ \emph {et~al.}(2012)\citenamefont {Maurer},
  \citenamefont {Kucsko}, \citenamefont {Latta}, \citenamefont {Jiang},
  \citenamefont {Yao}, \citenamefont {Bennett}, \citenamefont {Pastawski},
  \citenamefont {Hunger}, \citenamefont {Chisholm}, \citenamefont {Markham},
  \citenamefont {Twitchen}, \citenamefont {Cirac},\ and\ \citenamefont
  {Lukin}}]{Maurer2012}%
  \BibitemOpen
  \bibfield  {author} {\bibinfo {author} {\bibfnamefont {P.~C.}\ \bibnamefont
  {Maurer}}, \bibinfo {author} {\bibfnamefont {G.}~\bibnamefont {Kucsko}},
  \bibinfo {author} {\bibfnamefont {C.}~\bibnamefont {Latta}}, \bibinfo
  {author} {\bibfnamefont {L.}~\bibnamefont {Jiang}}, \bibinfo {author}
  {\bibfnamefont {N.~Y.}\ \bibnamefont {Yao}}, \bibinfo {author} {\bibfnamefont
  {S.~D.}\ \bibnamefont {Bennett}}, \bibinfo {author} {\bibfnamefont
  {F.}~\bibnamefont {Pastawski}}, \bibinfo {author} {\bibfnamefont
  {D.}~\bibnamefont {Hunger}}, \bibinfo {author} {\bibfnamefont
  {N.}~\bibnamefont {Chisholm}}, \bibinfo {author} {\bibfnamefont
  {M.}~\bibnamefont {Markham}}, \bibinfo {author} {\bibfnamefont {D.~J.}\
  \bibnamefont {Twitchen}}, \bibinfo {author} {\bibfnamefont {J.~I.}\
  \bibnamefont {Cirac}}, \ and\ \bibinfo {author} {\bibfnamefont {M.~D.}\
  \bibnamefont {Lukin}},\ }\bibfield  {title} {\enquote {\bibinfo {title}
  {{Room-Temperature Quantum Bit Memory Exceeding One Second}},}\ }\href
  {\doibase 10.1126/science.1220513} {\bibfield  {journal} {\bibinfo  {journal}
  {Science}\ }\textbf {\bibinfo {volume} {336}},\ \bibinfo {pages} {1283--1286}
  (\bibinfo {year} {2012})}\BibitemShut {NoStop}%
\bibitem [{\citenamefont {Bradley}\ \emph {et~al.}(2019)\citenamefont
  {Bradley}, \citenamefont {Randall}, \citenamefont {Abobeih}, \citenamefont
  {Berrevoets}, \citenamefont {Degen}, \citenamefont {Bakker}, \citenamefont
  {Markham}, \citenamefont {Twitchen},\ and\ \citenamefont
  {Taminiau}}]{Bradley2019}%
  \BibitemOpen
  \bibfield  {author} {\bibinfo {author} {\bibfnamefont {C.~E.}\ \bibnamefont
  {Bradley}}, \bibinfo {author} {\bibfnamefont {J.}~\bibnamefont {Randall}},
  \bibinfo {author} {\bibfnamefont {M.~H.}\ \bibnamefont {Abobeih}}, \bibinfo
  {author} {\bibfnamefont {R.~C.}\ \bibnamefont {Berrevoets}}, \bibinfo
  {author} {\bibfnamefont {M.~J.}\ \bibnamefont {Degen}}, \bibinfo {author}
  {\bibfnamefont {M.~A.}\ \bibnamefont {Bakker}}, \bibinfo {author}
  {\bibfnamefont {M.}~\bibnamefont {Markham}}, \bibinfo {author} {\bibfnamefont
  {D.~J.}\ \bibnamefont {Twitchen}}, \ and\ \bibinfo {author} {\bibfnamefont
  {T.~H.}\ \bibnamefont {Taminiau}},\ }\bibfield  {title} {\enquote {\bibinfo
  {title} {A ten-qubit solid-state spin register with quantum memory up to one
  minute},}\ }\href {\doibase 10.1103/PhysRevX.9.031045} {\bibfield  {journal}
  {\bibinfo  {journal} {Phys. Rev. X}\ }\textbf {\bibinfo {volume} {9}},\
  \bibinfo {pages} {031045} (\bibinfo {year} {2019})}\BibitemShut {NoStop}%
\bibitem [{\citenamefont {Bartling}\ \emph {et~al.}(2022)\citenamefont
  {Bartling}, \citenamefont {Abobeih}, \citenamefont {Pingault}, \citenamefont
  {Degen}, \citenamefont {Loenen}, \citenamefont {Bradley}, \citenamefont
  {Randall}, \citenamefont {Markham}, \citenamefont {Twitchen},\ and\
  \citenamefont {Taminiau}}]{Bartling2022}%
  \BibitemOpen
  \bibfield  {author} {\bibinfo {author} {\bibfnamefont {H.~P.}\ \bibnamefont
  {Bartling}}, \bibinfo {author} {\bibfnamefont {M.~H.}\ \bibnamefont
  {Abobeih}}, \bibinfo {author} {\bibfnamefont {B.}~\bibnamefont {Pingault}},
  \bibinfo {author} {\bibfnamefont {M.~J.}\ \bibnamefont {Degen}}, \bibinfo
  {author} {\bibfnamefont {S.~J.~H.}\ \bibnamefont {Loenen}}, \bibinfo {author}
  {\bibfnamefont {C.~E.}\ \bibnamefont {Bradley}}, \bibinfo {author}
  {\bibfnamefont {J.}~\bibnamefont {Randall}}, \bibinfo {author} {\bibfnamefont
  {M.}~\bibnamefont {Markham}}, \bibinfo {author} {\bibfnamefont {D.~J.}\
  \bibnamefont {Twitchen}}, \ and\ \bibinfo {author} {\bibfnamefont {T.~H.}\
  \bibnamefont {Taminiau}},\ }\bibfield  {title} {\enquote {\bibinfo {title}
  {Entanglement of spin-pair qubits with intrinsic dephasing times exceeding a
  minute},}\ }\href {\doibase 10.1103/PhysRevX.12.011048} {\bibfield  {journal}
  {\bibinfo  {journal} {Phys. Rev. X}\ }\textbf {\bibinfo {volume} {12}},\
  \bibinfo {pages} {011048} (\bibinfo {year} {2022})}\BibitemShut {NoStop}%
\bibitem [{\citenamefont {Xie}\ \emph {et~al.}(2023)\citenamefont {Xie},
  \citenamefont {Zhao}, \citenamefont {Xu}, \citenamefont {Kong}, \citenamefont
  {Yang}, \citenamefont {Wang}, \citenamefont {Wang}, \citenamefont {Shi},\
  and\ \citenamefont {Du}}]{Xie2023}%
  \BibitemOpen
  \bibfield  {author} {\bibinfo {author} {\bibfnamefont {Tianyu}\ \bibnamefont
  {Xie}}, \bibinfo {author} {\bibfnamefont {Zhiyuan}\ \bibnamefont {Zhao}},
  \bibinfo {author} {\bibfnamefont {Shaoyi}\ \bibnamefont {Xu}}, \bibinfo
  {author} {\bibfnamefont {Xi}~\bibnamefont {Kong}}, \bibinfo {author}
  {\bibfnamefont {Zhiping}\ \bibnamefont {Yang}}, \bibinfo {author}
  {\bibfnamefont {Mengqi}\ \bibnamefont {Wang}}, \bibinfo {author}
  {\bibfnamefont {Ya}~\bibnamefont {Wang}}, \bibinfo {author} {\bibfnamefont
  {Fazhan}\ \bibnamefont {Shi}}, \ and\ \bibinfo {author} {\bibfnamefont
  {Jiangfeng}\ \bibnamefont {Du}},\ }\bibfield  {title} {\enquote {\bibinfo
  {title} {99.92\%-fidelity cnot gates in solids by noise filtering},}\ }\href
  {\doibase 10.1103/PhysRevLett.130.030601} {\bibfield  {journal} {\bibinfo
  {journal} {Phys. Rev. Lett.}\ }\textbf {\bibinfo {volume} {130}},\ \bibinfo
  {pages} {030601} (\bibinfo {year} {2023})}\BibitemShut {NoStop}%
\bibitem [{\citenamefont {Wang}\ \emph {et~al.}(2023)\citenamefont {Wang},
  \citenamefont {Barr}, \citenamefont {Tang}, \citenamefont {Chen},
  \citenamefont {Li}, \citenamefont {Xu}, \citenamefont {Stasiuk},
  \citenamefont {Li},\ and\ \citenamefont {Cappellaro}}]{WangGuoqing2023}%
  \BibitemOpen
  \bibfield  {author} {\bibinfo {author} {\bibfnamefont {Guoqing}\ \bibnamefont
  {Wang}}, \bibinfo {author} {\bibfnamefont {Ariel~Rebekah}\ \bibnamefont
  {Barr}}, \bibinfo {author} {\bibfnamefont {Hao}\ \bibnamefont {Tang}},
  \bibinfo {author} {\bibfnamefont {Mo}~\bibnamefont {Chen}}, \bibinfo {author}
  {\bibfnamefont {Changhao}\ \bibnamefont {Li}}, \bibinfo {author}
  {\bibfnamefont {Haowei}\ \bibnamefont {Xu}}, \bibinfo {author} {\bibfnamefont
  {Andrew}\ \bibnamefont {Stasiuk}}, \bibinfo {author} {\bibfnamefont
  {Ju}~\bibnamefont {Li}}, \ and\ \bibinfo {author} {\bibfnamefont {Paola}\
  \bibnamefont {Cappellaro}},\ }\bibfield  {title} {\enquote {\bibinfo {title}
  {Characterizing temperature and strain variations with qubit ensembles for
  their robust coherence protection},}\ }\href {\doibase
  10.1103/PhysRevLett.131.043602} {\bibfield  {journal} {\bibinfo  {journal}
  {Phys. Rev. Lett.}\ }\textbf {\bibinfo {volume} {131}},\ \bibinfo {pages}
  {043602} (\bibinfo {year} {2023})}\BibitemShut {NoStop}%
\bibitem [{\citenamefont {Diniz}\ \emph {et~al.}(2011)\citenamefont {Diniz},
  \citenamefont {Portolan}, \citenamefont {Ferreira}, \citenamefont {G\'erard},
  \citenamefont {Bertet},\ and\ \citenamefont {Auff\`eves}}]{Diniz2011}%
  \BibitemOpen
  \bibfield  {author} {\bibinfo {author} {\bibfnamefont {I.}~\bibnamefont
  {Diniz}}, \bibinfo {author} {\bibfnamefont {S.}~\bibnamefont {Portolan}},
  \bibinfo {author} {\bibfnamefont {R.}~\bibnamefont {Ferreira}}, \bibinfo
  {author} {\bibfnamefont {J.~M.}\ \bibnamefont {G\'erard}}, \bibinfo {author}
  {\bibfnamefont {P.}~\bibnamefont {Bertet}}, \ and\ \bibinfo {author}
  {\bibfnamefont {A.}~\bibnamefont {Auff\`eves}},\ }\bibfield  {title}
  {\enquote {\bibinfo {title} {Strongly coupling a cavity to inhomogeneous
  ensembles of emitters: Potential for long-lived solid-state quantum
  memories},}\ }\href {\doibase 10.1103/PhysRevA.84.063810} {\bibfield
  {journal} {\bibinfo  {journal} {Phys. Rev. A}\ }\textbf {\bibinfo {volume}
  {84}},\ \bibinfo {pages} {063810} (\bibinfo {year} {2011})}\BibitemShut
  {NoStop}%
\bibitem [{\citenamefont {Zhu}\ \emph {et~al.}(2011)\citenamefont {Zhu},
  \citenamefont {Saito}, \citenamefont {Kemp}, \citenamefont {Kakuyanagi},
  \citenamefont {Karimoto}, \citenamefont {Nakano}, \citenamefont {Munro},
  \citenamefont {Tokura}, \citenamefont {Everitt}, \citenamefont {Nemoto},
  \citenamefont {Kasu}, \citenamefont {Mizuochi},\ and\ \citenamefont
  {Semba}}]{Zhu2011}%
  \BibitemOpen
  \bibfield  {author} {\bibinfo {author} {\bibfnamefont {Xiaobo}\ \bibnamefont
  {Zhu}}, \bibinfo {author} {\bibfnamefont {Shiro}\ \bibnamefont {Saito}},
  \bibinfo {author} {\bibfnamefont {Alexander}\ \bibnamefont {Kemp}}, \bibinfo
  {author} {\bibfnamefont {Kosuke}\ \bibnamefont {Kakuyanagi}}, \bibinfo
  {author} {\bibfnamefont {Shin-ichi}\ \bibnamefont {Karimoto}}, \bibinfo
  {author} {\bibfnamefont {Hayato}\ \bibnamefont {Nakano}}, \bibinfo {author}
  {\bibfnamefont {William~J.}\ \bibnamefont {Munro}}, \bibinfo {author}
  {\bibfnamefont {Yasuhiro}\ \bibnamefont {Tokura}}, \bibinfo {author}
  {\bibfnamefont {Mark~S.}\ \bibnamefont {Everitt}}, \bibinfo {author}
  {\bibfnamefont {Kae}\ \bibnamefont {Nemoto}}, \bibinfo {author}
  {\bibfnamefont {Makoto}\ \bibnamefont {Kasu}}, \bibinfo {author}
  {\bibfnamefont {Norikazu}\ \bibnamefont {Mizuochi}}, \ and\ \bibinfo {author}
  {\bibfnamefont {Kouichi}\ \bibnamefont {Semba}},\ }\bibfield  {title}
  {\enquote {\bibinfo {title} {Coherent coupling of a superconducting flux
  qubit to an electron spin ensemble in diamond},}\ }\href {\doibase
  10.1038/nature10462} {\bibfield  {journal} {\bibinfo  {journal} {Nature}\
  }\textbf {\bibinfo {volume} {478}},\ \bibinfo {pages} {221--224} (\bibinfo
  {year} {2011})}\BibitemShut {NoStop}%
\bibitem [{\citenamefont {Kubo}\ \emph {et~al.}(2011)\citenamefont {Kubo},
  \citenamefont {Grezes}, \citenamefont {Dewes}, \citenamefont {Umeda},
  \citenamefont {Isoya}, \citenamefont {Sumiya}, \citenamefont {Morishita},
  \citenamefont {Abe}, \citenamefont {Onoda}, \citenamefont {Ohshima},
  \citenamefont {Jacques}, \citenamefont {Dr\'eau}, \citenamefont {Roch},
  \citenamefont {Diniz}, \citenamefont {Auffeves}, \citenamefont {Vion},
  \citenamefont {Esteve},\ and\ \citenamefont {Bertet}}]{Kubo2011}%
  \BibitemOpen
  \bibfield  {author} {\bibinfo {author} {\bibfnamefont {Y.}~\bibnamefont
  {Kubo}}, \bibinfo {author} {\bibfnamefont {C.}~\bibnamefont {Grezes}},
  \bibinfo {author} {\bibfnamefont {A.}~\bibnamefont {Dewes}}, \bibinfo
  {author} {\bibfnamefont {T.}~\bibnamefont {Umeda}}, \bibinfo {author}
  {\bibfnamefont {J.}~\bibnamefont {Isoya}}, \bibinfo {author} {\bibfnamefont
  {H.}~\bibnamefont {Sumiya}}, \bibinfo {author} {\bibfnamefont
  {N.}~\bibnamefont {Morishita}}, \bibinfo {author} {\bibfnamefont
  {H.}~\bibnamefont {Abe}}, \bibinfo {author} {\bibfnamefont {S.}~\bibnamefont
  {Onoda}}, \bibinfo {author} {\bibfnamefont {T.}~\bibnamefont {Ohshima}},
  \bibinfo {author} {\bibfnamefont {V.}~\bibnamefont {Jacques}}, \bibinfo
  {author} {\bibfnamefont {A.}~\bibnamefont {Dr\'eau}}, \bibinfo {author}
  {\bibfnamefont {J.-F.}\ \bibnamefont {Roch}}, \bibinfo {author}
  {\bibfnamefont {I.}~\bibnamefont {Diniz}}, \bibinfo {author} {\bibfnamefont
  {A.}~\bibnamefont {Auffeves}}, \bibinfo {author} {\bibfnamefont
  {D.}~\bibnamefont {Vion}}, \bibinfo {author} {\bibfnamefont {D.}~\bibnamefont
  {Esteve}}, \ and\ \bibinfo {author} {\bibfnamefont {P.}~\bibnamefont
  {Bertet}},\ }\bibfield  {title} {\enquote {\bibinfo {title} {Hybrid quantum
  circuit with a superconducting qubit coupled to a spin ensemble},}\ }\href
  {\doibase 10.1103/PhysRevLett.107.220501} {\bibfield  {journal} {\bibinfo
  {journal} {Phys. Rev. Lett.}\ }\textbf {\bibinfo {volume} {107}},\ \bibinfo
  {pages} {220501} (\bibinfo {year} {2011})}\BibitemShut {NoStop}%
\bibitem [{\citenamefont {Kubo}\ \emph {et~al.}(2012)\citenamefont {Kubo},
  \citenamefont {Diniz}, \citenamefont {Dewes}, \citenamefont {Jacques},
  \citenamefont {Dr\'eau}, \citenamefont {Roch}, \citenamefont {Auffeves},
  \citenamefont {Vion}, \citenamefont {Esteve},\ and\ \citenamefont
  {Bertet}}]{Kubo2012}%
  \BibitemOpen
  \bibfield  {author} {\bibinfo {author} {\bibfnamefont {Y.}~\bibnamefont
  {Kubo}}, \bibinfo {author} {\bibfnamefont {I.}~\bibnamefont {Diniz}},
  \bibinfo {author} {\bibfnamefont {A.}~\bibnamefont {Dewes}}, \bibinfo
  {author} {\bibfnamefont {V.}~\bibnamefont {Jacques}}, \bibinfo {author}
  {\bibfnamefont {A.}~\bibnamefont {Dr\'eau}}, \bibinfo {author} {\bibfnamefont
  {J.-F.}\ \bibnamefont {Roch}}, \bibinfo {author} {\bibfnamefont
  {A.}~\bibnamefont {Auffeves}}, \bibinfo {author} {\bibfnamefont
  {D.}~\bibnamefont {Vion}}, \bibinfo {author} {\bibfnamefont {D.}~\bibnamefont
  {Esteve}}, \ and\ \bibinfo {author} {\bibfnamefont {P.}~\bibnamefont
  {Bertet}},\ }\bibfield  {title} {\enquote {\bibinfo {title} {Storage and
  retrieval of a microwave field in a spin ensemble},}\ }\href {\doibase
  10.1103/PhysRevA.85.012333} {\bibfield  {journal} {\bibinfo  {journal} {Phys.
  Rev. A}\ }\textbf {\bibinfo {volume} {85}},\ \bibinfo {pages} {012333}
  (\bibinfo {year} {2012})}\BibitemShut {NoStop}%
\bibitem [{\citenamefont {Heshami}\ \emph {et~al.}(2014)\citenamefont
  {Heshami}, \citenamefont {Santori}, \citenamefont {Khanaliloo}, \citenamefont
  {Healey}, \citenamefont {Acosta}, \citenamefont {Barclay},\ and\
  \citenamefont {Simon}}]{Heshami2014}%
  \BibitemOpen
  \bibfield  {author} {\bibinfo {author} {\bibfnamefont {Khabat}\ \bibnamefont
  {Heshami}}, \bibinfo {author} {\bibfnamefont {Charles}\ \bibnamefont
  {Santori}}, \bibinfo {author} {\bibfnamefont {Behzad}\ \bibnamefont
  {Khanaliloo}}, \bibinfo {author} {\bibfnamefont {Chris}\ \bibnamefont
  {Healey}}, \bibinfo {author} {\bibfnamefont {Victor~M.}\ \bibnamefont
  {Acosta}}, \bibinfo {author} {\bibfnamefont {Paul~E.}\ \bibnamefont
  {Barclay}}, \ and\ \bibinfo {author} {\bibfnamefont {Christoph}\ \bibnamefont
  {Simon}},\ }\bibfield  {title} {\enquote {\bibinfo {title} {Raman quantum
  memory based on an ensemble of nitrogen-vacancy centers coupled to a
  microcavity},}\ }\href {\doibase 10.1103/PhysRevA.89.040301} {\bibfield
  {journal} {\bibinfo  {journal} {Phys. Rev. A}\ }\textbf {\bibinfo {volume}
  {89}},\ \bibinfo {pages} {040301} (\bibinfo {year} {2014})}\BibitemShut
  {NoStop}%
\bibitem [{\citenamefont {Kalachev}\ \emph {et~al.}(2019)\citenamefont
  {Kalachev}, \citenamefont {Berezhnoi}, \citenamefont {Hemmer},\ and\
  \citenamefont {Kocharovskaya}}]{Kalachev2019}%
  \BibitemOpen
  \bibfield  {author} {\bibinfo {author} {\bibfnamefont {A}~\bibnamefont
  {Kalachev}}, \bibinfo {author} {\bibfnamefont {A}~\bibnamefont {Berezhnoi}},
  \bibinfo {author} {\bibfnamefont {P}~\bibnamefont {Hemmer}}, \ and\ \bibinfo
  {author} {\bibfnamefont {O}~\bibnamefont {Kocharovskaya}},\ }\bibfield
  {title} {\enquote {\bibinfo {title} {Raman quantum memory based on an
  ensemble of silicon-vacancy centers in diamond},}\ }\href {\doibase
  10.1088/1555-6611/ab4049} {\bibfield  {journal} {\bibinfo  {journal} {Laser
  Physics}\ }\textbf {\bibinfo {volume} {29}},\ \bibinfo {pages} {104001}
  (\bibinfo {year} {2019})}\BibitemShut {NoStop}%
\bibitem [{\citenamefont {Bhaskar}\ \emph {et~al.}(2020)\citenamefont
  {Bhaskar}, \citenamefont {Riedinger}, \citenamefont {Machielse},
  \citenamefont {Levonian}, \citenamefont {Nguyen}, \citenamefont {Knall},
  \citenamefont {Park}, \citenamefont {Englund}, \citenamefont {Lon{\v c}ar},
  \citenamefont {Sukachev},\ and\ \citenamefont {Lukin}}]{Bhaskar2020}%
  \BibitemOpen
  \bibfield  {author} {\bibinfo {author} {\bibfnamefont {M.~K.}\ \bibnamefont
  {Bhaskar}}, \bibinfo {author} {\bibfnamefont {R.}~\bibnamefont {Riedinger}},
  \bibinfo {author} {\bibfnamefont {B.}~\bibnamefont {Machielse}}, \bibinfo
  {author} {\bibfnamefont {D.~S.}\ \bibnamefont {Levonian}}, \bibinfo {author}
  {\bibfnamefont {C.~T.}\ \bibnamefont {Nguyen}}, \bibinfo {author}
  {\bibfnamefont {E.~N.}\ \bibnamefont {Knall}}, \bibinfo {author}
  {\bibfnamefont {H.}~\bibnamefont {Park}}, \bibinfo {author} {\bibfnamefont
  {D.}~\bibnamefont {Englund}}, \bibinfo {author} {\bibfnamefont
  {M.}~\bibnamefont {Lon{\v c}ar}}, \bibinfo {author} {\bibfnamefont {D.~D.}\
  \bibnamefont {Sukachev}}, \ and\ \bibinfo {author} {\bibfnamefont {M.~D.}\
  \bibnamefont {Lukin}},\ }\bibfield  {title} {\enquote {\bibinfo {title}
  {Experimental demonstration of memory-enhanced quantum communication},}\
  }\href {\doibase 10.1038/s41586-020-2103-5} {\bibfield  {journal} {\bibinfo
  {journal} {Nature}\ }\textbf {\bibinfo {volume} {580}},\ \bibinfo {pages}
  {60--64} (\bibinfo {year} {2020})}\BibitemShut {NoStop}%
\bibitem [{\citenamefont {Starling}\ \emph {et~al.}(2023)\citenamefont
  {Starling}, \citenamefont {Shtyrkova}, \citenamefont {Christen},
  \citenamefont {Murphy}, \citenamefont {Li}, \citenamefont {Chen},
  \citenamefont {Kharas}, \citenamefont {Zhang}, \citenamefont {Cummings},
  \citenamefont {Nowak}, \citenamefont {Bersin}, \citenamefont {Niffenegger},
  \citenamefont {Sutula}, \citenamefont {Englund}, \citenamefont {Hamilton},\
  and\ \citenamefont {Dixon}}]{Starling2023}%
  \BibitemOpen
  \bibfield  {author} {\bibinfo {author} {\bibfnamefont {David~J.}\
  \bibnamefont {Starling}}, \bibinfo {author} {\bibfnamefont {Katia}\
  \bibnamefont {Shtyrkova}}, \bibinfo {author} {\bibfnamefont {Ian}\
  \bibnamefont {Christen}}, \bibinfo {author} {\bibfnamefont {Ryan}\
  \bibnamefont {Murphy}}, \bibinfo {author} {\bibfnamefont {Linsen}\
  \bibnamefont {Li}}, \bibinfo {author} {\bibfnamefont {Kevin~C.}\ \bibnamefont
  {Chen}}, \bibinfo {author} {\bibfnamefont {Dave}\ \bibnamefont {Kharas}},
  \bibinfo {author} {\bibfnamefont {Xingyu}\ \bibnamefont {Zhang}}, \bibinfo
  {author} {\bibfnamefont {John}\ \bibnamefont {Cummings}}, \bibinfo {author}
  {\bibfnamefont {W.~John}\ \bibnamefont {Nowak}}, \bibinfo {author}
  {\bibfnamefont {Eric}\ \bibnamefont {Bersin}}, \bibinfo {author}
  {\bibfnamefont {Robert~J.}\ \bibnamefont {Niffenegger}}, \bibinfo {author}
  {\bibfnamefont {Madison}\ \bibnamefont {Sutula}}, \bibinfo {author}
  {\bibfnamefont {Dirk}\ \bibnamefont {Englund}}, \bibinfo {author}
  {\bibfnamefont {Scott}\ \bibnamefont {Hamilton}}, \ and\ \bibinfo {author}
  {\bibfnamefont {P.~Benjamin}\ \bibnamefont {Dixon}},\ }\bibfield  {title}
  {\enquote {\bibinfo {title} {Fully packaged multichannel cryogenic quantum
  memory module},}\ }\href {\doibase 10.1103/PhysRevApplied.19.064028}
  {\bibfield  {journal} {\bibinfo  {journal} {Phys. Rev. Appl.}\ }\textbf
  {\bibinfo {volume} {19}},\ \bibinfo {pages} {064028} (\bibinfo {year}
  {2023})}\BibitemShut {NoStop}%
\bibitem [{\citenamefont {Stas}\ \emph {et~al.}(2022)\citenamefont {Stas},
  \citenamefont {Huan}, \citenamefont {Machielse}, \citenamefont {Knall},
  \citenamefont {Suleymanzade}, \citenamefont {Pingault}, \citenamefont
  {Sutula}, \citenamefont {Ding}, \citenamefont {Knaut}, \citenamefont
  {Assumpcao}, \citenamefont {Wei}, \citenamefont {Bhaskar}, \citenamefont
  {Riedinger}, \citenamefont {Sukachev}, \citenamefont {Park}, \citenamefont
  {Lon{\v c}ar}, \citenamefont {Levonian},\ and\ \citenamefont
  {Lukin}}]{Stas2023}%
  \BibitemOpen
  \bibfield  {author} {\bibinfo {author} {\bibfnamefont {P.-J.}\ \bibnamefont
  {Stas}}, \bibinfo {author} {\bibfnamefont {Y.~Q.}\ \bibnamefont {Huan}},
  \bibinfo {author} {\bibfnamefont {B.}~\bibnamefont {Machielse}}, \bibinfo
  {author} {\bibfnamefont {E.~N.}\ \bibnamefont {Knall}}, \bibinfo {author}
  {\bibfnamefont {A.}~\bibnamefont {Suleymanzade}}, \bibinfo {author}
  {\bibfnamefont {B.}~\bibnamefont {Pingault}}, \bibinfo {author}
  {\bibfnamefont {M.}~\bibnamefont {Sutula}}, \bibinfo {author} {\bibfnamefont
  {S.~W.}\ \bibnamefont {Ding}}, \bibinfo {author} {\bibfnamefont {C.~M.}\
  \bibnamefont {Knaut}}, \bibinfo {author} {\bibfnamefont {D.~R.}\ \bibnamefont
  {Assumpcao}}, \bibinfo {author} {\bibfnamefont {Y.-C.}\ \bibnamefont {Wei}},
  \bibinfo {author} {\bibfnamefont {M.~K.}\ \bibnamefont {Bhaskar}}, \bibinfo
  {author} {\bibfnamefont {R.}~\bibnamefont {Riedinger}}, \bibinfo {author}
  {\bibfnamefont {D.~D.}\ \bibnamefont {Sukachev}}, \bibinfo {author}
  {\bibfnamefont {H.}~\bibnamefont {Park}}, \bibinfo {author} {\bibfnamefont
  {M.}~\bibnamefont {Lon{\v c}ar}}, \bibinfo {author} {\bibfnamefont {D.~S.}\
  \bibnamefont {Levonian}}, \ and\ \bibinfo {author} {\bibfnamefont {M.~D.}\
  \bibnamefont {Lukin}},\ }\bibfield  {title} {\enquote {\bibinfo {title}
  {Robust multi-qubit quantum network node with integrated error detection},}\
  }\href {\doibase 10.1126/science.add9771} {\bibfield  {journal} {\bibinfo
  {journal} {Science}\ }\textbf {\bibinfo {volume} {378}},\ \bibinfo {pages}
  {557--560} (\bibinfo {year} {2022})}\BibitemShut {NoStop}%
\bibitem [{\citenamefont {Siyushev}\ \emph {et~al.}(2017)\citenamefont
  {Siyushev}, \citenamefont {Metsch}, \citenamefont {Ijaz}, \citenamefont
  {Binder}, \citenamefont {Bhaskar}, \citenamefont {Sukachev}, \citenamefont
  {Sipahigil}, \citenamefont {Evans}, \citenamefont {Nguyen}, \citenamefont
  {Lukin}, \citenamefont {Hemmer}, \citenamefont {Palyanov}, \citenamefont
  {Kupriyanov}, \citenamefont {Borzdov}, \citenamefont {Rogers},\ and\
  \citenamefont {Jelezko}}]{Siyushev2017}%
  \BibitemOpen
  \bibfield  {author} {\bibinfo {author} {\bibfnamefont {Petr}\ \bibnamefont
  {Siyushev}}, \bibinfo {author} {\bibfnamefont {Mathias~H.}\ \bibnamefont
  {Metsch}}, \bibinfo {author} {\bibfnamefont {Aroosa}\ \bibnamefont {Ijaz}},
  \bibinfo {author} {\bibfnamefont {Jan~M.}\ \bibnamefont {Binder}}, \bibinfo
  {author} {\bibfnamefont {Mihir~K.}\ \bibnamefont {Bhaskar}}, \bibinfo
  {author} {\bibfnamefont {Denis~D.}\ \bibnamefont {Sukachev}}, \bibinfo
  {author} {\bibfnamefont {Alp}\ \bibnamefont {Sipahigil}}, \bibinfo {author}
  {\bibfnamefont {Ruffin~E.}\ \bibnamefont {Evans}}, \bibinfo {author}
  {\bibfnamefont {Christian~T.}\ \bibnamefont {Nguyen}}, \bibinfo {author}
  {\bibfnamefont {Mikhail~D.}\ \bibnamefont {Lukin}}, \bibinfo {author}
  {\bibfnamefont {Philip~R.}\ \bibnamefont {Hemmer}}, \bibinfo {author}
  {\bibfnamefont {Yuri~N.}\ \bibnamefont {Palyanov}}, \bibinfo {author}
  {\bibfnamefont {Igor~N.}\ \bibnamefont {Kupriyanov}}, \bibinfo {author}
  {\bibfnamefont {Yuri~M.}\ \bibnamefont {Borzdov}}, \bibinfo {author}
  {\bibfnamefont {Lachlan~J.}\ \bibnamefont {Rogers}}, \ and\ \bibinfo {author}
  {\bibfnamefont {Fedor}\ \bibnamefont {Jelezko}},\ }\bibfield  {title}
  {\enquote {\bibinfo {title} {Optical and microwave control of
  germanium-vacancy center spins in diamond},}\ }\href {\doibase
  10.1103/PhysRevB.96.081201} {\bibfield  {journal} {\bibinfo  {journal} {Phys.
  Rev. B}\ }\textbf {\bibinfo {volume} {96}},\ \bibinfo {pages} {081201}
  (\bibinfo {year} {2017})}\BibitemShut {NoStop}%
\bibitem [{\citenamefont {Debroux}\ \emph {et~al.}(2021)\citenamefont
  {Debroux}, \citenamefont {Michaels}, \citenamefont {Purser}, \citenamefont
  {Wan}, \citenamefont {Trusheim}, \citenamefont {Arjona~Mart\'{\i}nez},
  \citenamefont {Parker}, \citenamefont {Stramma}, \citenamefont {Chen},
  \citenamefont {de~Santis}, \citenamefont {Alexeev}, \citenamefont {Ferrari},
  \citenamefont {Englund}, \citenamefont {Gangloff},\ and\ \citenamefont
  {Atat\"ure}}]{Debroux2021}%
  \BibitemOpen
  \bibfield  {author} {\bibinfo {author} {\bibfnamefont {Romain}\ \bibnamefont
  {Debroux}}, \bibinfo {author} {\bibfnamefont {Cathryn~P.}\ \bibnamefont
  {Michaels}}, \bibinfo {author} {\bibfnamefont {Carola~M.}\ \bibnamefont
  {Purser}}, \bibinfo {author} {\bibfnamefont {Noel}\ \bibnamefont {Wan}},
  \bibinfo {author} {\bibfnamefont {Matthew~E.}\ \bibnamefont {Trusheim}},
  \bibinfo {author} {\bibfnamefont {Jes\'us}\ \bibnamefont
  {Arjona~Mart\'{\i}nez}}, \bibinfo {author} {\bibfnamefont {Ryan~A.}\
  \bibnamefont {Parker}}, \bibinfo {author} {\bibfnamefont {Alexander~M.}\
  \bibnamefont {Stramma}}, \bibinfo {author} {\bibfnamefont {Kevin~C.}\
  \bibnamefont {Chen}}, \bibinfo {author} {\bibfnamefont {Lorenzo}\
  \bibnamefont {de~Santis}}, \bibinfo {author} {\bibfnamefont {Evgeny~M.}\
  \bibnamefont {Alexeev}}, \bibinfo {author} {\bibfnamefont {Andrea~C.}\
  \bibnamefont {Ferrari}}, \bibinfo {author} {\bibfnamefont {Dirk}\
  \bibnamefont {Englund}}, \bibinfo {author} {\bibfnamefont {Dorian~A.}\
  \bibnamefont {Gangloff}}, \ and\ \bibinfo {author} {\bibfnamefont {Mete}\
  \bibnamefont {Atat\"ure}},\ }\bibfield  {title} {\enquote {\bibinfo {title}
  {Quantum control of the tin-vacancy spin qubit in diamond},}\ }\href
  {\doibase 10.1103/PhysRevX.11.041041} {\bibfield  {journal} {\bibinfo
  {journal} {Phys. Rev. X}\ }\textbf {\bibinfo {volume} {11}},\ \bibinfo
  {pages} {041041} (\bibinfo {year} {2021})}\BibitemShut {NoStop}%
\bibitem [{\citenamefont {Maurand}\ \emph {et~al.}(2016)\citenamefont
  {Maurand}, \citenamefont {Jehl}, \citenamefont {Kotekar-Patil}, \citenamefont
  {Corna}, \citenamefont {Bohuslavskyi}, \citenamefont {Lavi{\'e}ville},
  \citenamefont {Hutin}, \citenamefont {Barraud}, \citenamefont {Vinet},
  \citenamefont {Sanquer},\ and\ \citenamefont {De~Franceschi}}]{Maurand2016}%
  \BibitemOpen
  \bibfield  {author} {\bibinfo {author} {\bibfnamefont {R.}~\bibnamefont
  {Maurand}}, \bibinfo {author} {\bibfnamefont {X.}~\bibnamefont {Jehl}},
  \bibinfo {author} {\bibfnamefont {D.}~\bibnamefont {Kotekar-Patil}}, \bibinfo
  {author} {\bibfnamefont {A.}~\bibnamefont {Corna}}, \bibinfo {author}
  {\bibfnamefont {H.}~\bibnamefont {Bohuslavskyi}}, \bibinfo {author}
  {\bibfnamefont {R.}~\bibnamefont {Lavi{\'e}ville}}, \bibinfo {author}
  {\bibfnamefont {L.}~\bibnamefont {Hutin}}, \bibinfo {author} {\bibfnamefont
  {S.}~\bibnamefont {Barraud}}, \bibinfo {author} {\bibfnamefont
  {M.}~\bibnamefont {Vinet}}, \bibinfo {author} {\bibfnamefont
  {M.}~\bibnamefont {Sanquer}}, \ and\ \bibinfo {author} {\bibfnamefont
  {S.}~\bibnamefont {De~Franceschi}},\ }\bibfield  {title} {\enquote {\bibinfo
  {title} {A cmos silicon spin qubit},}\ }\href {\doibase 10.1038/ncomms13575}
  {\bibfield  {journal} {\bibinfo  {journal} {Nature Communications}\ }\textbf
  {\bibinfo {volume} {7}},\ \bibinfo {pages} {13575} (\bibinfo {year}
  {2016})}\BibitemShut {NoStop}%
\bibitem [{\citenamefont {Vandersypen}\ \emph {et~al.}(2017)\citenamefont
  {Vandersypen}, \citenamefont {Bluhm}, \citenamefont {Clarke}, \citenamefont
  {Dzurak}, \citenamefont {Ishihara}, \citenamefont {Morello}, \citenamefont
  {Reilly}, \citenamefont {Schreiber},\ and\ \citenamefont
  {Veldhorst}}]{Vandersypen2017}%
  \BibitemOpen
  \bibfield  {author} {\bibinfo {author} {\bibfnamefont {L.~M.~K.}\
  \bibnamefont {Vandersypen}}, \bibinfo {author} {\bibfnamefont
  {H.}~\bibnamefont {Bluhm}}, \bibinfo {author} {\bibfnamefont {J.~S.}\
  \bibnamefont {Clarke}}, \bibinfo {author} {\bibfnamefont {A.~S.}\
  \bibnamefont {Dzurak}}, \bibinfo {author} {\bibfnamefont {R.}~\bibnamefont
  {Ishihara}}, \bibinfo {author} {\bibfnamefont {A.}~\bibnamefont {Morello}},
  \bibinfo {author} {\bibfnamefont {D.~J.}\ \bibnamefont {Reilly}}, \bibinfo
  {author} {\bibfnamefont {L.~R.}\ \bibnamefont {Schreiber}}, \ and\ \bibinfo
  {author} {\bibfnamefont {M.}~\bibnamefont {Veldhorst}},\ }\bibfield  {title}
  {\enquote {\bibinfo {title} {Interfacing spin qubits in quantum dots and
  donors---hot, dense, and coherent},}\ }\href {\doibase
  10.1038/s41534-017-0038-y} {\bibfield  {journal} {\bibinfo  {journal} {npj
  Quantum Information}\ }\textbf {\bibinfo {volume} {3}},\ \bibinfo {pages}
  {34} (\bibinfo {year} {2017})}\BibitemShut {NoStop}%
\bibitem [{\citenamefont {Fedele}\ \emph {et~al.}(2021)\citenamefont {Fedele},
  \citenamefont {Chatterjee}, \citenamefont {Fallahi}, \citenamefont {Gardner},
  \citenamefont {Manfra},\ and\ \citenamefont {Kuemmeth}}]{Fedele2021}%
  \BibitemOpen
  \bibfield  {author} {\bibinfo {author} {\bibfnamefont {Federico}\
  \bibnamefont {Fedele}}, \bibinfo {author} {\bibfnamefont {Anasua}\
  \bibnamefont {Chatterjee}}, \bibinfo {author} {\bibfnamefont {Saeed}\
  \bibnamefont {Fallahi}}, \bibinfo {author} {\bibfnamefont {Geoffrey~C.}\
  \bibnamefont {Gardner}}, \bibinfo {author} {\bibfnamefont {Michael~J.}\
  \bibnamefont {Manfra}}, \ and\ \bibinfo {author} {\bibfnamefont {Ferdinand}\
  \bibnamefont {Kuemmeth}},\ }\bibfield  {title} {\enquote {\bibinfo {title}
  {Simultaneous operations in a two-dimensional array of singlet-triplet
  qubits},}\ }\href {\doibase 10.1103/PRXQuantum.2.040306} {\bibfield
  {journal} {\bibinfo  {journal} {PRX Quantum}\ }\textbf {\bibinfo {volume}
  {2}},\ \bibinfo {pages} {040306} (\bibinfo {year} {2021})}\BibitemShut
  {NoStop}%
\bibitem [{\citenamefont {Zajac}\ \emph {et~al.}(2018)\citenamefont {Zajac},
  \citenamefont {Sigillito}, \citenamefont {Russ}, \citenamefont {Borjans},
  \citenamefont {Taylor}, \citenamefont {Burkard},\ and\ \citenamefont
  {Petta}}]{Zajac2018}%
  \BibitemOpen
  \bibfield  {author} {\bibinfo {author} {\bibfnamefont {D.~M.}\ \bibnamefont
  {Zajac}}, \bibinfo {author} {\bibfnamefont {A.~J.}\ \bibnamefont
  {Sigillito}}, \bibinfo {author} {\bibfnamefont {M.}~\bibnamefont {Russ}},
  \bibinfo {author} {\bibfnamefont {F.}~\bibnamefont {Borjans}}, \bibinfo
  {author} {\bibfnamefont {J.~M.}\ \bibnamefont {Taylor}}, \bibinfo {author}
  {\bibfnamefont {G.}~\bibnamefont {Burkard}}, \ and\ \bibinfo {author}
  {\bibfnamefont {J.~R.}\ \bibnamefont {Petta}},\ }\bibfield  {title} {\enquote
  {\bibinfo {title} {{Resonantly driven CNOT gate for electron spins}},}\
  }\href {\doibase 10.1126/science.aao5965} {\bibfield  {journal} {\bibinfo
  {journal} {Science}\ }\textbf {\bibinfo {volume} {359}},\ \bibinfo {pages}
  {439--442} (\bibinfo {year} {2018})}\BibitemShut {NoStop}%
\bibitem [{\citenamefont {Mills}\ \emph {et~al.}(2022)\citenamefont {Mills},
  \citenamefont {Guinn}, \citenamefont {Gullans}, \citenamefont {Sigillito},
  \citenamefont {Feldman}, \citenamefont {Nielsen},\ and\ \citenamefont
  {Petta}}]{Mills2022}%
  \BibitemOpen
  \bibfield  {author} {\bibinfo {author} {\bibfnamefont {Adam~R.}\ \bibnamefont
  {Mills}}, \bibinfo {author} {\bibfnamefont {Charles~R.}\ \bibnamefont
  {Guinn}}, \bibinfo {author} {\bibfnamefont {Michael~J.}\ \bibnamefont
  {Gullans}}, \bibinfo {author} {\bibfnamefont {Anthony~J.}\ \bibnamefont
  {Sigillito}}, \bibinfo {author} {\bibfnamefont {Mayer~M.}\ \bibnamefont
  {Feldman}}, \bibinfo {author} {\bibfnamefont {Erik}\ \bibnamefont {Nielsen}},
  \ and\ \bibinfo {author} {\bibfnamefont {Jason~R.}\ \bibnamefont {Petta}},\
  }\bibfield  {title} {\enquote {\bibinfo {title} {Two-qubit silicon quantum
  processor with operation fidelity exceeding 99\%},}\ }\href {\doibase
  10.1126/sciadv.abn5130} {\bibfield  {journal} {\bibinfo  {journal} {Science
  Advances}\ }\textbf {\bibinfo {volume} {8}},\ \bibinfo {pages} {eabn5130}
  (\bibinfo {year} {2022})}\BibitemShut {NoStop}%
\bibitem [{\citenamefont {Burkard}\ \emph {et~al.}(2023)\citenamefont
  {Burkard}, \citenamefont {Ladd}, \citenamefont {Pan}, \citenamefont
  {Nichol},\ and\ \citenamefont {Petta}}]{Burkard2023}%
  \BibitemOpen
  \bibfield  {author} {\bibinfo {author} {\bibfnamefont {Guido}\ \bibnamefont
  {Burkard}}, \bibinfo {author} {\bibfnamefont {Thaddeus~D.}\ \bibnamefont
  {Ladd}}, \bibinfo {author} {\bibfnamefont {Andrew}\ \bibnamefont {Pan}},
  \bibinfo {author} {\bibfnamefont {John~M.}\ \bibnamefont {Nichol}}, \ and\
  \bibinfo {author} {\bibfnamefont {Jason~R.}\ \bibnamefont {Petta}},\
  }\bibfield  {title} {\enquote {\bibinfo {title} {Semiconductor spin
  qubits},}\ }\href {\doibase 10.1103/RevModPhys.95.025003} {\bibfield
  {journal} {\bibinfo  {journal} {Rev. Mod. Phys.}\ }\textbf {\bibinfo {volume}
  {95}},\ \bibinfo {pages} {025003} (\bibinfo {year} {2023})}\BibitemShut
  {NoStop}%
\bibitem [{\citenamefont {Loss}\ and\ \citenamefont
  {DiVincenzo}(1998)}]{Loss1998}%
  \BibitemOpen
  \bibfield  {author} {\bibinfo {author} {\bibfnamefont {Daniel}\ \bibnamefont
  {Loss}}\ and\ \bibinfo {author} {\bibfnamefont {David~P.}\ \bibnamefont
  {DiVincenzo}},\ }\bibfield  {title} {\enquote {\bibinfo {title} {Quantum
  computation with quantum dots},}\ }\href {\doibase 10.1103/PhysRevA.57.120}
  {\bibfield  {journal} {\bibinfo  {journal} {Phys. Rev. A}\ }\textbf {\bibinfo
  {volume} {57}},\ \bibinfo {pages} {120--126} (\bibinfo {year}
  {1998})}\BibitemShut {NoStop}%
\bibitem [{\citenamefont {Xue}\ \emph {et~al.}(2022)\citenamefont {Xue},
  \citenamefont {Russ}, \citenamefont {Samkharadze}, \citenamefont {Undseth},
  \citenamefont {Sammak}, \citenamefont {Scappucci},\ and\ \citenamefont
  {Vandersypen}}]{Xue2022}%
  \BibitemOpen
  \bibfield  {author} {\bibinfo {author} {\bibfnamefont {Xiao}\ \bibnamefont
  {Xue}}, \bibinfo {author} {\bibfnamefont {Maximilian}\ \bibnamefont {Russ}},
  \bibinfo {author} {\bibfnamefont {Nodar}\ \bibnamefont {Samkharadze}},
  \bibinfo {author} {\bibfnamefont {Brennan}\ \bibnamefont {Undseth}}, \bibinfo
  {author} {\bibfnamefont {Amir}\ \bibnamefont {Sammak}}, \bibinfo {author}
  {\bibfnamefont {Giordano}\ \bibnamefont {Scappucci}}, \ and\ \bibinfo
  {author} {\bibfnamefont {Lieven M.~K.}\ \bibnamefont {Vandersypen}},\
  }\bibfield  {title} {\enquote {\bibinfo {title} {Quantum logic with spin
  qubits crossing the surface code threshold},}\ }\href {\doibase
  10.1038/s41586-021-04273-w} {\bibfield  {journal} {\bibinfo  {journal}
  {Nature}\ }\textbf {\bibinfo {volume} {601}},\ \bibinfo {pages} {343--347}
  (\bibinfo {year} {2022})}\BibitemShut {NoStop}%
\bibitem [{\citenamefont {Yoneda}\ \emph {et~al.}(2018)\citenamefont {Yoneda},
  \citenamefont {Takeda}, \citenamefont {Otsuka}, \citenamefont {Nakajima},
  \citenamefont {Delbecq}, \citenamefont {Allison}, \citenamefont {Honda},
  \citenamefont {Kodera}, \citenamefont {Oda}, \citenamefont {Hoshi},
  \citenamefont {Usami}, \citenamefont {Itoh},\ and\ \citenamefont
  {Tarucha}}]{Yoneda2018}%
  \BibitemOpen
  \bibfield  {author} {\bibinfo {author} {\bibfnamefont {Jun}\ \bibnamefont
  {Yoneda}}, \bibinfo {author} {\bibfnamefont {Kenta}\ \bibnamefont {Takeda}},
  \bibinfo {author} {\bibfnamefont {Tomohiro}\ \bibnamefont {Otsuka}}, \bibinfo
  {author} {\bibfnamefont {Takashi}\ \bibnamefont {Nakajima}}, \bibinfo
  {author} {\bibfnamefont {Matthieu~R.}\ \bibnamefont {Delbecq}}, \bibinfo
  {author} {\bibfnamefont {Giles}\ \bibnamefont {Allison}}, \bibinfo {author}
  {\bibfnamefont {Takumu}\ \bibnamefont {Honda}}, \bibinfo {author}
  {\bibfnamefont {Tetsuo}\ \bibnamefont {Kodera}}, \bibinfo {author}
  {\bibfnamefont {Shunri}\ \bibnamefont {Oda}}, \bibinfo {author}
  {\bibfnamefont {Yusuke}\ \bibnamefont {Hoshi}}, \bibinfo {author}
  {\bibfnamefont {Noritaka}\ \bibnamefont {Usami}}, \bibinfo {author}
  {\bibfnamefont {Kohei~M.}\ \bibnamefont {Itoh}}, \ and\ \bibinfo {author}
  {\bibfnamefont {Seigo}\ \bibnamefont {Tarucha}},\ }\bibfield  {title}
  {\enquote {\bibinfo {title} {A quantum-dot spin qubit with coherence limited
  by charge noise and fidelity higher than 99.9{\%}},}\ }\href {\doibase
  10.1038/s41565-017-0014-x} {\bibfield  {journal} {\bibinfo  {journal} {Nature
  Nanotechnology}\ }\textbf {\bibinfo {volume} {13}},\ \bibinfo {pages}
  {102--106} (\bibinfo {year} {2018})}\BibitemShut {NoStop}%
\bibitem [{\citenamefont {Tanttu}\ \emph {et~al.}(2023)\citenamefont {Tanttu},
  \citenamefont {Lim}, \citenamefont {Huang}, \citenamefont {Stuyck},
  \citenamefont {Gilbert}, \citenamefont {Su}, \citenamefont {Feng},
  \citenamefont {Cifuentes}, \citenamefont {Seedhouse}, \citenamefont
  {Seritan}, \citenamefont {Ostrove}, \citenamefont {Rudinger}, \citenamefont
  {Leon}, \citenamefont {Huang}, \citenamefont {Escott}, \citenamefont {Itoh},
  \citenamefont {Abrosimov}, \citenamefont {Pohl}, \citenamefont {Thewalt},
  \citenamefont {Hudson}, \citenamefont {Blume-Kohout}, \citenamefont
  {Bartlett}, \citenamefont {Morello}, \citenamefont {Laucht}, \citenamefont
  {Yang}, \citenamefont {Saraiva},\ and\ \citenamefont {Dzurak}}]{Tanttu2023}%
  \BibitemOpen
  \bibfield  {author} {\bibinfo {author} {\bibfnamefont {Tuomo}\ \bibnamefont
  {Tanttu}}, \bibinfo {author} {\bibfnamefont {Wee~Han}\ \bibnamefont {Lim}},
  \bibinfo {author} {\bibfnamefont {Jonathan~Y.}\ \bibnamefont {Huang}},
  \bibinfo {author} {\bibfnamefont {Nard~Dumoulin}\ \bibnamefont {Stuyck}},
  \bibinfo {author} {\bibfnamefont {Will}\ \bibnamefont {Gilbert}}, \bibinfo
  {author} {\bibfnamefont {Rocky~Y.}\ \bibnamefont {Su}}, \bibinfo {author}
  {\bibfnamefont {MengKe}\ \bibnamefont {Feng}}, \bibinfo {author}
  {\bibfnamefont {Jesus~D.}\ \bibnamefont {Cifuentes}}, \bibinfo {author}
  {\bibfnamefont {Amanda~E.}\ \bibnamefont {Seedhouse}}, \bibinfo {author}
  {\bibfnamefont {Stefan~K.}\ \bibnamefont {Seritan}}, \bibinfo {author}
  {\bibfnamefont {Corey~I.}\ \bibnamefont {Ostrove}}, \bibinfo {author}
  {\bibfnamefont {Kenneth~M.}\ \bibnamefont {Rudinger}}, \bibinfo {author}
  {\bibfnamefont {Ross C.~C.}\ \bibnamefont {Leon}}, \bibinfo {author}
  {\bibfnamefont {Wister}\ \bibnamefont {Huang}}, \bibinfo {author}
  {\bibfnamefont {Christopher~C.}\ \bibnamefont {Escott}}, \bibinfo {author}
  {\bibfnamefont {Kohei~M.}\ \bibnamefont {Itoh}}, \bibinfo {author}
  {\bibfnamefont {Nikolay~V.}\ \bibnamefont {Abrosimov}}, \bibinfo {author}
  {\bibfnamefont {Hans-Joachim}\ \bibnamefont {Pohl}}, \bibinfo {author}
  {\bibfnamefont {Michael L.~W.}\ \bibnamefont {Thewalt}}, \bibinfo {author}
  {\bibfnamefont {Fay~E.}\ \bibnamefont {Hudson}}, \bibinfo {author}
  {\bibfnamefont {Robin}\ \bibnamefont {Blume-Kohout}}, \bibinfo {author}
  {\bibfnamefont {Stephen~D.}\ \bibnamefont {Bartlett}}, \bibinfo {author}
  {\bibfnamefont {Andrea}\ \bibnamefont {Morello}}, \bibinfo {author}
  {\bibfnamefont {Arne}\ \bibnamefont {Laucht}}, \bibinfo {author}
  {\bibfnamefont {Chih~Hwan}\ \bibnamefont {Yang}}, \bibinfo {author}
  {\bibfnamefont {Andre}\ \bibnamefont {Saraiva}}, \ and\ \bibinfo {author}
  {\bibfnamefont {Andrew~S.}\ \bibnamefont {Dzurak}},\ }\href@noop {} {\enquote
  {\bibinfo {title} {Stability of high-fidelity two-qubit operations in
  silicon},}\ } (\bibinfo {year} {2023}),\ \Eprint
  {http://arxiv.org/abs/2303.04090} {arXiv:2303.04090 [quant-ph]} \BibitemShut
  {NoStop}%
\bibitem [{\citenamefont {Noiri}\ \emph {et~al.}(2022)\citenamefont {Noiri},
  \citenamefont {Takeda}, \citenamefont {Nakajima}, \citenamefont {Kobayashi},
  \citenamefont {Sammak}, \citenamefont {Scappucci},\ and\ \citenamefont
  {Tarucha}}]{Noiri2022}%
  \BibitemOpen
  \bibfield  {author} {\bibinfo {author} {\bibfnamefont {Akito}\ \bibnamefont
  {Noiri}}, \bibinfo {author} {\bibfnamefont {Kenta}\ \bibnamefont {Takeda}},
  \bibinfo {author} {\bibfnamefont {Takashi}\ \bibnamefont {Nakajima}},
  \bibinfo {author} {\bibfnamefont {Takashi}\ \bibnamefont {Kobayashi}},
  \bibinfo {author} {\bibfnamefont {Amir}\ \bibnamefont {Sammak}}, \bibinfo
  {author} {\bibfnamefont {Giordano}\ \bibnamefont {Scappucci}}, \ and\
  \bibinfo {author} {\bibfnamefont {Seigo}\ \bibnamefont {Tarucha}},\
  }\bibfield  {title} {\enquote {\bibinfo {title} {Fast universal quantum gate
  above the fault-tolerance threshold in silicon},}\ }\href {\doibase
  10.1038/s41586-021-04182-y} {\bibfield  {journal} {\bibinfo  {journal}
  {Nature}\ }\textbf {\bibinfo {volume} {601}},\ \bibinfo {pages} {338--342}
  (\bibinfo {year} {2022})}\BibitemShut {NoStop}%
\bibitem [{\citenamefont {Cerfontaine}\ \emph {et~al.}(2020)\citenamefont
  {Cerfontaine}, \citenamefont {Botzem}, \citenamefont {Ritzmann},
  \citenamefont {Humpohl}, \citenamefont {Ludwig}, \citenamefont {Schuh},
  \citenamefont {Bougeard}, \citenamefont {Wieck},\ and\ \citenamefont
  {Bluhm}}]{Cerfontaine2020}%
  \BibitemOpen
  \bibfield  {author} {\bibinfo {author} {\bibfnamefont {Pascal}\ \bibnamefont
  {Cerfontaine}}, \bibinfo {author} {\bibfnamefont {Tim}\ \bibnamefont
  {Botzem}}, \bibinfo {author} {\bibfnamefont {Julian}\ \bibnamefont
  {Ritzmann}}, \bibinfo {author} {\bibfnamefont {Simon~Sebastian}\ \bibnamefont
  {Humpohl}}, \bibinfo {author} {\bibfnamefont {Arne}\ \bibnamefont {Ludwig}},
  \bibinfo {author} {\bibfnamefont {Dieter}\ \bibnamefont {Schuh}}, \bibinfo
  {author} {\bibfnamefont {Dominique}\ \bibnamefont {Bougeard}}, \bibinfo
  {author} {\bibfnamefont {Andreas~D.}\ \bibnamefont {Wieck}}, \ and\ \bibinfo
  {author} {\bibfnamefont {Hendrik}\ \bibnamefont {Bluhm}},\ }\bibfield
  {title} {\enquote {\bibinfo {title} {Closed-loop control of a gaas-based
  singlet-triplet spin qubit with 99.5{\%} gate fidelity and low leakage},}\
  }\href {\doibase 10.1038/s41467-020-17865-3} {\bibfield  {journal} {\bibinfo
  {journal} {Nature Communications}\ }\textbf {\bibinfo {volume} {11}},\
  \bibinfo {pages} {4144} (\bibinfo {year} {2020})}\BibitemShut {NoStop}%
\bibitem [{\citenamefont {Takeda}\ \emph {et~al.}(2020)\citenamefont {Takeda},
  \citenamefont {Noiri}, \citenamefont {Yoneda}, \citenamefont {Nakajima},\
  and\ \citenamefont {Tarucha}}]{Takeda2020}%
  \BibitemOpen
  \bibfield  {author} {\bibinfo {author} {\bibfnamefont {K.}~\bibnamefont
  {Takeda}}, \bibinfo {author} {\bibfnamefont {A.}~\bibnamefont {Noiri}},
  \bibinfo {author} {\bibfnamefont {J.}~\bibnamefont {Yoneda}}, \bibinfo
  {author} {\bibfnamefont {T.}~\bibnamefont {Nakajima}}, \ and\ \bibinfo
  {author} {\bibfnamefont {S.}~\bibnamefont {Tarucha}},\ }\bibfield  {title}
  {\enquote {\bibinfo {title} {Resonantly driven singlet-triplet spin qubit in
  silicon},}\ }\href {\doibase 10.1103/PhysRevLett.124.117701} {\bibfield
  {journal} {\bibinfo  {journal} {Phys. Rev. Lett.}\ }\textbf {\bibinfo
  {volume} {124}},\ \bibinfo {pages} {117701} (\bibinfo {year}
  {2020})}\BibitemShut {NoStop}%
\bibitem [{\citenamefont {Nichol}\ \emph {et~al.}(2017)\citenamefont {Nichol},
  \citenamefont {Orona}, \citenamefont {Harvey}, \citenamefont {Fallahi},
  \citenamefont {Gardner}, \citenamefont {Manfra},\ and\ \citenamefont
  {Yacoby}}]{Nichol2017}%
  \BibitemOpen
  \bibfield  {author} {\bibinfo {author} {\bibfnamefont {John~M.}\ \bibnamefont
  {Nichol}}, \bibinfo {author} {\bibfnamefont {Lucas~A.}\ \bibnamefont
  {Orona}}, \bibinfo {author} {\bibfnamefont {Shannon~P.}\ \bibnamefont
  {Harvey}}, \bibinfo {author} {\bibfnamefont {Saeed}\ \bibnamefont {Fallahi}},
  \bibinfo {author} {\bibfnamefont {Geoffrey~C.}\ \bibnamefont {Gardner}},
  \bibinfo {author} {\bibfnamefont {Michael~J.}\ \bibnamefont {Manfra}}, \ and\
  \bibinfo {author} {\bibfnamefont {Amir}\ \bibnamefont {Yacoby}},\ }\bibfield
  {title} {\enquote {\bibinfo {title} {High-fidelity entangling gate for
  double-quantum-dot spin qubits},}\ }\href {\doibase
  10.1038/s41534-016-0003-1} {\bibfield  {journal} {\bibinfo  {journal} {npj
  Quantum Information}\ }\textbf {\bibinfo {volume} {3}},\ \bibinfo {pages} {3}
  (\bibinfo {year} {2017})}\BibitemShut {NoStop}%
\bibitem [{\citenamefont {Weinstein}\ \emph {et~al.}(2023)\citenamefont
  {Weinstein}, \citenamefont {Reed}, \citenamefont {Jones}, \citenamefont
  {Andrews}, \citenamefont {Barnes}, \citenamefont {Blumoff}, \citenamefont
  {Euliss}, \citenamefont {Eng}, \citenamefont {Fong}, \citenamefont {Ha},
  \citenamefont {Hulbert}, \citenamefont {Jackson}, \citenamefont {Jura},
  \citenamefont {Keating}, \citenamefont {Kerckhoff}, \citenamefont {Kiselev},
  \citenamefont {Matten}, \citenamefont {Sabbir}, \citenamefont {Smith},
  \citenamefont {Wright}, \citenamefont {Rakher}, \citenamefont {Ladd},\ and\
  \citenamefont {Borselli}}]{Weinstein2023}%
  \BibitemOpen
  \bibfield  {author} {\bibinfo {author} {\bibfnamefont {Aaron~J.}\
  \bibnamefont {Weinstein}}, \bibinfo {author} {\bibfnamefont {Matthew~D.}\
  \bibnamefont {Reed}}, \bibinfo {author} {\bibfnamefont {Aaron~M.}\
  \bibnamefont {Jones}}, \bibinfo {author} {\bibfnamefont {Reed~W.}\
  \bibnamefont {Andrews}}, \bibinfo {author} {\bibfnamefont {David}\
  \bibnamefont {Barnes}}, \bibinfo {author} {\bibfnamefont {Jacob~Z.}\
  \bibnamefont {Blumoff}}, \bibinfo {author} {\bibfnamefont {Larken~E.}\
  \bibnamefont {Euliss}}, \bibinfo {author} {\bibfnamefont {Kevin}\
  \bibnamefont {Eng}}, \bibinfo {author} {\bibfnamefont {Bryan~H.}\
  \bibnamefont {Fong}}, \bibinfo {author} {\bibfnamefont {Sieu~D.}\
  \bibnamefont {Ha}}, \bibinfo {author} {\bibfnamefont {Daniel~R.}\
  \bibnamefont {Hulbert}}, \bibinfo {author} {\bibfnamefont {Clayton A.~C.}\
  \bibnamefont {Jackson}}, \bibinfo {author} {\bibfnamefont {Michael}\
  \bibnamefont {Jura}}, \bibinfo {author} {\bibfnamefont {Tyler~E.}\
  \bibnamefont {Keating}}, \bibinfo {author} {\bibfnamefont {Joseph}\
  \bibnamefont {Kerckhoff}}, \bibinfo {author} {\bibfnamefont {Andrey~A.}\
  \bibnamefont {Kiselev}}, \bibinfo {author} {\bibfnamefont {Justine}\
  \bibnamefont {Matten}}, \bibinfo {author} {\bibfnamefont {Golam}\
  \bibnamefont {Sabbir}}, \bibinfo {author} {\bibfnamefont {Aaron}\
  \bibnamefont {Smith}}, \bibinfo {author} {\bibfnamefont {Jeffrey}\
  \bibnamefont {Wright}}, \bibinfo {author} {\bibfnamefont {Matthew~T.}\
  \bibnamefont {Rakher}}, \bibinfo {author} {\bibfnamefont {Thaddeus~D.}\
  \bibnamefont {Ladd}}, \ and\ \bibinfo {author} {\bibfnamefont {Matthew~G.}\
  \bibnamefont {Borselli}},\ }\bibfield  {title} {\enquote {\bibinfo {title}
  {Universal logic with encoded spin qubits in silicon},}\ }\href {\doibase
  10.1038/s41586-023-05777-3} {\bibfield  {journal} {\bibinfo  {journal}
  {Nature}\ }\textbf {\bibinfo {volume} {615}},\ \bibinfo {pages} {817--822}
  (\bibinfo {year} {2023})}\BibitemShut {NoStop}%
\bibitem [{\citenamefont {DiVincenzo}\ \emph {et~al.}(2000)\citenamefont
  {DiVincenzo}, \citenamefont {Bacon}, \citenamefont {Kempe}, \citenamefont
  {Burkard},\ and\ \citenamefont {Whaley}}]{DiVincenzo2000}%
  \BibitemOpen
  \bibfield  {author} {\bibinfo {author} {\bibfnamefont {D.~P.}\ \bibnamefont
  {DiVincenzo}}, \bibinfo {author} {\bibfnamefont {D.}~\bibnamefont {Bacon}},
  \bibinfo {author} {\bibfnamefont {J.}~\bibnamefont {Kempe}}, \bibinfo
  {author} {\bibfnamefont {G.}~\bibnamefont {Burkard}}, \ and\ \bibinfo
  {author} {\bibfnamefont {K.~B.}\ \bibnamefont {Whaley}},\ }\bibfield  {title}
  {\enquote {\bibinfo {title} {Universal quantum computation with the exchange
  interaction},}\ }\href {\doibase 10.1038/35042541} {\bibfield  {journal}
  {\bibinfo  {journal} {Nature}\ }\textbf {\bibinfo {volume} {408}},\ \bibinfo
  {pages} {339--342} (\bibinfo {year} {2000})}\BibitemShut {NoStop}%
\bibitem [{\citenamefont {Bacon}\ \emph {et~al.}(2000)\citenamefont {Bacon},
  \citenamefont {Kempe}, \citenamefont {Lidar},\ and\ \citenamefont
  {Whaley}}]{Bacon2000}%
  \BibitemOpen
  \bibfield  {author} {\bibinfo {author} {\bibfnamefont {D.}~\bibnamefont
  {Bacon}}, \bibinfo {author} {\bibfnamefont {J.}~\bibnamefont {Kempe}},
  \bibinfo {author} {\bibfnamefont {D.~A.}\ \bibnamefont {Lidar}}, \ and\
  \bibinfo {author} {\bibfnamefont {K.~B.}\ \bibnamefont {Whaley}},\ }\bibfield
   {title} {\enquote {\bibinfo {title} {Universal fault-tolerant quantum
  computation on decoherence-free subspaces},}\ }\href {\doibase
  10.1103/PhysRevLett.85.1758} {\bibfield  {journal} {\bibinfo  {journal}
  {Phys. Rev. Lett.}\ }\textbf {\bibinfo {volume} {85}},\ \bibinfo {pages}
  {1758--1761} (\bibinfo {year} {2000})}\BibitemShut {NoStop}%
\bibitem [{\citenamefont {Stockklauser}\ \emph {et~al.}(2017)\citenamefont
  {Stockklauser}, \citenamefont {Scarlino}, \citenamefont {Koski},
  \citenamefont {Gasparinetti}, \citenamefont {Andersen}, \citenamefont
  {Reichl}, \citenamefont {Wegscheider}, \citenamefont {Ihn}, \citenamefont
  {Ensslin},\ and\ \citenamefont {Wallraff}}]{Stockklauser2017}%
  \BibitemOpen
  \bibfield  {author} {\bibinfo {author} {\bibfnamefont {A.}~\bibnamefont
  {Stockklauser}}, \bibinfo {author} {\bibfnamefont {P.}~\bibnamefont
  {Scarlino}}, \bibinfo {author} {\bibfnamefont {J.~V.}\ \bibnamefont {Koski}},
  \bibinfo {author} {\bibfnamefont {S.}~\bibnamefont {Gasparinetti}}, \bibinfo
  {author} {\bibfnamefont {C.~K.}\ \bibnamefont {Andersen}}, \bibinfo {author}
  {\bibfnamefont {C.}~\bibnamefont {Reichl}}, \bibinfo {author} {\bibfnamefont
  {W.}~\bibnamefont {Wegscheider}}, \bibinfo {author} {\bibfnamefont
  {T.}~\bibnamefont {Ihn}}, \bibinfo {author} {\bibfnamefont {K.}~\bibnamefont
  {Ensslin}}, \ and\ \bibinfo {author} {\bibfnamefont {A.}~\bibnamefont
  {Wallraff}},\ }\bibfield  {title} {\enquote {\bibinfo {title} {Strong
  coupling cavity qed with gate-defined double quantum dots enabled by a high
  impedance resonator},}\ }\href {\doibase 10.1103/PhysRevX.7.011030}
  {\bibfield  {journal} {\bibinfo  {journal} {Phys. Rev. X}\ }\textbf {\bibinfo
  {volume} {7}},\ \bibinfo {pages} {011030} (\bibinfo {year}
  {2017})}\BibitemShut {NoStop}%
\bibitem [{\citenamefont {Mi}\ \emph {et~al.}(2017)\citenamefont {Mi},
  \citenamefont {Cady}, \citenamefont {Zajac}, \citenamefont {Deelman},\ and\
  \citenamefont {Petta}}]{Mi2017}%
  \BibitemOpen
  \bibfield  {author} {\bibinfo {author} {\bibfnamefont {X.}~\bibnamefont
  {Mi}}, \bibinfo {author} {\bibfnamefont {J.~V.}\ \bibnamefont {Cady}},
  \bibinfo {author} {\bibfnamefont {D.~M.}\ \bibnamefont {Zajac}}, \bibinfo
  {author} {\bibfnamefont {P.~W.}\ \bibnamefont {Deelman}}, \ and\ \bibinfo
  {author} {\bibfnamefont {J.~R.}\ \bibnamefont {Petta}},\ }\bibfield  {title}
  {\enquote {\bibinfo {title} {Strong coupling of a single electron in silicon
  to a microwave photon},}\ }\href {\doibase 10.1126/science.aal2469}
  {\bibfield  {journal} {\bibinfo  {journal} {Science}\ }\textbf {\bibinfo
  {volume} {355}},\ \bibinfo {pages} {156--158} (\bibinfo {year}
  {2017})}\BibitemShut {NoStop}%
\bibitem [{\citenamefont {Scarlino}\ \emph {et~al.}(2019)\citenamefont
  {Scarlino}, \citenamefont {van Woerkom}, \citenamefont {Mendes},
  \citenamefont {Koski}, \citenamefont {Landig}, \citenamefont {Andersen},
  \citenamefont {Gasparinetti}, \citenamefont {Reichl}, \citenamefont
  {Wegscheider}, \citenamefont {Ensslin}, \citenamefont {Ihn}, \citenamefont
  {Blais},\ and\ \citenamefont {Wallraff}}]{Scarlino2019}%
  \BibitemOpen
  \bibfield  {author} {\bibinfo {author} {\bibfnamefont {P.}~\bibnamefont
  {Scarlino}}, \bibinfo {author} {\bibfnamefont {D.~J.}\ \bibnamefont {van
  Woerkom}}, \bibinfo {author} {\bibfnamefont {U.~C.}\ \bibnamefont {Mendes}},
  \bibinfo {author} {\bibfnamefont {J.~V.}\ \bibnamefont {Koski}}, \bibinfo
  {author} {\bibfnamefont {A.~J.}\ \bibnamefont {Landig}}, \bibinfo {author}
  {\bibfnamefont {C.~K.}\ \bibnamefont {Andersen}}, \bibinfo {author}
  {\bibfnamefont {S.}~\bibnamefont {Gasparinetti}}, \bibinfo {author}
  {\bibfnamefont {C.}~\bibnamefont {Reichl}}, \bibinfo {author} {\bibfnamefont
  {W.}~\bibnamefont {Wegscheider}}, \bibinfo {author} {\bibfnamefont
  {K.}~\bibnamefont {Ensslin}}, \bibinfo {author} {\bibfnamefont
  {T.}~\bibnamefont {Ihn}}, \bibinfo {author} {\bibfnamefont {A.}~\bibnamefont
  {Blais}}, \ and\ \bibinfo {author} {\bibfnamefont {A.}~\bibnamefont
  {Wallraff}},\ }\bibfield  {title} {\enquote {\bibinfo {title} {Coherent
  microwave-photon-mediated coupling between a semiconductor and a
  superconducting qubit},}\ }\href {\doibase 10.1038/s41467-019-10798-6}
  {\bibfield  {journal} {\bibinfo  {journal} {Nature Communications}\ }\textbf
  {\bibinfo {volume} {10}},\ \bibinfo {pages} {3011} (\bibinfo {year}
  {2019})}\BibitemShut {NoStop}%
\bibitem [{\citenamefont {Mi}\ \emph {et~al.}(2018)\citenamefont {Mi},
  \citenamefont {Benito}, \citenamefont {Putz}, \citenamefont {Zajac},
  \citenamefont {Taylor}, \citenamefont {Burkard},\ and\ \citenamefont
  {Petta}}]{Mi2018}%
  \BibitemOpen
  \bibfield  {author} {\bibinfo {author} {\bibfnamefont {X.}~\bibnamefont
  {Mi}}, \bibinfo {author} {\bibfnamefont {M.}~\bibnamefont {Benito}}, \bibinfo
  {author} {\bibfnamefont {S.}~\bibnamefont {Putz}}, \bibinfo {author}
  {\bibfnamefont {D.~M.}\ \bibnamefont {Zajac}}, \bibinfo {author}
  {\bibfnamefont {J.~M.}\ \bibnamefont {Taylor}}, \bibinfo {author}
  {\bibfnamefont {Guido}\ \bibnamefont {Burkard}}, \ and\ \bibinfo {author}
  {\bibfnamefont {J.~R.}\ \bibnamefont {Petta}},\ }\bibfield  {title} {\enquote
  {\bibinfo {title} {A coherent spin--photon interface in silicon},}\ }\href
  {\doibase 10.1038/nature25769} {\bibfield  {journal} {\bibinfo  {journal}
  {Nature}\ }\textbf {\bibinfo {volume} {555}},\ \bibinfo {pages} {599--603}
  (\bibinfo {year} {2018})}\BibitemShut {NoStop}%
\bibitem [{\citenamefont {Landig}\ \emph {et~al.}(2018)\citenamefont {Landig},
  \citenamefont {Koski}, \citenamefont {Scarlino}, \citenamefont {Mendes},
  \citenamefont {Blais}, \citenamefont {Reichl}, \citenamefont {Wegscheider},
  \citenamefont {Wallraff}, \citenamefont {Ensslin},\ and\ \citenamefont
  {Ihn}}]{Landig2018}%
  \BibitemOpen
  \bibfield  {author} {\bibinfo {author} {\bibfnamefont {A.~J.}\ \bibnamefont
  {Landig}}, \bibinfo {author} {\bibfnamefont {J.~V.}\ \bibnamefont {Koski}},
  \bibinfo {author} {\bibfnamefont {P.}~\bibnamefont {Scarlino}}, \bibinfo
  {author} {\bibfnamefont {U.~C.}\ \bibnamefont {Mendes}}, \bibinfo {author}
  {\bibfnamefont {A.}~\bibnamefont {Blais}}, \bibinfo {author} {\bibfnamefont
  {C.}~\bibnamefont {Reichl}}, \bibinfo {author} {\bibfnamefont
  {W.}~\bibnamefont {Wegscheider}}, \bibinfo {author} {\bibfnamefont
  {A.}~\bibnamefont {Wallraff}}, \bibinfo {author} {\bibfnamefont
  {K.}~\bibnamefont {Ensslin}}, \ and\ \bibinfo {author} {\bibfnamefont
  {T.}~\bibnamefont {Ihn}},\ }\bibfield  {title} {\enquote {\bibinfo {title}
  {Coherent spin--photon coupling using a resonant exchange qubit},}\ }\href
  {\doibase 10.1038/s41586-018-0365-y} {\bibfield  {journal} {\bibinfo
  {journal} {Nature}\ }\textbf {\bibinfo {volume} {560}},\ \bibinfo {pages}
  {179--184} (\bibinfo {year} {2018})}\BibitemShut {NoStop}%
\bibitem [{\citenamefont {Landig}\ \emph {et~al.}(2019)\citenamefont {Landig},
  \citenamefont {Koski}, \citenamefont {Scarlino}, \citenamefont {M{\"u}ller},
  \citenamefont {Abadillo-Uriel}, \citenamefont {Kratochwil}, \citenamefont
  {Reichl}, \citenamefont {Wegscheider}, \citenamefont {Coppersmith},
  \citenamefont {Friesen}, \citenamefont {Wallraff}, \citenamefont {Ihn},\ and\
  \citenamefont {Ensslin}}]{Landig2019}%
  \BibitemOpen
  \bibfield  {author} {\bibinfo {author} {\bibfnamefont {A.~J.}\ \bibnamefont
  {Landig}}, \bibinfo {author} {\bibfnamefont {J.~V.}\ \bibnamefont {Koski}},
  \bibinfo {author} {\bibfnamefont {P.}~\bibnamefont {Scarlino}}, \bibinfo
  {author} {\bibfnamefont {C.}~\bibnamefont {M{\"u}ller}}, \bibinfo {author}
  {\bibfnamefont {J.~C.}\ \bibnamefont {Abadillo-Uriel}}, \bibinfo {author}
  {\bibfnamefont {B.}~\bibnamefont {Kratochwil}}, \bibinfo {author}
  {\bibfnamefont {C.}~\bibnamefont {Reichl}}, \bibinfo {author} {\bibfnamefont
  {W.}~\bibnamefont {Wegscheider}}, \bibinfo {author} {\bibfnamefont {S.~N.}\
  \bibnamefont {Coppersmith}}, \bibinfo {author} {\bibfnamefont {Mark}\
  \bibnamefont {Friesen}}, \bibinfo {author} {\bibfnamefont {A.}~\bibnamefont
  {Wallraff}}, \bibinfo {author} {\bibfnamefont {T.}~\bibnamefont {Ihn}}, \
  and\ \bibinfo {author} {\bibfnamefont {K.}~\bibnamefont {Ensslin}},\
  }\bibfield  {title} {\enquote {\bibinfo {title} {Virtual-photon-mediated
  spin-qubit--transmon coupling},}\ }\href {\doibase
  10.1038/s41467-019-13000-z} {\bibfield  {journal} {\bibinfo  {journal}
  {Nature Communications}\ }\textbf {\bibinfo {volume} {10}},\ \bibinfo {pages}
  {5037} (\bibinfo {year} {2019})}\BibitemShut {NoStop}%
\bibitem [{\citenamefont {Samkharadze}\ \emph {et~al.}(2018)\citenamefont
  {Samkharadze}, \citenamefont {Zheng}, \citenamefont {Kalhor}, \citenamefont
  {Brousse}, \citenamefont {Sammak}, \citenamefont {Mendes}, \citenamefont
  {Blais}, \citenamefont {Scappucci},\ and\ \citenamefont
  {Vandersypen}}]{Samkharadze2018}%
  \BibitemOpen
  \bibfield  {author} {\bibinfo {author} {\bibfnamefont {N.}~\bibnamefont
  {Samkharadze}}, \bibinfo {author} {\bibfnamefont {G.}~\bibnamefont {Zheng}},
  \bibinfo {author} {\bibfnamefont {N.}~\bibnamefont {Kalhor}}, \bibinfo
  {author} {\bibfnamefont {D.}~\bibnamefont {Brousse}}, \bibinfo {author}
  {\bibfnamefont {A.}~\bibnamefont {Sammak}}, \bibinfo {author} {\bibfnamefont
  {U.~C.}\ \bibnamefont {Mendes}}, \bibinfo {author} {\bibfnamefont
  {A.}~\bibnamefont {Blais}}, \bibinfo {author} {\bibfnamefont
  {G.}~\bibnamefont {Scappucci}}, \ and\ \bibinfo {author} {\bibfnamefont
  {L.~M.~K.}\ \bibnamefont {Vandersypen}},\ }\bibfield  {title} {\enquote
  {\bibinfo {title} {Strong spin-photon coupling in silicon},}\ }\href
  {\doibase 10.1126/science.aar4054} {\bibfield  {journal} {\bibinfo  {journal}
  {Science}\ }\textbf {\bibinfo {volume} {359}},\ \bibinfo {pages} {1123--1127}
  (\bibinfo {year} {2018})}\BibitemShut {NoStop}%
\bibitem [{\citenamefont {Andrews}\ \emph {et~al.}(2014)\citenamefont
  {Andrews}, \citenamefont {Peterson}, \citenamefont {Purdy}, \citenamefont
  {Cicak}, \citenamefont {Simmonds}, \citenamefont {Regal},\ and\ \citenamefont
  {Lehnert}}]{Andrews2014}%
  \BibitemOpen
  \bibfield  {author} {\bibinfo {author} {\bibfnamefont {R.~W.}\ \bibnamefont
  {Andrews}}, \bibinfo {author} {\bibfnamefont {R.~W.}\ \bibnamefont
  {Peterson}}, \bibinfo {author} {\bibfnamefont {T.~P.}\ \bibnamefont {Purdy}},
  \bibinfo {author} {\bibfnamefont {K.}~\bibnamefont {Cicak}}, \bibinfo
  {author} {\bibfnamefont {R.~W.}\ \bibnamefont {Simmonds}}, \bibinfo {author}
  {\bibfnamefont {C.~A.}\ \bibnamefont {Regal}}, \ and\ \bibinfo {author}
  {\bibfnamefont {K.~W.}\ \bibnamefont {Lehnert}},\ }\bibfield  {title}
  {\enquote {\bibinfo {title} {Bidirectional and efficient conversion between
  microwave and optical light},}\ }\href {\doibase 10.1038/nphys2911}
  {\bibfield  {journal} {\bibinfo  {journal} {Nature Physics}\ }\textbf
  {\bibinfo {volume} {10}},\ \bibinfo {pages} {321--326} (\bibinfo {year}
  {2014})}\BibitemShut {NoStop}%
\bibitem [{\citenamefont {Bagci}\ \emph {et~al.}(2014)\citenamefont {Bagci},
  \citenamefont {Simonsen}, \citenamefont {Schmid}, \citenamefont {Villanueva},
  \citenamefont {Zeuthen}, \citenamefont {Appel}, \citenamefont {Taylor},
  \citenamefont {S{\o}rensen}, \citenamefont {Usami}, \citenamefont
  {Schliesser},\ and\ \citenamefont {Polzik}}]{Bagci2014}%
  \BibitemOpen
  \bibfield  {author} {\bibinfo {author} {\bibfnamefont {T.}~\bibnamefont
  {Bagci}}, \bibinfo {author} {\bibfnamefont {A.}~\bibnamefont {Simonsen}},
  \bibinfo {author} {\bibfnamefont {S.}~\bibnamefont {Schmid}}, \bibinfo
  {author} {\bibfnamefont {L.~G.}\ \bibnamefont {Villanueva}}, \bibinfo
  {author} {\bibfnamefont {E.}~\bibnamefont {Zeuthen}}, \bibinfo {author}
  {\bibfnamefont {J.}~\bibnamefont {Appel}}, \bibinfo {author} {\bibfnamefont
  {J.~M.}\ \bibnamefont {Taylor}}, \bibinfo {author} {\bibfnamefont
  {A.}~\bibnamefont {S{\o}rensen}}, \bibinfo {author} {\bibfnamefont
  {K.}~\bibnamefont {Usami}}, \bibinfo {author} {\bibfnamefont
  {A.}~\bibnamefont {Schliesser}}, \ and\ \bibinfo {author} {\bibfnamefont
  {E.~S.}\ \bibnamefont {Polzik}},\ }\bibfield  {title} {\enquote {\bibinfo
  {title} {Optical detection of radio waves through a nanomechanical
  transducer},}\ }\href {\doibase 10.1038/nature13029} {\bibfield  {journal}
  {\bibinfo  {journal} {Nature}\ }\textbf {\bibinfo {volume} {507}},\ \bibinfo
  {pages} {81--85} (\bibinfo {year} {2014})}\BibitemShut {NoStop}%
\bibitem [{\citenamefont {Shao}\ \emph {et~al.}(2019)\citenamefont {Shao},
  \citenamefont {Yu}, \citenamefont {Maity}, \citenamefont {Sinclair},
  \citenamefont {Zheng}, \citenamefont {Chia}, \citenamefont {Shams-Ansari},
  \citenamefont {Wang}, \citenamefont {Zhang}, \citenamefont {Lai},\ and\
  \citenamefont {Lon\v{c}ar}}]{Shao2019}%
  \BibitemOpen
  \bibfield  {author} {\bibinfo {author} {\bibfnamefont {Linbo}\ \bibnamefont
  {Shao}}, \bibinfo {author} {\bibfnamefont {Mengjie}\ \bibnamefont {Yu}},
  \bibinfo {author} {\bibfnamefont {Smarak}\ \bibnamefont {Maity}}, \bibinfo
  {author} {\bibfnamefont {Neil}\ \bibnamefont {Sinclair}}, \bibinfo {author}
  {\bibfnamefont {Lu}~\bibnamefont {Zheng}}, \bibinfo {author} {\bibfnamefont
  {Cleaven}\ \bibnamefont {Chia}}, \bibinfo {author} {\bibfnamefont
  {Amirhassan}\ \bibnamefont {Shams-Ansari}}, \bibinfo {author} {\bibfnamefont
  {Cheng}\ \bibnamefont {Wang}}, \bibinfo {author} {\bibfnamefont {Mian}\
  \bibnamefont {Zhang}}, \bibinfo {author} {\bibfnamefont {Keji}\ \bibnamefont
  {Lai}}, \ and\ \bibinfo {author} {\bibfnamefont {Marko}\ \bibnamefont
  {Lon\v{c}ar}},\ }\bibfield  {title} {\enquote {\bibinfo {title}
  {Microwave-to-optical conversion using lithium niobate thin-film acoustic
  resonators},}\ }\href {\doibase 10.1364/OPTICA.6.001498} {\bibfield
  {journal} {\bibinfo  {journal} {Optica}\ }\textbf {\bibinfo {volume} {6}},\
  \bibinfo {pages} {1498--1505} (\bibinfo {year} {2019})}\BibitemShut {NoStop}%
\bibitem [{\citenamefont {Forsch}\ \emph {et~al.}(2020)\citenamefont {Forsch},
  \citenamefont {Stockill}, \citenamefont {Wallucks}, \citenamefont
  {Marinkovi{\'c}}, \citenamefont {G{\"a}rtner}, \citenamefont {Norte},
  \citenamefont {van Otten}, \citenamefont {Fiore}, \citenamefont
  {Srinivasan},\ and\ \citenamefont {Gr{\"o}blacher}}]{Forsch2020}%
  \BibitemOpen
  \bibfield  {author} {\bibinfo {author} {\bibfnamefont {Moritz}\ \bibnamefont
  {Forsch}}, \bibinfo {author} {\bibfnamefont {Robert}\ \bibnamefont
  {Stockill}}, \bibinfo {author} {\bibfnamefont {Andreas}\ \bibnamefont
  {Wallucks}}, \bibinfo {author} {\bibfnamefont {Igor}\ \bibnamefont
  {Marinkovi{\'c}}}, \bibinfo {author} {\bibfnamefont {Claus}\ \bibnamefont
  {G{\"a}rtner}}, \bibinfo {author} {\bibfnamefont {Richard~A.}\ \bibnamefont
  {Norte}}, \bibinfo {author} {\bibfnamefont {Frank}\ \bibnamefont {van
  Otten}}, \bibinfo {author} {\bibfnamefont {Andrea}\ \bibnamefont {Fiore}},
  \bibinfo {author} {\bibfnamefont {Kartik}\ \bibnamefont {Srinivasan}}, \ and\
  \bibinfo {author} {\bibfnamefont {Simon}\ \bibnamefont {Gr{\"o}blacher}},\
  }\bibfield  {title} {\enquote {\bibinfo {title} {Microwave-to-optics
  conversion using a mechanical oscillator in its quantum ground state},}\
  }\href {\doibase 10.1038/s41567-019-0673-7} {\bibfield  {journal} {\bibinfo
  {journal} {Nature Physics}\ }\textbf {\bibinfo {volume} {16}},\ \bibinfo
  {pages} {69--74} (\bibinfo {year} {2020})}\BibitemShut {NoStop}%
\bibitem [{\citenamefont {Zhong}\ \emph {et~al.}(2020)\citenamefont {Zhong},
  \citenamefont {Wang}, \citenamefont {Zou}, \citenamefont {Zhang},
  \citenamefont {Han}, \citenamefont {Fu}, \citenamefont {Xu}, \citenamefont
  {Shankar}, \citenamefont {Devoret}, \citenamefont {Tang},\ and\ \citenamefont
  {Jiang}}]{Zhong2020}%
  \BibitemOpen
  \bibfield  {author} {\bibinfo {author} {\bibfnamefont {Changchun}\
  \bibnamefont {Zhong}}, \bibinfo {author} {\bibfnamefont {Zhixin}\
  \bibnamefont {Wang}}, \bibinfo {author} {\bibfnamefont {Changling}\
  \bibnamefont {Zou}}, \bibinfo {author} {\bibfnamefont {Mengzhen}\
  \bibnamefont {Zhang}}, \bibinfo {author} {\bibfnamefont {Xu}~\bibnamefont
  {Han}}, \bibinfo {author} {\bibfnamefont {Wei}\ \bibnamefont {Fu}}, \bibinfo
  {author} {\bibfnamefont {Mingrui}\ \bibnamefont {Xu}}, \bibinfo {author}
  {\bibfnamefont {S.}~\bibnamefont {Shankar}}, \bibinfo {author} {\bibfnamefont
  {Michel~H.}\ \bibnamefont {Devoret}}, \bibinfo {author} {\bibfnamefont
  {Hong~X.}\ \bibnamefont {Tang}}, \ and\ \bibinfo {author} {\bibfnamefont
  {Liang}\ \bibnamefont {Jiang}},\ }\bibfield  {title} {\enquote {\bibinfo
  {title} {Proposal for heralded generation and detection of entangled
  microwave--optical-photon pairs},}\ }\href {\doibase
  10.1103/PhysRevLett.124.010511} {\bibfield  {journal} {\bibinfo  {journal}
  {Phys. Rev. Lett.}\ }\textbf {\bibinfo {volume} {124}},\ \bibinfo {pages}
  {010511} (\bibinfo {year} {2020})}\BibitemShut {NoStop}%
\bibitem [{\citenamefont {Xu}\ \emph {et~al.}(2021{\natexlab{a}})\citenamefont
  {Xu}, \citenamefont {Sayem}, \citenamefont {Fan}, \citenamefont {Zou},
  \citenamefont {Wang}, \citenamefont {Cheng}, \citenamefont {Fu},
  \citenamefont {Yang}, \citenamefont {Xu},\ and\ \citenamefont
  {Tang}}]{Xu2021}%
  \BibitemOpen
  \bibfield  {author} {\bibinfo {author} {\bibfnamefont {Yuntao}\ \bibnamefont
  {Xu}}, \bibinfo {author} {\bibfnamefont {Ayed~Al}\ \bibnamefont {Sayem}},
  \bibinfo {author} {\bibfnamefont {Linran}\ \bibnamefont {Fan}}, \bibinfo
  {author} {\bibfnamefont {Chang-Ling}\ \bibnamefont {Zou}}, \bibinfo {author}
  {\bibfnamefont {Sihao}\ \bibnamefont {Wang}}, \bibinfo {author}
  {\bibfnamefont {Risheng}\ \bibnamefont {Cheng}}, \bibinfo {author}
  {\bibfnamefont {Wei}\ \bibnamefont {Fu}}, \bibinfo {author} {\bibfnamefont
  {Likai}\ \bibnamefont {Yang}}, \bibinfo {author} {\bibfnamefont {Mingrui}\
  \bibnamefont {Xu}}, \ and\ \bibinfo {author} {\bibfnamefont {Hong~X.}\
  \bibnamefont {Tang}},\ }\bibfield  {title} {\enquote {\bibinfo {title}
  {Bidirectional interconversion of microwave and light with thin-film lithium
  niobate},}\ }\href {\doibase 10.1038/s41467-021-24809-y} {\bibfield
  {journal} {\bibinfo  {journal} {Nature Communications}\ }\textbf {\bibinfo
  {volume} {12}},\ \bibinfo {pages} {4453} (\bibinfo {year}
  {2021}{\natexlab{a}})}\BibitemShut {NoStop}%
\bibitem [{\citenamefont {Mirhosseini}\ \emph {et~al.}(2020)\citenamefont
  {Mirhosseini}, \citenamefont {Sipahigil}, \citenamefont {Kalaee},\ and\
  \citenamefont {Painter}}]{Mirhosseini2020}%
  \BibitemOpen
  \bibfield  {author} {\bibinfo {author} {\bibfnamefont {Mohammad}\
  \bibnamefont {Mirhosseini}}, \bibinfo {author} {\bibfnamefont {Alp}\
  \bibnamefont {Sipahigil}}, \bibinfo {author} {\bibfnamefont {Mahmoud}\
  \bibnamefont {Kalaee}}, \ and\ \bibinfo {author} {\bibfnamefont {Oskar}\
  \bibnamefont {Painter}},\ }\bibfield  {title} {\enquote {\bibinfo {title}
  {Superconducting qubit to optical photon transduction},}\ }\href {\doibase
  10.1038/s41586-020-3038-6} {\bibfield  {journal} {\bibinfo  {journal}
  {Nature}\ }\textbf {\bibinfo {volume} {588}},\ \bibinfo {pages} {599--603}
  (\bibinfo {year} {2020})}\BibitemShut {NoStop}%
\bibitem [{\citenamefont {Vasileiadis}\ \emph {et~al.}(2021)\citenamefont
  {Vasileiadis}, \citenamefont {Varghese}, \citenamefont {Babacic},
  \citenamefont {Gomis-Bresco}, \citenamefont {Navarro~Urrios},\ and\
  \citenamefont {Graczykowski}}]{Vasileiadis2021}%
  \BibitemOpen
  \bibfield  {author} {\bibinfo {author} {\bibfnamefont {Thomas}\ \bibnamefont
  {Vasileiadis}}, \bibinfo {author} {\bibfnamefont {Jeena}\ \bibnamefont
  {Varghese}}, \bibinfo {author} {\bibfnamefont {Visnja}\ \bibnamefont
  {Babacic}}, \bibinfo {author} {\bibfnamefont {Jordi}\ \bibnamefont
  {Gomis-Bresco}}, \bibinfo {author} {\bibfnamefont {Daniel}\ \bibnamefont
  {Navarro~Urrios}}, \ and\ \bibinfo {author} {\bibfnamefont {Bartlomiej}\
  \bibnamefont {Graczykowski}},\ }\bibfield  {title} {\enquote {\bibinfo
  {title} {{Progress and perspectives on phononic crystals}},}\ }\href
  {\doibase 10.1063/5.0042337} {\bibfield  {journal} {\bibinfo  {journal}
  {Journal of Applied Physics}\ }\textbf {\bibinfo {volume} {129}},\ \bibinfo
  {pages} {160901} (\bibinfo {year} {2021})}\BibitemShut {NoStop}%
\bibitem [{\citenamefont {Bachtold}\ \emph {et~al.}(2022)\citenamefont
  {Bachtold}, \citenamefont {Moser},\ and\ \citenamefont
  {Dykman}}]{Bachtold2022}%
  \BibitemOpen
  \bibfield  {author} {\bibinfo {author} {\bibfnamefont {Adrian}\ \bibnamefont
  {Bachtold}}, \bibinfo {author} {\bibfnamefont {Joel}\ \bibnamefont {Moser}},
  \ and\ \bibinfo {author} {\bibfnamefont {M.~I.}\ \bibnamefont {Dykman}},\
  }\bibfield  {title} {\enquote {\bibinfo {title} {Mesoscopic physics of
  nanomechanical systems},}\ }\href {\doibase 10.1103/RevModPhys.94.045005}
  {\bibfield  {journal} {\bibinfo  {journal} {Rev. Mod. Phys.}\ }\textbf
  {\bibinfo {volume} {94}},\ \bibinfo {pages} {045005} (\bibinfo {year}
  {2022})}\BibitemShut {NoStop}%
\bibitem [{\citenamefont {Heinrich}\ \emph {et~al.}(2021)\citenamefont
  {Heinrich}, \citenamefont {Oliver}, \citenamefont {Vandersypen},
  \citenamefont {Ardavan}, \citenamefont {Sessoli}, \citenamefont {Loss},
  \citenamefont {Jayich}, \citenamefont {Fernandez-Rossier}, \citenamefont
  {Laucht},\ and\ \citenamefont {Morello}}]{Heinrich2021}%
  \BibitemOpen
  \bibfield  {author} {\bibinfo {author} {\bibfnamefont {Andreas~J.}\
  \bibnamefont {Heinrich}}, \bibinfo {author} {\bibfnamefont {William~D.}\
  \bibnamefont {Oliver}}, \bibinfo {author} {\bibfnamefont {Lieven M.~K.}\
  \bibnamefont {Vandersypen}}, \bibinfo {author} {\bibfnamefont {Arzhang}\
  \bibnamefont {Ardavan}}, \bibinfo {author} {\bibfnamefont {Roberta}\
  \bibnamefont {Sessoli}}, \bibinfo {author} {\bibfnamefont {Daniel}\
  \bibnamefont {Loss}}, \bibinfo {author} {\bibfnamefont {Ania~Bleszynski}\
  \bibnamefont {Jayich}}, \bibinfo {author} {\bibfnamefont {Joaquin}\
  \bibnamefont {Fernandez-Rossier}}, \bibinfo {author} {\bibfnamefont {Arne}\
  \bibnamefont {Laucht}}, \ and\ \bibinfo {author} {\bibfnamefont {Andrea}\
  \bibnamefont {Morello}},\ }\bibfield  {title} {\enquote {\bibinfo {title}
  {Quantum-coherent nanoscience},}\ }\href {\doibase
  10.1038/s41565-021-00994-1} {\bibfield  {journal} {\bibinfo  {journal}
  {Nature Nanotechnology}\ }\textbf {\bibinfo {volume} {16}},\ \bibinfo {pages}
  {1318--1329} (\bibinfo {year} {2021})}\BibitemShut {NoStop}%
\bibitem [{\citenamefont {Seis}\ \emph {et~al.}(2022)\citenamefont {Seis},
  \citenamefont {Capelle}, \citenamefont {Langman}, \citenamefont {Saarinen},
  \citenamefont {Planz},\ and\ \citenamefont {Schliesser}}]{Seis2022}%
  \BibitemOpen
  \bibfield  {author} {\bibinfo {author} {\bibfnamefont {Yannick}\ \bibnamefont
  {Seis}}, \bibinfo {author} {\bibfnamefont {Thibault}\ \bibnamefont
  {Capelle}}, \bibinfo {author} {\bibfnamefont {Eric}\ \bibnamefont {Langman}},
  \bibinfo {author} {\bibfnamefont {Sampo}\ \bibnamefont {Saarinen}}, \bibinfo
  {author} {\bibfnamefont {Eric}\ \bibnamefont {Planz}}, \ and\ \bibinfo
  {author} {\bibfnamefont {Albert}\ \bibnamefont {Schliesser}},\ }\bibfield
  {title} {\enquote {\bibinfo {title} {Ground state cooling of an ultracoherent
  electromechanical system},}\ }\href {\doibase 10.1038/s41467-022-29115-9}
  {\bibfield  {journal} {\bibinfo  {journal} {Nature Communications}\ }\textbf
  {\bibinfo {volume} {13}},\ \bibinfo {pages} {1507} (\bibinfo {year}
  {2022})}\BibitemShut {NoStop}%
\bibitem [{\citenamefont {Beccari}\ \emph {et~al.}(2022)\citenamefont
  {Beccari}, \citenamefont {Visani}, \citenamefont {Fedorov}, \citenamefont
  {Bereyhi}, \citenamefont {Boureau}, \citenamefont {Engelsen},\ and\
  \citenamefont {Kippenberg}}]{Beccari2022}%
  \BibitemOpen
  \bibfield  {author} {\bibinfo {author} {\bibfnamefont {A.}~\bibnamefont
  {Beccari}}, \bibinfo {author} {\bibfnamefont {D.~A.}\ \bibnamefont {Visani}},
  \bibinfo {author} {\bibfnamefont {S.~A.}\ \bibnamefont {Fedorov}}, \bibinfo
  {author} {\bibfnamefont {M.~J.}\ \bibnamefont {Bereyhi}}, \bibinfo {author}
  {\bibfnamefont {V.}~\bibnamefont {Boureau}}, \bibinfo {author} {\bibfnamefont
  {N.~J.}\ \bibnamefont {Engelsen}}, \ and\ \bibinfo {author} {\bibfnamefont
  {T.~J.}\ \bibnamefont {Kippenberg}},\ }\bibfield  {title} {\enquote {\bibinfo
  {title} {Strained crystalline nanomechanical resonators with quality factors
  above 10 billion},}\ }\href {\doibase 10.1038/s41567-021-01498-4} {\bibfield
  {journal} {\bibinfo  {journal} {Nature Physics}\ }\textbf {\bibinfo {volume}
  {18}},\ \bibinfo {pages} {436--441} (\bibinfo {year} {2022})}\BibitemShut
  {NoStop}%
\bibitem [{\citenamefont {MacCabe}\ \emph {et~al.}(2020)\citenamefont
  {MacCabe}, \citenamefont {Ren}, \citenamefont {Luo}, \citenamefont {Cohen},
  \citenamefont {Zhou}, \citenamefont {Sipahigil}, \citenamefont
  {Mirhosseini},\ and\ \citenamefont {Painter}}]{MacCabe2020}%
  \BibitemOpen
  \bibfield  {author} {\bibinfo {author} {\bibfnamefont {Gregory~S.}\
  \bibnamefont {MacCabe}}, \bibinfo {author} {\bibfnamefont {Hengjiang}\
  \bibnamefont {Ren}}, \bibinfo {author} {\bibfnamefont {Jie}\ \bibnamefont
  {Luo}}, \bibinfo {author} {\bibfnamefont {Justin~D.}\ \bibnamefont {Cohen}},
  \bibinfo {author} {\bibfnamefont {Hengyun}\ \bibnamefont {Zhou}}, \bibinfo
  {author} {\bibfnamefont {Alp}\ \bibnamefont {Sipahigil}}, \bibinfo {author}
  {\bibfnamefont {Mohammad}\ \bibnamefont {Mirhosseini}}, \ and\ \bibinfo
  {author} {\bibfnamefont {Oskar}\ \bibnamefont {Painter}},\ }\bibfield
  {title} {\enquote {\bibinfo {title} {Nano-acoustic resonator with ultralong
  phonon lifetime},}\ }\href {\doibase 10.1126/science.abc7312} {\bibfield
  {journal} {\bibinfo  {journal} {Science}\ }\textbf {\bibinfo {volume}
  {370}},\ \bibinfo {pages} {840--843} (\bibinfo {year} {2020})}\BibitemShut
  {NoStop}%
\bibitem [{\citenamefont {Teufel}\ \emph {et~al.}(2011)\citenamefont {Teufel},
  \citenamefont {Donner}, \citenamefont {Li}, \citenamefont {Harlow},
  \citenamefont {Allman}, \citenamefont {Cicak}, \citenamefont {Sirois},
  \citenamefont {Whittaker}, \citenamefont {Lehnert},\ and\ \citenamefont
  {Simmonds}}]{Teufel2011}%
  \BibitemOpen
  \bibfield  {author} {\bibinfo {author} {\bibfnamefont {J.~D.}\ \bibnamefont
  {Teufel}}, \bibinfo {author} {\bibfnamefont {T.}~\bibnamefont {Donner}},
  \bibinfo {author} {\bibfnamefont {Dale}\ \bibnamefont {Li}}, \bibinfo
  {author} {\bibfnamefont {J.~W.}\ \bibnamefont {Harlow}}, \bibinfo {author}
  {\bibfnamefont {M.~S.}\ \bibnamefont {Allman}}, \bibinfo {author}
  {\bibfnamefont {K.}~\bibnamefont {Cicak}}, \bibinfo {author} {\bibfnamefont
  {A.~J.}\ \bibnamefont {Sirois}}, \bibinfo {author} {\bibfnamefont {J.~D.}\
  \bibnamefont {Whittaker}}, \bibinfo {author} {\bibfnamefont {K.~W.}\
  \bibnamefont {Lehnert}}, \ and\ \bibinfo {author} {\bibfnamefont {R.~W.}\
  \bibnamefont {Simmonds}},\ }\bibfield  {title} {\enquote {\bibinfo {title}
  {Sideband cooling of micromechanical motion to the quantum ground state},}\
  }\href {\doibase 10.1038/nature10261} {\bibfield  {journal} {\bibinfo
  {journal} {Nature}\ }\textbf {\bibinfo {volume} {475}},\ \bibinfo {pages}
  {359--363} (\bibinfo {year} {2011})}\BibitemShut {NoStop}%
\bibitem [{\citenamefont {Wollack}\ \emph {et~al.}(2022)\citenamefont
  {Wollack}, \citenamefont {Cleland}, \citenamefont {Gruenke}, \citenamefont
  {Wang}, \citenamefont {Arrangoiz-Arriola},\ and\ \citenamefont
  {Safavi-Naeini}}]{Wollack2022}%
  \BibitemOpen
  \bibfield  {author} {\bibinfo {author} {\bibfnamefont {E.~Alex}\ \bibnamefont
  {Wollack}}, \bibinfo {author} {\bibfnamefont {Agnetta~Y.}\ \bibnamefont
  {Cleland}}, \bibinfo {author} {\bibfnamefont {Rachel~G.}\ \bibnamefont
  {Gruenke}}, \bibinfo {author} {\bibfnamefont {Zhaoyou}\ \bibnamefont {Wang}},
  \bibinfo {author} {\bibfnamefont {Patricio}\ \bibnamefont
  {Arrangoiz-Arriola}}, \ and\ \bibinfo {author} {\bibfnamefont {Amir~H.}\
  \bibnamefont {Safavi-Naeini}},\ }\bibfield  {title} {\enquote {\bibinfo
  {title} {Quantum state preparation and tomography of entangled mechanical
  resonators},}\ }\href {\doibase 10.1038/s41586-022-04500-y} {\bibfield
  {journal} {\bibinfo  {journal} {Nature}\ }\textbf {\bibinfo {volume} {604}},\
  \bibinfo {pages} {463--467} (\bibinfo {year} {2022})}\BibitemShut {NoStop}%
\bibitem [{\citenamefont {von L{\"u}pke}\ \emph {et~al.}(2022)\citenamefont
  {von L{\"u}pke}, \citenamefont {Yang}, \citenamefont {Bild}, \citenamefont
  {Michaud}, \citenamefont {Fadel},\ and\ \citenamefont {Chu}}]{Lupke2022}%
  \BibitemOpen
  \bibfield  {author} {\bibinfo {author} {\bibfnamefont {Uwe}\ \bibnamefont
  {von L{\"u}pke}}, \bibinfo {author} {\bibfnamefont {Yu}~\bibnamefont {Yang}},
  \bibinfo {author} {\bibfnamefont {Marius}\ \bibnamefont {Bild}}, \bibinfo
  {author} {\bibfnamefont {Laurent}\ \bibnamefont {Michaud}}, \bibinfo {author}
  {\bibfnamefont {Matteo}\ \bibnamefont {Fadel}}, \ and\ \bibinfo {author}
  {\bibfnamefont {Yiwen}\ \bibnamefont {Chu}},\ }\bibfield  {title} {\enquote
  {\bibinfo {title} {Parity measurement in the strong dispersive regime of
  circuit quantum acoustodynamics},}\ }\href {\doibase
  10.1038/s41567-022-01591-2} {\bibfield  {journal} {\bibinfo  {journal}
  {Nature Physics}\ }\textbf {\bibinfo {volume} {18}},\ \bibinfo {pages}
  {794--799} (\bibinfo {year} {2022})}\BibitemShut {NoStop}%
\bibitem [{\citenamefont {Stockill}\ \emph {et~al.}(2022)\citenamefont
  {Stockill}, \citenamefont {Forsch}, \citenamefont {Hijazi}, \citenamefont
  {Beaudoin}, \citenamefont {Pantzas}, \citenamefont {Sagnes}, \citenamefont
  {Braive},\ and\ \citenamefont {Gr{\"o}blacher}}]{Stockill2022}%
  \BibitemOpen
  \bibfield  {author} {\bibinfo {author} {\bibfnamefont {Robert}\ \bibnamefont
  {Stockill}}, \bibinfo {author} {\bibfnamefont {Moritz}\ \bibnamefont
  {Forsch}}, \bibinfo {author} {\bibfnamefont {Frederick}\ \bibnamefont
  {Hijazi}}, \bibinfo {author} {\bibfnamefont {Gr{\'e}goire}\ \bibnamefont
  {Beaudoin}}, \bibinfo {author} {\bibfnamefont {Konstantinos}\ \bibnamefont
  {Pantzas}}, \bibinfo {author} {\bibfnamefont {Isabelle}\ \bibnamefont
  {Sagnes}}, \bibinfo {author} {\bibfnamefont {R{\'e}my}\ \bibnamefont
  {Braive}}, \ and\ \bibinfo {author} {\bibfnamefont {Simon}\ \bibnamefont
  {Gr{\"o}blacher}},\ }\bibfield  {title} {\enquote {\bibinfo {title}
  {Ultra-low-noise microwave to optics conversion in gallium phosphide},}\
  }\href {\doibase 10.1038/s41467-022-34338-x} {\bibfield  {journal} {\bibinfo
  {journal} {Nature Communications}\ }\textbf {\bibinfo {volume} {13}},\
  \bibinfo {pages} {6583} (\bibinfo {year} {2022})}\BibitemShut {NoStop}%
\bibitem [{\citenamefont {Bravyi}\ and\ \citenamefont
  {Kitaev}(1998)}]{bravyi1998quantum}%
  \BibitemOpen
  \bibfield  {author} {\bibinfo {author} {\bibfnamefont {S.~B.}\ \bibnamefont
  {Bravyi}}\ and\ \bibinfo {author} {\bibfnamefont {A.~Yu.}\ \bibnamefont
  {Kitaev}},\ }\href@noop {} {\enquote {\bibinfo {title} {Quantum codes on a
  lattice with boundary},}\ } (\bibinfo {year} {1998}),\ \Eprint
  {http://arxiv.org/abs/quant-ph/9811052} {arXiv:quant-ph/9811052 [quant-ph]}
  \BibitemShut {NoStop}%
\bibitem [{\citenamefont {Dennis}\ \emph {et~al.}(2002)\citenamefont {Dennis},
  \citenamefont {Kitaev}, \citenamefont {Landahl},\ and\ \citenamefont
  {Preskill}}]{Dennis2002}%
  \BibitemOpen
  \bibfield  {author} {\bibinfo {author} {\bibfnamefont {Eric}\ \bibnamefont
  {Dennis}}, \bibinfo {author} {\bibfnamefont {Alexei}\ \bibnamefont {Kitaev}},
  \bibinfo {author} {\bibfnamefont {Andrew}\ \bibnamefont {Landahl}}, \ and\
  \bibinfo {author} {\bibfnamefont {John}\ \bibnamefont {Preskill}},\
  }\bibfield  {title} {\enquote {\bibinfo {title} {{Topological quantum
  memory}},}\ }\href {\doibase 10.1063/1.1499754} {\bibfield  {journal}
  {\bibinfo  {journal} {Journal of Mathematical Physics}\ }\textbf {\bibinfo
  {volume} {43}},\ \bibinfo {pages} {4452--4505} (\bibinfo {year}
  {2002})}\BibitemShut {NoStop}%
\bibitem [{\citenamefont {Brown}\ \emph {et~al.}(2016)\citenamefont {Brown},
  \citenamefont {Loss}, \citenamefont {Pachos}, \citenamefont {Self},\ and\
  \citenamefont {Wootton}}]{Brown2012016}%
  \BibitemOpen
  \bibfield  {author} {\bibinfo {author} {\bibfnamefont {Benjamin~J.}\
  \bibnamefont {Brown}}, \bibinfo {author} {\bibfnamefont {Daniel}\
  \bibnamefont {Loss}}, \bibinfo {author} {\bibfnamefont {Jiannis~K.}\
  \bibnamefont {Pachos}}, \bibinfo {author} {\bibfnamefont {Chris~N.}\
  \bibnamefont {Self}}, \ and\ \bibinfo {author} {\bibfnamefont {James~R.}\
  \bibnamefont {Wootton}},\ }\bibfield  {title} {\enquote {\bibinfo {title}
  {Quantum memories at finite temperature},}\ }\href {\doibase
  10.1103/RevModPhys.88.045005} {\bibfield  {journal} {\bibinfo  {journal}
  {Rev. Mod. Phys.}\ }\textbf {\bibinfo {volume} {88}},\ \bibinfo {pages}
  {045005} (\bibinfo {year} {2016})}\BibitemShut {NoStop}%
\bibitem [{\citenamefont {Alicea}\ \emph {et~al.}(2011)\citenamefont {Alicea},
  \citenamefont {Oreg}, \citenamefont {Refael}, \citenamefont {von Oppen},\
  and\ \citenamefont {Fisher}}]{Alicea2011}%
  \BibitemOpen
  \bibfield  {author} {\bibinfo {author} {\bibfnamefont {Jason}\ \bibnamefont
  {Alicea}}, \bibinfo {author} {\bibfnamefont {Yuval}\ \bibnamefont {Oreg}},
  \bibinfo {author} {\bibfnamefont {Gil}\ \bibnamefont {Refael}}, \bibinfo
  {author} {\bibfnamefont {Felix}\ \bibnamefont {von Oppen}}, \ and\ \bibinfo
  {author} {\bibfnamefont {Matthew P.~A.}\ \bibnamefont {Fisher}},\ }\bibfield
  {title} {\enquote {\bibinfo {title} {Non-abelian statistics and topological
  quantum information processing in 1d wire networks},}\ }\href {\doibase
  10.1038/nphys1915} {\bibfield  {journal} {\bibinfo  {journal} {Nature
  Physics}\ }\textbf {\bibinfo {volume} {7}},\ \bibinfo {pages} {412--417}
  (\bibinfo {year} {2011})}\BibitemShut {NoStop}%
\bibitem [{\citenamefont {Sarma}\ \emph {et~al.}(2015)\citenamefont {Sarma},
  \citenamefont {Freedman},\ and\ \citenamefont {Nayak}}]{Sarma2015}%
  \BibitemOpen
  \bibfield  {author} {\bibinfo {author} {\bibfnamefont {Sankar~Das}\
  \bibnamefont {Sarma}}, \bibinfo {author} {\bibfnamefont {Michael}\
  \bibnamefont {Freedman}}, \ and\ \bibinfo {author} {\bibfnamefont {Chetan}\
  \bibnamefont {Nayak}},\ }\bibfield  {title} {\enquote {\bibinfo {title}
  {Majorana zero modes and topological quantum computation},}\ }\href {\doibase
  10.1038/npjqi.2015.1} {\bibfield  {journal} {\bibinfo  {journal} {npj Quantum
  Information}\ }\textbf {\bibinfo {volume} {1}},\ \bibinfo {pages} {15001}
  (\bibinfo {year} {2015})}\BibitemShut {NoStop}%
\bibitem [{\citenamefont {Chirolli}\ \emph {et~al.}(2022)\citenamefont
  {Chirolli}, \citenamefont {Yao},\ and\ \citenamefont {Moore}}]{Chirolli2022}%
  \BibitemOpen
  \bibfield  {author} {\bibinfo {author} {\bibfnamefont {Luca}\ \bibnamefont
  {Chirolli}}, \bibinfo {author} {\bibfnamefont {Norman~Y.}\ \bibnamefont
  {Yao}}, \ and\ \bibinfo {author} {\bibfnamefont {Joel~E.}\ \bibnamefont
  {Moore}},\ }\bibfield  {title} {\enquote {\bibinfo {title} {Swap gate between
  a majorana qubit and a parity-protected superconducting qubit},}\ }\href
  {\doibase 10.1103/PhysRevLett.129.177701} {\bibfield  {journal} {\bibinfo
  {journal} {Phys. Rev. Lett.}\ }\textbf {\bibinfo {volume} {129}},\ \bibinfo
  {pages} {177701} (\bibinfo {year} {2022})}\BibitemShut {NoStop}%
\bibitem [{\citenamefont {Roy}\ and\ \citenamefont
  {DiVincenzo}(2017)}]{roy2017topological}%
  \BibitemOpen
  \bibfield  {author} {\bibinfo {author} {\bibfnamefont {Ananda}\ \bibnamefont
  {Roy}}\ and\ \bibinfo {author} {\bibfnamefont {David~P.}\ \bibnamefont
  {DiVincenzo}},\ }\href@noop {} {\enquote {\bibinfo {title} {Topological
  quantum computing},}\ } (\bibinfo {year} {2017}),\ \Eprint
  {http://arxiv.org/abs/1701.05052} {arXiv:1701.05052 [quant-ph]} \BibitemShut
  {NoStop}%
\bibitem [{\citenamefont {Lahtinen}\ and\ \citenamefont
  {Pachos}(2017)}]{Lahtinen2017}%
  \BibitemOpen
  \bibfield  {author} {\bibinfo {author} {\bibfnamefont {Ville}\ \bibnamefont
  {Lahtinen}}\ and\ \bibinfo {author} {\bibfnamefont {Jiannis~K.}\ \bibnamefont
  {Pachos}},\ }\bibfield  {title} {\enquote {\bibinfo {title} {{A Short
  Introduction to Topological Quantum Computation}},}\ }\href {\doibase
  10.21468/SciPostPhys.3.3.021} {\bibfield  {journal} {\bibinfo  {journal}
  {SciPost Phys.}\ }\textbf {\bibinfo {volume} {3}},\ \bibinfo {pages} {021}
  (\bibinfo {year} {2017})}\BibitemShut {NoStop}%
\bibitem [{\citenamefont {Liu}\ \emph {et~al.}(2022)\citenamefont {Liu},
  \citenamefont {Hann},\ and\ \citenamefont {Jiang}}]{Liu2022quantum}%
  \BibitemOpen
  \bibfield  {author} {\bibinfo {author} {\bibfnamefont {Junyu}\ \bibnamefont
  {Liu}}, \bibinfo {author} {\bibfnamefont {Connor~T.}\ \bibnamefont {Hann}}, \
  and\ \bibinfo {author} {\bibfnamefont {Liang}\ \bibnamefont {Jiang}},\
  }\href@noop {} {\enquote {\bibinfo {title} {Quantum data center: Theories and
  applications},}\ } (\bibinfo {year} {2022}),\ \Eprint
  {http://arxiv.org/abs/2207.14336} {arXiv:2207.14336 [quant-ph]} \BibitemShut
  {NoStop}%
\bibitem [{\citenamefont {Chang}\ \emph {et~al.}(2019)\citenamefont {Chang},
  \citenamefont {Li}, \citenamefont {Wu}, \citenamefont {Jiang}, \citenamefont
  {Zhang}, \citenamefont {Pu}, \citenamefont {Chang},\ and\ \citenamefont
  {Duan}}]{Chang2019}%
  \BibitemOpen
  \bibfield  {author} {\bibinfo {author} {\bibfnamefont {W.}~\bibnamefont
  {Chang}}, \bibinfo {author} {\bibfnamefont {C.}~\bibnamefont {Li}}, \bibinfo
  {author} {\bibfnamefont {Y.-K.}\ \bibnamefont {Wu}}, \bibinfo {author}
  {\bibfnamefont {N.}~\bibnamefont {Jiang}}, \bibinfo {author} {\bibfnamefont
  {S.}~\bibnamefont {Zhang}}, \bibinfo {author} {\bibfnamefont {Y.-F.}\
  \bibnamefont {Pu}}, \bibinfo {author} {\bibfnamefont {X.-Y.}\ \bibnamefont
  {Chang}}, \ and\ \bibinfo {author} {\bibfnamefont {L.-M.}\ \bibnamefont
  {Duan}},\ }\bibfield  {title} {\enquote {\bibinfo {title} {Long-distance
  entanglement between a multiplexed quantum memory and a telecom photon},}\
  }\href {\doibase 10.1103/PhysRevX.9.041033} {\bibfield  {journal} {\bibinfo
  {journal} {Phys. Rev. X}\ }\textbf {\bibinfo {volume} {9}},\ \bibinfo {pages}
  {041033} (\bibinfo {year} {2019})}\BibitemShut {NoStop}%
\bibitem [{\citenamefont {O'Sullivan}\ \emph {et~al.}(2022)\citenamefont
  {O'Sullivan}, \citenamefont {Kennedy}, \citenamefont {Debnath}, \citenamefont
  {Alexander}, \citenamefont {Zollitsch}, \citenamefont {\ifmmode
  \check{S}\else \v{S}\fi{}im\ifmmode~\dot{e}\else \.{e}\fi{}nas},
  \citenamefont {Hashim}, \citenamefont {Thomas}, \citenamefont {Withington},
  \citenamefont {Siddiqi}, \citenamefont {M\o{}lmer},\ and\ \citenamefont
  {Morton}}]{OSullivan2022}%
  \BibitemOpen
  \bibfield  {author} {\bibinfo {author} {\bibfnamefont {James}\ \bibnamefont
  {O'Sullivan}}, \bibinfo {author} {\bibfnamefont {Oscar~W.}\ \bibnamefont
  {Kennedy}}, \bibinfo {author} {\bibfnamefont {Kamanasish}\ \bibnamefont
  {Debnath}}, \bibinfo {author} {\bibfnamefont {Joseph}\ \bibnamefont
  {Alexander}}, \bibinfo {author} {\bibfnamefont {Christoph~W.}\ \bibnamefont
  {Zollitsch}}, \bibinfo {author} {\bibfnamefont {Mantas}\ \bibnamefont
  {\ifmmode \check{S}\else \v{S}\fi{}im\ifmmode~\dot{e}\else \.{e}\fi{}nas}},
  \bibinfo {author} {\bibfnamefont {Akel}\ \bibnamefont {Hashim}}, \bibinfo
  {author} {\bibfnamefont {Christopher~N.}\ \bibnamefont {Thomas}}, \bibinfo
  {author} {\bibfnamefont {Stafford}\ \bibnamefont {Withington}}, \bibinfo
  {author} {\bibfnamefont {Irfan}\ \bibnamefont {Siddiqi}}, \bibinfo {author}
  {\bibfnamefont {Klaus}\ \bibnamefont {M\o{}lmer}}, \ and\ \bibinfo {author}
  {\bibfnamefont {John J.~L.}\ \bibnamefont {Morton}},\ }\bibfield  {title}
  {\enquote {\bibinfo {title} {Random-access quantum memory using chirped pulse
  phase encoding},}\ }\href {\doibase 10.1103/PhysRevX.12.041014} {\bibfield
  {journal} {\bibinfo  {journal} {Phys. Rev. X}\ }\textbf {\bibinfo {volume}
  {12}},\ \bibinfo {pages} {041014} (\bibinfo {year} {2022})}\BibitemShut
  {NoStop}%
\bibitem [{\citenamefont {Besse}\ \emph {et~al.}(2020)\citenamefont {Besse},
  \citenamefont {Reuer}, \citenamefont {Collodo}, \citenamefont {Wulff},
  \citenamefont {Wernli}, \citenamefont {Copetudo}, \citenamefont {Malz},
  \citenamefont {Magnard}, \citenamefont {Akin}, \citenamefont {Gabureac},
  \citenamefont {Norris}, \citenamefont {Cirac}, \citenamefont {Wallraff},\
  and\ \citenamefont {Eichler}}]{Besse2020}%
  \BibitemOpen
  \bibfield  {author} {\bibinfo {author} {\bibfnamefont {Jean-Claude}\
  \bibnamefont {Besse}}, \bibinfo {author} {\bibfnamefont {Kevin}\ \bibnamefont
  {Reuer}}, \bibinfo {author} {\bibfnamefont {Michele~C.}\ \bibnamefont
  {Collodo}}, \bibinfo {author} {\bibfnamefont {Arne}\ \bibnamefont {Wulff}},
  \bibinfo {author} {\bibfnamefont {Lucien}\ \bibnamefont {Wernli}}, \bibinfo
  {author} {\bibfnamefont {Adrian}\ \bibnamefont {Copetudo}}, \bibinfo {author}
  {\bibfnamefont {Daniel}\ \bibnamefont {Malz}}, \bibinfo {author}
  {\bibfnamefont {Paul}\ \bibnamefont {Magnard}}, \bibinfo {author}
  {\bibfnamefont {Abdulkadir}\ \bibnamefont {Akin}}, \bibinfo {author}
  {\bibfnamefont {Mihai}\ \bibnamefont {Gabureac}}, \bibinfo {author}
  {\bibfnamefont {Graham~J.}\ \bibnamefont {Norris}}, \bibinfo {author}
  {\bibfnamefont {J.~Ignacio}\ \bibnamefont {Cirac}}, \bibinfo {author}
  {\bibfnamefont {Andreas}\ \bibnamefont {Wallraff}}, \ and\ \bibinfo {author}
  {\bibfnamefont {Christopher}\ \bibnamefont {Eichler}},\ }\bibfield  {title}
  {\enquote {\bibinfo {title} {Realizing a deterministic source of
  multipartite-entangled photonic qubits},}\ }\href {\doibase
  10.1038/s41467-020-18635-x} {\bibfield  {journal} {\bibinfo  {journal}
  {Nature Communications}\ }\textbf {\bibinfo {volume} {11}},\ \bibinfo {pages}
  {4877} (\bibinfo {year} {2020})}\BibitemShut {NoStop}%
\bibitem [{\citenamefont {Reuer}\ \emph {et~al.}(2022)\citenamefont {Reuer},
  \citenamefont {Besse}, \citenamefont {Wernli}, \citenamefont {Magnard},
  \citenamefont {Kurpiers}, \citenamefont {Norris}, \citenamefont {Wallraff},\
  and\ \citenamefont {Eichler}}]{Reuer2022}%
  \BibitemOpen
  \bibfield  {author} {\bibinfo {author} {\bibfnamefont {Kevin}\ \bibnamefont
  {Reuer}}, \bibinfo {author} {\bibfnamefont {Jean-Claude}\ \bibnamefont
  {Besse}}, \bibinfo {author} {\bibfnamefont {Lucien}\ \bibnamefont {Wernli}},
  \bibinfo {author} {\bibfnamefont {Paul}\ \bibnamefont {Magnard}}, \bibinfo
  {author} {\bibfnamefont {Philipp}\ \bibnamefont {Kurpiers}}, \bibinfo
  {author} {\bibfnamefont {Graham~J.}\ \bibnamefont {Norris}}, \bibinfo
  {author} {\bibfnamefont {Andreas}\ \bibnamefont {Wallraff}}, \ and\ \bibinfo
  {author} {\bibfnamefont {Christopher}\ \bibnamefont {Eichler}},\ }\bibfield
  {title} {\enquote {\bibinfo {title} {Realization of a universal quantum gate
  set for itinerant microwave photons},}\ }\href {\doibase
  10.1103/PhysRevX.12.011008} {\bibfield  {journal} {\bibinfo  {journal} {Phys.
  Rev. X}\ }\textbf {\bibinfo {volume} {12}},\ \bibinfo {pages} {011008}
  (\bibinfo {year} {2022})}\BibitemShut {NoStop}%
\bibitem [{\citenamefont {Delaney}\ \emph {et~al.}(2022)\citenamefont
  {Delaney}, \citenamefont {Urmey}, \citenamefont {Mittal}, \citenamefont
  {Brubaker}, \citenamefont {Kindem}, \citenamefont {Burns}, \citenamefont
  {Regal},\ and\ \citenamefont {Lehnert}}]{Delaney2022}%
  \BibitemOpen
  \bibfield  {author} {\bibinfo {author} {\bibfnamefont {RD}~\bibnamefont
  {Delaney}}, \bibinfo {author} {\bibfnamefont {MD}~\bibnamefont {Urmey}},
  \bibinfo {author} {\bibfnamefont {S}~\bibnamefont {Mittal}}, \bibinfo
  {author} {\bibfnamefont {BM}~\bibnamefont {Brubaker}}, \bibinfo {author}
  {\bibfnamefont {JM}~\bibnamefont {Kindem}}, \bibinfo {author} {\bibfnamefont
  {PS}~\bibnamefont {Burns}}, \bibinfo {author} {\bibfnamefont
  {CA}~\bibnamefont {Regal}}, \ and\ \bibinfo {author} {\bibfnamefont
  {KW}~\bibnamefont {Lehnert}},\ }\bibfield  {title} {\enquote {\bibinfo
  {title} {Superconducting-qubit readout via low-backaction electro-optic
  transduction},}\ }\href@noop {} {\bibfield  {journal} {\bibinfo  {journal}
  {Nature}\ }\textbf {\bibinfo {volume} {606}},\ \bibinfo {pages} {489--493}
  (\bibinfo {year} {2022})}\BibitemShut {NoStop}%
\bibitem [{\citenamefont {Sahu}\ \emph {et~al.}(2022)\citenamefont {Sahu},
  \citenamefont {Hease}, \citenamefont {Rueda}, \citenamefont {Arnold},
  \citenamefont {Qiu},\ and\ \citenamefont {Fink}}]{Sahu2022}%
  \BibitemOpen
  \bibfield  {author} {\bibinfo {author} {\bibfnamefont {Rishabh}\ \bibnamefont
  {Sahu}}, \bibinfo {author} {\bibfnamefont {William}\ \bibnamefont {Hease}},
  \bibinfo {author} {\bibfnamefont {Alfredo}\ \bibnamefont {Rueda}}, \bibinfo
  {author} {\bibfnamefont {Georg}\ \bibnamefont {Arnold}}, \bibinfo {author}
  {\bibfnamefont {Liu}\ \bibnamefont {Qiu}}, \ and\ \bibinfo {author}
  {\bibfnamefont {Johannes~M}\ \bibnamefont {Fink}},\ }\bibfield  {title}
  {\enquote {\bibinfo {title} {Quantum-enabled operation of a microwave-optical
  interface},}\ }\href@noop {} {\bibfield  {journal} {\bibinfo  {journal} {Nat.
  Commun.}\ }\textbf {\bibinfo {volume} {13}},\ \bibinfo {pages} {1276}
  (\bibinfo {year} {2022})}\BibitemShut {NoStop}%
\bibitem [{\citenamefont {Xu}\ \emph {et~al.}(2021{\natexlab{b}})\citenamefont
  {Xu}, \citenamefont {Sayem}, \citenamefont {Fan}, \citenamefont {Zou},
  \citenamefont {Wang}, \citenamefont {Cheng}, \citenamefont {Fu},
  \citenamefont {Yang}, \citenamefont {Xu},\ and\ \citenamefont
  {Tang}}]{xu2021bidirectional}%
  \BibitemOpen
  \bibfield  {author} {\bibinfo {author} {\bibfnamefont {Yuntao}\ \bibnamefont
  {Xu}}, \bibinfo {author} {\bibfnamefont {Ayed~Al}\ \bibnamefont {Sayem}},
  \bibinfo {author} {\bibfnamefont {Linran}\ \bibnamefont {Fan}}, \bibinfo
  {author} {\bibfnamefont {Chang-Ling}\ \bibnamefont {Zou}}, \bibinfo {author}
  {\bibfnamefont {Sihao}\ \bibnamefont {Wang}}, \bibinfo {author}
  {\bibfnamefont {Risheng}\ \bibnamefont {Cheng}}, \bibinfo {author}
  {\bibfnamefont {Wei}\ \bibnamefont {Fu}}, \bibinfo {author} {\bibfnamefont
  {Likai}\ \bibnamefont {Yang}}, \bibinfo {author} {\bibfnamefont {Mingrui}\
  \bibnamefont {Xu}}, \ and\ \bibinfo {author} {\bibfnamefont {Hong~X}\
  \bibnamefont {Tang}},\ }\bibfield  {title} {\enquote {\bibinfo {title}
  {Bidirectional interconversion of microwave and light with thin-film lithium
  niobate},}\ }\href@noop {} {\bibfield  {journal} {\bibinfo  {journal} {Nat.
  Commun.}\ }\textbf {\bibinfo {volume} {12}},\ \bibinfo {pages} {4453}
  (\bibinfo {year} {2021}{\natexlab{b}})}\BibitemShut {NoStop}%
\bibitem [{\citenamefont {Ang}\ \emph {et~al.}(2022)\citenamefont {Ang},
  \citenamefont {Carini}, \citenamefont {Chen}, \citenamefont {Chuang},
  \citenamefont {DeMarco}, \citenamefont {Economou}, \citenamefont {Eickbusch},
  \citenamefont {Faraon}, \citenamefont {Fu}, \citenamefont {Girvin},
  \citenamefont {Hatridge}, \citenamefont {Houck}, \citenamefont {Hilaire},
  \citenamefont {Krsulich}, \citenamefont {Li}, \citenamefont {Liu},
  \citenamefont {Liu}, \citenamefont {Martonosi}, \citenamefont {McKay},
  \citenamefont {Misewich}, \citenamefont {Ritter}, \citenamefont {Schoelkopf},
  \citenamefont {Stein}, \citenamefont {Sussman}, \citenamefont {Tang},
  \citenamefont {Tang}, \citenamefont {Tomesh}, \citenamefont {Tubman},
  \citenamefont {Wang}, \citenamefont {Wiebe}, \citenamefont {Yao},
  \citenamefont {Yost},\ and\ \citenamefont {Zhou}}]{ang2022architectures}%
  \BibitemOpen
  \bibfield  {author} {\bibinfo {author} {\bibfnamefont {James}\ \bibnamefont
  {Ang}}, \bibinfo {author} {\bibfnamefont {Gabriella}\ \bibnamefont {Carini}},
  \bibinfo {author} {\bibfnamefont {Yanzhu}\ \bibnamefont {Chen}}, \bibinfo
  {author} {\bibfnamefont {Isaac}\ \bibnamefont {Chuang}}, \bibinfo {author}
  {\bibfnamefont {Michael~Austin}\ \bibnamefont {DeMarco}}, \bibinfo {author}
  {\bibfnamefont {Sophia~E.}\ \bibnamefont {Economou}}, \bibinfo {author}
  {\bibfnamefont {Alec}\ \bibnamefont {Eickbusch}}, \bibinfo {author}
  {\bibfnamefont {Andrei}\ \bibnamefont {Faraon}}, \bibinfo {author}
  {\bibfnamefont {Kai-Mei}\ \bibnamefont {Fu}}, \bibinfo {author}
  {\bibfnamefont {Steven~M.}\ \bibnamefont {Girvin}}, \bibinfo {author}
  {\bibfnamefont {Michael}\ \bibnamefont {Hatridge}}, \bibinfo {author}
  {\bibfnamefont {Andrew}\ \bibnamefont {Houck}}, \bibinfo {author}
  {\bibfnamefont {Paul}\ \bibnamefont {Hilaire}}, \bibinfo {author}
  {\bibfnamefont {Kevin}\ \bibnamefont {Krsulich}}, \bibinfo {author}
  {\bibfnamefont {Ang}\ \bibnamefont {Li}}, \bibinfo {author} {\bibfnamefont
  {Chenxu}\ \bibnamefont {Liu}}, \bibinfo {author} {\bibfnamefont {Yuan}\
  \bibnamefont {Liu}}, \bibinfo {author} {\bibfnamefont {Margaret}\
  \bibnamefont {Martonosi}}, \bibinfo {author} {\bibfnamefont {David~C.}\
  \bibnamefont {McKay}}, \bibinfo {author} {\bibfnamefont {James}\ \bibnamefont
  {Misewich}}, \bibinfo {author} {\bibfnamefont {Mark}\ \bibnamefont {Ritter}},
  \bibinfo {author} {\bibfnamefont {Robert~J.}\ \bibnamefont {Schoelkopf}},
  \bibinfo {author} {\bibfnamefont {Samuel~A.}\ \bibnamefont {Stein}}, \bibinfo
  {author} {\bibfnamefont {Sara}\ \bibnamefont {Sussman}}, \bibinfo {author}
  {\bibfnamefont {Hong~X.}\ \bibnamefont {Tang}}, \bibinfo {author}
  {\bibfnamefont {Wei}\ \bibnamefont {Tang}}, \bibinfo {author} {\bibfnamefont
  {Teague}\ \bibnamefont {Tomesh}}, \bibinfo {author} {\bibfnamefont {Norm~M.}\
  \bibnamefont {Tubman}}, \bibinfo {author} {\bibfnamefont {Chen}\ \bibnamefont
  {Wang}}, \bibinfo {author} {\bibfnamefont {Nathan}\ \bibnamefont {Wiebe}},
  \bibinfo {author} {\bibfnamefont {Yong-Xin}\ \bibnamefont {Yao}}, \bibinfo
  {author} {\bibfnamefont {Dillon~C.}\ \bibnamefont {Yost}}, \ and\ \bibinfo
  {author} {\bibfnamefont {Yiyu}\ \bibnamefont {Zhou}},\ }\href@noop {}
  {\enquote {\bibinfo {title} {Architectures for multinode superconducting
  quantum computers},}\ } (\bibinfo {year} {2022}),\ \Eprint
  {http://arxiv.org/abs/2212.06167} {arXiv:2212.06167 [quant-ph]} \BibitemShut
  {NoStop}%
\bibitem [{\citenamefont {Giovannetti}\ \emph
  {et~al.}(2008{\natexlab{a}})\citenamefont {Giovannetti}, \citenamefont
  {Lloyd},\ and\ \citenamefont {Maccone}}]{Giovannetti2008}%
  \BibitemOpen
  \bibfield  {author} {\bibinfo {author} {\bibfnamefont {Vittorio}\
  \bibnamefont {Giovannetti}}, \bibinfo {author} {\bibfnamefont {Seth}\
  \bibnamefont {Lloyd}}, \ and\ \bibinfo {author} {\bibfnamefont {Lorenzo}\
  \bibnamefont {Maccone}},\ }\bibfield  {title} {\enquote {\bibinfo {title}
  {Quantum random access memory},}\ }\href {\doibase
  10.1103/PhysRevLett.100.160501} {\bibfield  {journal} {\bibinfo  {journal}
  {Phys. Rev. Lett.}\ }\textbf {\bibinfo {volume} {100}},\ \bibinfo {pages}
  {160501} (\bibinfo {year} {2008}{\natexlab{a}})}\BibitemShut {NoStop}%
\bibitem [{\citenamefont {Giovannetti}\ \emph
  {et~al.}(2008{\natexlab{b}})\citenamefont {Giovannetti}, \citenamefont
  {Lloyd},\ and\ \citenamefont {Maccone}}]{Giovannetti2008PRA}%
  \BibitemOpen
  \bibfield  {author} {\bibinfo {author} {\bibfnamefont {Vittorio}\
  \bibnamefont {Giovannetti}}, \bibinfo {author} {\bibfnamefont {Seth}\
  \bibnamefont {Lloyd}}, \ and\ \bibinfo {author} {\bibfnamefont {Lorenzo}\
  \bibnamefont {Maccone}},\ }\bibfield  {title} {\enquote {\bibinfo {title}
  {Architectures for a quantum random access memory},}\ }\href {\doibase
  10.1103/PhysRevA.78.052310} {\bibfield  {journal} {\bibinfo  {journal} {Phys.
  Rev. A}\ }\textbf {\bibinfo {volume} {78}},\ \bibinfo {pages} {052310}
  (\bibinfo {year} {2008}{\natexlab{b}})}\BibitemShut {NoStop}%
\bibitem [{\citenamefont {Hong}\ \emph {et~al.}(2012)\citenamefont {Hong},
  \citenamefont {Xiang}, \citenamefont {Zhu}, \citenamefont {Jiang},\ and\
  \citenamefont {Wu}}]{Hong2012}%
  \BibitemOpen
  \bibfield  {author} {\bibinfo {author} {\bibfnamefont {Fang-Yu}\ \bibnamefont
  {Hong}}, \bibinfo {author} {\bibfnamefont {Yang}\ \bibnamefont {Xiang}},
  \bibinfo {author} {\bibfnamefont {Zhi-Yan}\ \bibnamefont {Zhu}}, \bibinfo
  {author} {\bibfnamefont {Li-zhen}\ \bibnamefont {Jiang}}, \ and\ \bibinfo
  {author} {\bibfnamefont {Liang-neng}\ \bibnamefont {Wu}},\ }\bibfield
  {title} {\enquote {\bibinfo {title} {Robust quantum random access memory},}\
  }\href {\doibase 10.1103/PhysRevA.86.010306} {\bibfield  {journal} {\bibinfo
  {journal} {Phys. Rev. A}\ }\textbf {\bibinfo {volume} {86}},\ \bibinfo
  {pages} {010306} (\bibinfo {year} {2012})}\BibitemShut {NoStop}%
\bibitem [{\citenamefont {Harrow}\ \emph {et~al.}(2009)\citenamefont {Harrow},
  \citenamefont {Hassidim},\ and\ \citenamefont {Lloyd}}]{Harrow2009}%
  \BibitemOpen
  \bibfield  {author} {\bibinfo {author} {\bibfnamefont {Aram~W.}\ \bibnamefont
  {Harrow}}, \bibinfo {author} {\bibfnamefont {Avinatan}\ \bibnamefont
  {Hassidim}}, \ and\ \bibinfo {author} {\bibfnamefont {Seth}\ \bibnamefont
  {Lloyd}},\ }\bibfield  {title} {\enquote {\bibinfo {title} {Quantum algorithm
  for linear systems of equations},}\ }\href {\doibase
  10.1103/PhysRevLett.103.150502} {\bibfield  {journal} {\bibinfo  {journal}
  {Phys. Rev. Lett.}\ }\textbf {\bibinfo {volume} {103}},\ \bibinfo {pages}
  {150502} (\bibinfo {year} {2009})}\BibitemShut {NoStop}%
\bibitem [{\citenamefont {Hann}\ \emph {et~al.}(2019)\citenamefont {Hann},
  \citenamefont {Zou}, \citenamefont {Zhang}, \citenamefont {Chu},
  \citenamefont {Schoelkopf}, \citenamefont {Girvin},\ and\ \citenamefont
  {Jiang}}]{Hann2019}%
  \BibitemOpen
  \bibfield  {author} {\bibinfo {author} {\bibfnamefont {Connor~T.}\
  \bibnamefont {Hann}}, \bibinfo {author} {\bibfnamefont {Chang-Ling}\
  \bibnamefont {Zou}}, \bibinfo {author} {\bibfnamefont {Yaxing}\ \bibnamefont
  {Zhang}}, \bibinfo {author} {\bibfnamefont {Yiwen}\ \bibnamefont {Chu}},
  \bibinfo {author} {\bibfnamefont {Robert~J.}\ \bibnamefont {Schoelkopf}},
  \bibinfo {author} {\bibfnamefont {S.~M.}\ \bibnamefont {Girvin}}, \ and\
  \bibinfo {author} {\bibfnamefont {Liang}\ \bibnamefont {Jiang}},\ }\bibfield
  {title} {\enquote {\bibinfo {title} {Hardware-efficient quantum random access
  memory with hybrid quantum acoustic systems},}\ }\href {\doibase
  10.1103/PhysRevLett.123.250501} {\bibfield  {journal} {\bibinfo  {journal}
  {Phys. Rev. Lett.}\ }\textbf {\bibinfo {volume} {123}},\ \bibinfo {pages}
  {250501} (\bibinfo {year} {2019})}\BibitemShut {NoStop}%
\bibitem [{\citenamefont {Chen}\ \emph
  {et~al.}(2021{\natexlab{b}})\citenamefont {Chen}, \citenamefont {Dai},
  \citenamefont {Errando-Herranz}, \citenamefont {Lloyd},\ and\ \citenamefont
  {Englund}}]{Chen2021}%
  \BibitemOpen
  \bibfield  {author} {\bibinfo {author} {\bibfnamefont {K.~C.}\ \bibnamefont
  {Chen}}, \bibinfo {author} {\bibfnamefont {W.}~\bibnamefont {Dai}}, \bibinfo
  {author} {\bibfnamefont {C.}~\bibnamefont {Errando-Herranz}}, \bibinfo
  {author} {\bibfnamefont {S.}~\bibnamefont {Lloyd}}, \ and\ \bibinfo {author}
  {\bibfnamefont {D.}~\bibnamefont {Englund}},\ }\bibfield  {title} {\enquote
  {\bibinfo {title} {Scalable and high-fidelity quantum random access memory in
  spin-photon networks},}\ }\href {\doibase 10.1103/PRXQuantum.2.030319}
  {\bibfield  {journal} {\bibinfo  {journal} {PRX Quantum}\ }\textbf {\bibinfo
  {volume} {2}},\ \bibinfo {pages} {030319} (\bibinfo {year}
  {2021}{\natexlab{b}})}\BibitemShut {NoStop}%
\bibitem [{\citenamefont {Weiss}\ \emph {et~al.}(2023)\citenamefont {Weiss},
  \citenamefont {Puri},\ and\ \citenamefont {Girvin}}]{weiss2023qram}%
  \BibitemOpen
  \bibfield  {author} {\bibinfo {author} {\bibfnamefont {D.~K.}\ \bibnamefont
  {Weiss}}, \bibinfo {author} {\bibfnamefont {Shruti}\ \bibnamefont {Puri}}, \
  and\ \bibinfo {author} {\bibfnamefont {S.~M.}\ \bibnamefont {Girvin}},\
  }\href@noop {} {\enquote {\bibinfo {title} {Qram architectures using
  superconducting cavities},}\ } (\bibinfo {year} {2023}),\ \Eprint
  {http://arxiv.org/abs/2310.08288} {arXiv:2310.08288 [quant-ph]} \BibitemShut
  {NoStop}%
\bibitem [{\citenamefont {Asaka}\ \emph {et~al.}(2021)\citenamefont {Asaka},
  \citenamefont {Sakai},\ and\ \citenamefont {Yahagi}}]{Asaka2021}%
  \BibitemOpen
  \bibfield  {author} {\bibinfo {author} {\bibfnamefont {Ryo}\ \bibnamefont
  {Asaka}}, \bibinfo {author} {\bibfnamefont {Kazumitsu}\ \bibnamefont
  {Sakai}}, \ and\ \bibinfo {author} {\bibfnamefont {Ryoko}\ \bibnamefont
  {Yahagi}},\ }\bibfield  {title} {\enquote {\bibinfo {title} {Quantum random
  access memory via quantum walk},}\ }\href {\doibase 10.1088/2058-9565/abf484}
  {\bibfield  {journal} {\bibinfo  {journal} {Quantum Science and Technology}\
  }\textbf {\bibinfo {volume} {6}},\ \bibinfo {pages} {035004} (\bibinfo {year}
  {2021})}\BibitemShut {NoStop}%
\bibitem [{\citenamefont {Asaka}\ \emph
  {et~al.}(2023{\natexlab{a}})\citenamefont {Asaka}, \citenamefont {Sakai},\
  and\ \citenamefont {Yahagi}}]{Asaka2023I}%
  \BibitemOpen
  \bibfield  {author} {\bibinfo {author} {\bibfnamefont {Ryo}\ \bibnamefont
  {Asaka}}, \bibinfo {author} {\bibfnamefont {Kazumitsu}\ \bibnamefont
  {Sakai}}, \ and\ \bibinfo {author} {\bibfnamefont {Ryoko}\ \bibnamefont
  {Yahagi}},\ }\bibfield  {title} {\enquote {\bibinfo {title} {Two-level
  quantum walkers on directed graphs. i. universal quantum computing},}\ }\href
  {\doibase 10.1103/PhysRevA.107.022415} {\bibfield  {journal} {\bibinfo
  {journal} {Phys. Rev. A}\ }\textbf {\bibinfo {volume} {107}},\ \bibinfo
  {pages} {022415} (\bibinfo {year} {2023}{\natexlab{a}})}\BibitemShut
  {NoStop}%
\bibitem [{\citenamefont {Asaka}\ \emph
  {et~al.}(2023{\natexlab{b}})\citenamefont {Asaka}, \citenamefont {Sakai},\
  and\ \citenamefont {Yahagi}}]{Asaka2023II}%
  \BibitemOpen
  \bibfield  {author} {\bibinfo {author} {\bibfnamefont {Ryo}\ \bibnamefont
  {Asaka}}, \bibinfo {author} {\bibfnamefont {Kazumitsu}\ \bibnamefont
  {Sakai}}, \ and\ \bibinfo {author} {\bibfnamefont {Ryoko}\ \bibnamefont
  {Yahagi}},\ }\bibfield  {title} {\enquote {\bibinfo {title} {Two-level
  quantum walkers on directed graphs. ii. application to quantum random access
  memory},}\ }\href {\doibase 10.1103/PhysRevA.107.022416} {\bibfield
  {journal} {\bibinfo  {journal} {Phys. Rev. A}\ }\textbf {\bibinfo {volume}
  {107}},\ \bibinfo {pages} {022416} (\bibinfo {year}
  {2023}{\natexlab{b}})}\BibitemShut {NoStop}%
\bibitem [{\citenamefont {Park}\ \emph {et~al.}(2019)\citenamefont {Park},
  \citenamefont {Petruccione},\ and\ \citenamefont {Rhee}}]{Park2019}%
  \BibitemOpen
  \bibfield  {author} {\bibinfo {author} {\bibfnamefont {Daniel~K.}\
  \bibnamefont {Park}}, \bibinfo {author} {\bibfnamefont {Francesco}\
  \bibnamefont {Petruccione}}, \ and\ \bibinfo {author} {\bibfnamefont
  {June-Koo~Kevin}\ \bibnamefont {Rhee}},\ }\bibfield  {title} {\enquote
  {\bibinfo {title} {Circuit-based quantum random access memory for classical
  data},}\ }\href {\doibase 10.1038/s41598-019-40439-3} {\bibfield  {journal}
  {\bibinfo  {journal} {Scientific Reports}\ }\textbf {\bibinfo {volume} {9}},\
  \bibinfo {pages} {3949} (\bibinfo {year} {2019})}\BibitemShut {NoStop}%
\bibitem [{\citenamefont {Matteo}\ \emph {et~al.}(2020)\citenamefont {Matteo},
  \citenamefont {Gheorghiu},\ and\ \citenamefont {Mosca}}]{Matteo2020}%
  \BibitemOpen
  \bibfield  {author} {\bibinfo {author} {\bibfnamefont {Olivia~Di}\
  \bibnamefont {Matteo}}, \bibinfo {author} {\bibfnamefont {Vlad}\ \bibnamefont
  {Gheorghiu}}, \ and\ \bibinfo {author} {\bibfnamefont {Michele}\ \bibnamefont
  {Mosca}},\ }\bibfield  {title} {\enquote {\bibinfo {title} {Fault-tolerant
  resource estimation of quantum random-access memories},}\ }\href {\doibase
  10.1109/TQE.2020.2965803} {\bibfield  {journal} {\bibinfo  {journal} {IEEE
  Transactions on Quantum Engineering}\ }\textbf {\bibinfo {volume} {1}},\
  \bibinfo {pages} {1--13} (\bibinfo {year} {2020})}\BibitemShut {NoStop}%
\bibitem [{\citenamefont {Paler}\ \emph {et~al.}(2020)\citenamefont {Paler},
  \citenamefont {Oumarou},\ and\ \citenamefont {Basmadjian}}]{Alexandru2020}%
  \BibitemOpen
  \bibfield  {author} {\bibinfo {author} {\bibfnamefont {Alexandru}\
  \bibnamefont {Paler}}, \bibinfo {author} {\bibfnamefont {Oumarou}\
  \bibnamefont {Oumarou}}, \ and\ \bibinfo {author} {\bibfnamefont {Robert}\
  \bibnamefont {Basmadjian}},\ }\bibfield  {title} {\enquote {\bibinfo {title}
  {Parallelizing the queries in a bucket-brigade quantum random access
  memory},}\ }\href {\doibase 10.1103/PhysRevA.102.032608} {\bibfield
  {journal} {\bibinfo  {journal} {Phys. Rev. A}\ }\textbf {\bibinfo {volume}
  {102}},\ \bibinfo {pages} {032608} (\bibinfo {year} {2020})}\BibitemShut
  {NoStop}%
\bibitem [{\citenamefont {Arunachalam}\ \emph {et~al.}(2015)\citenamefont
  {Arunachalam}, \citenamefont {Gheorghiu}, \citenamefont {Jochym-O'Connor},
  \citenamefont {Mosca},\ and\ \citenamefont {Srinivasan}}]{Arunachalam2015}%
  \BibitemOpen
  \bibfield  {author} {\bibinfo {author} {\bibfnamefont {Srinivasan}\
  \bibnamefont {Arunachalam}}, \bibinfo {author} {\bibfnamefont {Vlad}\
  \bibnamefont {Gheorghiu}}, \bibinfo {author} {\bibfnamefont {Tomas}\
  \bibnamefont {Jochym-O'Connor}}, \bibinfo {author} {\bibfnamefont {Michele}\
  \bibnamefont {Mosca}}, \ and\ \bibinfo {author} {\bibfnamefont
  {Priyaa~Varshinee}\ \bibnamefont {Srinivasan}},\ }\bibfield  {title}
  {\enquote {\bibinfo {title} {On the robustness of bucket brigade quantum
  ram},}\ }\href {\doibase 10.1088/1367-2630/17/12/123010} {\bibfield
  {journal} {\bibinfo  {journal} {New Journal of Physics}\ }\textbf {\bibinfo
  {volume} {17}},\ \bibinfo {pages} {123010} (\bibinfo {year}
  {2015})}\BibitemShut {NoStop}%
\bibitem [{\citenamefont {Hann}\ \emph {et~al.}(2021)\citenamefont {Hann},
  \citenamefont {Lee}, \citenamefont {Girvin},\ and\ \citenamefont
  {Jiang}}]{Hann2021}%
  \BibitemOpen
  \bibfield  {author} {\bibinfo {author} {\bibfnamefont {Connor~T.}\
  \bibnamefont {Hann}}, \bibinfo {author} {\bibfnamefont {Gideon}\ \bibnamefont
  {Lee}}, \bibinfo {author} {\bibfnamefont {S.M.}\ \bibnamefont {Girvin}}, \
  and\ \bibinfo {author} {\bibfnamefont {Liang}\ \bibnamefont {Jiang}},\
  }\bibfield  {title} {\enquote {\bibinfo {title} {Resilience of quantum random
  access memory to generic noise},}\ }\href {\doibase
  10.1103/PRXQuantum.2.020311} {\bibfield  {journal} {\bibinfo  {journal} {PRX
  Quantum}\ }\textbf {\bibinfo {volume} {2}},\ \bibinfo {pages} {020311}
  (\bibinfo {year} {2021})}\BibitemShut {NoStop}%
\bibitem [{\citenamefont {Chen}\ \emph {et~al.}(2023)\citenamefont {Chen},
  \citenamefont {Xue}, \citenamefont {Wang}, \citenamefont {Sun}, \citenamefont
  {Liu}, \citenamefont {Zhuang}, \citenamefont {Dou}, \citenamefont {Zou},
  \citenamefont {Fang}, \citenamefont {Wu},\ and\ \citenamefont
  {Guo}}]{chen2023efficient}%
  \BibitemOpen
  \bibfield  {author} {\bibinfo {author} {\bibfnamefont {Zhao-Yun}\
  \bibnamefont {Chen}}, \bibinfo {author} {\bibfnamefont {Cheng}\ \bibnamefont
  {Xue}}, \bibinfo {author} {\bibfnamefont {Yun-Jie}\ \bibnamefont {Wang}},
  \bibinfo {author} {\bibfnamefont {Tai-Ping}\ \bibnamefont {Sun}}, \bibinfo
  {author} {\bibfnamefont {Huan-Yu}\ \bibnamefont {Liu}}, \bibinfo {author}
  {\bibfnamefont {Xi-Ning}\ \bibnamefont {Zhuang}}, \bibinfo {author}
  {\bibfnamefont {Meng-Han}\ \bibnamefont {Dou}}, \bibinfo {author}
  {\bibfnamefont {Tian-Rui}\ \bibnamefont {Zou}}, \bibinfo {author}
  {\bibfnamefont {Yuan}\ \bibnamefont {Fang}}, \bibinfo {author} {\bibfnamefont
  {Yu-Chun}\ \bibnamefont {Wu}}, \ and\ \bibinfo {author} {\bibfnamefont
  {Guo-Ping}\ \bibnamefont {Guo}},\ }\href@noop {} {\enquote {\bibinfo {title}
  {Efficient and error-resilient data access protocols for a limited-sized
  quantum random access memory},}\ } (\bibinfo {year} {2023}),\ \Eprint
  {http://arxiv.org/abs/2303.05207} {arXiv:2303.05207 [quant-ph]} \BibitemShut
  {NoStop}%
\bibitem [{\citenamefont {Xu}\ \emph {et~al.}(2023)\citenamefont {Xu},
  \citenamefont {Hann}, \citenamefont {Foxman}, \citenamefont {Girvin},\ and\
  \citenamefont {Ding}}]{xu2023systems}%
  \BibitemOpen
  \bibfield  {author} {\bibinfo {author} {\bibfnamefont {Shifan}\ \bibnamefont
  {Xu}}, \bibinfo {author} {\bibfnamefont {Connor~T.}\ \bibnamefont {Hann}},
  \bibinfo {author} {\bibfnamefont {Ben}\ \bibnamefont {Foxman}}, \bibinfo
  {author} {\bibfnamefont {Steven~M.}\ \bibnamefont {Girvin}}, \ and\ \bibinfo
  {author} {\bibfnamefont {Yongshan}\ \bibnamefont {Ding}},\ }\href@noop {}
  {\enquote {\bibinfo {title} {Systems architecture for quantum random access
  memory},}\ } (\bibinfo {year} {2023}),\ \Eprint
  {http://arxiv.org/abs/2306.03242} {arXiv:2306.03242 [quant-ph]} \BibitemShut
  {NoStop}%
\bibitem [{\citenamefont {Hann}(2021)}]{HannThesis}%
  \BibitemOpen
  \bibfield  {author} {\bibinfo {author} {\bibfnamefont {Connor}\ \bibnamefont
  {Hann}},\ }\href
  {https://elischolar.library.yale.edu/gsas_dissertations/346/} {\enquote
  {\bibinfo {title} {Practicality of quantum random access memory},}\ }
  (\bibinfo {year} {2021}),\ \bibinfo {note} {{Ph.D. Dissertation}}\BibitemShut
  {NoStop}%
\bibitem [{\citenamefont {Phalak}\ \emph {et~al.}(2023)\citenamefont {Phalak},
  \citenamefont {Chatterjee},\ and\ \citenamefont {Ghosh}}]{phalak2023quantum}%
  \BibitemOpen
  \bibfield  {author} {\bibinfo {author} {\bibfnamefont {Koustubh}\
  \bibnamefont {Phalak}}, \bibinfo {author} {\bibfnamefont {Avimita}\
  \bibnamefont {Chatterjee}}, \ and\ \bibinfo {author} {\bibfnamefont
  {Swaroop}\ \bibnamefont {Ghosh}},\ }\href@noop {} {\enquote {\bibinfo {title}
  {Quantum random access memory for dummies},}\ } (\bibinfo {year} {2023}),\
  \Eprint {http://arxiv.org/abs/2305.01178} {arXiv:2305.01178 [quant-ph]}
  \BibitemShut {NoStop}%
\bibitem [{\citenamefont {Jaques}\ and\ \citenamefont
  {Rattew}(2023)}]{jaques2023qram}%
  \BibitemOpen
  \bibfield  {author} {\bibinfo {author} {\bibfnamefont {Samuel}\ \bibnamefont
  {Jaques}}\ and\ \bibinfo {author} {\bibfnamefont {Arthur~G.}\ \bibnamefont
  {Rattew}},\ }\href@noop {} {\enquote {\bibinfo {title} {Qram: A survey and
  critique},}\ } (\bibinfo {year} {2023}),\ \Eprint
  {http://arxiv.org/abs/2305.10310} {arXiv:2305.10310 [quant-ph]} \BibitemShut
  {NoStop}%
\bibitem [{\citenamefont {Grover}(1996)}]{Grover1996}%
  \BibitemOpen
  \bibfield  {author} {\bibinfo {author} {\bibfnamefont {Lov~K.}\ \bibnamefont
  {Grover}},\ }\bibfield  {title} {\enquote {\bibinfo {title} {A fast quantum
  mechanical algorithm for database search},}\ }in\ \href {\doibase
  10.1145/237814.237866} {\emph {\bibinfo {booktitle} {Proceedings of the
  Twenty-eighth Annual ACM Symposium on Theory of Computing}}},\ \bibinfo
  {series and number} {STOC '96}\ (\bibinfo  {publisher} {ACM},\ \bibinfo
  {address} {New York, NY, USA},\ \bibinfo {year} {1996})\ pp.\ \bibinfo
  {pages} {212--219}\BibitemShut {NoStop}%
\bibitem [{\citenamefont {Wu}\ \emph {et~al.}(2022)\citenamefont {Wu},
  \citenamefont {Ding},\ and\ \citenamefont {Li}}]{wu2022collcomm}%
  \BibitemOpen
  \bibfield  {author} {\bibinfo {author} {\bibfnamefont {Anbang}\ \bibnamefont
  {Wu}}, \bibinfo {author} {\bibfnamefont {Yufei}\ \bibnamefont {Ding}}, \ and\
  \bibinfo {author} {\bibfnamefont {Ang}\ \bibnamefont {Li}},\ }\href@noop {}
  {\enquote {\bibinfo {title} {Collcomm: Enabling efficient collective quantum
  communication based on epr buffering},}\ } (\bibinfo {year} {2022}),\ \Eprint
  {http://arxiv.org/abs/2208.06724} {arXiv:2208.06724 [quant-ph]} \BibitemShut
  {NoStop}%
\bibitem [{\citenamefont {Cross}\ \emph {et~al.}(2017)\citenamefont {Cross},
  \citenamefont {Bishop}, \citenamefont {Smolin},\ and\ \citenamefont
  {Gambetta}}]{cross2017open}%
  \BibitemOpen
  \bibfield  {author} {\bibinfo {author} {\bibfnamefont {Andrew~W.}\
  \bibnamefont {Cross}}, \bibinfo {author} {\bibfnamefont {Lev~S.}\
  \bibnamefont {Bishop}}, \bibinfo {author} {\bibfnamefont {John~A.}\
  \bibnamefont {Smolin}}, \ and\ \bibinfo {author} {\bibfnamefont {Jay~M.}\
  \bibnamefont {Gambetta}},\ }\href@noop {} {\enquote {\bibinfo {title} {Open
  quantum assembly language},}\ } (\bibinfo {year} {2017}),\ \Eprint
  {http://arxiv.org/abs/1707.03429} {arXiv:1707.03429 [quant-ph]} \BibitemShut
  {NoStop}%
\bibitem [{\citenamefont {Cross}\ \emph {et~al.}(2022)\citenamefont {Cross},
  \citenamefont {Javadi-Abhari}, \citenamefont {Alexander}, \citenamefont
  {De~Beaudrap}, \citenamefont {Bishop}, \citenamefont {Heidel}, \citenamefont
  {Ryan}, \citenamefont {Sivarajah}, \citenamefont {Smolin}, \citenamefont
  {Gambetta},\ and\ \citenamefont {Johnson}}]{qasm3}%
  \BibitemOpen
  \bibfield  {author} {\bibinfo {author} {\bibfnamefont {Andrew}\ \bibnamefont
  {Cross}}, \bibinfo {author} {\bibfnamefont {Ali}\ \bibnamefont
  {Javadi-Abhari}}, \bibinfo {author} {\bibfnamefont {Thomas}\ \bibnamefont
  {Alexander}}, \bibinfo {author} {\bibfnamefont {Niel}\ \bibnamefont
  {De~Beaudrap}}, \bibinfo {author} {\bibfnamefont {Lev~S.}\ \bibnamefont
  {Bishop}}, \bibinfo {author} {\bibfnamefont {Steven}\ \bibnamefont {Heidel}},
  \bibinfo {author} {\bibfnamefont {Colm~A.}\ \bibnamefont {Ryan}}, \bibinfo
  {author} {\bibfnamefont {Prasahnt}\ \bibnamefont {Sivarajah}}, \bibinfo
  {author} {\bibfnamefont {John}\ \bibnamefont {Smolin}}, \bibinfo {author}
  {\bibfnamefont {Jay~M.}\ \bibnamefont {Gambetta}}, \ and\ \bibinfo {author}
  {\bibfnamefont {Blake~R.}\ \bibnamefont {Johnson}},\ }\bibfield  {title}
  {\enquote {\bibinfo {title} {Openqasm 3: A broader and deeper quantum
  assembly language},}\ }\href {\doibase 10.1145/3505636} {\bibfield  {journal}
  {\bibinfo  {journal} {ACM Transactions on Quantum Computing}\ }\textbf
  {\bibinfo {volume} {3}} (\bibinfo {year} {2022}),\
  10.1145/3505636}\BibitemShut {NoStop}%
\bibitem [{\citenamefont {{Qiskit contributors}}(2023)}]{Qiskit}%
  \BibitemOpen
  \bibfield  {author} {\bibinfo {author} {\bibnamefont {{Qiskit
  contributors}}},\ }\href {\doibase 10.5281/zenodo.2573505} {\enquote
  {\bibinfo {title} {Qiskit: An open-source framework for quantum computing},}\
  } (\bibinfo {year} {2023})\BibitemShut {NoStop}%
\bibitem [{\citenamefont {Biamonte}\ \emph {et~al.}(2017)\citenamefont
  {Biamonte}, \citenamefont {Wittek}, \citenamefont {Pancotti}, \citenamefont
  {Rebentrost}, \citenamefont {Wiebe},\ and\ \citenamefont
  {Lloyd}}]{Biamonte2017}%
  \BibitemOpen
  \bibfield  {author} {\bibinfo {author} {\bibfnamefont {Jacob}\ \bibnamefont
  {Biamonte}}, \bibinfo {author} {\bibfnamefont {Peter}\ \bibnamefont
  {Wittek}}, \bibinfo {author} {\bibfnamefont {Nicola}\ \bibnamefont
  {Pancotti}}, \bibinfo {author} {\bibfnamefont {Patrick}\ \bibnamefont
  {Rebentrost}}, \bibinfo {author} {\bibfnamefont {Nathan}\ \bibnamefont
  {Wiebe}}, \ and\ \bibinfo {author} {\bibfnamefont {Seth}\ \bibnamefont
  {Lloyd}},\ }\bibfield  {title} {\enquote {\bibinfo {title} {Quantum machine
  learning},}\ }\href {\doibase 10.1038/nature23474} {\bibfield  {journal}
  {\bibinfo  {journal} {Nature}\ }\textbf {\bibinfo {volume} {549}},\ \bibinfo
  {pages} {195--202} (\bibinfo {year} {2017})}\BibitemShut {NoStop}%
\bibitem [{\citenamefont {Cerezo}\ \emph {et~al.}(2022)\citenamefont {Cerezo},
  \citenamefont {Verdon}, \citenamefont {Huang}, \citenamefont {Cincio},\ and\
  \citenamefont {Coles}}]{Cerezo2022}%
  \BibitemOpen
  \bibfield  {author} {\bibinfo {author} {\bibfnamefont {M.}~\bibnamefont
  {Cerezo}}, \bibinfo {author} {\bibfnamefont {Guillaume}\ \bibnamefont
  {Verdon}}, \bibinfo {author} {\bibfnamefont {Hsin-Yuan}\ \bibnamefont
  {Huang}}, \bibinfo {author} {\bibfnamefont {Lukasz}\ \bibnamefont {Cincio}},
  \ and\ \bibinfo {author} {\bibfnamefont {Patrick~J.}\ \bibnamefont {Coles}},\
  }\bibfield  {title} {\enquote {\bibinfo {title} {Challenges and opportunities
  in quantum machine learning},}\ }\href {\doibase 10.1038/s43588-022-00311-3}
  {\bibfield  {journal} {\bibinfo  {journal} {Nature Computational Science}\
  }\textbf {\bibinfo {volume} {2}},\ \bibinfo {pages} {567--576} (\bibinfo
  {year} {2022})}\BibitemShut {NoStop}%
\bibitem [{\citenamefont {Buhrman}\ \emph {et~al.}(2001)\citenamefont
  {Buhrman}, \citenamefont {Cleve}, \citenamefont {Watrous},\ and\
  \citenamefont {de~Wolf}}]{Buhrman2001}%
  \BibitemOpen
  \bibfield  {author} {\bibinfo {author} {\bibfnamefont {Harry}\ \bibnamefont
  {Buhrman}}, \bibinfo {author} {\bibfnamefont {Richard}\ \bibnamefont
  {Cleve}}, \bibinfo {author} {\bibfnamefont {John}\ \bibnamefont {Watrous}}, \
  and\ \bibinfo {author} {\bibfnamefont {Ronald}\ \bibnamefont {de~Wolf}},\
  }\bibfield  {title} {\enquote {\bibinfo {title} {Quantum fingerprinting},}\
  }\href {\doibase 10.1103/PhysRevLett.87.167902} {\bibfield  {journal}
  {\bibinfo  {journal} {Phys. Rev. Lett.}\ }\textbf {\bibinfo {volume} {87}},\
  \bibinfo {pages} {167902} (\bibinfo {year} {2001})}\BibitemShut {NoStop}%
\bibitem [{\citenamefont {Rosset}\ \emph {et~al.}(2018)\citenamefont {Rosset},
  \citenamefont {Buscemi},\ and\ \citenamefont {Liang}}]{Rosset2018}%
  \BibitemOpen
  \bibfield  {author} {\bibinfo {author} {\bibfnamefont {Denis}\ \bibnamefont
  {Rosset}}, \bibinfo {author} {\bibfnamefont {Francesco}\ \bibnamefont
  {Buscemi}}, \ and\ \bibinfo {author} {\bibfnamefont {Yeong-Cherng}\
  \bibnamefont {Liang}},\ }\bibfield  {title} {\enquote {\bibinfo {title}
  {Resource theory of quantum memories and their faithful verification with
  minimal assumptions},}\ }\href {\doibase 10.1103/PhysRevX.8.021033}
  {\bibfield  {journal} {\bibinfo  {journal} {Phys. Rev. X}\ }\textbf {\bibinfo
  {volume} {8}},\ \bibinfo {pages} {021033} (\bibinfo {year}
  {2018})}\BibitemShut {NoStop}%
\bibitem [{\citenamefont {Pusey}(2015)}]{Pusey2015}%
  \BibitemOpen
  \bibfield  {author} {\bibinfo {author} {\bibfnamefont {Matthew~F.}\
  \bibnamefont {Pusey}},\ }\bibfield  {title} {\enquote {\bibinfo {title}
  {Verifying the quantumness of a channel with an untrusted device},}\ }\href
  {\doibase 10.1364/JOSAB.32.000A56} {\bibfield  {journal} {\bibinfo  {journal}
  {J. Opt. Soc. Am. B}\ }\textbf {\bibinfo {volume} {32}},\ \bibinfo {pages}
  {A56--A63} (\bibinfo {year} {2015})}\BibitemShut {NoStop}%
\bibitem [{\citenamefont {Graffitti}\ \emph {et~al.}(2020)\citenamefont
  {Graffitti}, \citenamefont {Pickston}, \citenamefont {Barrow}, \citenamefont
  {Proietti}, \citenamefont {Kundys}, \citenamefont {Rosset}, \citenamefont
  {Ringbauer},\ and\ \citenamefont {Fedrizzi}}]{Graffitti2020}%
  \BibitemOpen
  \bibfield  {author} {\bibinfo {author} {\bibfnamefont {Francesco}\
  \bibnamefont {Graffitti}}, \bibinfo {author} {\bibfnamefont {Alexander}\
  \bibnamefont {Pickston}}, \bibinfo {author} {\bibfnamefont {Peter}\
  \bibnamefont {Barrow}}, \bibinfo {author} {\bibfnamefont {Massimiliano}\
  \bibnamefont {Proietti}}, \bibinfo {author} {\bibfnamefont {Dmytro}\
  \bibnamefont {Kundys}}, \bibinfo {author} {\bibfnamefont {Denis}\
  \bibnamefont {Rosset}}, \bibinfo {author} {\bibfnamefont {Martin}\
  \bibnamefont {Ringbauer}}, \ and\ \bibinfo {author} {\bibfnamefont
  {Alessandro}\ \bibnamefont {Fedrizzi}},\ }\bibfield  {title} {\enquote
  {\bibinfo {title} {Measurement-device-independent verification of quantum
  channels},}\ }\href {\doibase 10.1103/PhysRevLett.124.010503} {\bibfield
  {journal} {\bibinfo  {journal} {Phys. Rev. Lett.}\ }\textbf {\bibinfo
  {volume} {124}},\ \bibinfo {pages} {010503} (\bibinfo {year}
  {2020})}\BibitemShut {NoStop}%
\bibitem [{\citenamefont {Mao}\ \emph {et~al.}(2020)\citenamefont {Mao},
  \citenamefont {Zhen}, \citenamefont {Liu}, \citenamefont {Zou}, \citenamefont
  {Tang}, \citenamefont {Zhang}, \citenamefont {Wang}, \citenamefont {Liang},
  \citenamefont {Zhang}, \citenamefont {Li}, \citenamefont {You}, \citenamefont
  {Wang}, \citenamefont {Li}, \citenamefont {Liu}, \citenamefont {Chen},
  \citenamefont {Chen},\ and\ \citenamefont {Pan}}]{Mao2020}%
  \BibitemOpen
  \bibfield  {author} {\bibinfo {author} {\bibfnamefont {Yingqiu}\ \bibnamefont
  {Mao}}, \bibinfo {author} {\bibfnamefont {Yi-Zheng}\ \bibnamefont {Zhen}},
  \bibinfo {author} {\bibfnamefont {Hui}\ \bibnamefont {Liu}}, \bibinfo
  {author} {\bibfnamefont {Mi}~\bibnamefont {Zou}}, \bibinfo {author}
  {\bibfnamefont {Qi-Jie}\ \bibnamefont {Tang}}, \bibinfo {author}
  {\bibfnamefont {Si-Jie}\ \bibnamefont {Zhang}}, \bibinfo {author}
  {\bibfnamefont {Jian}\ \bibnamefont {Wang}}, \bibinfo {author} {\bibfnamefont
  {Hao}\ \bibnamefont {Liang}}, \bibinfo {author} {\bibfnamefont {Weijun}\
  \bibnamefont {Zhang}}, \bibinfo {author} {\bibfnamefont {Hao}\ \bibnamefont
  {Li}}, \bibinfo {author} {\bibfnamefont {Lixing}\ \bibnamefont {You}},
  \bibinfo {author} {\bibfnamefont {Zhen}\ \bibnamefont {Wang}}, \bibinfo
  {author} {\bibfnamefont {Li}~\bibnamefont {Li}}, \bibinfo {author}
  {\bibfnamefont {Nai-Le}\ \bibnamefont {Liu}}, \bibinfo {author}
  {\bibfnamefont {Kai}\ \bibnamefont {Chen}}, \bibinfo {author} {\bibfnamefont
  {Teng-Yun}\ \bibnamefont {Chen}}, \ and\ \bibinfo {author} {\bibfnamefont
  {Jian-Wei}\ \bibnamefont {Pan}},\ }\bibfield  {title} {\enquote {\bibinfo
  {title} {Experimentally verified approach to nonentanglement-breaking channel
  certification},}\ }\href {\doibase 10.1103/PhysRevLett.124.010502} {\bibfield
   {journal} {\bibinfo  {journal} {Phys. Rev. Lett.}\ }\textbf {\bibinfo
  {volume} {124}},\ \bibinfo {pages} {010502} (\bibinfo {year}
  {2020})}\BibitemShut {NoStop}%
\bibitem [{\citenamefont {Yu}\ \emph {et~al.}(2021)\citenamefont {Yu},
  \citenamefont {Sun}, \citenamefont {Zhang}, \citenamefont {Bai},
  \citenamefont {Fang}, \citenamefont {Luo}, \citenamefont {An}, \citenamefont
  {Li}, \citenamefont {Zhang}, \citenamefont {Xu}, \citenamefont {Bao},\ and\
  \citenamefont {Pan}}]{Yu2021}%
  \BibitemOpen
  \bibfield  {author} {\bibinfo {author} {\bibfnamefont {Yong}\ \bibnamefont
  {Yu}}, \bibinfo {author} {\bibfnamefont {Peng-Fei}\ \bibnamefont {Sun}},
  \bibinfo {author} {\bibfnamefont {Yu-Zhe}\ \bibnamefont {Zhang}}, \bibinfo
  {author} {\bibfnamefont {Bing}\ \bibnamefont {Bai}}, \bibinfo {author}
  {\bibfnamefont {Yu-Qiang}\ \bibnamefont {Fang}}, \bibinfo {author}
  {\bibfnamefont {Xi-Yu}\ \bibnamefont {Luo}}, \bibinfo {author} {\bibfnamefont
  {Zi-Ye}\ \bibnamefont {An}}, \bibinfo {author} {\bibfnamefont {Jun}\
  \bibnamefont {Li}}, \bibinfo {author} {\bibfnamefont {Jun}\ \bibnamefont
  {Zhang}}, \bibinfo {author} {\bibfnamefont {Feihu}\ \bibnamefont {Xu}},
  \bibinfo {author} {\bibfnamefont {Xiao-Hui}\ \bibnamefont {Bao}}, \ and\
  \bibinfo {author} {\bibfnamefont {Jian-Wei}\ \bibnamefont {Pan}},\ }\bibfield
   {title} {\enquote {\bibinfo {title} {Measurement-device-independent
  verification of a quantum memory},}\ }\href {\doibase
  10.1103/PhysRevLett.127.160502} {\bibfield  {journal} {\bibinfo  {journal}
  {Phys. Rev. Lett.}\ }\textbf {\bibinfo {volume} {127}},\ \bibinfo {pages}
  {160502} (\bibinfo {year} {2021})}\BibitemShut {NoStop}%
\bibitem [{\citenamefont {Abiuso}(2023)}]{abiuso2023verification}%
  \BibitemOpen
  \bibfield  {author} {\bibinfo {author} {\bibfnamefont {Paolo}\ \bibnamefont
  {Abiuso}},\ }\href@noop {} {\enquote {\bibinfo {title} {Verification of
  continuous-variable quantum memories},}\ } (\bibinfo {year} {2023}),\ \Eprint
  {http://arxiv.org/abs/2305.07513} {arXiv:2305.07513 [quant-ph]} \BibitemShut
  {NoStop}%
\end{thebibliography}%
\end{document}